\documentclass[useAMS,usenatbib]{mn2e}
\usepackage{lscape,graphicx,pifont}
\usepackage{times}
\usepackage{xtab}
\usepackage{longtable}
\usepackage{amsmath}

\newcommand{\spitzer}{{\it Spitzer}}
\newcommand{\spec}{SAGE-Spec}
\newcommand{\lsim}{\vcenter{\hbox{$<$}\offinterlineskip\hbox{$\sim$}}}
\newcommand{\gsim}{\vcenter{\hbox{$>$}\offinterlineskip\hbox{$\sim$}}}
\newcommand{\hho}{H$_2$O}
\newcommand{\coo}{CO$_2$}
\newcommand{\al}{\llap}
\newcommand{\ar}{\rlap}
\newcommand{\iras}{IRAS}
\newcommand{\msx}{MSX\,SMC}
\newcommand{\mum}{$\mu$m}
\newcommand{\tick}{\ding{51}}
\newcommand{\dtick}{\ding{51}\hspace{-1mm}\ding{51}}
\newcommand{\wtick}{{\tiny\ding{51}}}
\newcommand{\cross}{\ding{53}}
\newcommand{\emm}{$\bigwedge$}

\title[Young stellar objects in the SMC]{Early-stage young stellar objects in
the Small Magellanic Cloud}
\author[Oliveira et al.]
{J.M. Oliveira$^{1}$\thanks{j.oliveira@keele.ac.uk},
J.Th. van Loon$^{1}$,
G.C. Sloan$^{2}$,
M. Sewi\l{}o$^{3}$, 
K.E. Kraemer$^{4}$, 
P.R. Wood$^{5}$, \and 
R. Indebetouw$^{6,7},$
M.D. Filipovi\'c$^{8}$, 
E.J. Crawford$^{8}$, 
G.F. Wong$^{8}$,
J.L. Hora$^{9}$,
M. Meixner$^{10}$, \and
T.P. Robitaille$^{11}$,
B. Shiao$^{10}$,
J.D. Simon$^{12}$
\\
$^{1}$ School of Physical \& Geographical Sciences, Lennard-Jones 
       Laboratories, Keele University, Staffordshire ST5 5BG, UK\\
$^{2}$ Department of Astronomy, Cornell University, Ithaca, NY 14853, USA\\
$^{3}$ Department of Physics \& Astronomy, The Johns Hopkins University, 3400 
N. Charles Street, Baltimore, MD 21218, USA\\
$^{4}$ Boston College, Institute for Scientific Research, 140 Commonwealth 
       Avenue, Chestnut Hill, MA 02467, USA\\
$^{5}$ Research School of Astronomy and Astrophysics, Australian National 
       University, Cotter Road, Weston Creek, ACT 2611, Australia\\
$^{6}$ Department of Astronomy, University of Virginia, PO Box 3818, 
       Charlottesville, VA 22903, USA\\
$^{7}$ National Radio Astronomical Observatory, Charlottesville, VA 22904, USA\\
$^{8}$ University of Western Sydney, Locked Bag 1797, Penrith South DC, NSW 1797, 
Australia\\
$^{9}$ Harvard-Smithsonian Center for Astrophysics, 60 Garden Street, MS-65
Cambridge, MA 02138-1516, USA\\
$^{10}$ Space Telescope Science Institute, 3700 San Martin Drive, Baltimore, MD 
       21218, USA\\
$^{11}$ Max-Planck-Institut f\"ur Astronomie, K\"onigstuhl 17, D-69117 Heidelberg,
Germany\\
$^{12}$ Observatories of the Carnegie Institution of Washington, 813 Santa 
Barbara St., Pasadena, CA 91101, USA}
\begin{document}

\date{Accepted 2012 October 18. Received 2012 October 18; in original form 
2012 September 18}

\pagerange{\pageref{firstpage}--\pageref{lastpage}} \pubyear{2012}

\maketitle

\label{firstpage}

\begin{abstract}

We present new observations of 34 Young Stellar Object (YSO) candidates in the Small
Magellanic Cloud (SMC). The photometric selection required sources to be bright at 24 
and 70\,\mum\ (to exclude evolved stars and galaxies). The anchor of the analysis is a
set of \spitzer-IRS spectra, supplemented by groundbased 3$-$5\,\mum\ spectra, 
\spitzer\ IRAC and MIPS photometry, near-IR imaging and photometry, optical 
spectroscopy and radio data. 
The sources' spectral energy distributions (SEDs) and spectral indices are consistent 
with embedded YSOs; prominent silicate absorption is observed in the spectra of at 
least ten sources, silicate emission is observed towards four sources. Polycyclic 
Aromatic Hydrocarbon
(PAH) emission is detected towards all but two sources. Based on band ratios (in 
particular the strength of the 11.3-\mum\ and the weakness of the 8.6-\mum\ bands) PAH 
emission towards SMC YSOs is dominated by predominantly small neutral grains. Ice 
absorption is observed towards fourteen sources in the SMC. The comparison of \hho\ 
and \coo\ ice column densities for SMC, Large Magellanic Cloud (LMC) and Galactic 
samples suggests that there is a significant \hho\ column density threshold for the 
detection of \coo\ ice. This supports the scenario proposed by \citet{oliveira11}, 
where the reduced shielding in metal-poor environments depletes the \hho\ column 
density in the outer regions of the YSO envelopes. No CO ice is detected towards the 
SMC sources. Emission due to pure-rotational $0-0$ transitions of molecular hydrogen 
is detected towards the majority of SMC sources, allowing us to estimate rotational 
temperatures and H$_2$ column densities. 
All but one source are spectroscopically confirmed as SMC YSOs. 
Based on the presence of ice absorption, silicate emission or absorption, and PAH
emission, the sources are classified and placed in an evolutionary sequence. Of the 33
YSOs identified in the SMC, 30 sources populate different stages of massive stellar 
evolution. The presence of ice- and/or silicate-absorption features indicates sources 
in the early embedded stages; as a source evolves, a compact H\,{\sc ii} region starts to
emerge, and at the later stages the source's IR spectrum is completely dominated by 
PAH and fine-structure emission. The remaining three sources are classified as 
intermediate-mass YSOs with a thick dusty disc and a tenuous envelope still present.
We propose one of the SMC sources is a D-type symbiotic system, based on the 
presence of Raman, H and He emission lines in the optical spectrum, and silicate 
emission in the IRS-spectrum. This would be the first dust-rich symbiotic system
identified in the SMC.

\end{abstract}


\section{Introduction}

Star formation is a complex interplay of various chemo-physical processes. During the 
onset of gravitational collapse of a dense cloud, dense cores can only develop if heat 
can be dissipated. The most efficient cooling mechanisms are via radiation through 
fine-structure lines of carbon and oxygen, and rotational transitions in abundant 
molecules such as water \citep[e.g.,][]{vandishoeck04}. These cooling agents all 
contain at least one metallic atom. Furthermore, dust grains are crucial in driving 
molecular cloud chemistry, as dust opacity shields cores from radiation, and grain 
surfaces enable chemical reactions to occur that would not happen in the gas phase. 
These facts suggest that metallicity is an important parameter to explore in 
understanding star formation. However, most of what is known about the physics of star 
formation is deduced from observations in solar-metallicity Galactic star forming 
regions. The nearest templates for the detailed study of star formation under 
metal-poor conditions are the Magellanic Clouds (MCs), with ISM metallicities of 
$Z_{\rm SMC}\sim 0.2 \,Z_\odot$ and $Z_{\rm LMC}\sim 0.4 \,Z_\odot$, respectively 
for the SMC and LMC \citep[e.g.,][]{russell92}. Even over the metallicity range covered
by the Galaxy and the MCs, cooling and chemistry rates, and star formation timescales 
could be affected \citep{banerji09}. The low metallicity of the SMC in particular is 
typical of galaxies during the early phases of their assembly, and studies of star 
formation in the SMC provide a stepping stone to understanding star formation at high 
redshift where these processes cannot be directly observed.

The {\it Spitzer Space Telescope} \citep[\spitzer,][]{werner04} finally 
allowed the identification of sizable samples of Young Stellar Objects (YSOs) in the
MCs. The \spitzer\ Legacy Programs (SAGE, \citealt{meixner06}; SAGE-SMC,
\citealt{gordon11}) have identified 1000s of previously unknown YSOs, both in the LMC 
and the SMC (\citealt{whitney08,gruendl09}; Sewi\l{}o et al in preparation). Follow-up 
spectroscopic programmes have provided unique insight into the abundances of ices in 
Magellanic YSOs, revealing differences in the composition of circumstellar material at 
lower metallicity \citep{oliveira09,oliveira11,shimonishi10,seale11}. 

In this paper we present a sample of 34 photometrically selected YSO candidates in the 
SMC, observed with \spitzer's InfraRed Spectrograph \citep[IRS,][]{houck04}. Based on 
the properties of their IRS spectra and a variety of multi-wavelength diagnostics, we 
classify 33 sources as YSOs in the SMC (27 previously unknown) and 1 source as a 
symbiotic system. This paper is organised as follows. After describing the sample 
selection and the different observations, the spectral properties of the SMC sources 
are analysed in detail in Section 4, with the aim of assessing the sources' 
evolutionary stage. We describe the SED properties and modelling, dust and ice 
absorption, polycyclic aromatic hydrocarbon (PAH), ionic fine structure and H$_2$ 
emission, and the properties of optical counterparts to the infrared (IR) sources. The 
source classification and their evolutionary status is then discussed in Section 5. 
Section 6 describes the process that allowed us to identify one of the sources as the 
first D-type symbiotic system in the SMC. 

\section{Defining the YSO sample}

\begin{figure*}
\includegraphics[scale=0.73,angle=-90]{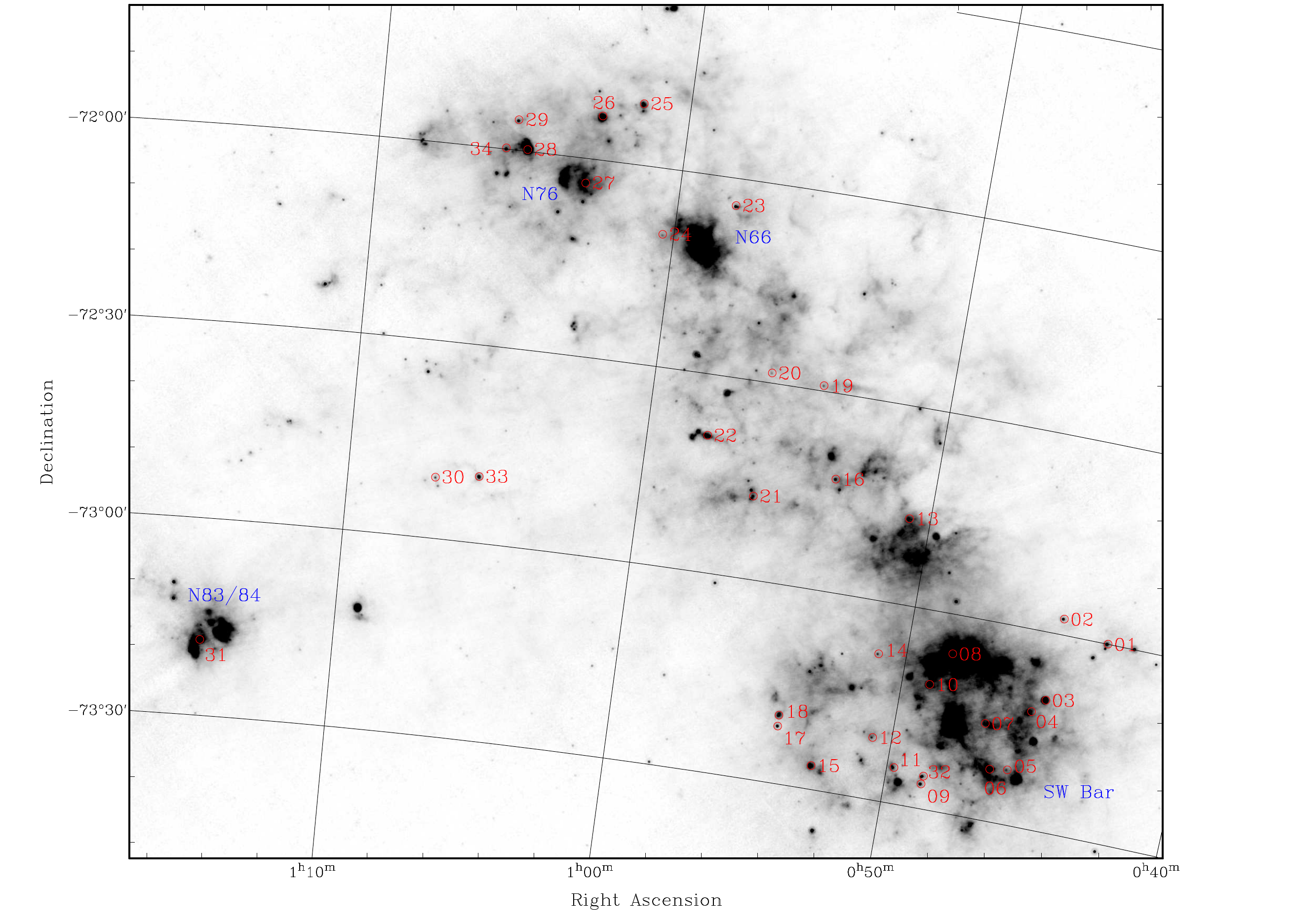}
\caption{SAGE-SMC MIPS 70-\mum\ image showing the location of the 34 YSO candidates
(red circles and labels); sources \#01 to 31 were first identified in this survey 
\citep[see also][]{oliveira11,vanloon10b}, while sources \#32 (MSX\,SMC\,79), \#33 
(IRAS\,01039$-$7305) and \#34 (IRAS\,01042$-$7215) were identified by 
\citet{vanloon08}. The star forming regions discussed in Section 4.3 are labelled in
blue.}
\label{image}
\end{figure*}

The S$^3$MC survey \citep{bolatto07} imaged the main body of the SMC (central 3 
square degrees) in all seven \spitzer\ photometric bands, using the Infrared Array 
Camera \citep[IRAC,][]{fazio04} and the Multiband Imaging Photometer for \spitzer\ 
\citep[MIPS,][]{rieke04}. The resulting photometric catalogue was used to select the 
present YSO sample as described below.

\begin{figure*}
\includegraphics[scale=0.42]{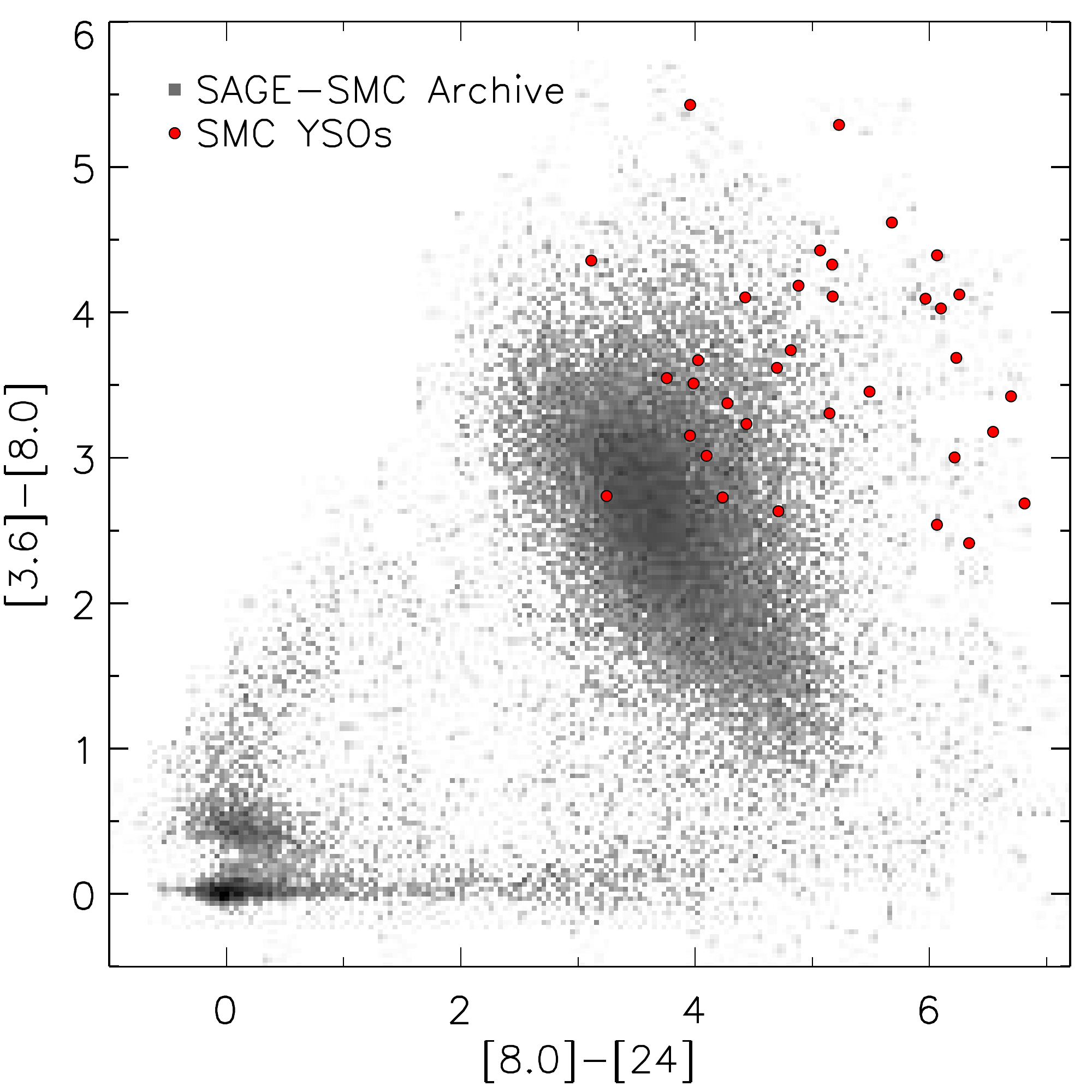}
\includegraphics[scale=0.42]{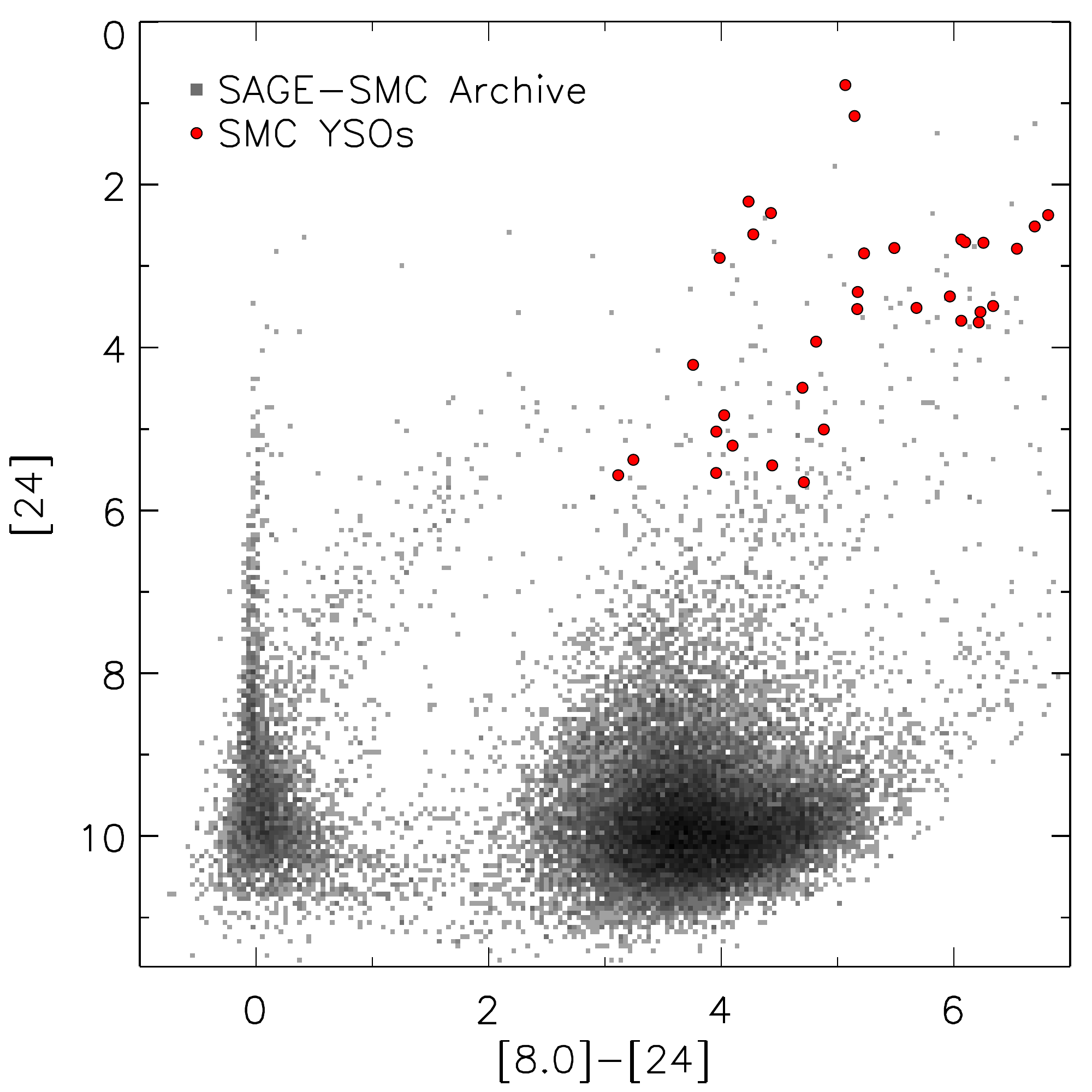}
\caption{Colour--colour and colour--magnitude diagrams for the YSO candidates (filled 
red circles), as well as the field SMC population (grey scale).}
\label{ccd_cmd}
\end{figure*}

The near- and mid-IR colours of evolved stars can be as red as embedded YSOs 
(especially carbon stars, which are common in the SMC). This makes the 70\,\mum\ 
photometry critical in identifying YSOs with cold dust (as opposed to the warm dust 
surrounding evolved stars). Thus, we selected targets with detections in all IRAC 
bands and at 24 and 70\,\mum. We also imposed a 10\,mJy lower limit to the 8 and 
24\,\mum\ fluxes so that the sources are suitable for \spitzer\ spectroscopy. From the 
160 objects with 70\,\mum\ detections, we were left with 46 candidates. 
By cross-correlating their positions with the SIMBAD database and the \spitzer's 
Reserved Object Catalogue, we rejected 15 objects which might be (or are near to) 
evolved stars or planetary nebulae. The 31 remaining targets have colours 
$[3.6]-[5.8] > 1.35$\,mag and $[8]-[24] > 2.75$\,mag, typical for embedded YSOs 
\citep{rho06}. Background galaxies can also have colours similar to those of YSOs
 \citep*{eisenhardt04}. However, the selected YSO candidates are bright, sit well
above the bulk of background galaxy contamination \citep{lee06,jorgensen06}, 
and most are located in active regions of star formation (Fig.\ref{image}). Thus 
they are likely genuine YSOs and not more uniformly distributed evolved stars 
or background galaxies. YSO candidates separate better from the bulk SMC
population in colour--magnitude diagrams, rather than colour--colour diagrams
(\citealt{whitney08}; Sewi\l{}o et al. in preparation), as demonstrated in 
Fig.\,\ref{ccd_cmd}. Our selection criteria imply the final sample is composed 
of massive YSOs ($M \ga 10 M_{\odot}$). The sample is obviously not complete but
it is ideally designed to identify massive YSOs that can be the subjects of a 
{\it detailed} spectroscopic analysis.

Our strategy has proved very successful, with all but one of the 31 candidates 
confirmed as an embedded YSO or compact H\,{\sc ii} region (Section 4). We add to 
the SMC sample the three YSOs identified by \citet{vanloon08}, MSX\,SMC\,79, 
IRAS\,01039$-$7305 and IRAS\,01042$-$7215, based on the appearance of their 
3$-$4\,\mum\ spectra(Section 3.3). The locations of these 34 YSO candidates 
superposed on a 70-\mum\ MIPS image of the SMC are shown in Fig.\,\ref{image}. 
Fig.\,\ref{ccd_cmd} shows $[3.6]-[8.0]\,vs.\,[8.0]-[24]$ colour--colour and 
$[8.0]\,vs\,[8.0]-[24]$ colour--magnitude diagrams for the 34 YSO candidates, 
together with the field SMC population \citep{gordon11}. 

\section{Observations and catalogues}

\subsection{Spitzer-IRS and MIPS-SED spectroscopy}

The mid-IR, low-resolution spectra of the 31 candidate YSOs were obtained 
using the \spitzer-IRS and the spectral energy distribution (SED) mode of MIPS 
\citep{rieke04} on board \spitzer, taken as part of the SMC-Spec program (PID: 50240, 
P.I.: G.C.\ Sloan). This GTO programme aimed at providing a comprehensive 
spectroscopic survey of the SMC. Scientifically its goals were to study dust in nearly 
every stage of its life cycle in the SMC in order to assess how the interactions of 
dust and its host galaxy differ from more metal-rich systems like the Galaxy and the
LMC. The programme was designed to provide the SMC counterpart of the LMC 
\spec\ programme \citep{kemper10}, albeit at a smaller scale. The SMC observations 
targeted important classes generally underrepresented in the \spitzer\ archive of SMC 
sources, such as YSOs and compact H\,{\sc ii} regions, objects in transition to and 
from the asymptotic giant branch, and supergiants. The aforementioned three additional 
objects from \citet{vanloon08} were initially thought to be evolved stars; their 
\spitzer-IRS spectra were obtained as part of a cycle\,1 GTO programme (PID: 3277, 
P.I.: M.\ Egan).

The IRS observations of the point sources in SMC-Spec were performed in 
staring mode, using the Short-Low and Long-Low modules (SL and LL, 
respectively). The 8- and 24-\mum\ fluxes were used to set exposure times, 
aiming at signal-to-noise ratios of $\sim60$ in SL and $\sim30$ in LL. 
The archival spectra mentioned previously were also obtained in the SL and LL staring 
modes. All spectra were reduced following standard reduction techniques. 
Flat-fielded images from the standard \spitzer\ reduction pipeline (version 
S18.18) were background subtracted and cleaned of ``rogue'' pixels and 
artefacts. Spectra were extracted individually for each data collection event 
(DCE) and co-added to produce one spectrum per nod position. Individual spectra
are extracted both using optimal and tapered extraction; optimal extraction
uses super-sampled point-spread function (PSF) profiles to weight the pixels from the 
spatial profiles, while tapered column extraction integrates the flux in a spectral 
window that expands with wavelength \citep{lebouteiller10}. For each extraction, the 
nods were then combined to produce a single spectrum per order, rejecting ``spikes'' 
that appear in only one of the nod positions. Finally, the spectra of all the segments 
were combined including the two bonus orders that are useful in correcting for 
discontinuities between the orders. 

While for point sources optimal extraction produces the best signal-to-noise 
ratio, tapered extraction performs better if the source is extended 
\citep{lebouteiller10}. Some of the sources in our sample could be marginally extended;
we opt to use optimal extracted spectra but check the veracity and strength of the 
spectral features against the tapered extraction spectra. The strength of relevant 
emission features (Sections 4.3 and 4.4) is measured by first fitting a series of line 
segments to the continua on either side of feature and then integrating in 
$F_{\lambda}$ space.

For 10 objects in the sample (see Table \ref{table4}) we also have MIPS-SED spectra, 
covering the wavelength range 52$-$90\,\mum, as described in \citet{vanloon10b}. The IRS 
and MIPS-SED \spitzer\ spectra are shown in Fig.\ \ref{seds} (selected examples) and 
Appendix B (complete sample). 

\subsection{SAGE-SMC photometric data}

The SAGE-SMC (Surveying the Agents of Galaxy Evolution in the 
Tidally-Stripped, Low Metallicity Small Magellanic Cloud) \spitzer\ Legacy 
programme (PID: 40245, P.I.: K.\ Gordon) mapped almost the entire SMC using IRAC and
MIPS. \citet{gordon11} provide a full discussion of the observations, data 
reduction and catalogue generation; we highlight here only the more relevant 
details.

In the overlap region, the SAGE-SMC and (reprocessed) S$^3$MC images were 
combined to produce mosaic images from which photometric catalogues were created
(we use the ``Single Frame + Mosaic Photometry'' catalogue). There is a systematic 
offset between the IRAC photometry in the SAGE-SMC and S$^3$MC catalogues 
\citep{gordon11}; since there is an excellent agreement between the SAGE-SMC fluxes 
and those predicted from the IRAC calibration stars we adopt the IRAC SAGE-SMC 
catalogue fluxes. Due to updated processing and improved calibrations, the S$^3$MC and 
SAGE-SMC MIPS fluxes are also different \citep[see][for full details]{gordon11}; we 
also adopt the MIPS SAGE-SMC catalogue fluxes. Catalogue fluxes are listed in Table 
\ref{table1}. 

Due to the complexity of the multiple datasets, the SAGE-SMC catalogue relies on
a stringent set of rules to create the final catalogues, designed to maximise
reliability rather than completeness \citep{gordon11b}. Therefore, the SAGE-SMC 
catalogues do not provide fluxes for all the IR sources in our target list: even
though the images clearly show a point source, a few sources are missing the shortest 
or all IRAC band fluxes for a variety of reasons. Some YSOs sit in regions where 
there is a complex structure of extended emission (characteristic of star forming 
environments), and may be also marginally extended. Other issues can arise 
during the source extraction and band merging processes causing a particular 
source not to make it into the final IRAC photometric catalogue, for instance 
variability and source confusion. 
As part of their paper on SMC YSOs identified using SAGE-SMC images and catalogues, 
Sewi\l{}o et al. (in preparation) visually inspected all the images of the 34 YSO 
candidates in this sample, and performed aperture photometry on the mosaic images. When
SAGE-SMC catalogue fluxes are unavailable, Table \ref{table1} lists extracted aperture 
fluxes (identified by $*$). The adopted aperture sizes are $5\arcsec$ for the IRAC 
bands, and $8\arcsec$ and $15\arcsec$ respectively for MIPS 24 and 70\,\mum. 

\subsection{Thermal-infrared ground-based spectroscopy}

L-band (2.8$-$4.1\,\mum) spectra of eleven bright sources in the SMC sample 
(Table \ref{table4}) were obtained with the Infrared Spectrometer And Array Camera 
(ISAAC) at the European Southern Observatory (ESO) Very Large Telescope (VLT) at 
Paranal, on the nights of 2006 October 28 and 29 (ESO Programme 078.C-0338, P.I.: 
J.M.\ Oliveira) to search for \hho\ ice absorption at 3\,\mum. The standard IR 
technique of chopping and nodding was employed to cancel the high background. The 
resolving power was $\lambda/\Delta\lambda\approx500$. Exposure times varied between
60 and 105\,min. The hydrogen lines in the standard stars left remnants of at most 
a few per cent of the continuum level. Telluric lines were used to calibrate the 
wavelength scale to an accuracy of $\Delta\lambda\approx 0.002$\,\mum. 

M-band spectra of five bright SMC objects (Table \ref{table4}) were also obtained 
with ISAAC at the VLT, on the nights of 2009 November 4 and 5 (ESO Programme 
084.C-0955, P.I.: J.M.\ Oliveira) to search for the CO ice feature at 4.67\,\mum. 
Exposure times varied between 45 and 90\,min. The M-band spectra were obtained 
and reduced in the same way as the L-band spectra. Telluric lines were used to 
calibrate the wavelength scale to an accuracy of $\Delta\lambda\approx0.003$\,\mum. 

Acquisition for the L-band spectroscopy was done using high spatial resolution 
L$^{\prime}$-band images ($\lambda = 3.8$\,\mum). Magnitudes were obtained using 
aperture photometry on the targets and spectral standard stars observed at regular
intervals during each night. Magnitudes were converted to fluxes using the 
following conversion: a 16-mag star has a flux of $9.414 \times 10^{-5}$\,Jy. 
These resulting fluxes (Table \ref{table1}) were used to flux-calibrate the 
L-band spectra.

Some of these spectra have been published previously; \citet{vanloon08} first 
identified MSX\,SMC\,79 (source \#32), IRAS\,01039$-$7305 (\#33) and IRAS\,01042$-$7215
(\#34) as YSOs based on their L-band spectra; \citet{oliveira11} discussed the ice 
features of sources \#03, 17, 18 and 34; they did not discuss the spectra of source 
\#33 since the ice detections are uncertain. 

\subsection{Near- and mid-infrared ground-based imaging}

The near-infrared imaging observations were performed with the SOFI (Son OF ISAAC)
imager at the New Technology Telescope (NTT) at ESO La Silla, between 2006 October 1 
and 2006 November 10 in service mode (ESO Programme 078.C-0319, P.I.: J.Th.\ van 
Loon). Images were obtained with the $J_s$ and $K_s$ filters using the 
$0\rlap{.}^{\prime\prime}288$ pixel$^{-1}$ plate scale, with a total integration time
of 11\,min per filter. The total integration time was split into jittered 1\,min 
exposures to allow for efficient sky removal. For each filter, the jittered images
were reduced and combined using standard infrared reduction steps implemented 
within ESO's data reduction pipelines: detector cross-talk correction, flatfield 
correction, sky subtraction and shift-addition of jittered frames. Photometric 
calibration was done using dedicated standard star observations; these 
were observed several times per night in order to estimate magnitude 
zeropoints. Since zeropoints were well behaved, we adopted a single value to 
calibrate all observations: $z_{J_s} = 23.09 \pm 0.02$ mag and $z_{K_s} = 22.37
\pm 0.04$ mag. PSF photometry was performed using the {\sc daophot} package 
\citep{stetson87} within {\sc iraf}. Typical full-width at half-maximum (FWHM) 
of the stellar profiles were 3.8 and 3.6 pixels, corresponding to respectively 
$1\rlap{.}^{\prime\prime}1$ and $1\rlap{.}^{\prime\prime}0$ in the $J_s$- and 
$K_s$-bands. Aperture correction was performed using bright PSF 
stars. Magnitudes were converted to fluxes using the following 
conversion: a 16-mag star has fluxes of $6.112 \times 10^{-4}$\,Jy and $2.584 
\times 10^{-4}$\,Jy, respectively in $J_s$ and $K_s$. Fluxes are listed in 
Table\,\ref{table1}.
In Appendix C, $J_sK_s$ colour composite images are shown for each target 
(Fig.\,\ref{thumbs}), together with \spitzer/IRAC [3.6]-[5.8]-[8] colour 
composites. Some sources sit in complex cluster-like environments; we carefully 
investigated each image to identify the redder source as the counterpart for 
the mid-IR sources (these identifications were usually very obvious).

SOFI imaging was obtained only for the original 31 objects, not for the 
additional objects identified by \citet{vanloon08}. However, source \#32 is in 
the SOFI field-of-view of source \#09 and $J_s$ and $K_s$ fluxes were also 
measured. For the remaining two sources (\#33 and 34) we use $JHK_s$ fluxes and 
images from the IR Survey Facility \citep[IRSF,][]{kato07}.

The 21 brightest mid-IR objects were imaged with the VLT Imager and 
Spectrograph for the IR (VISIR) at ESO, Paranal, in service mode over the 
period 2006 September 4 to October 3 (ESO Programme 078.C-0319, P.I.: J.Th.\ 
van Loon). Images were obtained through the narrow-band PAH2 filter (centered 
at $\lambda=11.25$\,\mum, half-band width $\Delta\lambda=0.59$\,\mum). Eight 
standard stars with flux densities in the range 5$-$11\,Jy were observed for 
photometric calibration. The plate scale was $0\rlap{.}^{\prime\prime}075$,
providing a field-of-view of $19\arcsec\times19\arcsec$. The standard IR 
technique of chopping and nodding was employed to cancel the high background. 
Integration times were 24 minutes, split in 16 exposures of 23 chop cycles each.
The individual exposures were combined and corrected for instrumental effects 
using version 1.5.0 of the VISIR pipeline. Photometry was performed by 
collecting the counts within a circular aperture centered on the zero-order 
maximum of the diffraction pattern, avoiding the first Airy ring; if the source 
was not detected upper limits were calculated. Table\,\ref{table1} lists the 
fluxes.

\subsection{Ancillary multi-wavelength data}

\subsubsection{Optical spectroscopy}

Our optical spectroscopy was obtained using the Double-Beam Spectrograph (DBS) 
mounted on the Nasmyth focus of the Australian National University 2.3-m 
telescope at Siding Spring Observatory. More information on the instrument can 
be found in \citet*{rodgers88}. Standard {\sc iraf} routines were used for the 
data reduction (bias subtraction, flat fielding, wavelength calibration). The 
blue and red spectra were joined in the interval 6160$-$6300\,\AA; the joined 
spectra cover the wavelength range 3600$-$9500\,\AA. Spectral standard stars, 
observed with the same instrumental setup as the programme objects, were used 
to provide a relative flux calibration for each object. Telluric features were 
removed by observing white dwarf standards with few intrinsic spectral 
features. The final spectra have an effective resolution of $\sim4.5$\,\AA.

Optical spectra were obtained for 32 objects (Appendix D, Fig.\,\ref{optical}). 
At the position of the IR source, source \#24 shows no point source at optical
wavelengths, only extended emission. There is clearly a point source in the 
$J_s$-band (Fig.\,\ref{thumbs}) but it is the faintest object in the sample 
(Table\ \ref{table1}); thus it was too faint to obtain an optical spectrum. The 
position of IR source \#31 sits between two bright optical sources; in 
Fig.\,\ref{thumbs} it is clear that this region is very complex with many bright
sources at short wavelengths as well as bright IR sources. Thus no optical 
spectrum was obtained for this source.

For sources \#05, 06 and 29 the optical spectra exhibit only Balmer absorption 
lines, i.e.\ the spectra are typical of more evolved stars. Since these objects 
are also very faint in the $J_s$-band (Table \ref{table1}) it is unlikely the 
obtained optical spectra are associated with the IR source. For sources \#02, 10, 11, 
14, 16, 17, 18, 23 and 34 the spectra only show H$\alpha$ emission thus we perform no 
further analysis of their optical spectra. For the remaining 20 sources we analyse 
the optical spectra in detail (Section 4.5).

\subsubsection{Radio data}

Radio free-free emission from (ultra-)compact H\,{\sc ii} regions is a common 
signpost of massive star formation.\footnotemark\footnotetext{By (ultra-)compact 
H\,{\sc ii} region we mean a H\,{\sc ii} region associated with the early stages 
of massive star formation, without making any statement about the degree of 
compactness or physical properties \citep{hoare07}; this is to distinguish
these objects from ``classical'' evolved H\,{\sc ii} regions.} Using all available
archival data from the Australia Telescope Compact Array (ATCA) and the
Parkes radio telescope, \citet{wong11a} and \citet{crawford11} created new 
high-sensitivity, high-resolution radio-continuum images of the SMC at 1.42, 
2.37, 4.80 and 8.64\,GHz (wavelengths respectively 20, 13, 6 and 3\,cm), as well as
point source catalogues \citep{wong11b,wong12}. Of the 34 objects in the sample, 11 
sources have radio-continuum detections at at least 2 frequencies. Of these, 8 are radio 
point sources found within $4\arcsec$ of the IR source (the positional accuracy at 
20 and 13\,cm). Sources \#08 and 31 are in complex and extended H\,{\sc ii} 
regions and therefore the source position is less certain; there is no point 
source at the position of source \#15 but we clearly see an extended 
radio-continuum object. Radio spectral indices ($S_{\nu} \propto \nu^{\alpha}$) 
for these sources are $\alpha\ \gsim-0.3$, consistent with their classification 
as compact H\,{\sc ii} regions \citep[e.g.,][]{filipovic98} --- background 
galaxies are dominated by synchrotron emission with a steeper spectral index. 
The sources detected at radio wavelengths are the most luminous in our sample 
(Section 4.1).

\onecolumn

\setlongtables
\LTcapwidth=9.3in
{\small
\begin{landscape}
\begin{longtable}{l@{\hspace{2.mm}}c@{\hspace{2mm}}c@{\hspace{2mm}}c@{\hspace{2.5mm}}c@{\hspace{2.5mm}}c@{\hspace{3.5mm}}ccc@{\hspace{3.5mm}}c@{\hspace{5mm}}c@{\hspace{5mm}}c@{\hspace{1.5mm}}c@{\hspace{2mm}}l}
\caption{\normalsize Infrared fluxes (in mJy) for the SMC YSO candidates. 
{\it Spitzer} photometry is from the SAGE-SMC catalogue \citep{gordon11}; when 
catalogue photometry is unavailable, aperture photometry fluxes are used 
(indicated by $*$). Listed $J_sK_s$ fluxes are from this work,
except for fluxes of sources \#32 and 33 that are from the IRSF catalogue
\citep{kato07}. Source IDs are from the S$^3$MC catalogue, unless stated 
otherwise. S$^3$MC objects were identified by \citet{bolatto07}. Some sources 
were identified spectroscopically using IRS and MIPS-SED spectra by 
\citet[][1]{vanloon08}, \citet[][2]{vanloon10b} and \citet[][3]{oliveira11};  
\citet[][4]{martayan07} identified source \#20 as a candidate Herbig B[e] star. 
Source classification is discussed in Sections 5 and 6. \label{table1}}\\
\hline
\#  &RA \& Dec (J2000)              &$J_s$                  &$K_s$       
                &$L^{\prime}$                &$F_{3.6}$              &$F_{4.5}$    
		         &$F_{5.8}$          &$F_{8}$                
			 &$F_{11.3}$         &$F_{24}$               &$F_{70}$  			    &source ID&Ref.\\
\hline 
\endfirsthead
\endhead
\hline
\endfoot
\hline
\endlastfoot
01         &00\,43\,12.86\,$-$72\,59\,58.3 &0.339\,$\pm$\,0.029&      0.341\,$\pm$\,0.030&                  \null  &\al{*}2.56\,$\pm$\,0.30&\al{*}1.98\,$\pm$\,0.27&\al{*1}0.35\,$\pm$\,0.62 &\al{*2}5.74\,$\pm$\, 0.98&              $<$\,6& \al{33}7.6\,$\pm$\, 1.7&	\al{3}316\,$\pm$\,21&004312.85$-$725958.30&\\
	&&&&&&&&&&&&IRAS\,00413$-$7316&\\
02         &00\,44\,51.87\,$-$72\,57\,34.2 &0.258\,$\pm$\,0.016&      0.937\,$\pm$\,0.047&      7.76\,$\pm$\,0.75  &      5.76\,$\pm$\,0.17&       9.92\,$\pm$\,0.23& \al{1}5.73\,$\pm$\,0.36 & \al{3}1.69\,$\pm$\, 0.45& \al{3}3\,$\pm$\, 4& \al{55}5.4\,$\pm$\, 4.9& \al{1}620\,$\pm$\,18&IRAS\,00429$-$7313          & 2\\
03         &00\,44\,56.30\,$-$73\,10\,11.8 &0.210\,$\pm$\,0.030&      3.583\,$\pm$\,0.171&\al{8}2.75\,$\pm$\,7.98  &\al{4}0.27\,$\pm$\,0.66&\al{*8}2.85\,$\pm$\,1.00 &\al{15}0.60\,$\pm$\,1.60 &\al{19}3.10\,$\pm$\, 2.46&\al{16}4\,$\pm$\,10&\al{247}0.0\,$\pm$\,12.4&\al{12}440\,$\pm$\,60&IRAS\,00430$-$7326          &2, 3\\
04         &00\,45\,21.26\,$-$73\,12\,18.7 &0.977\,$\pm$\,0.028&      1.010\,$\pm$\,0.051&                  \null  &      1.78\,$\pm$\,0.18&       2.17\,$\pm$\,0.06&       2.92\,$\pm$\,0.19 &       4.59\,$\pm$\, 0.63&              \null&  \al{3}9.3\,$\pm$\, 0.4&       917\,$\pm$\,11&004521.26$-$731218.68&\\
05         &00\,45\,47.51\,$-$73\,21\,42.4 &0.081\,$\pm$\,0.004&      0.527\,$\pm$\,0.024&                  \null  &      3.33\,$\pm$\,0.08&       5.81\,$\pm$\,0.07&       8.18\,$\pm$\,0.14 & \al{1}2.20\,$\pm$\, 0.24&              \null&  \al{5}9.4\,$\pm$\, 0.5&       281\,$\pm$\, 6\,&\null&\\
06	   &00\,46\,24.45\,$-$73\,22\,07.1 &0.077\,$\pm$\,0.006&      0.363\,$\pm$\,0.024&	3.89\,$\pm$\,0.38  &	  2.85\,$\pm$\,0.14&	   8.82\,$\pm$\,0.17& \al{1}4.40\,$\pm$\,0.31 & \al{2}0.40\,$\pm$\, 0.48& 	      $<$\,6& \al{19}2.5\,$\pm$\, 0.9& \al{2}154\,$\pm$\,20&004624.46$-$732207.30&2\\
07	   &00\,46\,51.72\,$-$73\,15\,25.3 &0.078\,$\pm$\,0.008&      0.101\,$\pm$\,0.012&		    \null  &\al{*}1.71\,$\pm$\,0.25& \al{*}1.08\,$\pm$\,0.20& \al{*}9.04\,$\pm$\,0.58 &\al{*2}1.60\,$\pm$\, 1.00& 	       \null&  \al{4}2.5\,$\pm$\, 0.4&       776\,$\pm$\,11&004651.71$-$731525.34&\\
08	   &00\,48\,25.83\,$-$73\,05\,57.3 &0.166\,$\pm$\,0.014&      0.316\,$\pm$\,0.023&		    \null  &	  2.07\,$\pm$\,0.08&	   3.25\,$\pm$\,0.06&	    8.76\,$\pm$\,0.16 & \al{1}9.30\,$\pm$\, 0.44& \al{1}9\,$\pm$\, 2& \al{59}2.8\,$\pm$\, 5.2& \al{7}779\,$\pm$\,55&004825.83$-$730557.29&\\
09	   &00\,48\,41.78\,$-$73\,26\,15.3 &0.141\,$\pm$\,0.015&      0.146\,$\pm$\,0.028&		    \null  &\al{*}1.93\,$\pm$\,0.26&	   0.98\,$\pm$\,0.04&	    3.01\,$\pm$\,0.10 &       7.00\,$\pm$\, 0.43& 	      $<$\,6& \al{23}9.4\,$\pm$\, 1.0& \al{1}875\,$\pm$\,18&004841.77$-$732615.25&\\
10	   &00\,49\,01.64\,$-$73\,11\,09.6 &0.086\,$\pm$\,0.008&      0.163\,$\pm$\,0.016&		    \null  &\al{*}2.49\,$\pm$\,0.30&	   0.80\,$\pm$\,0.13&	    5.06\,$\pm$\,0.27 & \al{1}3.30\,$\pm$\, 0.95& 	      $<$\,6& \al{70}8.7\,$\pm$\, 4.0& \al{3}109\,$\pm$\,25&004901.63$-$731109.60&\\
11	   &00\,49\,44.57\,$-$73\,24\,32.8 &0.121\,$\pm$\,0.009&      0.166\,$\pm$\,0.015&		    \null  &\al{*}1.59\,$\pm$\,0.24&	   0.52\,$\pm$\,0.03&	    3.39\,$\pm$\,0.17 &       7.13\,$\pm$\, 0.46& 	       \null&  \al{4}7.5\,$\pm$\, 0.5& \al{*}900\,$\pm$\,73&004944.57$-$732432.75&\\
12	   &00\,50\,40.25\,$-$73\,20\,37.0 &0.090\,$\pm$\,0.007&      0.124\,$\pm$\,0.011&		    \null  &	  0.66\,$\pm$\,0.03&	   0.46\,$\pm$\,0.01&	    3.02\,$\pm$\,0.07 &       7.11\,$\pm$\, 0.19& 	       \null&  \al{7}1.3\,$\pm$\, 0.5&       452\,$\pm$\, 6\,&005040.24$-$732036.99&\\
13	   &00\,50\,43.24\,$-$72\,46\,56.2 &0.252\,$\pm$\,0.021&      0.336\,$\pm$\,0.024&		    \null  &	  1.63\,$\pm$\,0.27&	   1.27\,$\pm$\,0.05&	    5.89\,$\pm$\,0.18 & \al{1}6.60\,$\pm$\, 0.54&       8\,$\pm$\, 1& \al{58}9.2\,$\pm$\, 3.6& \al{2}762\,$\pm$\,28&005043.23$-$724656.24&\\
14	   &00\,50\,58.09\,$-$73\,07\,56.8 &0.391\,$\pm$\,0.019&      0.817\,$\pm$\,0.048&		    \null  &	  2.11\,$\pm$\,0.04&	   2.42\,$\pm$\,0.03&	    4.95\,$\pm$\,0.07 & \al{1}3.50\,$\pm$\, 0.19& 	       \null& \al{11}4.3\,$\pm$\, 1.1&       461\,$\pm$\, 6\,&005058.09$-$730756.78&\\
15	   &00\,52\,38.84\,$-$73\,26\,23.9 &0.227\,$\pm$\,0.013&      0.246\,$\pm$\,0.020&		    \null  &	  1.57\,$\pm$\,0.20&	   1.04\,$\pm$\,0.06&	    6.98\,$\pm$\,0.22 & \al{2}0.50\,$\pm$\, 0.51& \al{1}4\,$\pm$\, 6& \al{61}0.7\,$\pm$\, 3.4& \al{2}695\,$\pm$\,19&005238.84$-$732623.92&\\
&&&&&&&&&&&&IRAS\,00509$-$7342&\\
16	   &00\,53\,25.36\,$-$72\,42\,53.2 &0.364\,$\pm$\,0.022&      0.369\,$\pm$\,0.025&		    \null  &	  0.84\,$\pm$\,0.10&	   0.67\,$\pm$\,0.02&	    4.03\,$\pm$\,0.19 & \al{1}3.50\,$\pm$\, 0.57& \al{1}9\,$\pm$\, 4& \al{28}2.0\,$\pm$\, 1.5& \al{1}226\,$\pm$\,14&005325.36$-$724253.20&\\
&&&&&&&&&&&&IRAS\,00516$-$7259&\\
17	   &00\,54\,02.31\,$-$73\,21\,18.6 &0.357\,$\pm$\,0.022&      2.244\,$\pm$\,0.120&\al{2}6.41\,$\pm$\,2.55  &\al{1}9.50\,$\pm$\,0.25& \al{4}6.30\,$\pm$\,0.66& \al{7}8.40\,$\pm$\,0.71 &\al{11}3.00\,$\pm$\, 0.91& \al{6}9\,$\pm$\,10& \al{49}6.5\,$\pm$\, 3.5& \al{1}669\,$\pm$\,14&005402.30$-$732118.70&2, 3\\					 
18	   &00\,54\,03.36\,$-$73\,19\,38.4 &0.247\,$\pm$\,0.022&      1.104\,$\pm$\,0.060&\al{1}8.61\,$\pm$\,1.80  &\al{1}2.50\,$\pm$\,0.18& \al{3}8.60\,$\pm$\,0.37& \al{8}4.20\,$\pm$\,0.78 &\al{12}5.00\,$\pm$\, 1.02& \al{9}1\,$\pm$\, 7& \al{82}4.6\,$\pm$\, 5.6& \al{3}987\,$\pm$\,25&005403.36$-$731938.30&2, 3\\
19	   &00\,54\,19.16\,$-$72\,29\,09.6 &0.249\,$\pm$\,0.009&      0.287\,$\pm$\,0.012&		    \null  &	  1.27\,$\pm$\,0.03&	   3.72\,$\pm$\,0.05& \al{1}0.80\,$\pm$\,0.15 & \al{3}7.90\,$\pm$\, 0.34& \al{7}6\,$\pm$\, 3& \al{52}2.6\,$\pm$\, 3.1&       319\,$\pm$\, 5\,&005419.16$-$722909.63&\\
20	   &00\,56\,06.38\,$-$72\,28\,28.1 &0.731\,$\pm$\,0.048&      0.877\,$\pm$\,0.045&		    \null  &	  2.74\,$\pm$\,0.06&	   4.28\,$\pm$\,0.07&	    6.81\,$\pm$\,0.11 & \al{1}8.40\,$\pm$\, 0.20& 	       \null&  \al{8}3.8\,$\pm$\, 0.7&       221\,$\pm$\, 3\,&005606.37$-$722828.05&4\\
21	   &00\,56\,06.50\,$-$72\,47\,22.7 &0.075\,$\pm$\,0.013&      0.167\,$\pm$\,0.021&		    \null  &	  1.14\,$\pm$\,0.05&	   1.15\,$\pm$\,0.02&	    3.75\,$\pm$\,0.11 &       7.77\,$\pm$\, 0.27& 	      $<$\,6& \al{26}9.3\,$\pm$\, 1.8& \al{1}081\,$\pm$\,13&005606.49$-$724722.66&\\
22	   &00\,57\,57.11\,$-$72\,39\,15.4 &0.330\,$\pm$\,0.021&      0.655\,$\pm$\,0.048&	2.77\,$\pm$\,0.27  &	  3.46\,$\pm$\,0.63&	   3.80\,$\pm$\,0.12&	    6.04\,$\pm$\,0.23 &       8.19\,$\pm$\, 0.66& 	      $<$\,6& \al{24}3.8\,$\pm$\, 1.6& \al{3}276\,$\pm$\,29&005757.10$-$723915.40&\\
&&&&&&&&&&&&IRAS\,00562$-$7255&\\
23         &00\,58\,06.41\,$-$72\,04\,07.3 &0.347\,$\pm$\,0.015&      0.385\,$\pm$\,0.020&                  \null  &      1.19\,$\pm$\,0.14&       0.86\,$\pm$\,0.03&       4.89\,$\pm$\,0.17 & \al{1}1.80\,$\pm$\, 0.67&             $<$\,6& \al{32}1.1\,$\pm$\, 2.5& \al{1}442\,$\pm$\,17&005806.41$-$720407.32&\\
&&&&&&&&&&&&IRAS\,00563$-$7220&\\
24	   &01\,00\,22.32\,$-$72\,09\,58.1 &0.072\,$\pm$\,0.003&      0.469\,$\pm$\,0.023&		    \null  &	  2.45\,$\pm$\,0.05&	   4.04\,$\pm$\,0.06&	    5.90\,$\pm$\,0.08 & \al{1}0.20\,$\pm$\, 0.13& 	       \null&  \al{4}3.6\,$\pm$\, 0.5&       202\,$\pm$\, 4\,&\null&\\
25	   &01\,01\,31.70\,$-$71\,50\,40.3 &0.348\,$\pm$\,0.027&      0.386\,$\pm$\,0.047&		    \null  &\al{*}2.79\,$\pm$\,0.32&	   0.91\,$\pm$\,0.17&	    4.55\,$\pm$\,0.26 & \al{1}1.90\,$\pm$\, 1.39& 	      $<$\,6& \al{55}0.7\,$\pm$\, 2.3& \al{3}875\,$\pm$\,19&010131.69$-$715040.30&\\
26	   &01\,02\,48.54\,$-$71\,53\,18.0 &0.167\,$\pm$\,0.011&      0.482\,$\pm$\,0.028&		    \null  &\al{*}3.57\,$\pm$\,0.36& \al{*}4.58\,$\pm$\,0.41&	    5.07\,$\pm$\,0.24 &       7.52\,$\pm$\, 0.64& 	      $<$\,6& \al{28}8.2\,$\pm$\, 2.1& \al{5}616\,$\pm$\,30&010248.54$-$715317.98&\\
27	   &01\,03\,06.14\,$-$72\,03\,44.0 &0.073\,$\pm$\,0.005&      0.154\,$\pm$\,0.009&		    \null  &	  0.48\,$\pm$\,0.09& 	   0.37\,$\pm$\,0.01&	    2.01\,$\pm$\,0.12 &\al{*1}6.26\,$\pm$\, 1.00&   	       \null&  \al{6}9.6\,$\pm$\, 0.6& \al{1}123\,$\pm$\,11&010306.13$-$720343.95&\\
28	   &01\,05\,07.26\,$-$71\,59\,42.7 &0.325\,$\pm$\,0.022&      1.503\,$\pm$\,0.082&\al{3}2.34\,$\pm$\,3.12  &\al{2}1.90\,$\pm$\,0.37& \al{5}5.40\,$\pm$\,0.85&\al{12}6.00\,$\pm$\,1.31 &\al{29}5.00\,$\pm$\, 3.46&\al{39}6\,$\pm$\,22&\al{350}7.0\,$\pm$\,22.8&\al{10}960\,$\pm$\,67&010507.25$-$715942.70&2\\
29	   &01\,05\,30.71\,$-$71\,55\,21.3 &0.074\,$\pm$\,0.005&      0.156\,$\pm$\,0.023&		    \null  &	  1.73\,$\pm$\,0.07& 	   4.48\,$\pm$\,0.06& \al{1}1.30\,$\pm$\,0.16 & \al{2}1.30\,$\pm$\, 0.20& \al{2}1\,$\pm$\, 6& \al{27}8.3\,$\pm$\, 2.2& \al{1}467\,$\pm$\,14&010530.71$-$715521.25&\\
30	   &01\,06\,59.67\,$-$72\,50\,43.1 &0.699\,$\pm$\,0.020&      2.282\,$\pm$\,0.101&	8.75\,$\pm$\,0.84  &	  8.03\,$\pm$\,0.13& \al{1}0.90\,$\pm$\,0.12& \al{1}6.30\,$\pm$\,0.18 & \al{2}2.80\,$\pm$\, 0.22& 	       \null&  \al{5}0.6\,$\pm$\, 0.6&       536\,$\pm$\, 5\,&010659.66$-$725043.10&2\\
31         &01\,14\,39.38\,$-$73\,18\,29.3 &0.137\,$\pm$\,0.015&      0.310\,$\pm$\,0.023&                  \null  &\al{*}5.02\,$\pm$\,0.42&       2.96\,$\pm$\,0.14&       5.71\,$\pm$\,0.43 & \al{1}3.60\,$\pm$\, 2.56& \al{1}2\,$\pm$\, 3&\al{*80}5.9\,$\pm$\,14.9& \al{9}351\,$\pm$\,66&011439.38$-$731829.26&2\\
32	   &00\,48\,39.64\,$-$73\,25\,01.0 &0.282\,$\pm$\,0.008&      1.010\,$\pm$\,0.017&\al{1}1.11\,$\pm$\,1.02  &      6.94\,$\pm$\,0.08& \al{1}2.80\,$\pm$\,0.13& \al{2}3.40\,$\pm$\,0.24 & \al{4}1.60\,$\pm$\, 0.34&              \null& \al{14}8.1\,$\pm$\, 1.7&\al{*1}271\,$\pm$\,45&{\msx}\,79                  &1\\
&&&&&&&&&&&&004839.63-732500.98&\\
33         &01\,05\,30.22\,$-$72\,49\,53.9 &1.064\,$\pm$\,0.049&\al{1}1.063\,$\pm$\,0.214&\al{8}8.26\,$\pm$\,8.13  &\al{6}0.40\,$\pm$\,1.92& \al{9}3.90\,$\pm$\,2.65&\al{12}2.00\,$\pm$\,1.21 &\al{17}0.00\,$\pm$\, 3.71&              \null& \al{93}9.1\,$\pm$\, 4.9& \al{3}208\,$\pm$\,17&{\iras}\,01039$-$7305       &1, 2\\
34         &01\,05\,49.29\,$-$71\,59\,48.8 &0.344\,$\pm$\,0.013&      2.578\,$\pm$\,0.071&\al{4}7.62\,$\pm$\,4.39  &\al{2}2.10\,$\pm$\,0.28& \al{4}8.30\,$\pm$\,0.65& \al{8}0.10\,$\pm$\,0.81 &\al{11}3.00\,$\pm$\, 0.88&              \null& \al{64}8.0\,$\pm$\, 3.4& \al{1}998\,$\pm$\,18&{\iras}\,01042$-$7215       &1, 2, 3
\end{longtable}

\end{landscape}}

\twocolumn

\subsection{Resolving YSOs at the distance of the SMC}

The \spitzer\ photometry has a spatial resolution of 2$\arcsec$, 6$\arcsec$, 18$\arcsec$ 
respectively for the IRAC bands and at 24 and 70\,\mum, while the resolution of the IRS 
spectra varies between $2\rlap{.}^{\prime\prime}5$ and 10$\arcsec$. At the distance of 
the SMC \citep[60\,pc,][]{szewczyk09}, this corresponds to $\sim 0.6-5$\,pc for the 
photometry and $\sim 0.8-3$\,pc for the spectra. For comparison the Trapezium core of the
Orion Nebula cluster extends $\sim0.5$\,pc. Using the near-IR photometry (FWHM 
typically 1$\arcsec$ or $\sim 0.3$\,pc) we show that the \spitzer\ fluxes are 
dominated by the contribution of a single bright IR source. However, it is likely that 
some IR sources encase a massive binary or a dominant massive star surrounded by 
low-mass siblings that we are unable to resolve. Most of our analysis focuses 
on the YSO chemistry and other envelope properties therefore the detailed
luminosity distribution within the YSO source is less important. Another
important issue to bear in mind is that the different datasets sample 
physically distinct spatial scales due to their different spatial resolutions
(see discussion in Section 5.1).  

\begin{figure}
\includegraphics[scale=0.91]{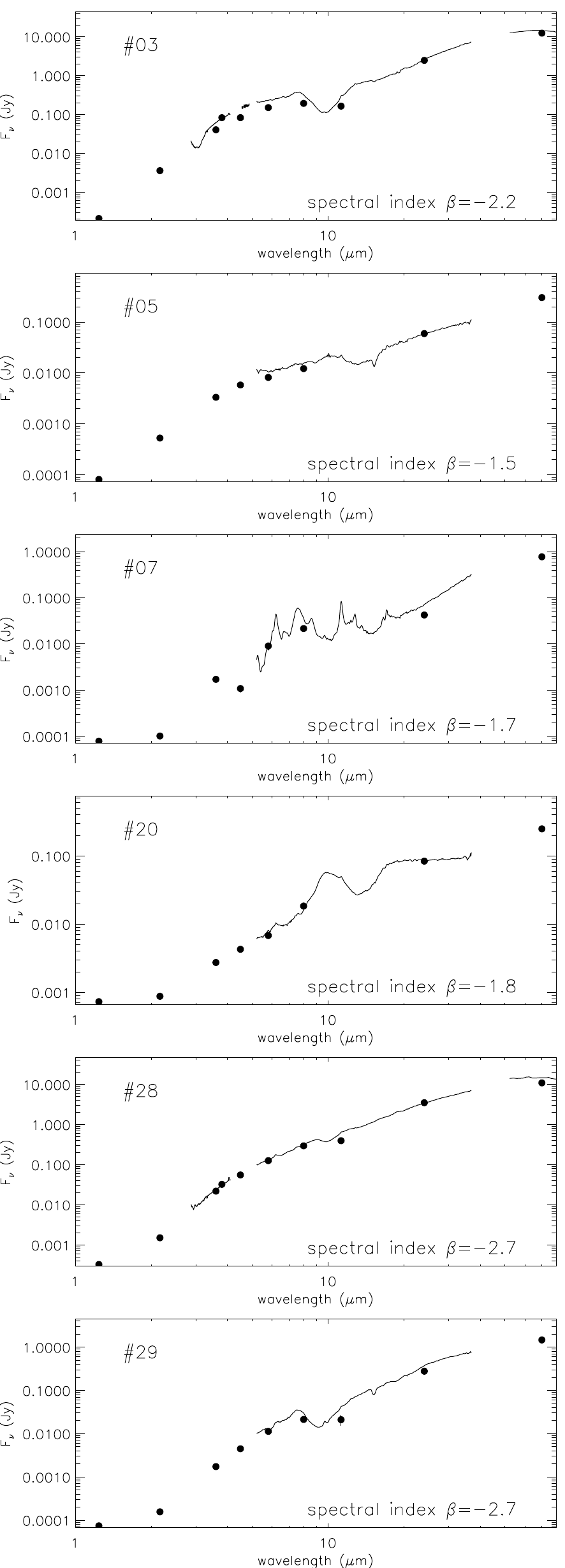}
\caption{Selected examples of SEDs of YSO candidates in the SMC. Spectral 
indices calculated in the range 3.6$-$24\,\mum\ are given in each panel. The 
SEDs of all objects in the sample are provided in Appendix B.}
\label{seds}
\end{figure}

\section{Spectral properties}

The spectra of YSOs are characterised by a cold dust continuum, and often 
exhibit strong silicate features at 10 and 18\,\mum. Silicate absorption 
superimposed on a very red continuum is indicative of embedded protostellar 
objects \citep[e.g.,][]{furlan08}. Ice absorption features are another common 
feature in spectra of embedded YSOs. These objects are traditionally 
classified as Class I sources \citep[based on their IR spectral 
index,][]{lada87}, or as Stage I sources \citep[based on their modelled spectral 
energy distributions,][]{robitaille06}. As the nascent massive star becomes hotter 
and excites its environment (i.e.\ it develops a compact H\,{\sc ii} region), 
emission features attributed to PAHs, and atomic fine-structure and H$_2$ emission
lines also become common. Thus the infrared spectra of YSOs can show a superposition 
of ice, dust and PAH features that can be difficult to disentangle, in particular in 
extra-galactic sources \citep[e.g.,][]{oliveira09}. 

For intermediate-mass YSOs in the later stages the dust cocoons dissipate and 
become hotter, and eventually IR emission from a circumstellar disc dominates, 
with prominent silicate emission. Such objects are usually classified as Class 
II or Stage II objects. Amongst such objects are Herbig Ae/Be (HAeBe) stars, that are 
often hot enough to excite PAH emission \citep[e.g.,][]{keller08}.

Figures \ref{seds} and \ref{seds_ap} show all available IR photometry and
spectroscopy for the 34 YSO candidates. In this section we discuss the main 
spectral features and their properties and in the next section we discuss object
classification.

\subsection{Spectral energy distribution and dust features}

%We use the IRS spectra to measure the IR spectral index between 20 and 32\,\mum. 
%This interval was selected to help identify possible evolved star interlopers in the 
%sample: generally speaking, if a spectrum turns over in this wavelength range the 
%emitting object is an evolved star with only moderate amounts of cold dust 
%\citep{vanloon10a}. 

The spectral index in frequency space defined as $F_{\nu} \propto \nu^\beta$
corresponds to $\beta = -\alpha -1$, where $\alpha$ is the spectral index 
defined in wavelength space: 
$\alpha = \dfrac{d\,\log(\lambda F_{\lambda})}{d\,\log \lambda}$ \citep{andre94}. As
determined using \spitzer\ photometry of Galactic YSO samples, Class I objects 
have $\alpha \ga -0.5$, Class II YSOs have $-0.5 \ga \alpha \ga -1.8$ and Class III
(disc-less YSOs) have $\alpha \la -1.8$ \citep[e.g.,][]{muench07}. This corresponds to 
$\beta \la -0.5$ (Class I), $-0.5 \la \beta \la 0.8$ (Class II) and finally 
$\beta \ga 0.8$ (Class III) sources. These classes are more commonly used to describe the
early stages of intermediate- and low-mass stellar evolution. 

We calculate the spectral indices of the objects in the sample between 3.6 and 24\,\mum. 
All objects have steep spectra with $\beta < -1.0$, consistent with a 
Class I classification. Closer inspection of the SEDs (Fig.\,\ref{seds_ap}) reveals that
for \#19 and 20 the SED flattens considerably above 20\,\mum, with respectively  
$\beta \sim -0.6$ and $\beta \sim -0.2$ between 20 and 32\,\mum. This suggests the
presence of only moderate amounts of cold dust. 

%Seven of these objects exhibit some 
%degree of flattening in their SEDs ($-1.6 \la \beta \la -1.3$), suggestive of a somewhat
%less embedded source. The two objects with $\beta \ga -1.3$, sources \#19 and 20 
%(respectively $\beta=-0.6$ and $\beta=-0.2$) show typical silicate dust emission, 
%consistent with a Class II classification.

As part of their survey of photometric YSO candidates, Sewi\l{}o et al. (in 
preparation) have fitted the SEDs of all the sources in our sample, using the
grid of models and fitting tool developed by \citet{robitaille06,robitaille07}. Based 
on model parameters (mass accretion rate, stellar mass and disc mass) the fitting tool 
classifies objects as Stage I, II and III. The SED models do not include the 
contribution from PAH emission. Good SED fits are achieved for the majority of objects,
and all objects are classified as Stage I sources. Integrated SED luminosities and 
stellar masses are in the ranges $1.5 \times 10^{3}-1.4\times 10^{5}$\,L$_{\odot}$ and 
$8.2-30$\,M$_{\odot}$; sources \#20 and 28 are the least and most luminous 
respectively. We should point out that this grid of models assumes that a single YSO 
source is responsible for the fluxes measured at different wavelengths. As discussed in 
Section\,3.6, some of the SMC sources may be binaries or small unresolved clusters; we
assume that the central source can be represented by the equivalent of a single 
luminous object. 
 
Prominent silicate absorption is observed in the IRS spectra of at least ten YSOs 
(indicated by \tick\ in the relevant column of Table \ref{table3}). A further two 
objects have tentative detections of silicate absorption (\tick? in Table 
\ref{table3}): source \#30 shows strong PAH emission while the spectrum of source \#24 
shows only weak features. As discussed by \citet{oliveira09}, without 
detailed modelling of {\it all} the spectral features in the spectrum it is often 
difficult to distinguish weak silicate absorption from PAH emission complexes between 
7 and 12\,\mum. Indeed, silicate absorption is conclusively identified only in objects 
where PAH emission is relatively weak or non existent. The red wing of the silicate 
absorption band can be deformed by the presence of the 13-\mum\ libration mode of \hho\
ice. This might explain the particularly extended red wing of the silicate feature for 
source \#06, the source with strongest \hho\ and \coo\ ice in our sample (see 
Section\,\ref{ices_section}). Recent modelling work \citep*{robinson12} suggests that 
constraining envelope optical depth and dust properties is crucial in asserting whether
the libration mode is present in a spectrum. The 18-\mum\ silicate absorption feature
seems to be present in the spectrum of some sources, namely \#03, 17, 18, 30, and maybe
\#29 as well.

The shape of the silicate emission features is a reflection of the composition of
the dust grains. Amorphous olivine grains create a typical sawtooth-shaped 
profile peaking at 9.8\,\mum, characteristic of small interstellar dust grains. 
The presence of processed (crystalline) grains is revealed by a shift in the 
feature's peak to longer wavelengths, and a significant population of larger 
grains broadens the profile \citep[e.g.,][]{kessler06}. However, as pointed out by
\citet{sargent09}, the presence of a population of large grains is difficult to identify
unambiguously. %It is often be challenging to distinguish YSOs with silicate emission 
%from oxygen-rich evolved stars \citep[e.g.,][]{woods11}.

Four objects in the sample exhibit silicates in emission (indicated by \emm\ in 
Table\,\ref{table3}). None of the spectra exhibits crystalline features. The profile of
source \#19 is narrow with a sharp 9.8-\mum\ peak, indicating unprocessed small dust 
grains. This object is not a YSO, as detailed in Section\,\ref{symbiotic}. Sources 
\#05, 14, and 20 exhibit broader profiles that could suggest larger grains or mixed 
grain populations.  The spectra of these three objects also show PAH emission at 
11.3\,\mum\ that, in particular for source \#14, masks the underlying profile shape. 
Sources \#05 and 14 exhibit a red underlying continuum while the SED of \#20 is 
essentially flat at wavelengths longer than 20\,\mum\ (see above). These objects are 
discussed in more detail in Section\,\ref{class}.3.

\subsection{Ice absorption bands}
\label{ices_section}

\begin{figure*}
\includegraphics[scale=0.42]{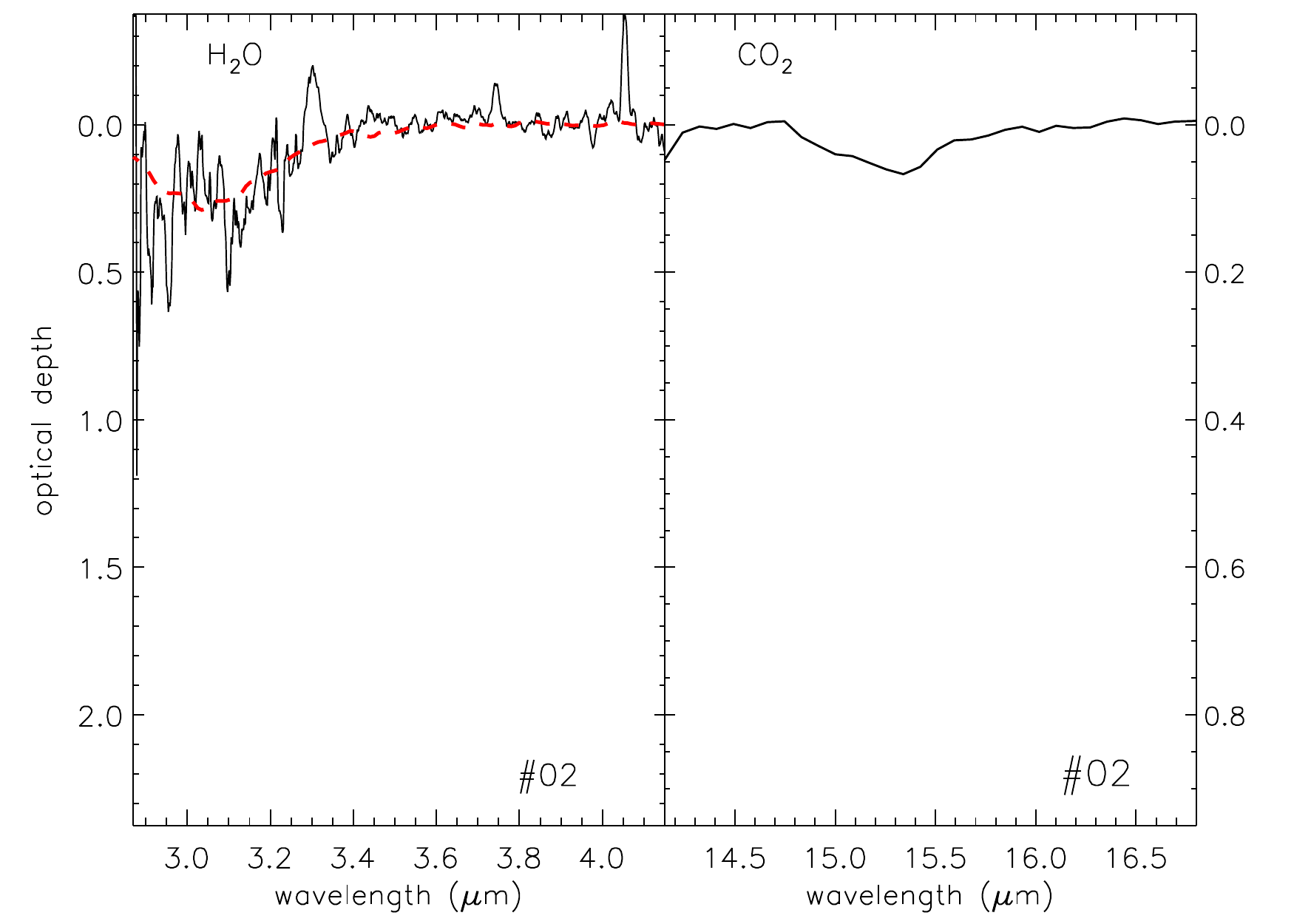}
\includegraphics[scale=0.42]{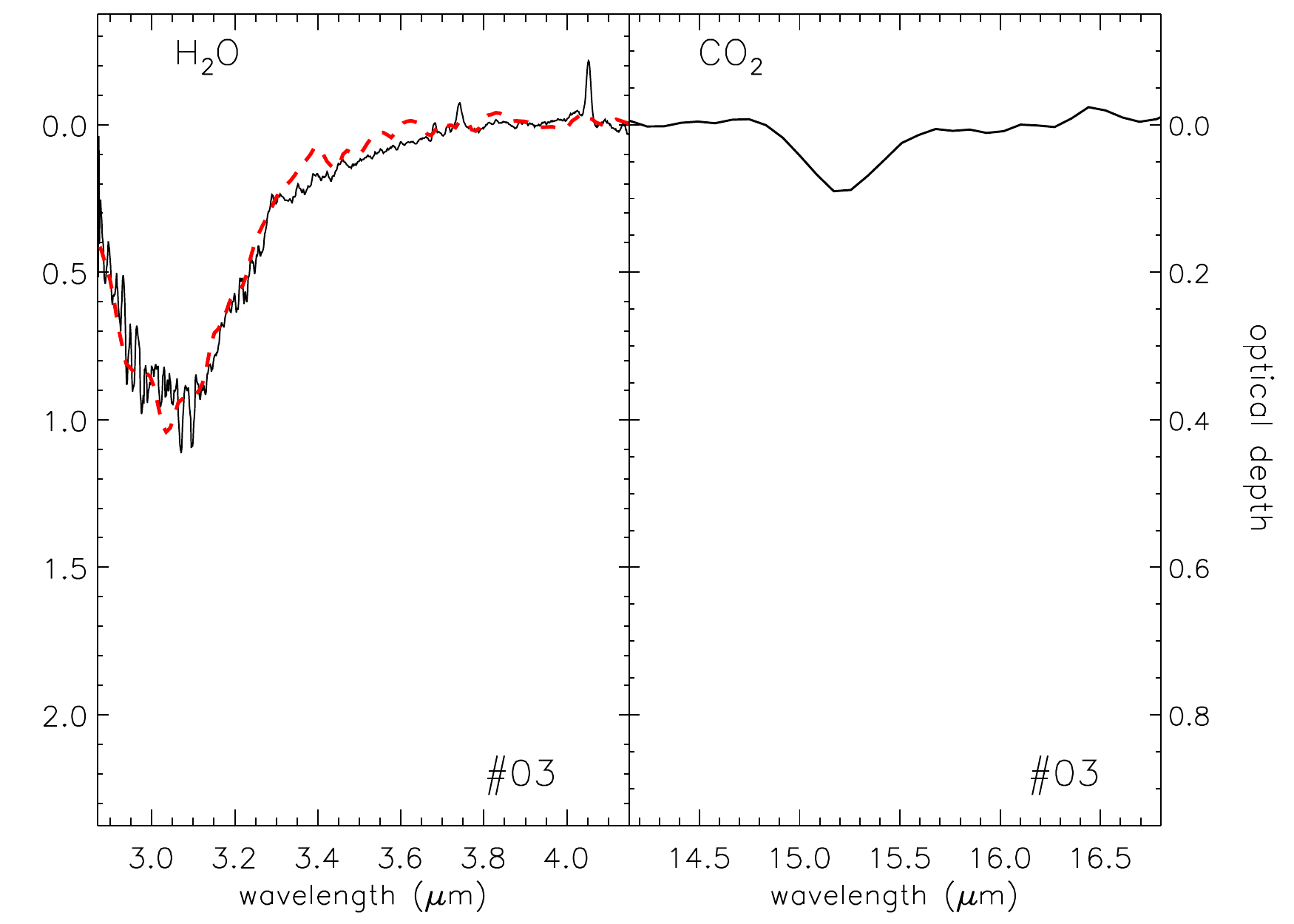}
\includegraphics[scale=0.42]{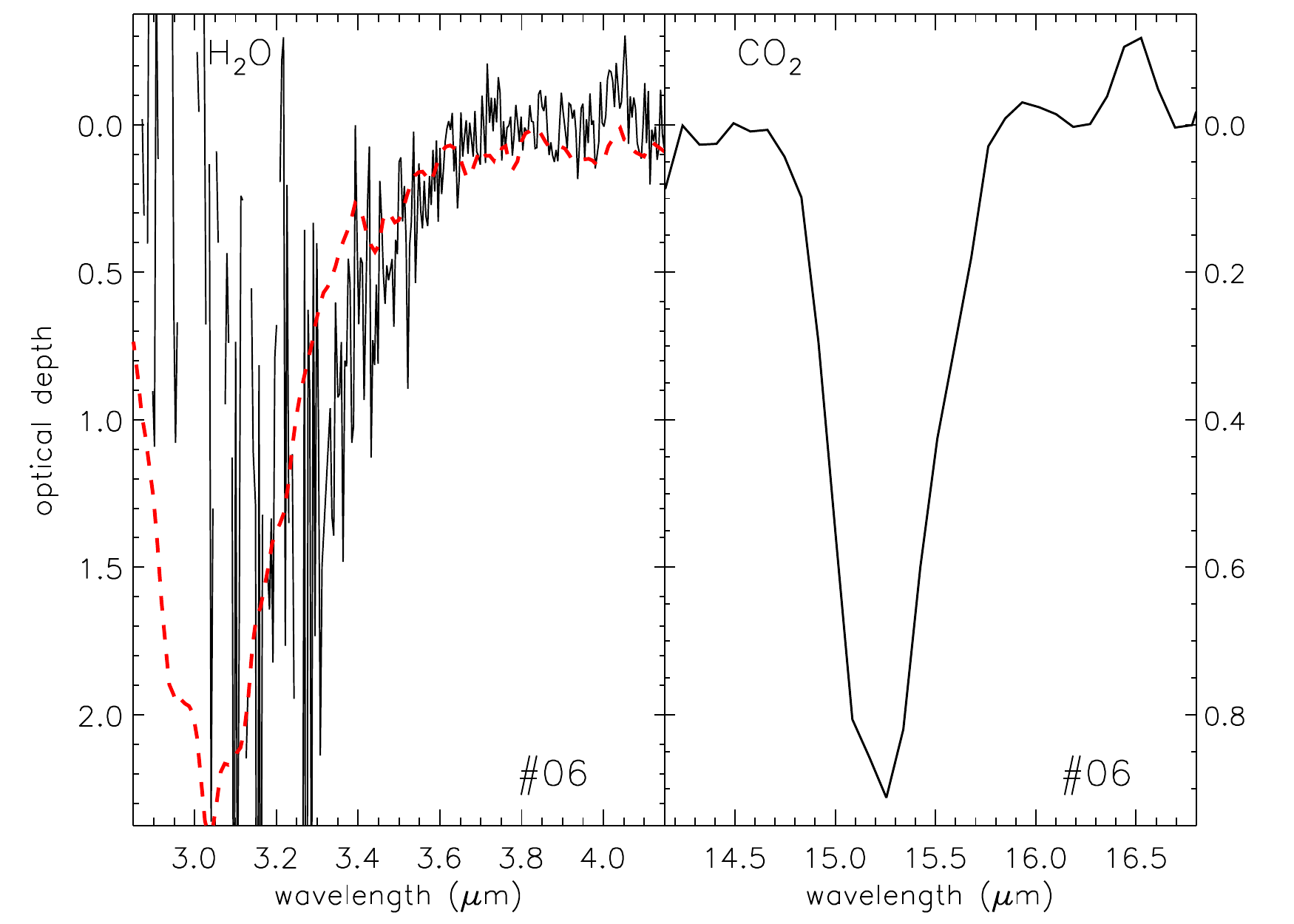}
\includegraphics[scale=0.42]{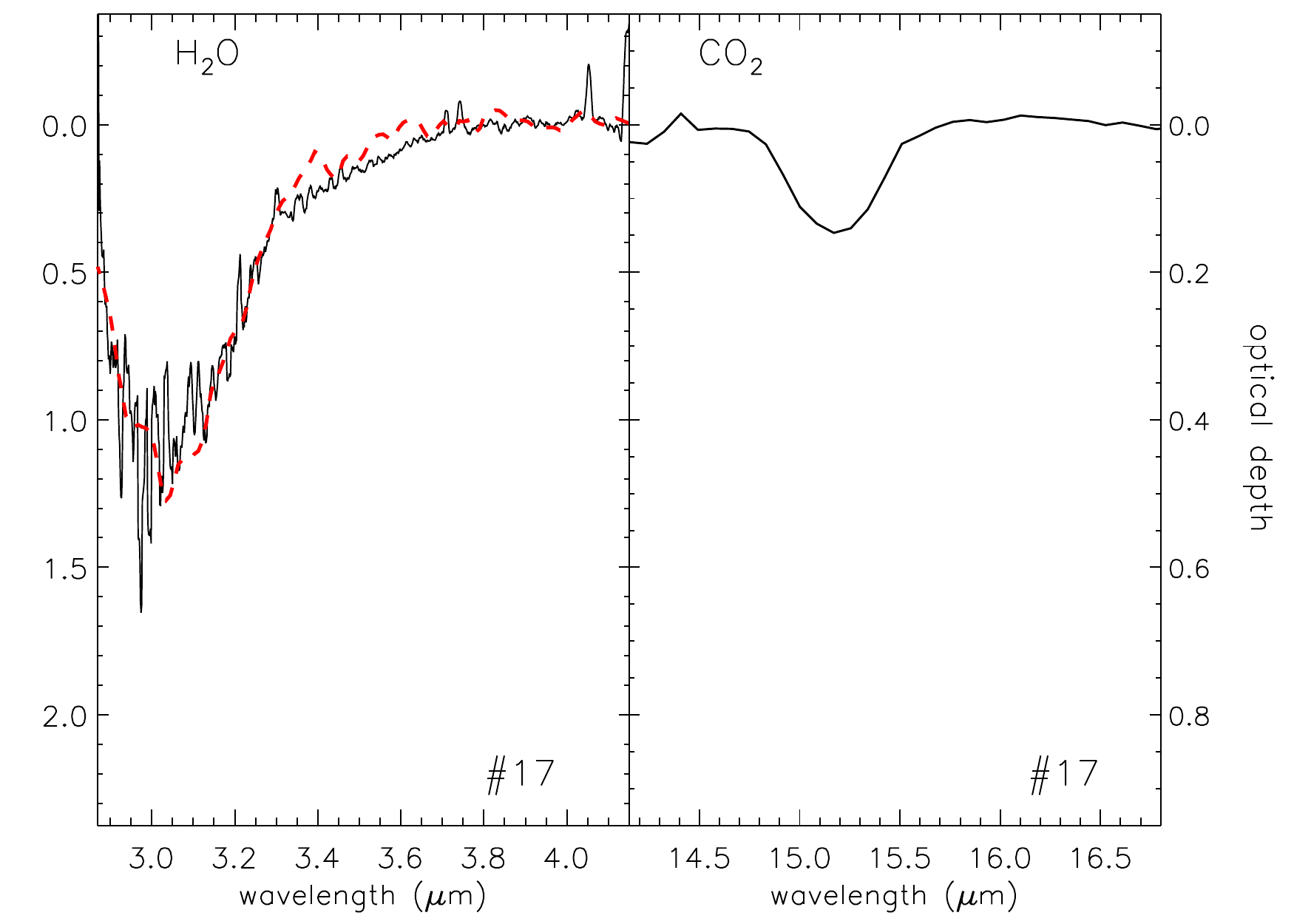}
\includegraphics[scale=0.42]{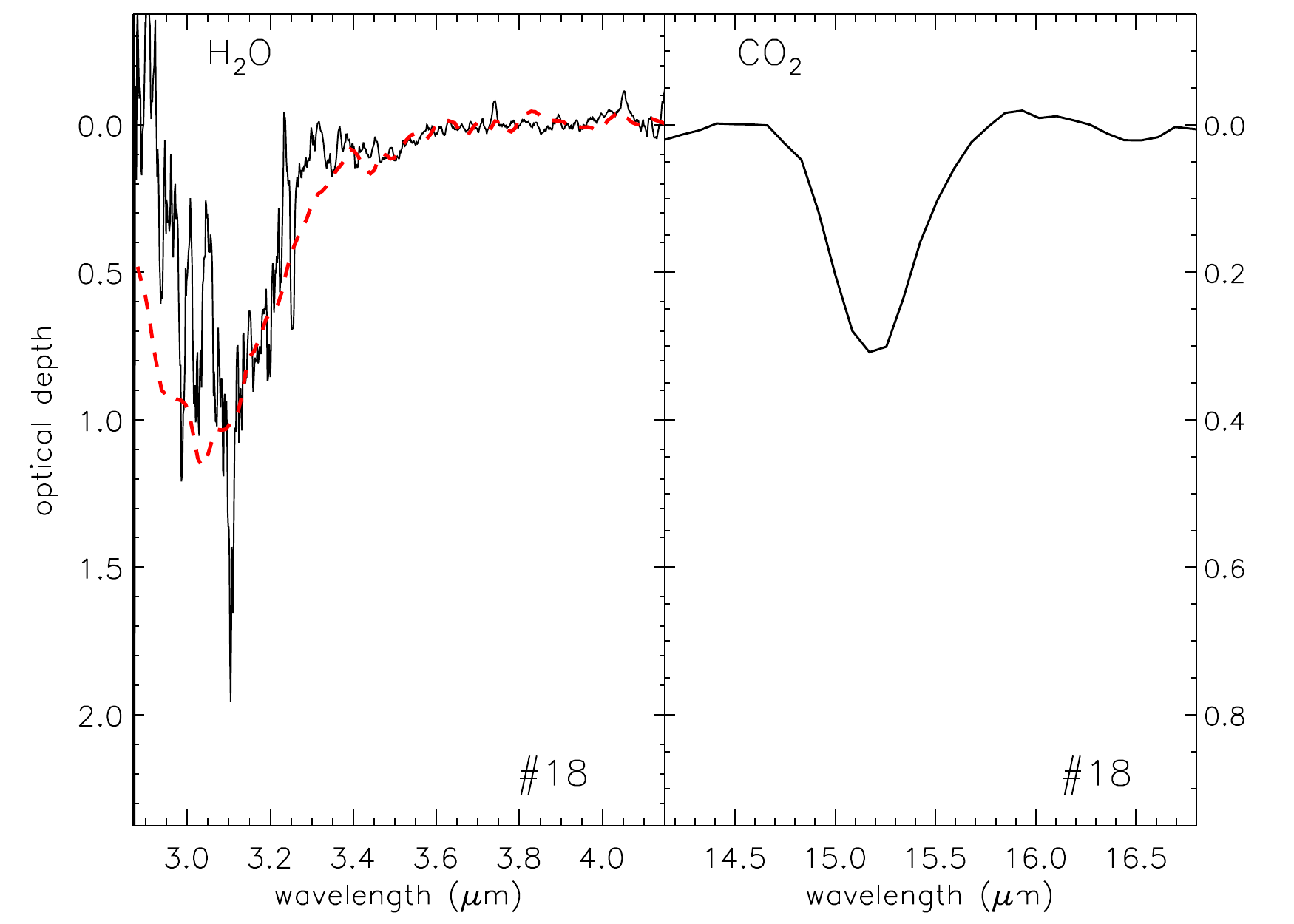}
\includegraphics[scale=0.42]{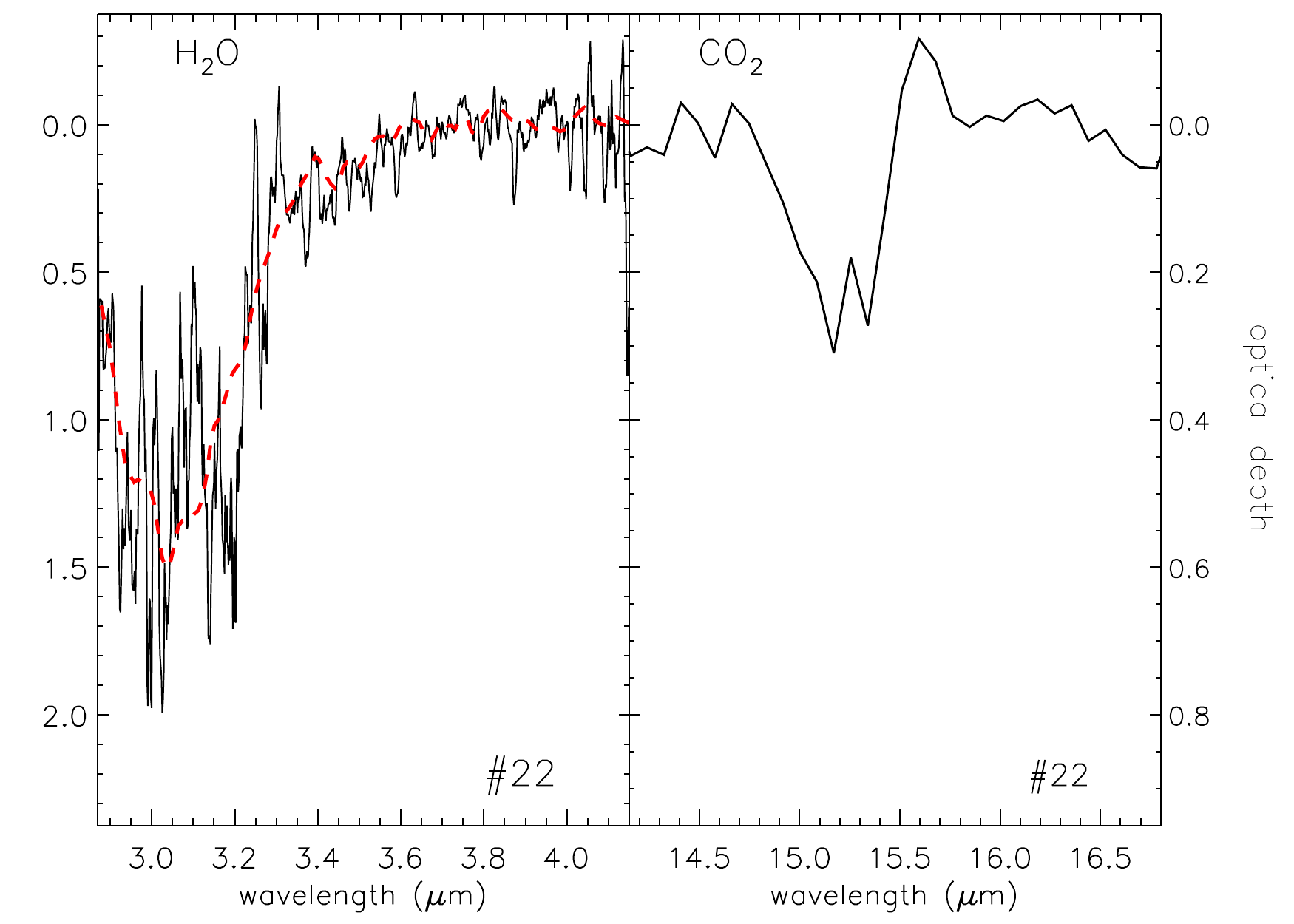}
\includegraphics[scale=0.42]{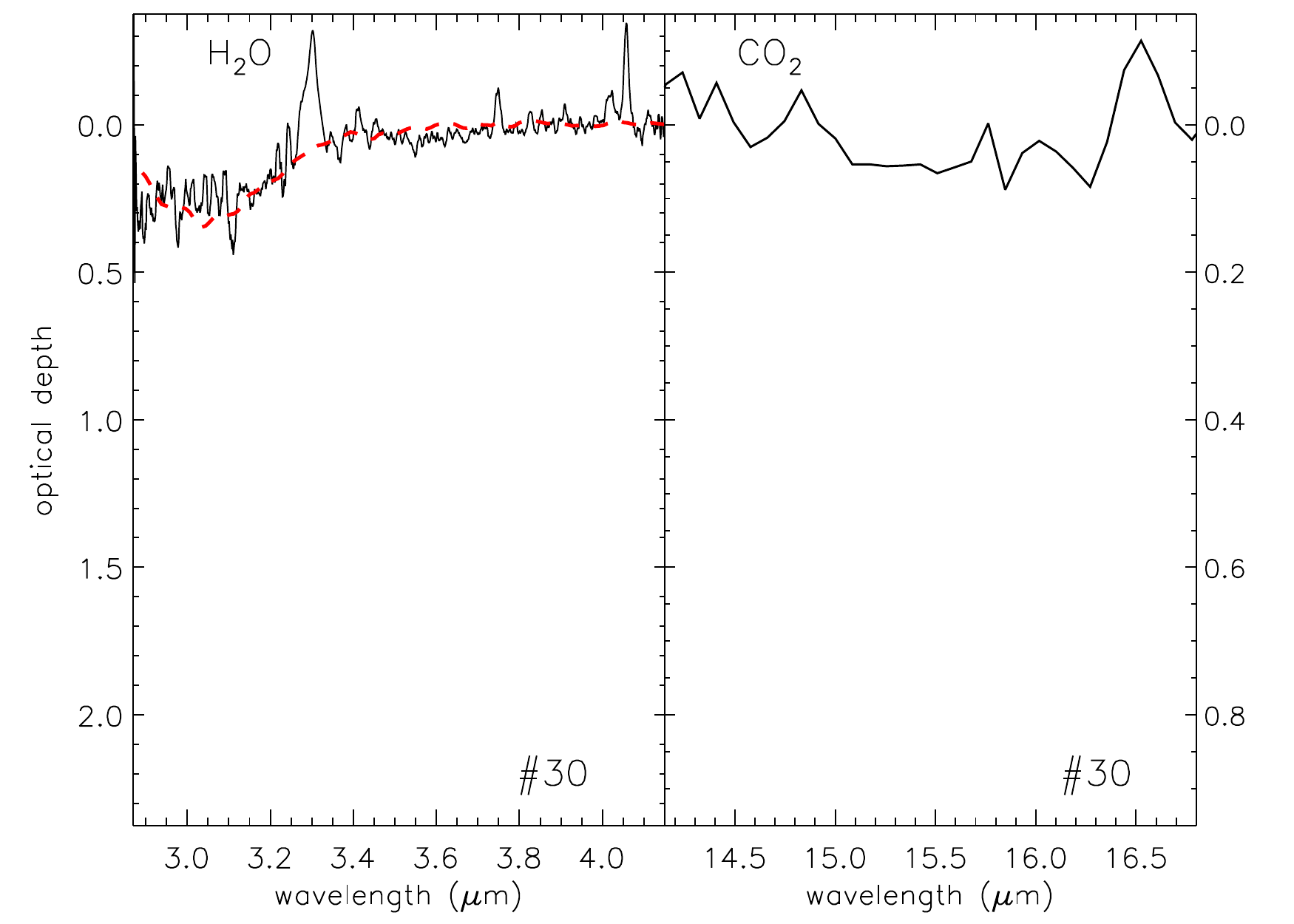}
\includegraphics[scale=0.42]{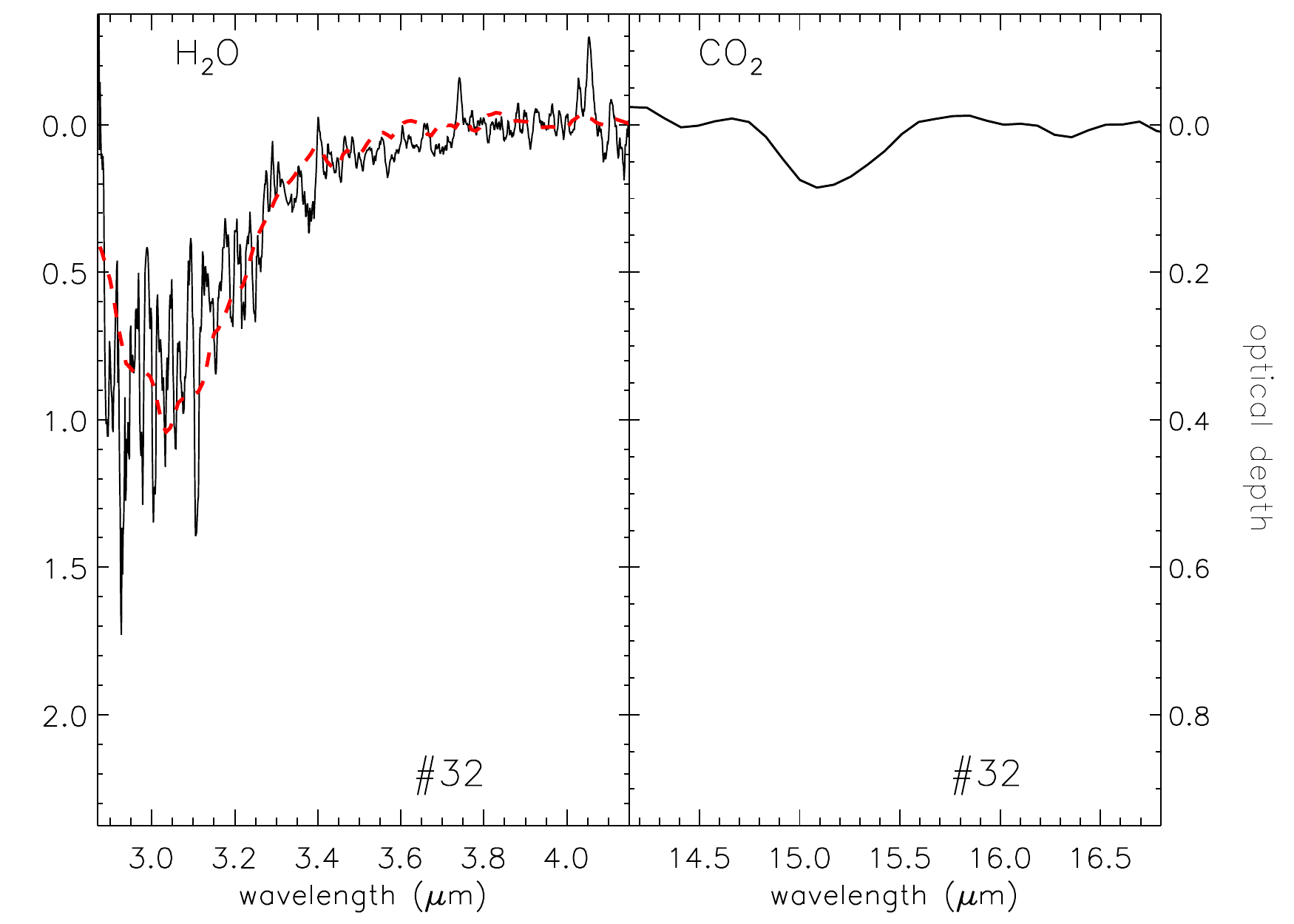}
\caption{SMC sources with \hho\ ice (left) and \coo\ ice (right) detections. 
Dashed (red) lines show a low-resolution AKARI spectrum of a LMC YSO 
\citep{shimonishi10} to help constrain the blue wing of the \hho\ ice feature.
Source \#30 shows \hho\ ice but not \coo\ ice. Many sources show hydrogen emission at 
3.74\,\mum\ (Pf$\gamma$) and 4.05\,\mum\ (Br$\alpha$); sources \#02 and 30 exhibit 
PAH emission at 3.3\,\mum, and \#06 and 30 at 16.45\,\mum. Source \#22 shows 
[Ne\,{\sc iii}] emission at 15.6\,\mum.\label{ices}}
\end{figure*}

\begin{figure*}
\begin{center}
\includegraphics[scale=0.42]{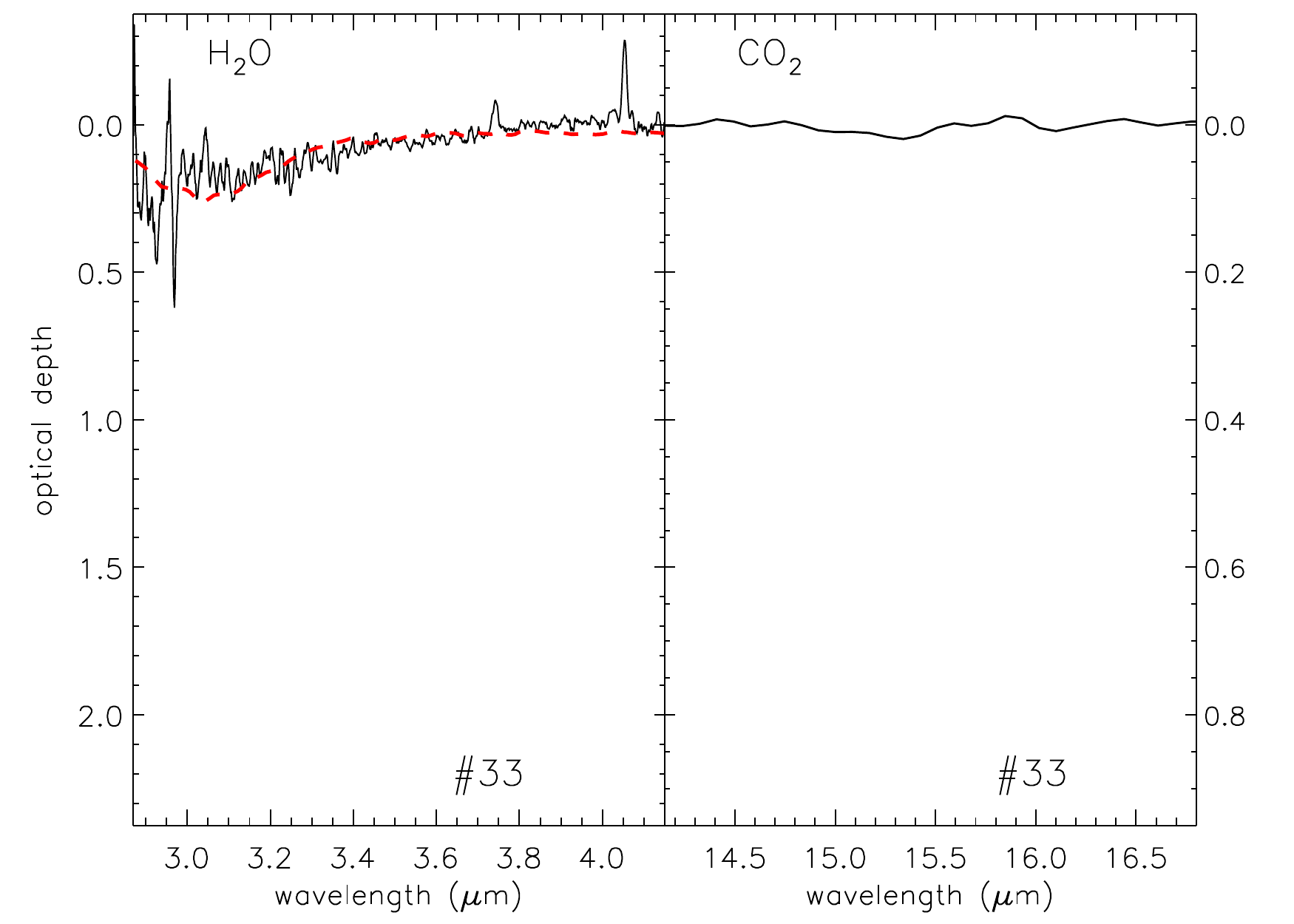}
\includegraphics[scale=0.42]{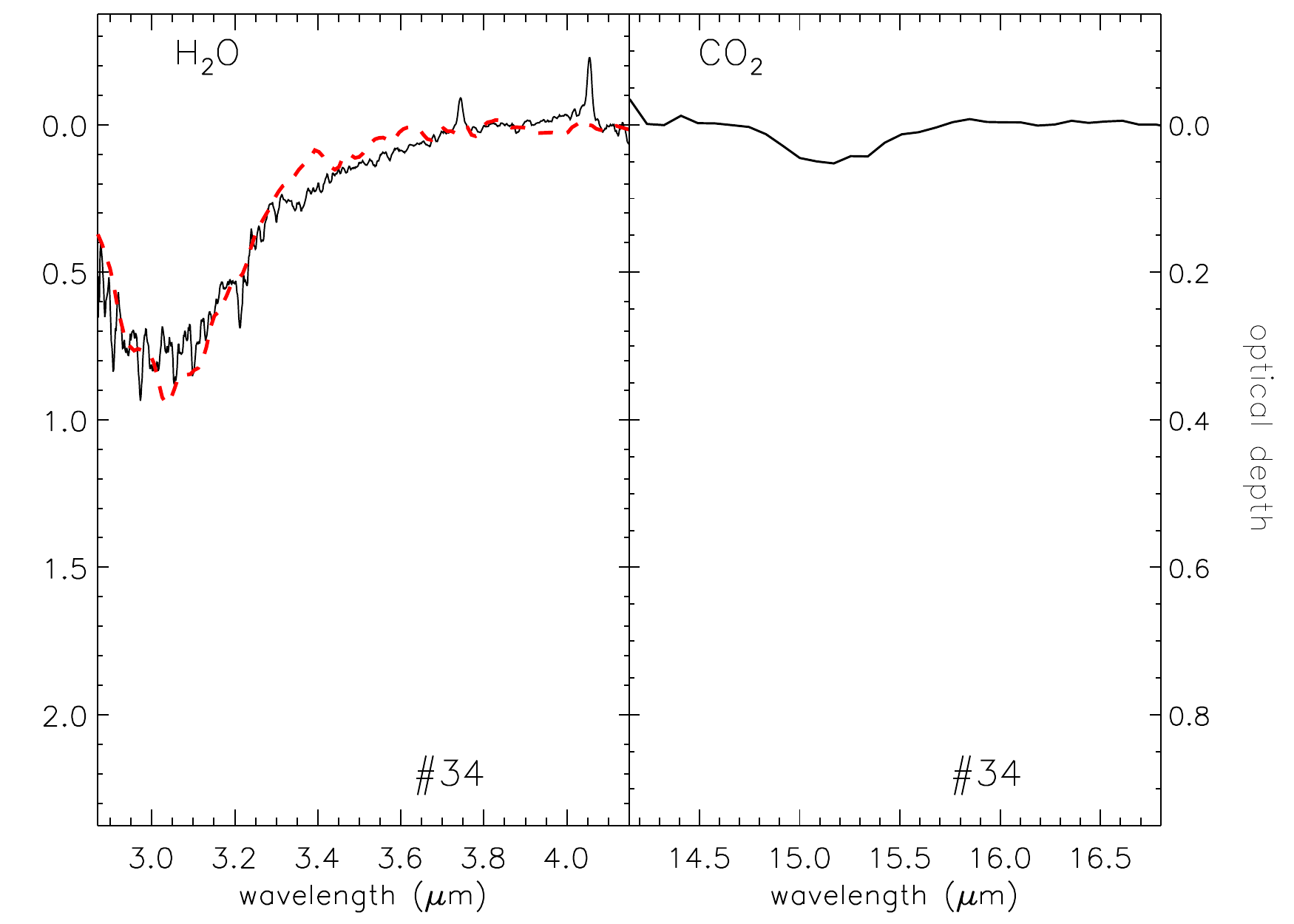}
\end{center}
\contcaption{The \hho\ ice detection for source \#33 is marginal. The 
\coo\ ice detection for source \#34 is weak but significant.}
\end{figure*}

\begin{figure*}
\includegraphics[scale=0.83]{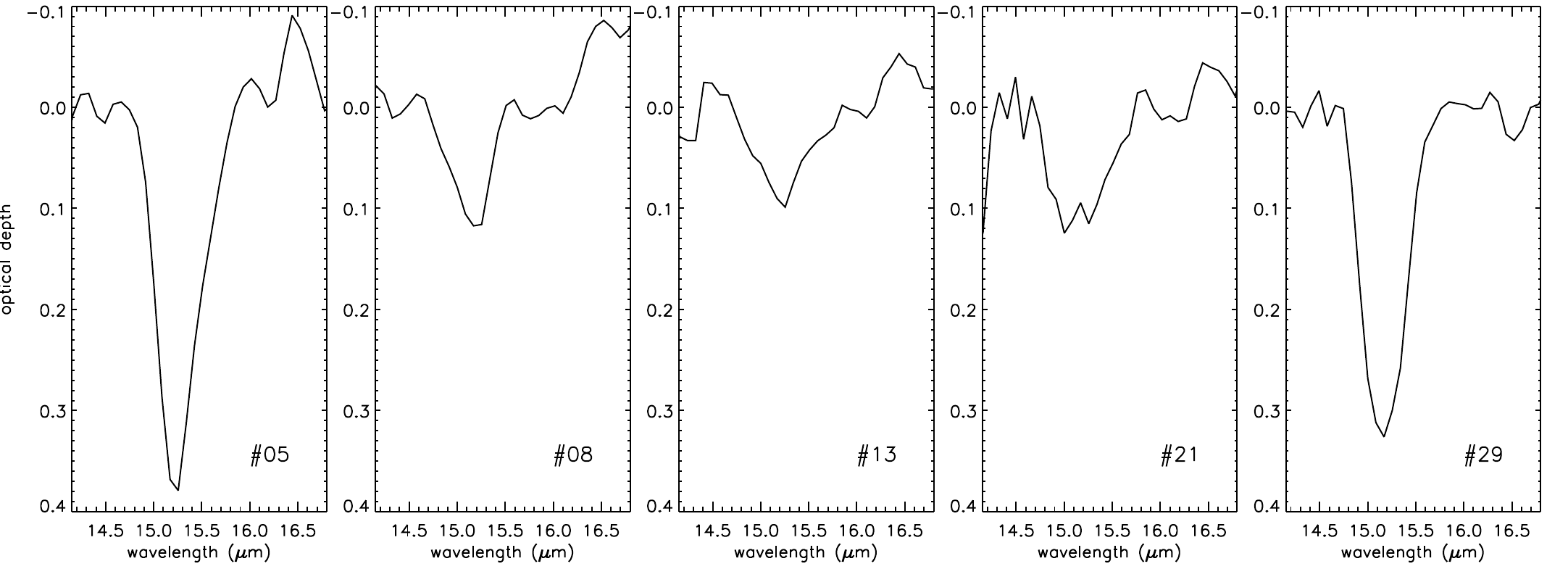}
\caption{Additional SMC sources with \coo\ ice detections. No 3$-$4\,\mum\ spectrum 
is available for these sources. The \coo\ ice profile of source \#21 is distorted 
by a couple of pixels with large errors, shifting the peak position. \label{ices2}}
\end{figure*}

\begin{figure*}
\includegraphics[scale=0.85]{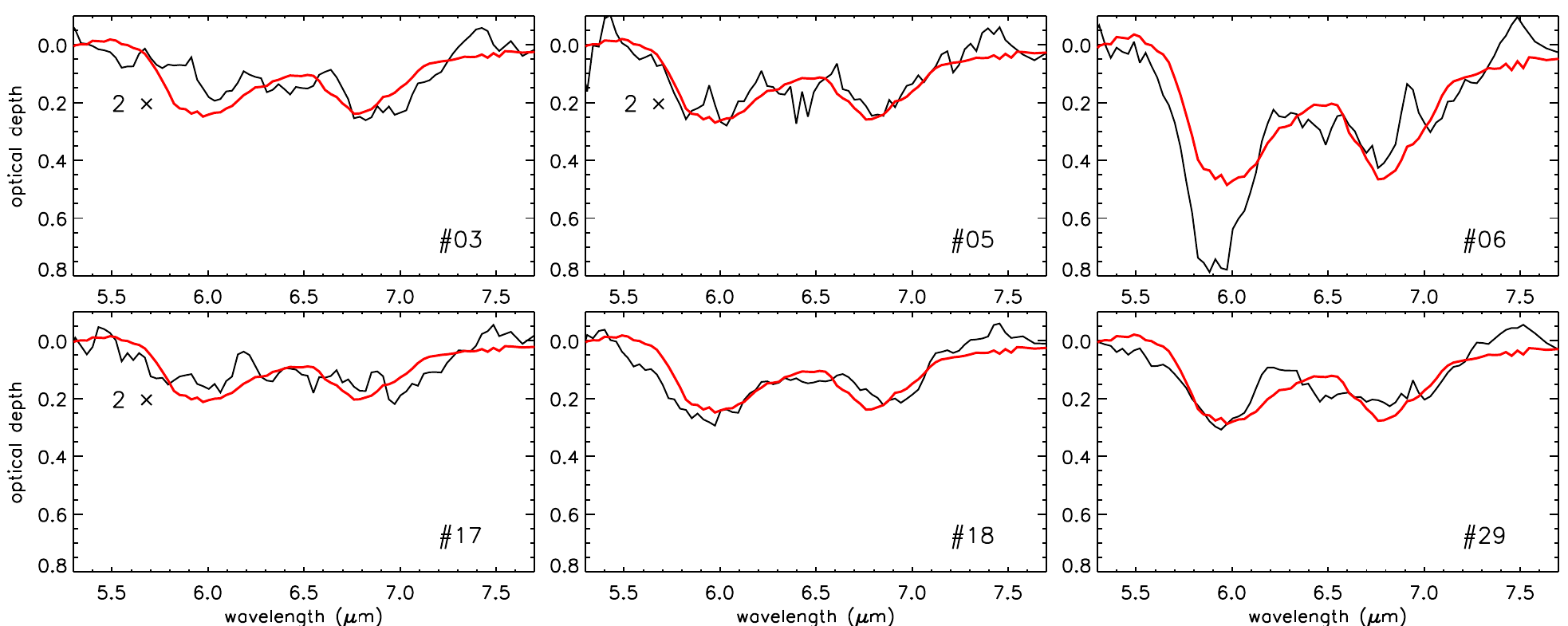}
\caption{SMC sources with ice detections in the 5$-$7\,\mum\ range. This band is
attributed to \hho\ ice at 6\,\mum\ with smaller contributions from more complex 
molecules (HCOOH, H$_2$CO and NH$_3$), and ammonium (NH$_4^+$) with a small 
contribution from methanol (CH$_3$OH) at 6.85\,\mum\ \citep{boogert08}. This ice 
complex can only be clearly identified when the 6.22-\mum\ PAH feature is weak or
absent. Some optical depth spectra are multiplied by a factor 2 (as indicated) 
to improve feature visibility. The red line is an optical depth spectrum of a
Galactic YSO for comparison \citep{zasowski09}.\label{ices3}}
\end{figure*}

In cold molecular clouds, layers of ice form on the surface of dust grains. 
\hho\ is by far the most abundant ice (typically $10^{-5}-10^{-4}$ with 
respect to H$_2$), followed by \coo\ and CO, with a combined abundance of 
10$-$30\% with respect to \hho\ ice \citep[e.g.,][]{vandishoeck04}. Understanding ice 
chemistry is crucial to understanding the gas-phase chemistry and to probing the 
physical conditions within molecular clouds. Surface chemistry and UV and 
cosmic-ray processing of the ice mantles are thought to play an important role 
not only in the formation of more complex ice species but also \hho, O$_2$ and 
gas-phase organic molecules. Prominent ice features in the IRS spectral range 
are found between 5 and 7\,\mum\ (attributed to a mixture of \hho, NH$_3$, CH$_3$OH, 
HCOOH and H$_2$CO ices, e.g., \citealt{boogert08}), and at 15.2\,\mum\ (attributed to 
\coo\ ice, e.g., \citealt{gerakines99,pontoppidan08}). The ice complex at 5$-$7\,\mum\ 
can be difficult to identify due to the superposition of numerous PAH emission features 
\citep{spoon02}. At shorter wavelengths, ice features of \hho\ and CO are found in 
the 3$-$5\,\mum\ range \citep[e.g.,][]{oliveira11}. Circumstellar ices have been 
detected in massive YSO environments in the Galaxy \citep[e.g.,][]
{gerakines99,gibb04} and the LMC 
\citep{vanloon05,vanloon08,oliveira09,oliveira11,seale11,shimonishi10}.

In the SMC, \hho\ and \coo\ ices (at 3\,\mum\ and 15.2\,\mum\ respectively) have been
detected in the environments of sources \#03, 17, 18, 32 and 34 
\citep{vanloon08,oliveira11}. Even though CO ice at 4.67\,\mum\ is detected towards 
YSOs in the Galaxy and the LMC \citep{gibb04,oliveira11}, no CO ice is detected
towards sources \#03, 17, 18, 33 and 34 (the only SMC sources observed at relevant
wavelengths). \citet{oliveira11} interpreted this as a metallicity effect. Sources
\#03, 06 and 17 exhibit a broad feature between 60$-$70\,\mum\ in the MIPS-SED 
spectrum, attributed to crystalline \hho\ ice \citep{vanloon10b}.

In this work we newly identify ice features towards a number of SMC YSO
candidates. We detect \hho\ and \coo\ ice absorption in the environments of sources 
\#02, 05, 06, 22 and 29 (Figs.\,\ref{ices}, \ref{ices2} and \ref{ices3}). For sources 
\#05 and 29 we do not have 3$-$4\,\mum\ spectra to probe the strongest 3-\mum\ \hho\ 
band, but we detect the 6-\mum\ \hho\ band --- also identified in sources \#03, 06, 17 
and 18 (Fig.\ref{ices3}). Additionally we detect only \coo\ ice towards
sources \#08, 13 and 21 (Fig.\,\ref{ices2}); no 3-\mum\ spectra are available and no ice 
is detected at 5$-$7\,\mum\ due to the presence of strong PAH emission 
(Table\,\ref{table3}).

The blue edge of the 3-\mum\ \hho\ ice feature is set by the Earth's atmospheric 
cut-off. Therefore we use the sources' $K_{\rm s}$-band magnitudes to help constrain the 
continuum bluewards of the feature;  the red continuum is constrained avoiding hydrogen
emission at 3.74\,\mum\ (Pf$\gamma$) and 4.05\,\mum\ (Br$\alpha$). For the \coo\ ice we
use a narrow wavelength interval surrounding the feature to constrain the continuum. 
Optical depth spectra are determined by subtracting a polynomial fitted to this 
local pseudo-continuum from each spectrum. 

The L-band spectrum of source \#06 is very red and the signal-to-noise ratio in the
blue wing of the 3-\mum\ \hho\ feature is poor. Therefore we do not measure 
the column density directly from the ISAAC spectrum. Instead we scaled a 
low-resolution AKARI spectrum of a LMC YSO \citep[SSTISAGE1C\,J051449.41$-$671221.5,][]
{shimonishi10} to match the red wing of the feature in \#06 (as shown in 
Fig.\,\ref{ices}), and measure the column density of the scaled spectrum. This 
measurement is indicated by $*$ in Table\,\ref{table2}. We validate this technique by 
checking that for the other sources the two methods provide consistent column density 
measurements. The \hho\ column density measurements for sources \#02 and 30 are 
uncertain, since the detection is weak and PAH emission at 3.3\,\mum\ fills-in the red 
wing of the feature. For source \#33 the \hho\ ice detection is very tentative 
(4-$\sigma$ detection); we do not consider it any further.

The 6-\mum\ \hho\ ice feature is very complex; the contributions from other ice
species can lead to overestimates of the \hho\ column density \citep[see discussion in]
[and references therein]{oliveira09}. Therefore we do not measure the \hho\ ice column
density directly from the 6-\mum\ feature. Since the peak optical depth $\tau_6$ 
correlates with the column density, we use the $\tau_6$ measurements and the column 
densities derived from the 3-\mum\ feature for sources \#03, 06, 17 and 18 to estimate 
the \hho\ column densities for sources \#05 and 29 (assuming a simple linear 
correlation). These measurements are indicated by $**$ in Table\,\ref{table2}.

\begin{table}
\begin{center}
\caption{\normalsize Column density measurements (in $10^{17}$ molecules 
cm$^{-2}$) for \hho\ and \coo\ ices in SMC YSO candidates. For source \#06 the 
\hho\ column density measurement (indicated by $*$) was performed by scaling a 
better quality spectrum (see text). For sources \#05 and 29 \hho\ column 
densities (indicated by $**$) are estimated from the 6-\mum\ feature, rather 
than the 3-\mum\ feature (see text).}\label{table2}
\begin{tabular}{c|c|c}
\hline
source \#        & N(\hho)&N(\coo)\\
\hline
02&      7.2$\pm$\,2.2&      1.6$\pm$0.2\\
03&\al{1}7.7$\pm$\,0.7&      1.7$\pm$0.1\\%improved wing
05&\al{**1}9\,$\pm$\,5\,&      7.3$\pm$0.9\\
06&\al{*4}6.3$\pm$\,7\,\,\,\,\,&\al{1}9.2$\pm$0.5\\
%08&                 &      2.3$\pm$0.2\\changed to improve fit due to 15.55 micron feature
08&                 &      2.3$\pm$0.2\\%stays the same
13&                 &      2.2$\pm$0.2\\
17&\al{2}1.6$\pm$\,0.8&      2.8$\pm$0.1\\
18&\al{2}2.3$\pm$\,1.2&      6.0$\pm$0.2\\
21&                 &      3.0$\pm$0.3\\
%22&\al{2}7.7$\pm$3.3&      8.6$\pm$1.0\\changed to improve fit due to 15.55 micron feature
22&\al{2}7.7$\pm$\,3.3&      6.0$\pm$1.0\\
29&\al{**3}2 $\pm$\,5\,\,&      6.7$\pm$0.1\\
30&      4.9$\pm$\,1.0&$\lsim$2         \\
32&\al{1}8.8$\pm$\,1.5&      1.5$\pm$0.2\\
33&      3.9$\pm$\,1.0& $\lsim$0.3      \\
34&\al{1}6.6$\pm$\,0.7&      1.0$\pm$0.2\\
\hline
\end{tabular}
\end{center}
\end{table}

The red wing of the \coo\ ice feature for sources \#08 and 22 is affected by weak
[Ne\,{\sc iii}] emission at 15.6\,\mum\ (see next subsection). Weak \hho\ ice is 
detected towards source \#30 but no \coo\ ice is identified (an upper limit is 
listed in Table\,\ref{table2}). The \coo\ ice detection towards source \#34 is weak
but significant, since clear \hho\ ice is detected. For source \#33 no \coo\ ice is
detected. 

Calculated ice column densities are listed in Table\,\ref{table2}. The adopted band 
strengths are $2.0 \times 10^{16}$ and $1.1 \times 10^{17}$\,cm$^2$ per molecule, 
respectively for \hho\ and \coo\ \citep{gerakines95}. The quoted uncertainties for 
measurements in Table\,\ref{table2} do not reflect uncertainties in continuum 
determination. We have re-calculated \coo\ column densities for sources \#03, 17, 18 
and 34 using optimally extracted IRS spectra (\citealt{oliveira11} used tapered-column 
extraction spectra; the optimal spectra provide better feature contrast due to improved
signal-to-noise ratio). Sources with stronger ice absorption tend to be the most 
embedded (as measured by the spectral index), as observed also in Galactic samples 
\citep[see for instance][]{forbrich10}. However, a steep spectral index does not imply
the presence of ice absorption. There is weak anti-correlation between ice column density
and the source luminosity.

\begin{figure}
\includegraphics[scale=0.85]{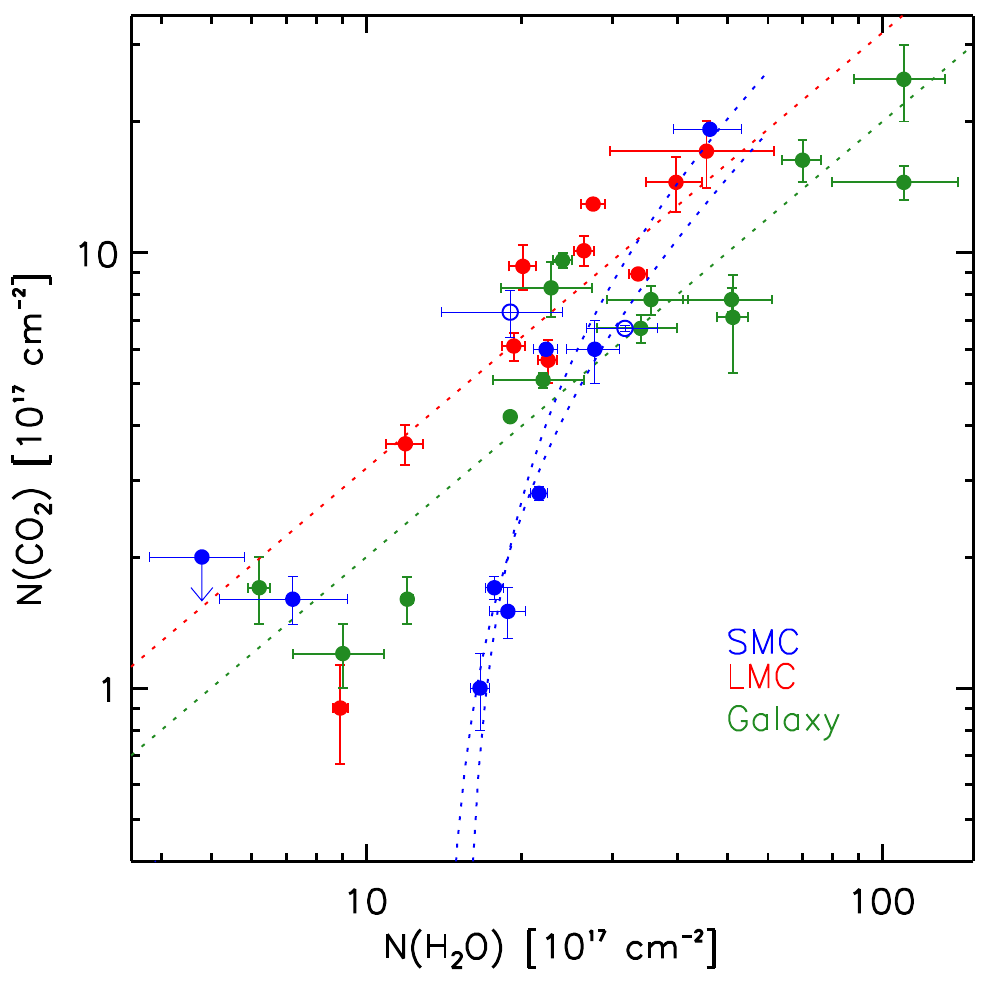}
\caption{Column densities for \hho\ and \coo\ ices. Red and green filled
circles represent respectively LMC and Galactic YSOs; the dashed red and green 
lines represent the estimated N(\coo)/N(\hho) ratios, respectively 0.32 and 0.2 
\citep[][and references therein]{oliveira11}. SMC measurements are represented 
by blue circles. Two fits to the SMC data, of the type $N({\rm CO}_2) = m \times
N({\rm H}_2{\rm O}) + N_0({\rm H}_2{\rm O})$ ($m$ is the slope and 
$N_0({\rm H}_2{\rm O})$ is the \hho\ column density threshold), are shown 
(blue dashed lines), including and excluding the 6-\mum\ \hho\ measurements 
(open blue circles). 
\label{ices_ratio}}
\end{figure}

\citet{oliveira11} compared the column densities for \hho\ and \coo\ ices for
Galactic and LMC samples. They found $N$(\coo)/$N$(\hho) ratios for the LMC and the 
Galaxy $\sim 0.32$ and $\sim 0.2$ respectively, consistent with previous 
determinations. For the SMC, rather than a constant ratio, Fig.\,\ref{ices_ratio} 
suggests that there is a \hho\ column density threshold for the detection of \coo\ 
ice, something not observed in either the LMC or the Galaxy. Source \#02 is the 
only SMC source with $N$(\hho)\,$< 1 \times 10^{18}$\,cm$^{-2}$ and a \coo\ ice 
detection, and as explained above $N$(\hho) may be underestimated for this source. We 
perform linear fits to the SMC data, of the form 
$N({\rm CO}_2) = m \times N({\rm H}_2{\rm O}) + N_0({\rm H}_2{\rm O})$ where $m$
is the slope and $N_0$(\hho) is the \hho\ ice column density threshold for
the detection of \coo\ ice. Excluding sources \#02 (see above) and \#30 (upper
limit), the fitted slope is $\sim 0.4-0.6$ and $N_0$(\hho)\,$\sim 1.4-1.55 
\times 10^{18}$\,cm$^{-2}$ (depending on whether the 6-\mum\ \hho\ measurements are
included, open blue circles in Fig.\,\ref{ices_ratio}).

In the LMC, \coo\ ice column densities are enhanced with respect to \hho\ ice, while 
the relative CO-to-\coo\ abundances are unchanged. \coo\ production could be increased 
due to the stronger UV field and/or higher dust temperatures in the LMC. However such 
harsher conditions would also destroy CO ice (the most volatile ice species), something 
that is not observed. Instead \citet{oliveira11} suggest that \hho\ ice is depleted
due to the combined effects of a lower gas-to-dust ratio and stronger UV radiation field.
This would push the onset of water ice freeze-out deeper into the YSO envelope therefore
reducing the observed column density (see their Figure 3). Forming deeper in the YSO 
envelope, \coo\ ice would remain unaffected.

In the Galaxy, the $A_{\rm V}$-thresholds for the detection of \hho\ and \coo\ ices are 
statistically indistinguishable \citep[see][and references therein]{oliveira11}, 
suggesting that the two ices species are co-spatial in YSO envelopes. However, in the 
SMC the present observations suggest a column density threshold $N_0$(\hho)\,$\sim 1.5 
\times 10^{18}$\,cm$^{-2}$ for the detection of \coo\ ice; even the LMC measurements are
consistent with a small threshold $N_0$(\hho)\,$\sim 3 \times 10^{17}$\,cm$^{-2}$. This 
suggests that in metal-poor environments part of the envelope may have \hho\ ice but not 
\coo\ ice, supporting the scenario proposed by \citet{oliveira11}. The role of reduced 
shielding in regulating the ice chemistry mentioned above is also supported by the 
non-detection (with high confidence) of CO ice absorption in the spectra of five SMC 
sources (the only sources observed at 4.67\,\mum).

\subsection{PAH and fine structure emission}

\begin{figure*}
\includegraphics[scale=0.75]{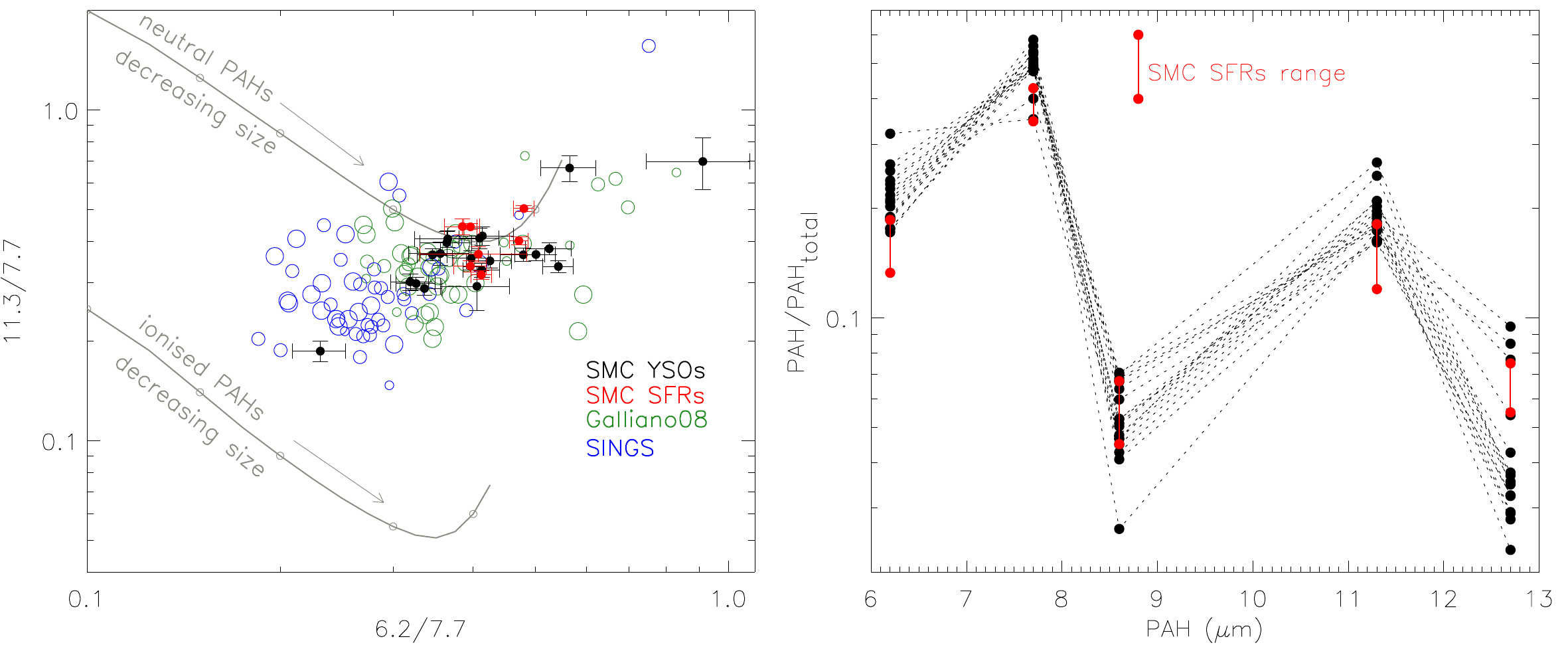}
\caption{PAH properties towards SMC sources. Left: 11.3/7.7 intensity ratio against
6.2/7.7 ratio. The different samples are: SMC YSO candidates (this work, filled black
circles), SMC SFRs \citep[][filled red circles]{sandstrom12}, a sample of galaxies
and individual SFRs \citep[][green circles]{galliano08} and finally the SINGS
galaxies \citep[][blue circles]{smith07}. Symbol sizes represent three metallicity 
ranges: $12+\log(\rm{O/H})=7.7-8.2$, $8.2-8.7$, and $8.7-9.2$\,dex. Solid grey lines 
indicate model trends for neutral and ionised PAHs of varying grain sizes adapted from 
\citet{draine01}. Right: PAH intensity carried in each band normalised to total PAH 
intensity. \label{pahs}}
\end{figure*}

Numerous emission features originate from the C--C and C--H stretching and 
bending modes of PAH molecules, excited by UV radiation. The best studied 
emission bands are at 6.2, 7.7, 8.6, 11.3 and 12.7\,\mum, ubiquitous in the 
spectra of compact H\,{\sc ii} regions and planetary nebulae. Many sources also exhibit
a PAH emission complex at 17\,\mum. Often these emission bands are accompanied by 
fine-structure emission lines such as [S\,{\sc iv}] at 10.5\,\mum, [Ne\,{\sc ii}] at 
12.8\,\mum, [Ne\,{\sc iii}] at 15.6\,\mum, [Si\,{\sc ii}] at 34.8\,\mum, and 
[S\,{\sc iii}] at 18.7 and 33.5\,\mum. These lines (unresolved in the low-resolution IRS 
modes) originate from ionised gas. The presence of both PAH and fine-structure line 
emission clearly suggests the presence of an ionising source of UV radiation.

In the sample of 34 objects, all but two exhibit PAH emission. Even when other 
PAH bands are weak, the relatively isolated 11.3-\mum\ band can be easily 
identified. Thus we use the relative peak strength of this feature to assess how
dominant PAH emission is in shaping the IRS spectrum. We fit a local 
continuum using spectral points to the left and right of the feature, and 
measure the peak strength of the feature with respect to the underlying 
continuum, $F_{11.3}/F_{c}$. Sources with $F_{11.3}/F_{c} \le 1.05$ essentially 
show no PAH emission; this is the case of sources \#18 and 19. For sources with 
$1.05<F_{11.3}/F_{c}\le 1.30$ only the 11.3-\mum\ band can be easily identified,
other bands are not clearly visible (typical examples are sources \#03 and 20). 
Sources with $F_{11.3}/F_{c}>1.3$ clearly exhibit the complete zoo of PAH 
features (e.g., \#08), and for some sources the spectra are completely dominated
by PAH emission ($F_{11.3}/F_{c}>3$, e.g., \#07). The different PAH groups are 
listed in Table\,\ref{table3} as \cross, \wtick, \tick, \dtick\ from absent to 
very strong, and are used in the source classification in Section 5. 

Recently, \citet{sandstrom12} analysed the properties of PAH emission observed
towards a sample of six diverse star forming regions (SFRs) in the SMC, namely N76
and N66 in the northeast SMC bar, three regions in the southwest bar including 
N22, and N83/84 in the SMC Wing. Our sample comprises sources in the {\it same six 
regions}, and further extends the spatial coverage by sampling for instance the region 
in the SMC body that connects N\,66 to the southwest bar (see Fig.\,\ref{image}). 
Fig.\,\ref{pahs} summarises PAH properties in the SMC. On the left we show intensity 
ratios for the YSO candidates (black filled circles) and the averages for the 
aforementioned SMC SFRs (red filled circles). On the right we show the intensity of each
main band normalised to total PAH intensity. In our sample the 11.3-\mum\ band is 
strong compared to the 7.7- and 6.2-\mum\ bands, and the 8.6-\mum\ band is very weak 
compared to total PAH strength, consistent with the \citet{sandstrom12} results. Our 
analysis does not reveal changes to PAH emission properties that would be evidence for
the YSO's irradiation of its environment. Thus the PAH emission observed towards at 
least some YSO candidates may have an important environmental contribution. 
Nevertheless, for sources with strong emission, total PAH intensity correlates with
the luminosity of the source.

Model PAH band ratios \citep[adapted from][as described by \citealt{sandstrom12}]
{draine01} show the regions of the ratio diagram occupied by neutral and ionised 
PAHs, indicating also the effect of decreasing grain size. The strength of the 
radiation field also has a slight effect on PAH ratios. As proposed by
\citet{sandstrom12}, the comparison with these models suggests that SMC PAHs are 
predominantly small and neutral.

By comparing PAH properties in the SMC and the high-metallicity SINGS galaxies 
\citep{smith07}, \citet{sandstrom12} explain the observed differences as a metallicity
effect: they speculate that, in the metal-poor SMC, PAHs {\it form} preferably in
smaller grains and mostly neutral PAHs {\it survive} in the ISM. To further
investigate whether the observed PAH ratios are directly related to metallicity we 
extend the comparison to also include the measurements compiled by \citet{galliano08}
for a diverse sample of galaxies and individual SFRs of a range of metallicities. These 
data are show in Fig.\,\ref{pahs} (left): the SINGS and Galliano samples are represented 
by open blue and green circles respectively, and the symbol size reflects each region's 
gas-phase oxygen abundances, in the intervals $12+\log(\rm{O/H})=7.7-8.2$, $8.2-8.7$, 
and $8.7-9.2$. We adopt $12+\log(\rm{O/H})=8.0$ for the SMC. Both the SINGS and Galliano 
samples cover a similar metallicity range, with the majority of objects in each sample 
above 8.3 and 8.8\,dex respectively. However, these two samples occupy different regions 
in the PAH ratio diagram (with some overlap) despite their similar metallicities. The 
SMC measurements sit in a similar region to the Galliano sample, despite its lower 
metallicity. In summary our analysis supports the suggestion that PAHs throughout the 
whole SMC are indeed predominantly small and neutral; however the sample comparisons 
we describe do not support a simple metallicity explanation; other global environmental
parameters should also play a role \citep[e.g.,][]{haynes10}.

Fine-structure emission from atomic ions can be intrinsic to YSO sources but it can 
also be a result of contamination from the diffuse gas or nearby H\,{\sc ii} 
regions. Furthermore, at this resolution, it is difficult to separate from PAH 
emission, particularly for the 12.8-\mum\ [Ne\,{\sc ii}] line. While the IRS spectra of
8 objects exhibit the 18.7- and 33.5-\mum\ [S\,{\sc iii}] emission lines 
(Table\,\ref{table3}), only sources \#26 and 31 show strong 8.99-\mum\ [Ar\,{\sc iii}],
10.5-\mum\ [S\,{\sc iv}] and 15.6-\mum\ [Ne\,{\sc iii}] emission lines, also present 
but weaker in the spectra of \#08 and 22. Spectral contamination from the environment 
is likely given the location of many targets in the vicinity of known H\,{\sc ii} 
regions (Fig.\,\ref{image} and further discussion in Section\,5.1).

%==============================================================================

\subsection{H$_2$ emission}

H$_2$ emission is expected to be ubiquitous in YSO environments. However, only
when the molecular gas is heated to a few hundred K is H$_2$ emission observable.
Both UV radiation (released by the accretion process, the emerging star itself, 
or from nearby environment) and shocks (created as outflows interact with the 
quiescent molecular cloud) can produce warm H$_2$ gas \citep[for a review see][]
{habart05}. Extinction-corrected excitation diagrams can be used to diagnose the 
gas conditions and constrain the excitation mechanism of massive YSOs 
\citep*[e.g.,][]{vandenancker00}. 

Several emission lines due to molecular hydrogen are included in the IRS range.
In particular pure-rotational $0-0$ transitions occur at 5.51 S(7), 6.10 S(6),
6.91 S(5), 8.03 S(4), 9.66 S(3), 12.28 S(2), 17.03 S(1) and 28.22\,\mum\ S(0). The
S(6) and S(4) transitions are difficult to disentangle from the PAH emission, the S(5) 
line can be contaminated by [Ar\,{\sc iii}] at 6.99\,\mum, and the S/N ratio in the
region of the S(7) line is sometimes low. Nevertheless we have identified H$_2$ emission 
in the spectra of the majority of objects in our sample: for 24 sources three or more 
unblended emission lines are measured (\tick\ in Table\,\ref{table3}), while for eight 
sources only two weak lines are identified (\tick? in Table\,\ref{table3}). There are 
two sources with no detectable H$_2$ emission lines (\cross\ in Table\,\ref{table3}).

\begin{figure*}
\includegraphics[scale=0.88]{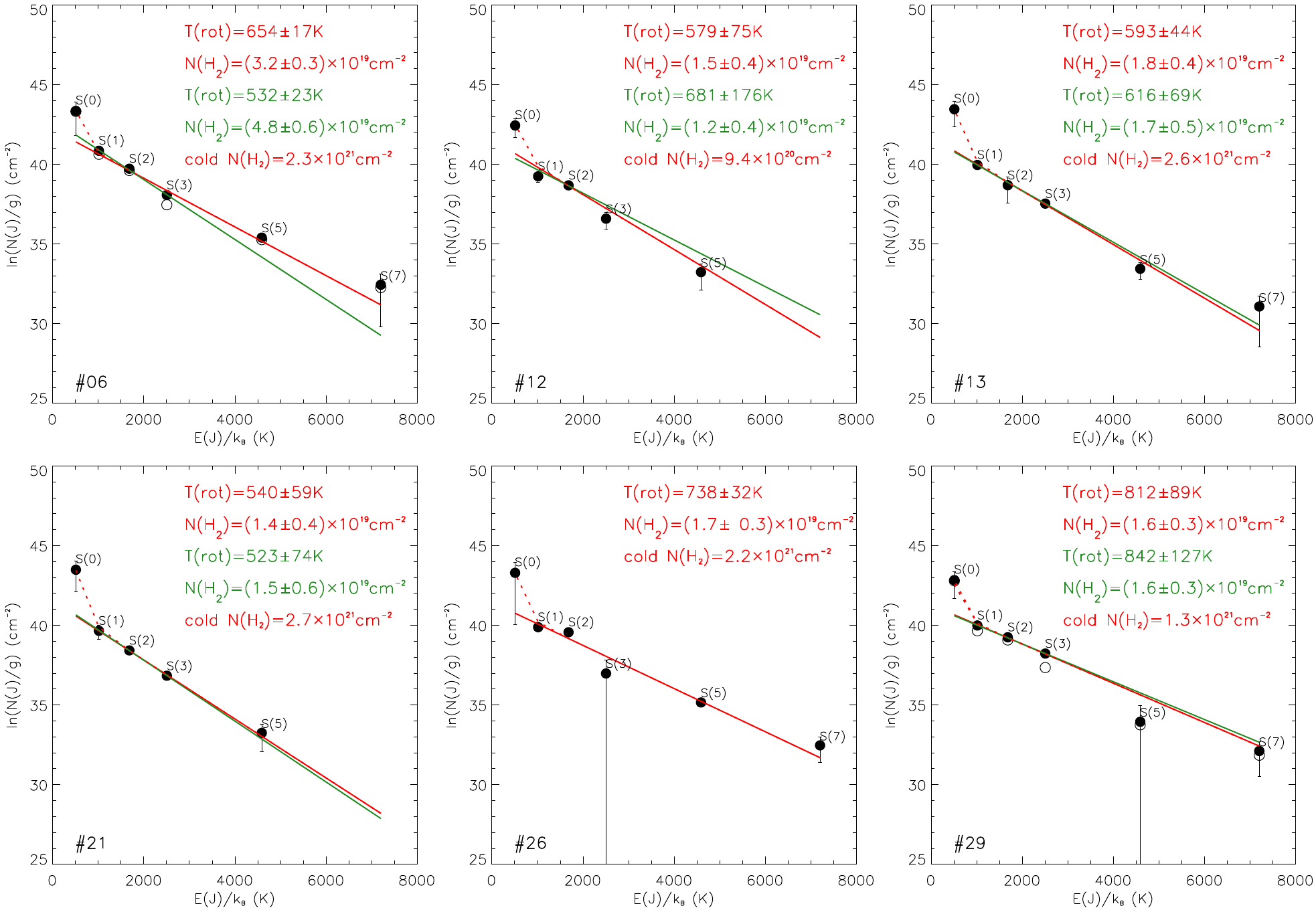}
\caption{H$_2$ excitation diagrams for selected SMC YSO candidates. For sources \#06 
and 29, the line intensities are corrected for extinction (see text for details). The 
results of two fits are displayed when available: a fit to all measured lines 
(red), or a fit to only S(1) to S(3) (green). For each fit, the derived 
rotational temperature $T_{\rm rot}$ and column density $N({\rm H_2})$ are 
indicated. The dashed red line shows the contribution of a cold H$_2$
component, with fixed temperature $T_{\rm rot}=100$\,K. \label{h2_rot}}
\end{figure*}

Since the lines discussed here are optically thin \citep*{parmar91}, the 
measured line intensities can be used to derive the total number of molecules 
and the rotational temperature. For excited H$_2$ gas in local thermodynamic 
equilibrium (LTE) at a single temperature, the density $N(\rm{H_2})$ and 
temperature $T_{\rm rot}$ are described by the Boltzmann distribution 
(in logarithmic form):
\begin{equation}\ln\left(\frac{N(J)}{g_J}\right)= - \frac{E(J)}{k_B\,T_{\rm rot}} + 
\ln\left(N(H_2)\frac{h\,c\,B}{2\,k_B\,T_{\rm rot}}\right)\end{equation}
where $g_J$, $E(J)$ and $N(J)$ are respectively the statistical weight, energy 
and column density of the upper $J$ level \footnotemark, $B=59.33$\,cm$^{-1}$ is
the H$_2$ molecular constant. The statistical weight is given by 
$g_J=g_S(2J+1)$, where the nuclear spin weight $g_S=3$ for $ortho$-states (odd 
$J$) and $g_S=1$ for $para$-states (even $J$) --- we assume an equilibrium 
{\it ortho}-{\it to}-{\it para} ratio $opr=3$, appropriate for gas with $T>300$\,K
\citep[e.g.,][]{sternberg99}. The column density $N(J)$ is derived from the 
measured extinction-corrected line intensities $I(J)$ as follows:
\begin{equation}N(J) = 4\,\pi\,\frac{\lambda}{h\,c} \frac{I(J)}{A_J \Omega}
\end{equation}
where $A_J$ are the transition probabilities and $\Omega$ is the beamsize in
steradian. More details on the method and values of relevant constants and 
energies can be found in the literature \citep[e.g.,][]
{parmar91, bernard-salas05,barsony10}.
\footnotetext{In the notation 0\,-\,0\,S($J'$), $J'$ indicates the lower 
rotational state.}

To estimate the beamsize $\Omega$ we multiply the FWHM of the PSF by the slit width. 
Slit widths are $3\rlap{.}^{\prime\prime}6$ and $10\rlap{.}^{\prime\prime}6$ 
respectively for SL and LL. The FWHM is measured from the two-dimensional PSF as a 
function of wavelength, ranging from $1\rlap{.}^{\prime\prime}9$ at 5.51\,\mum\ to 
7$^{\prime\prime}$ at 28.2\,\mum. The PSFs are those used for the optimal extraction 
of the spectra \citep[Section 3.1 and][]{lebouteiller10}. We choose to have a beamsize
that increases with wavelength since some objects are marginally resolved, and in 
such complex environments at these large distances, the IRS spectra do sample 
different sized emission regions.

To estimate the extinction, we use {\sc pahfit} \citep{smith07} since it 
disentangles the contributions from the dust and PAH emission; the code returns 
fully mixed relative extinctions that can be used to correct the observed line 
intensities. The extinction corrections are largest for the S(3) 9.66-\mum\ line, 
but still usually small for the targets in the present sample. 

Figure\,\ref{h2_rot} shows examples of H$_2$ excitation diagrams for several 
SMC sources, open and full circles represent respectively uncorrected and 
extinction-corrected measurements. For each object with more than two H$_2$ 
emission line measurements a straight line can be fitted to the data points, 
allowing the determination of the temperature and column density, using Eq.\,(1).
It can be seen from the examples in Fig.\,\ref{h2_rot} that the weighted column 
density for S(0) is always high compared to the measurements for other 
lines. This is to be expected since S(0) has the lowest energy, therefore it
probes the reservoir of H$_2$ that is too cool to excite any other transitions.
We discuss S(0) emission later. 

Whenever possible we perform two fits; the first makes use of all good 
measurements available except S(0), the second makes use only of S(1), S(2) and
S(3) (red and green lines in Fig.\,\ref{h2_rot} respectively). The reason to 
perform the second fit to the three transitions closest in energy level is to 
investigate possible deviations from the simple single temperature scenario, as 
it could be expected if a smaller column of hotter H$_2$ is also present 
\citep{vandenancker00}. In general, a single-temperature Boltzmann 
distribution provides good fits to the weighted column densities within the 
measurement uncertainties. Firstly this validates the adopted 
wavelength-dependent beamsizes, rather than a single beamsize. Furthermore the
column density of the S(2) line, the only $para$-state we are able to measure, does 
not deviate from those of the $ortho$-states, suggesting that the adopted LTE value 
of $opr=3$ is adequate. The two fits described above (see examples in 
Fig.\,\ref{h2_rot}) result in temperatures that agree within 1.5$\sigma$ for all
objects except for source \#06 (5-$\sigma$ difference). This again shows that a single 
temperature fit is appropriate to describe the molecular gas responsible for 
most of the emission. The exception is source \#06, for which the excitation diagram 
and fitted temperatures suggest the presence of a hotter H$_2$ contribution 
(top left panel in Fig.\,\ref{h2_rot}). 

For the fits to 4 or 5 data points the median temperature is $T_{\rm rot}=594\pm99$\,K 
(range $439-812$\,K), while for the 3-measurement fits the median temperature is 
$T_{\rm rot}=562\pm100$\,K (range $439-843$\,K). The median densities are respectively
$N({\rm H_2}) = 1.8\pm0.9 \times 10^{19}$\,cm$^{-2}$ (range 
$0.6-3.8\times 10^{19}$\,cm$^{-2}$) and 
$N({\rm H_2}) = 1.6\pm1.5 \times 10^{19}$\,cm$^{-2}$ (range $0.7-4.8\times 
10^{19}$\,cm$^{-2}$). When compared to a sample of Galactic YSOs
\citep{vandenancker99} covering the same luminosity range, the rotational temperatures 
are similar. However, the column densities for the SMC YSO candidates tend to be 
smaller by a factor three, typically $\sim 1.7 \times 10^{19}$\,cm$^{-2}$ rather than 
$\sim 5 \times 10^{19}$\,cm$^{-2}$.

Neither H$_2$ temperature or column density seem correlated with source luminosity for 
the SMC and Galactic samples. For sources with strong PAH emission, H$_2$ intensity 
(using S(3) as a proxy) is largest for sources with the largest total PAH intensity. 
However, there is no distinction between the derived H$_2$ properties for sources with 
ice and silicate absorption, and PAH emission. Since, as discussed in the next section, 
these features trace the YSO evolution, this implies that H$_2$ emission does not seem to
correlate with evolutionary stage, neither in SMC nor Galactic samples 
\citep[see also][]{forbrich10}.

As already mentioned, the weighted column density for S(0) is too large when
compared to the other transitions, suggesting a reservoir of quiescent cold gas. 
Since this cold component is poorly constrained, we opt to fix its temperature at 
100\,K \citep[e.g.,][]{lahuis10} and simply adjust the column density to match the 
observed $N(0)/g(0)$. We do not constrain $opr$, expected to be $\sim1.6$ for this 
temperature \citep{sternberg99}. The contribution of this cold component, added to the 
main warm component, is shown in Fig.\,\ref{h2_rot} (dashed red line). The fitted column 
densities are in the range $N({\rm H_2}) = 0.4-3.8\times 10^{21}$\,cm$^{-2}$ (median 
density $2.3\times 10^{21}$\,cm$^{-2}$). Therefore the contribution of the cold molecular
gas reservoir is substantial, even though its signature is only observed in the S(0) 
emission. 

On their own the column densities and temperatures derived in this way cannot
constrain the excitation mechanism: there is significant overlap in the parameter range
predicted for shocked and photo-dissociated gas \citep{habart05,bernard-salas05}, even 
though higher temperatures are suggestive of shocked gas. Other diagnostics are
available: PAH emission suggests photo-dissociated gas while [S\,{\sc i}] emission at 
25.25\,\mum\ indicates shocked gas. As already discussed, the majority of the 
sources in the sample exhibit PAH emission. We do not detect the [S\,{\sc i}] line in 
any of the spectra. Even though it is likely that both mechanisms contribute to H$_2$ 
excitation, the available evidence suggests that radiation is the dominant excitation 
mechanism in these SMC sources. Since H$_2$ emission is also excited in 
photo-dissociation regions \citep[e.g.,][]{habart05}, it is possible that there is  
an environmental contribution to emission observed towards the YSO candidates. This may 
be true in particular for the warmer component, the cool component originating from the 
denser more shielded regions.

%==============================================================================

\subsection{Optical spectra}

In this section we discuss the optical emission-line spectra of 20 SMC sources --- for 
the remaining 14 sources either no optical spectrum could be obtained, the spectrum 
is not associated with the IR source, or only H$\alpha$ emission is detected (Section 
3.5.1). In H\,{\sc ii} regions and their precursors surrounding young massive stars, 
common optical emission lines are due to permitted hydrogen (Balmer
and Paschen) and O\,{\sc i} emission (8446 \AA) and numerous forbidden emission 
lines: [O\,{\sc i}] (6300\,\AA), [O\,{\sc ii}] (3727, 7322, 7332\,\AA), 
[N\,{\sc ii}] (6548, 6583\,\AA), [S\,{\sc ii}] (6717, 6731\,\AA), as well as 
[O\,{\sc iii}] (4363, 4959, 5007\,\AA) and [S\,{\sc iii}] (9068, 9530\,\AA).
If the object is massive enough (early O-type source) an appreciable He$^+$ ionisation 
zone develops \citep[e.g.,][]{draine11} and He\,{\sc i} recombination emission is 
detected (at 3888, 4471, 5875, 6678, 7065\,\AA). Hydrogen emission in YSOs originates 
both from the accretion columns and outflows. Forbidden line emission can also 
originate in the relatively low-density environments of outflowing jets or winds 
\citep[e.g.,][]{white04} but it is also observed in PDR-like environments 
\citep{storzer00}. Velocity information can distinguish between the two excitation 
scenarios \citep[e.g.,][]{storzer00} but our spectra have insufficient velocity 
resolution (e.g., $\sim 205$\,km\,s$^{-1}$ at the position of H$\alpha$).

Figure \ref{optical} shows the optical spectra of the 32 SMC sources; the last 
panel shows the rich spectrum of source \#26 with the emission lines identified. We 
implement a classification scheme that relies on the detection of progressively higher 
excitation energy emission lines. Thus the classification reflects the harshness of the
near-YSO environment. Type I objects exhibit emission from the Balmer and Paschen 
series and O\,{\sc i}, Type II objects add collisionally excited lines of 
[O\,{\sc ii}], [N\,{\sc ii}] and [S\,{\sc ii}]. Type III objects exhibit also 
[O\,{\sc iii}] emission, Type IV objects add [S\,{\sc iii}], and finally Type V objects 
show prominent He\,{\sc i} recombination emission. The type breakdown of the sample of 
twenty sources is as follows: two Type I, two Type I/II (only H$\alpha$ and 
[S\,{\sc ii}] emission, see below), four Type II, one Type III, two Type IV, four Type 
IV/V (a single He\,{\sc i} line identified), and four Type V; the optical spectrum of
source \#19 is discussed in Section 5. 

Of the ten sources classified as Type IV or V, all sources exhibit PAH emission 
and seven show IR fine-structure emission. Most are also detected at radio wavelengths 
(see Table \ref{table3}). Conversely, all the objects that show IR fine-structure 
emission, and for which we have an optical spectrum, are classified as Type IV$-$V. 
These ten sources have the highest luminosities ($L\,{\gsim}\,10^4\,L_{\odot}$), as 
determined from SED fits (Table\,\ref{table3}). This builds a consistent picture of 
these objects representing more evolved YSO candidates, i.e. ultracompact H\,{\sc ii} 
regions. However, four of these sources do exhibit ice features in their IRS spectrum 
(see next section), suggesting deeply embedded objects. 

As mentioned above the optical spectra do not have enough resolution to investigate the
origin of the emission features. However, we have looked for broadening of the line 
profiles that would indicate infall and/or outflow activity. In fact, a number of 
sources show evidence of broadened H$\alpha$ profiles: 13 sources (indicated in 
Table\,\ref{table3}) have FWHM in the range 300$-$440\,km\,s$^{-1}$. Furthermore, two 
sources (\#32 and 33) exhibit extremely broad profiles (FWHM$> 600$\,km\,s$^{-1}$) with
line centroids shifted to $\sim-200$\,km\,s$^{-1}$. These two sources are classified as
Type I/II, since besides H$\alpha$ only [S\,{\sc ii}] emission is detected (equally 
broad). The profiles clearly suggest an origin in optically thick winds in the 
environments of these two YSO candidates. 

\section{Source classification}
\label{class}
%\onecolumn

\newcommand{\mtc}{\multicolumn}
\newcommand{\m}{$\mu$m}
\renewcommand{\bottomfraction}{1.}

{\scriptsize
\begin{table*}
\caption{Detailed properties of SMC YSO candidates. We use the IRS spectra to 
investigate PAH, fine-structure and H$_2$ emission and silicate 
absorption/emission, and calculate the spectral index between 3.6 and 24\,\m. 
The presence or absence of a feature is indicated by \tick\ and \cross, ? 
indicating doubt (see discussion in the text). For PAH emission the 
contrast of the 11.3-\m\ feature with respect to the continuum is analysed
in more detail: \dtick, \tick, \wtick\ indicate very strong, strong and 
weak emission respectively (see text). Four objects with silicate emission
are identified with \emm. For H$_2$ emission, \tick? indicates objects for
which only two emission lines were detected, rather than three to five. Ice
species investigated are \hho\ (at 3.1, 6 and 60\,\m), CO (4.67\,\m) and \coo\ 
(15.2\,\m). The optical ionisation classes are defined according to the 
emission lines present in the spectrum: Type I objects exhibit Balmer, Paschen
and O\,{\sc i}, Type II objects show hydrogen, O\,{\sc i}, [N\,{\sc ii}], 
[O\,{\sc ii}] and [S\,{\sc ii}] emission, Type III objects show the same 
lines plus [O\,{\sc iii}], Type IV objects further add [S\,{\sc iii}],
and finally Type V objects show all these lines plus He\,{\sc i} emission. 
Some objects exhibit a stellar absorption spectrum and others only 
H$\alpha$ in emission (see text for discussion). Objects that exhibit H$\alpha$
emission broader than 300\,km\,s$^{-1}$ are also indicated. The next column 
indicates whether the object has been detected at radio wavelengths; $*$ 
signals an extended source. The sources are classified using features in their
IRS spectrum, according to two classification schemes previously applied to 
samples of LMC YSOs \citep{seale09,woods11}. The last 
column provides the luminosities determined from the SED fits. All 
sources are classified as YSOs (Section 5), except for source \#19 that is a 
D-type symbiotic system (Section 6).}
\label{table3}
\begin{tabular}{l@{\hspace{0.5mm}}|c@{\hspace{1.2mm}}c@{\hspace{1.2mm}}c@{\hspace{1.2mm}}c@{\hspace{0mm}}c|@{\hspace{-1mm}}c@{\hspace{1.2mm}}c@{\hspace{1.2mm}}c@{\hspace{1.2mm}}c@{\hspace{1.5mm}}c@{\hspace{-2.mm}}c@{\hspace{-5mm}}c@{\hspace{1.mm}}c@{\hspace{1mm}}c@{\hspace{1mm}}c@{\hspace{0.5mm}}c}
\hline
\#   &PAH&silicate&H$_{2}$ &fine struct.& $\beta$  &\mtc{3}{c}{H$_2$O ice}&CO ice &CO$_2$ ice&optical&broad     &radio &\multicolumn{2}{@{\hspace{-1mm}}c}{YSO class.}&$L$\\	      
     &emission&    &emission&emission    &3.6$-$24\m &3\,\m&6\,\m&60\,\m    &4.67\m  &15.2\m    &Type   &H$\alpha$&source&S09&W11&$(10^3$L$_{\odot})$\\
\hline
01 &\tick &\cross&\tick &\tick      &$-$2.6&\null &\cross&\null &\null &\cross&IV/V&n&y&PE&G3&16\\
02 &\wtick&\tick &\tick &\cross     &$-$2.4&\tick &\cross&\cross&\null &\tick &\mtc{1}{c}{only H$\alpha$ emission}&y&y&S&G1&19\\
03 &\wtick&\tick &\tick\ar{?}&\cross&$-$2.2&\tick &\tick &\tick &\cross&\tick &V&n&y&S&G1&61\\
04 &\tick &\cross&\tick &\cross     &$-$1.6&\null &\cross&\null &\null &\cross&II&y&n&P&G3&2.3\\
05 &\wtick&\emm  &\tick &\cross     &$-$1.5&\null &\tick &\null &\null &\tick &\mtc{1}{c}{absorption lines}&\null&n&O&G1&1.6\\
06 &\tick &\tick &\tick &\cross     &$-$2.2&\tick &\tick &\tick &\null &\tick &\mtc{1}{c}{absorption lines}&\null&n&S&G1&5.8\\
07 &\dtick&\cross&\tick &\cross     &$-$1.7&\null &\cross&\null &\null &\cross&II&n&n&P&G3&4.2\\
08 &\tick &\cross&\tick &\tick      &$-$3.0&\null &\cross&\null &\null &\tick &V&n&y\ar{$^{*}$}&PE&G1&1.9\\
09 &\tick &\cross&\tick &\tick      &$-$2.5&\null &\cross&\null &\null &\cross&IV/V&n&y&PE&G3&7.9\\
10 &\tick &\cross&\tick &\cross     &$-$3.0&\null &\cross&\null &\null &\cross&\mtc{1}{c}{only H$\alpha$ emission}&y&y&P&G3&33\\
11 &\dtick&\cross&\tick &\cross     &$-$1.8&\null &\cross&\null &\null &\cross&\mtc{1}{c}{only H$\alpha$ emission}&y&n&P&G3&2.2\\
12 &\dtick&\cross&\tick &\cross     &$-$2.5&\null &\cross&\null &\null &\cross&III&n&n&P&G3&2.3\\
13 &\tick &\cross&\tick &\tick      &$-$3.1&\null &\cross&\null &\null &\tick &IV&n&n&PE&G1&22\\
14 &\tick &\emm  &\tick\ar{?}&\cross&$-$2.1&\null &\cross&\null &\null &\cross&\mtc{1}{c}{only H$\alpha$ emission}&y&n&O&G4&1.8\\
15 &\tick &\cross&\tick &\tick      &$-$3.1&\null &\cross&\null &\null &\cross&V&n&y\ar{$^{*}$}&PE&G3&21\\
16 &\tick &\cross&\tick &\cross     &$-$3.1&\null &\cross&\null &\null &\cross&\mtc{1}{c}{only H$\alpha$ emission}&y&n&P&G3&12\\
17 &\wtick&\tick &\tick\ar{?}&\cross&$-$1.7&\tick &\tick &\tick &\cross&\tick &\mtc{1}{c}{only H$\alpha$ emission}&y&n&S&G1&22\\
18 &\cross&\tick &\cross&\cross     &$-$2.2&\tick &\tick &\cross&\cross&\tick &\mtc{1}{c}{only H$\alpha$ emission}&y&n&S&G1&28\\
19 &\cross&\emm  &\cross&\cross     &$-$3.2&\null &\cross&\null &\null &\cross&III-V&y&n&\mtc{2}{c}{not a YSO}&34\\
20 &\wtick&\emm  &\tick?&\cross     &$-$1.8&\null &\cross&\null &\null &\cross&I&y&n&O&G4&1.5\\
21 &\tick &\cross&\tick &\cross     &$-$2.9&\null &\cross&\null &\null &\tick &II&n&n&P&G1&11\\
22 &\tick &\cross&\tick &\tick      &$-$2.2&\tick &\cross&\null &\null &\tick &IV/V&y&n&PE&G1&9.1\\
23 &\dtick&\cross&\tick &\cross     &$-$2.9&\null &\cross&\null &\null &\cross&\mtc{1}{c}{only H$\alpha$ emission}&n&n&P&G3&14\\
24 &\wtick &\tick\ar{?}&\tick&\cross&$-$1.5&\null &\cross&\null &\null &\cross&\mtc{1}{c}{no spectrum}&\null&n&P&G2/G3&4.5\\
25 &\tick &\cross&\tick &\cross     &$-$2.8&\null &\cross&\null &\null &\cross&IV&n&y&P&G3&17\\
26 &\tick &\cross&\tick &\tick      &$-$2.3&\null &\cross&\null &\null &\cross&V&n&y&PE&G3&12\\
27 &\dtick&\cross&\tick &\cross     &$-$2.6&\null &\cross&\null &\null &\cross&II&n&n&P&G3&3.3\\
28 &\wtick&\tick &\tick\ar{?}&\cross&$-$2.7&\cross&\cross&\cross&\null &\cross&IV/V&n&y&S&G2&140\\
29 &\wtick&\tick &\tick &\cross     &$-$2.7&\null &\tick &\null &\null &\tick &\mtc{1}{c}{absorption lines}&\null&n&S&G1&10\\
30 &\tick &\tick\ar{?}&\tick &\cross&$-$1.0&\tick &\cross&\cross&\null &\cross&I&y&n&P&G1&7.9\\
31 &\tick &\cross&\tick\ar{?}&\tick &$-$2.7&\null &\cross&\cross&\null &\cross&\mtc{1}{c}{no spectrum}&\null&y\ar{$^{*}$}&PE&G3&6.7\\
32 &\wtick&\tick &\tick &\cross     &$-$1.6&\tick &\cross&\null &\null &\tick &I/II&y&n&S&G1&3.5\\
33 &\wtick&\tick &\tick\ar{?}&\cross&$-$1.4&?	  &\cross&\cross&\cross&\cross&I/II&y&n&S&G2&26\\
34 &\wtick&\tick &\tick\ar{?}&\cross&$-$1.8&\tick &\cross&\cross&\cross&\tick &\mtc{1}{c}{only H$\alpha$ emission}&y&n&S&G1&23\\
\hline
\end{tabular}
\end{table*}}

%\end{landscape}

All but one of the sources in the SMC sample are YSOs, based on the IR spectral 
properties analysed in the previous section: ice absorption, silicate absorption 
or emission, PAH and H$_2$ emission, red continuum and SED fitting. \citet{boyer11}
tentatively proposed that nine of these sources are very dusty evolved stars, based
on \spitzer\ photometric criteria. Seven of those sources show ice absorption and 
two others show silicate absorption, and are thus clearly SMC YSOs. The only non-YSO
source in our sample (\#19) is discussed in Section\,\ref{symbiotic}.

Two YSO spectral classification schemes have been recently developed and applied to 
YSO samples in the LMC. \citet{seale09} performed an automated spectral classification
of YSOs using principal component analysis to identify the spectral features that 
dominate the spectra; the following classes are relevant for the SMC sample: S
objects show spectra dominated by silicate absorption, P and PE objects show PAH 
emission and IR fine-structure emission and O objects show silicate emission. There
is an evolutionary sequence associated with this classification in the sense that
objects with S spectra are more embedded while P and PE spectra are associated 
with more evolved compact H\,{\sc ii} regions. We have classified the objects in the 
YSO sample using this classification scheme (Table\,\ref{table3}).

The other classification scheme was introduced by \citet{woods11}: objects are
classified in groups depending on spectral features present in the spectrum, from 
G1 (ice absorption), G2 (silicate absorption), G3 (PAH emission) and G4 (silicate 
emission). The main difference from the Seale classification is the usage of ice 
absorption as a clear indicator of a very cool envelope, identifying the earliest 
embedded sources. Thus we discuss the classification of the SMC YSOs using this
method in more detail (Table\,\ref{table3}).

\subsection{Embedded YSOs}

There are fourteen SMC sources classified as belonging to group G1, i.e. exhibiting ice
absorption. Most G1 sources (eight) also exhibit silicate absorption, but this is not 
always the case: source \#05 shows a silicate emission feature (see Section 5.3), while
sources \#08, 13, 21, 22 and 30 show strong PAH emission that could mask weak silicate 
absorption --- we suspect this to be the case particularly for \#30. Of the 
aforementioned fourteen sources, ten show definite H$_2$ emission, three have weak 
H$_2$ emission, and one source shows no H$_2$ emission. Eight G1 sources have weak or no 
PAH emission features while six sources show strong emission (three of which also show 
fine-structure emission). Of the fourteen G1 sources, three have no optical spectra and 
seven exhibit either only H$\alpha$ or low-excitation emission lines. Another four G1 
sources, namely \#03, 08, 13 and 22 are classified as Type IV$-$V based on the 
counterpart optical spectrum; except for \#03, these sources also show strong PAH and 
fine-structure emission.

As already mentioned the presence of ice and silicate absorption features indicates
the presence of an embedded YSO while PAH, and forbidden and fine-structure emission 
hint at a compact H\,{\sc ii} region. However, many sources exhibit both ice absorption 
and emission features. Galactic YSOs can exhibit both PAH and fine structure
emission and ice absorption \citep[e.g., W3\,IRS5 and MonR2\,IRS2;][]{gibb04}, and 
many LMC YSOs with \coo\ ice signatures also exhibit PAH emission 
\citep{oliveira09,seale11}. The same is true for the SMC sample, with half the sources
exhibiting mixed-property spectra. Dust and ice features originate from the cooler 
regions of the embedded YSO envelope. On the other hand PAH and fine-structure emission 
can be excited not only by the emerging YSO itself but also by neighbouring massive 
stars, and more generally in the larger H\,{\sc ii} complexes in which many YSOs reside. 
At the LMC and SMC distances \citep[respectively $\sim$\,50 and 60\,kpc,][]
{ngeow08,szewczyk09}, it becomes impossible to disentangle the contributions of the
different physical environments. The spatial resolution of a IRS spectrum varies 
typically from 3$\arcsec$ to 10$\arcsec$, corresponding to 1 to 3\,pc in the SMC, 
implying that very different spatial scales are sampled at different wavelengths.
This makes it difficult to use spectral features to unequivocally constrain the 
evolutionary stage of the object. 

Nevertheless the six G1 YSOs with ice and silicate absorption, weak or absent PAH 
emission, no fine-structure emission and quiescent optical spectra are the more 
embedded, earliest YSOs (sources \#02, 17, 18, 29, 32 and 34). Sources \#28 and 
33 (the two G2 sources) and \#24 (G2/G3 source) are still relatively embedded 
(silicate is seen in absorption) but no ice absorption is detected, suggesting warmer 
envelopes. At these early stages the YSO has little influence on the physical conditions 
in its envelope. The remaining G1 YSOs (sources \#03, 06, 08, 13, 21, 22 and 30) show
both ice absorption and PAH emission in their spectra. These objects are likely more 
evolved since the PAH and fine-structure emission indicates that the YSO is able to emit 
copious amounts of UV radiation, while still retaining enough of its cold envelope 
responsible for the absorption features (with the possible caveat that \#06, 08 and 13 
sit in particularly complex environments, see Fig.\,\ref{image}). At this stage a 
(ultra-)compact H\,{\sc ii} region is emerging.

\subsection{PAH dominated YSOs}

There are fourteen sources in the sample that we classify as G3, indicating that PAH
emission is the noticeable feature in their steep IRS spectra. Five of the sources have 
fine-structure emission as well. In terms of counterpart optical spectra, there does 
not seem to be a strong relation between PAH emission and optical type (Section 4.5),
since four sources only exhibit H$\alpha$ emission, three sources are Type II, one
source is Type III and five sources are Type IV$-$V (for one further source we have no 
spectrum).

These fourteen G3 sources are the most evolved in the sample, with the IR spectral 
features clearly indicating the presence of a well developed compact H\,{\sc ii} 
region, in the process of clearing out the remnant dusty envelope.

\subsection{Intermediate-mass YSOs}

In this subsection we discuss the three YSOs that exhibit silicates in emission 
in their IRS spectrum, classified as G4 (sources \#14 and 20) and source \#05 that also
shows ice absorption. The SED of \#20 is steep but flattens considerably above 20\,\mum. 
In fact, its IRS spectrum is very similar to that of the Galactic HAeBe star 
IRAS\,04101+3103 \citep{furlan06}, including the weak PAH emission. A spectrum and 
photometry of the optical counterpart have been analysed by \citet[]
[identified as SMC5$\_037102$]{martayan07}; they suggest that this source is a Herbig 
B[e] star, based on the presence of numerous Fe\,{\sc ii} and [Fe\,{\sc ii}] emission 
lines and accretion signatures in the Balmer lines. We cannot detect the iron lines in 
our low-resolution optical spectrum, but we confirm that H$\alpha$ is broad. The fact 
that the continuum flux rises from the $J_s$-band to 70\,\mum\ suggests that a tenuous 
dusty envelope remains. 

The SEDs of sources \#05 and 14 both have silicate emission and steep IRS spectra
throughout. Source \#14 shows strong PAH emission, while \#05 shows weak PAH and H$_2$ 
emission as well as \hho\ and \coo\ ice absorption. \citet{furlan08} describes a number
of Galactic objects that they classify as evolved Class I YSOs; these objects still 
retain a low-density envelope, tenuous enough to reveal the thick accretion disc and 
the central star. The observed properties of sources \#05, 14 and 20 suggests the same 
classification, with \#20 the more evolved of the three sources. Note that the 
evolutionary scenario discussed in the previous subsection refers to massive star 
formation, while the present discussion addresses the evolution of intermediate-mass 
stars. 

\section{A D-type symbiotic star in the SMC}
\label{symbiotic}

\begin{figure}
\includegraphics[scale=0.93]{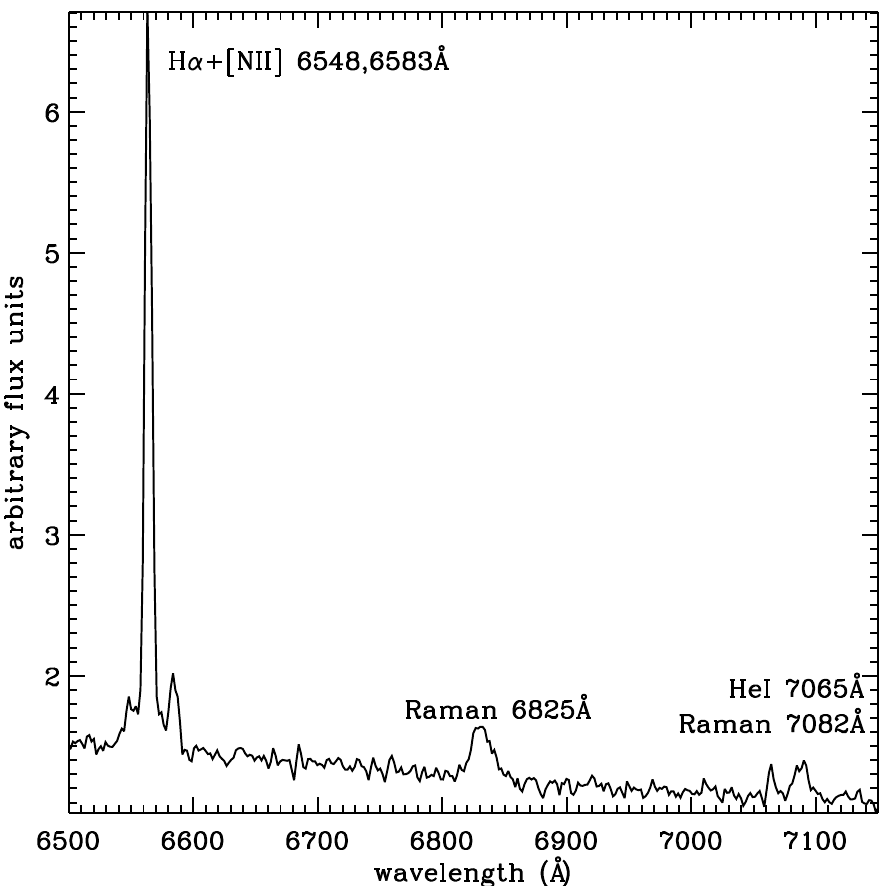}
\caption{Detail of the optical spectrum of \#19, showing the prominent Raman emission
lines.\label{raman}}
\end{figure}

Source \#19 is a clear oddity since its optical spectrum (Figs.\ref{raman}, 
\ref{optical}) shows a blue continuum with broad emission lines at 6825 and 7082\,\AA, 
as well as H$\alpha$, [N\,{\sc ii}] and [O\,{\sc iii}] emission, with weak He\,{\sc i} 
emission at 5875 and 7065\,\AA, and a tentative detection of He\,{\sc ii} emission at 
4686\,\AA. Figure \ref{thumbs} shows that next to the IR source there is a bright blue 
star. By looking at the 2-D spectrum (before extraction) we see that the continuum and 
emission line contributions are offset by about $2^{\prime\prime}$, in the sense that 
the continuum originates from the bright blue source, while the emission lines originate 
from the IR-bright source. 

The emission features at 6825 and 7082\,\AA\ are identified as Raman scattering
of the O\,{\sc vi} resonance photons at 1032 and 1038\,\AA\ by neutral hydrogen; these
lines are usually observed in the spectra of symbiotic binary systems \citep[e.g.,][]
{schmid89}. Symbiotic objects are interacting binaries, in which an evolved giant 
transfers material to a much hotter, compact companion; according to the classification
criteria of \citet{belczynski00}, the presence of the Raman scattering lines and
optical emission lines is enough to identify a symbiotic system, even if no features of
the cool giant are found. Recently it has been proposed that massive luminous B[e] stars
may also exhibit Raman scattering lines \citep{torres12}. 

Source \#19 has a steep SED from 2 to 20\,\mum\ (Fig.\,\ref{seds}). Above 20\,\mum\ 
the IRS spectrum flattens considerably and the source becomes fainter at 70\,\mum. This 
indicates the presence of a significant amount of (not very cold) dust. It shows a 
prominent silicate emission feature, suggestive of unprocessed dust, and no PAH emission.
This is consistent with this object being a dusty (D-type) symbiotic star  
\citep{angeloni07}. The optical spectrum shows no
molecular TiO bands that would indicate the presence of the red giant. This is another 
common feature of D-type symbiotic stars: the reddened asymptotic giant branch star is 
not detected at optical wavelengths but it reveals itself at IR wavelengths \citep[e.g.,]
[]{corradi10}. In terms of photometry, source \#19 has a colour $J_s-K_s\sim 1.1$\,mag, 
and it has not been identified as variable. Given the presence of Raman scattering 
and H and He emission, its brightness at 70\,\mum, and the properties of
its IRS spectrum, we propose that \#19 is a D-type symbiotic system in the SMC. This 
adds to the six S-type (dust-less) symbiotic systems already confirmed in the SMC 
\citep{mikolajewska04}.

\section{Summary}

We present a multi-wavelength characterisation of 34 YSO candidates in the SMC. The 
target objects are bright in the 70-\mum\ MIPS band, and the selection strategy aims at
excluding both evolved star and bright galaxy interlopers. The basis of the analysis 
described here are low-resolution IR spectra obtained with \spitzer-IRS, supported by 
\spitzer\ photometry (IRAC and MIPS), near-IR photometry, 3$-$5\,\mum\ spectroscopy, 
low-resolution optical spectroscopy and radio data. The objective is to confirm 
the YSO nature of these SMC sources and characterise them. We summarise here our most 
important results.
\begin{itemize}

\item Of the 34 sources 33 are spectroscopically identified as YSOs in the SMC. This 
now sizable sample adds to the SMC YSOs previously identified by 
\citet{vanloon08,vanloon10b} and \citet{oliveira11}.

\item One object (source \#19) is identified as a D-type symbiotic system, based on the
presence of Raman emission at 6825 and 7082\,\AA\ and nebular emission lines, as well 
as prominent silicate emission. This is the first D-type symbiotic identified in the 
SMC. 

\item Fourteen YSOs exhibit ice absorption in their spectra. We analyse \hho\ and \coo\
ice column densities; we suggest the presence of a significant \hho\ column density
threshold for the detection of \coo\ ice in the SMC. The observed differences between 
Galactic, LMC and SMC samples can be explained as due to metallicity.

\item We analyse PAH emission, which is ubiquitous in the sample. We confirm previous 
results from \citet{sandstrom12} who propose that the grains responsible for PAH
emission in the SMC are mostly small and neutral. Based on the comparison of different 
published samples, the observed PAH properties cannot be solely determined by 
metallicity. 

\item Many objects show narrow emission lines in the IRS spectra attributed to 
molecular hydrogen. Excitation diagrams constrain the rotational temperature and
H$_2$ column density of the bulk of the gas responsible to the emission. When 
compared to Galactic sources \citep{vandenancker99} the rotational temperatures 
are similar, but the H$_2$ column densities in the SMC are generally smaller. 
Photo-dissociation is the dominant excitation mechanism. There does not seem to
be a clear correlation between the detection of H$_2$ emission and its derived 
properties and the evolutionary stage of the YSO. For most sources there is also
a significant reservoir of colder molecular gas ($T_{\rm rot} \sim 100$\,K).

\item Of the 33 YSOs in our sample, 30 cover the main stages of massive YSO evolution. 
Based on the presence of ice and silicate absorption and weak/absent PAH emission, six 
YSOs are still deeply embedded. Three other embedded sources have envelopes already too 
warm for significant amount of ice to survive on the dust grains. A further seven 
sources, while still embedded (i.e. showing IR molecular absorption features), already 
present evidence of an emerging H\,{\sc ii} region. Finally, fourteen sources exhibit 
spectra with the hallmarks of compact H\,{\sc ii} regions (strong PAH and fine-structure 
emission). Three other sources exhibit silicate emission (one shows ice absorption as 
well); these sources are probably sources in transition between Class I and Class II, 
i.e. precursors to intermediate-mass HAeBe stars that still retain a tenuous envelope.

Scheduled ground-based observations will target more YSOs with ice signatures; column
density measurements and modelling of the ice profiles will help constrain the reason 
for the observed differences in ice properties for samples of sub-solar metallicity. 
Some of the SMC YSOs are targets of an ongoing Herschel spectroscopy program that aims 
to constrain the role of the main cooling agents (gas-phase CO and water, OH, 
[C\,{\sc ii}] and [O\,{\sc i}]) and obtain an inventory of the species present both
in gas and solid phases.

\end{itemize}

\section*{Acknowledgments}
We thank the staff at ESO's Paranal and La Silla Observatories for their support 
during the observations. This work is based on observations made with the \spitzer\
Space Telescope, which is operated by the Jet Propulsion Laboratory, California 
Institute of Technology under contract with NASA. This research has made use of the
SIMBAD database, operated at CDS, Strasbourg, France. We thank the referee for helpful
comments.

\clearpage

\begin{table}
\begin{center}
\caption{\normalsize Summary of spectroscopic observations available for the SMC
sources.}
\label{table4}
\begin{tabular}{l|c|c|c|c|c}
\hline
\#& optical&3-4\mum&4-5\mum&IRS   &MIPS-SED\\
\hline
01       &\tick   &\cross &\cross &\tick &\cross\\
02       &\tick   &\tick  &\cross &\tick &\tick \\
03       &\tick   &\tick  &\tick  &\tick &\tick \\
04       &\tick   &\cross &\cross &\tick &\cross\\
05       &\cross  &\cross &\cross &\tick &\cross\\
06       &\cross  &\tick  &\cross &\tick &\tick \\
07       &\tick   &\cross &\cross &\tick &\cross\\
08       &\tick   &\cross &\cross &\tick &\cross\\
09       &\tick   &\cross &\cross &\tick &\cross\\
10       &\tick   &\cross &\cross &\tick &\cross\\
11       &\tick   &\cross &\cross &\tick &\cross\\
12       &\tick   &\cross &\cross &\tick &\cross\\
13       &\tick   &\cross &\cross &\tick &\cross\\
14       &\tick   &\cross &\cross &\tick &\cross\\
15       &\tick   &\cross &\cross &\tick &\cross\\
16       &\tick   &\cross &\cross &\tick &\cross\\
17       &\tick   &\tick  &\tick  &\tick &\tick \\
18       &\tick   &\tick  &\tick  &\tick &\tick \\
19       &\tick   &\cross &\cross &\tick &\cross\\
20       &\tick   &\cross &\cross &\tick &\cross\\
21       &\tick   &\cross &\cross &\tick &\cross\\
22       &\tick   &\tick  &\cross &\tick &\cross \\
23       &\tick   &\cross &\cross &\tick &\cross\\
24       &\cross  &\cross &\cross &\tick &\cross\\
25       &\tick   &\cross &\cross &\tick &\cross\\
26       &\tick   &\cross &\cross &\tick &\cross\\
27       &\tick   &\cross &\cross &\tick &\cross\\
28       &\tick   &\tick  &\cross &\tick &\tick \\
29       &\cross  &\cross &\cross &\tick &\cross \\
30       &\tick   &\tick  &\cross &\tick &\tick \\
31       &\cross  &\cross &\cross &\tick &\tick \\
32       &\tick   &\tick  &\cross &\tick &\cross\\
33       &\tick   &\tick  &\tick  &\tick &\tick\\
34       &\tick   &\tick  &\tick  &\tick &\tick\\
\hline
\end{tabular}
\end{center}
\end{table}

\appendix

\section{Summary of spectroscopic observations}

Table\,\ref{table4} lists the spectroscopic observations available for each target.
The details of the observations are described in Section 3.

%\clearpage

\section{Spectral energy distributions}

In this Appendix we show the SEDs of the 34 sources in the SMC sample
(Fig.\,\ref{seds_ap}). The photometric data are compiled in Table\,\ref{table1}, and 
spectra were obtained at the VLT (3$-$5\,\mum) and with \spitzer-IRS and MIPS-SED
(Table\,\ref{table4}). The derived spectral indices $\beta$ are indicated in each panel.

\begin{figure*}
\includegraphics[scale=0.9]{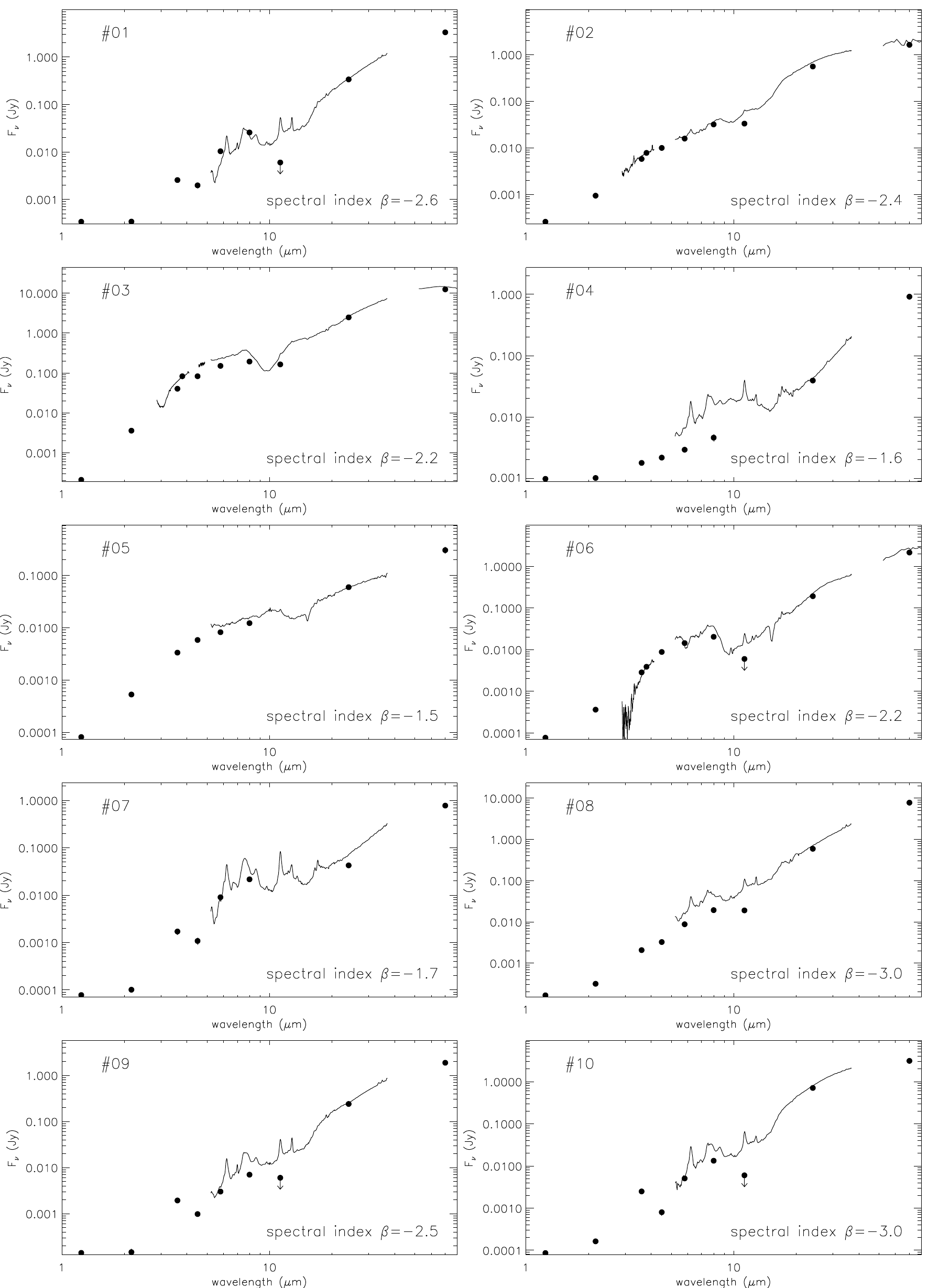}
\caption{SEDs of the 34 SMC sources analysed in this work, showing all available IR
photometry and spectroscopy. Spectral indices calculated in the range 3.6$-$24\,\mum\ 
are provided.}
\label{seds_ap}
\end{figure*}
\begin{figure*}
\includegraphics[scale=0.9]{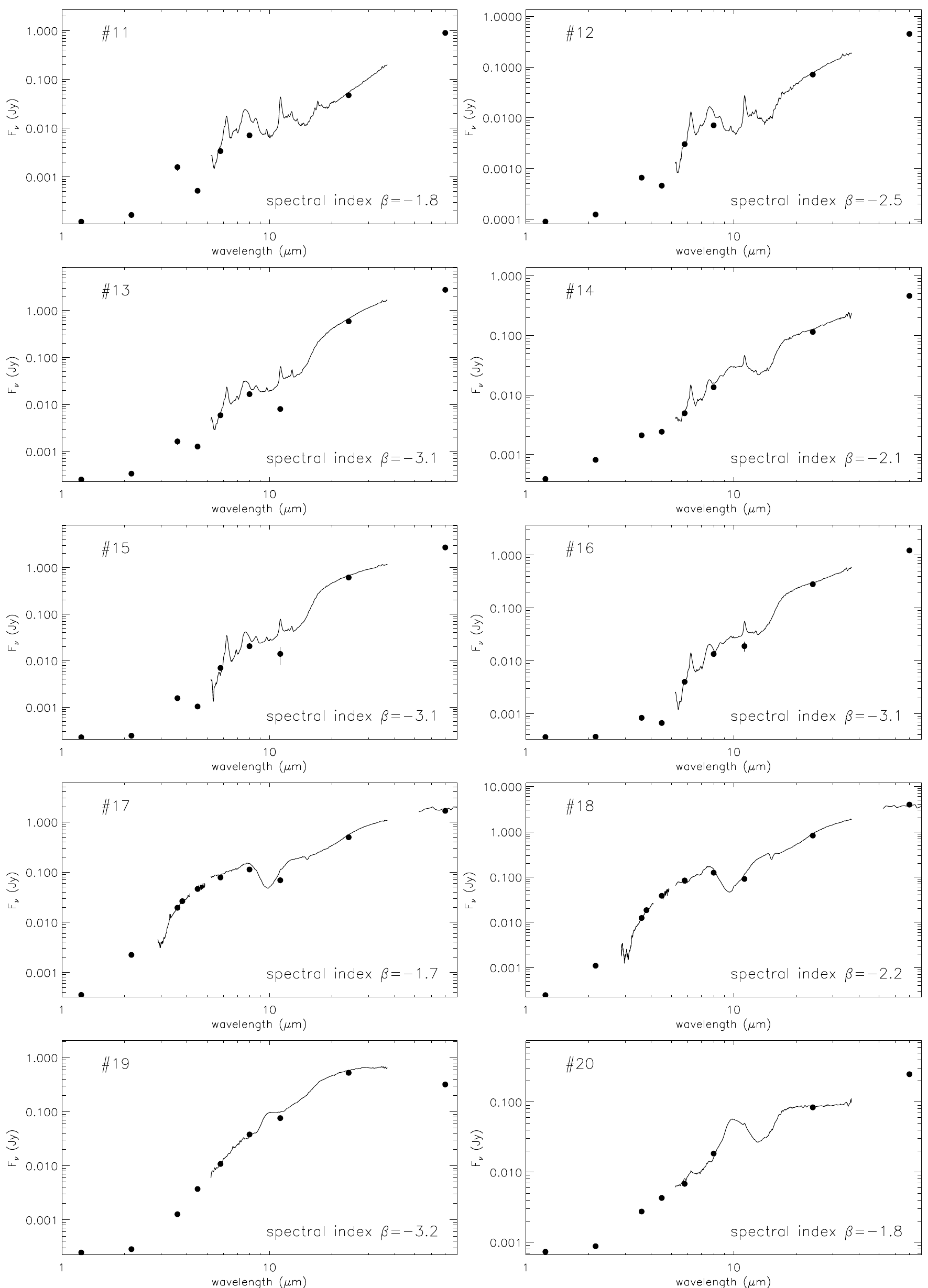}
\contcaption{}
\end{figure*}
\begin{figure*}
\includegraphics[scale=0.9]{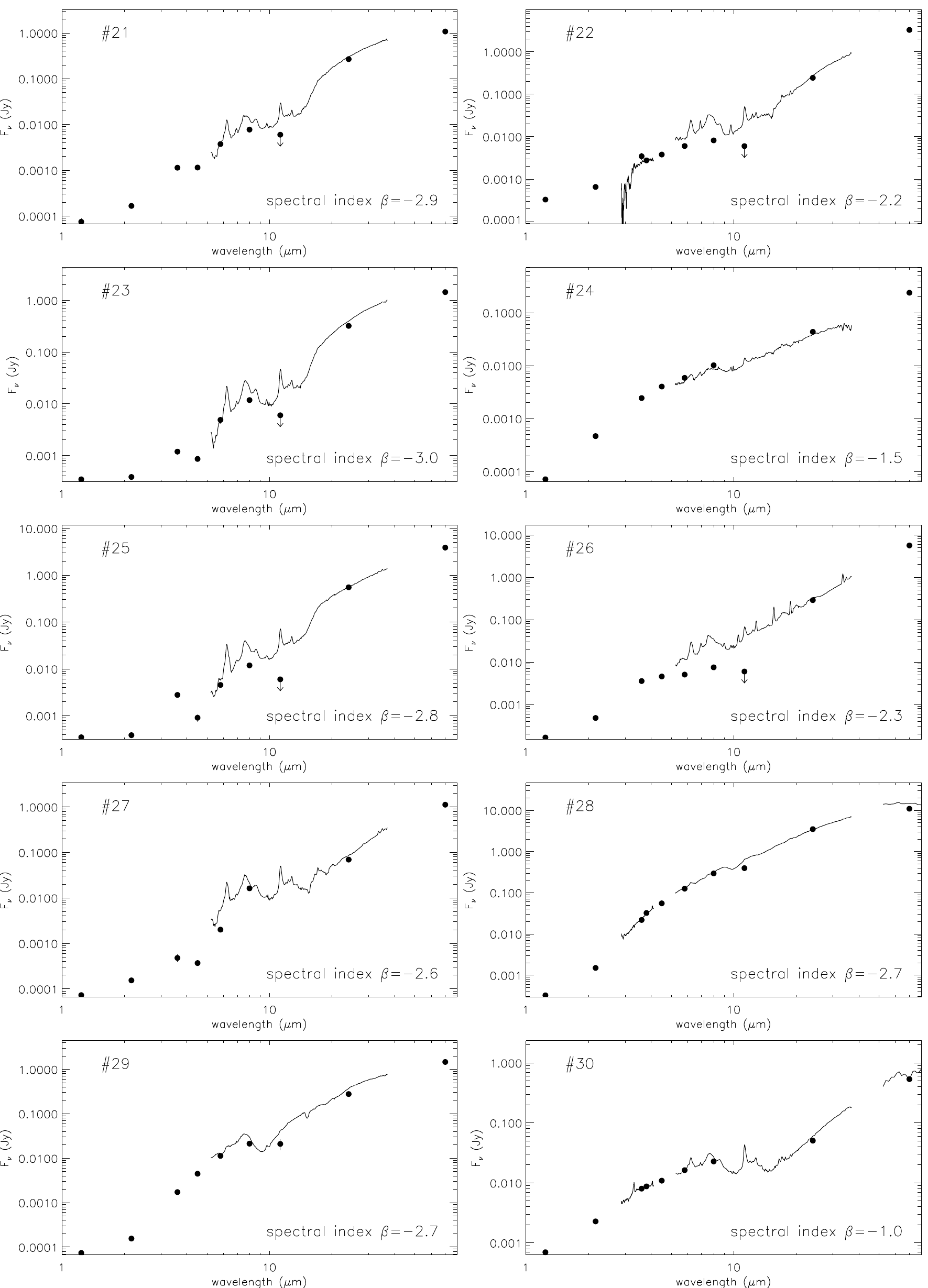}
\contcaption{}
\end{figure*}
\begin{figure*}
\includegraphics*[scale=0.9]{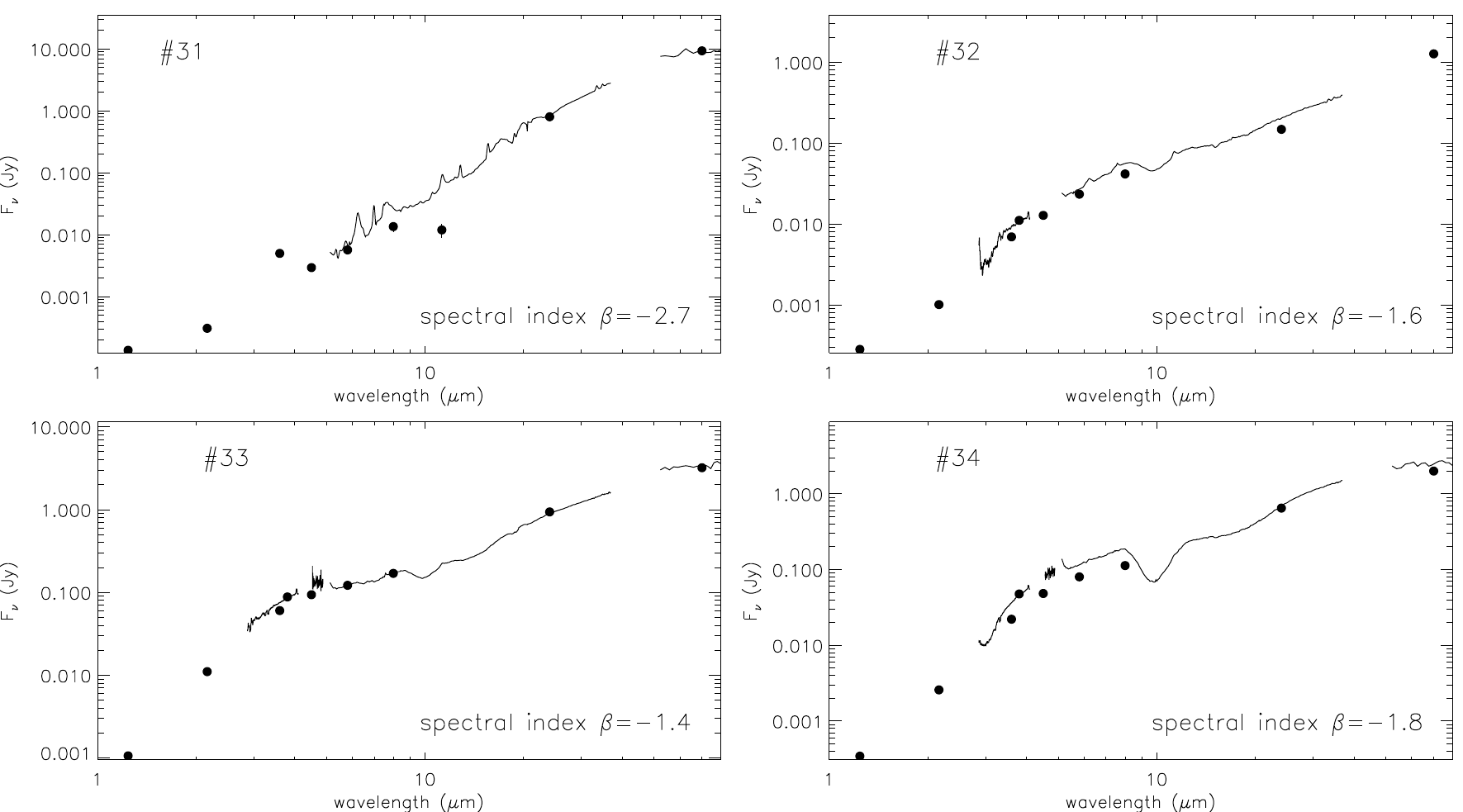}
\contcaption{}
\end{figure*}

\section{NIR thumbnail images}

In this appendix we show thumbnail images for all sources in the SMC sample 
(Fig.\,\ref{thumbs}). For sources \#01$-$32, the $J_{\rm s}K_{\rm s}$ composite is 
shown on the left and the \spitzer\ [3.6]-[5.8]-[8] composite is shown on the right 
--- see Section 3 for details on these observations. For sources \#33 and 34 we show a 
$JHK$ colour-composite, constructed using IRSF images \citep{kato07}. These images 
provide evidence of the complexity of some of the YSO candidate environments.

\begin{figure*}
\includegraphics[scale=0.363]{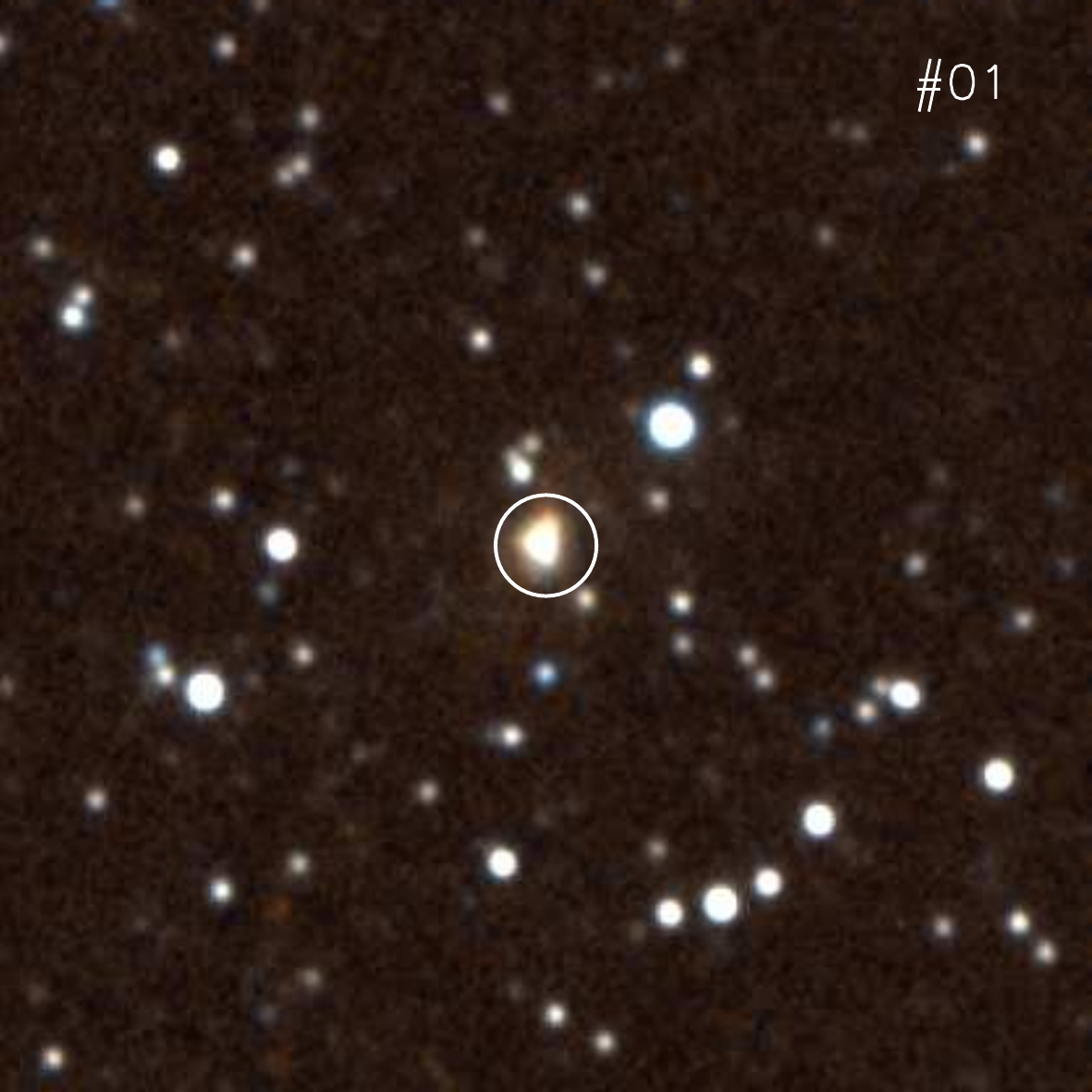}
\includegraphics[scale=0.363]{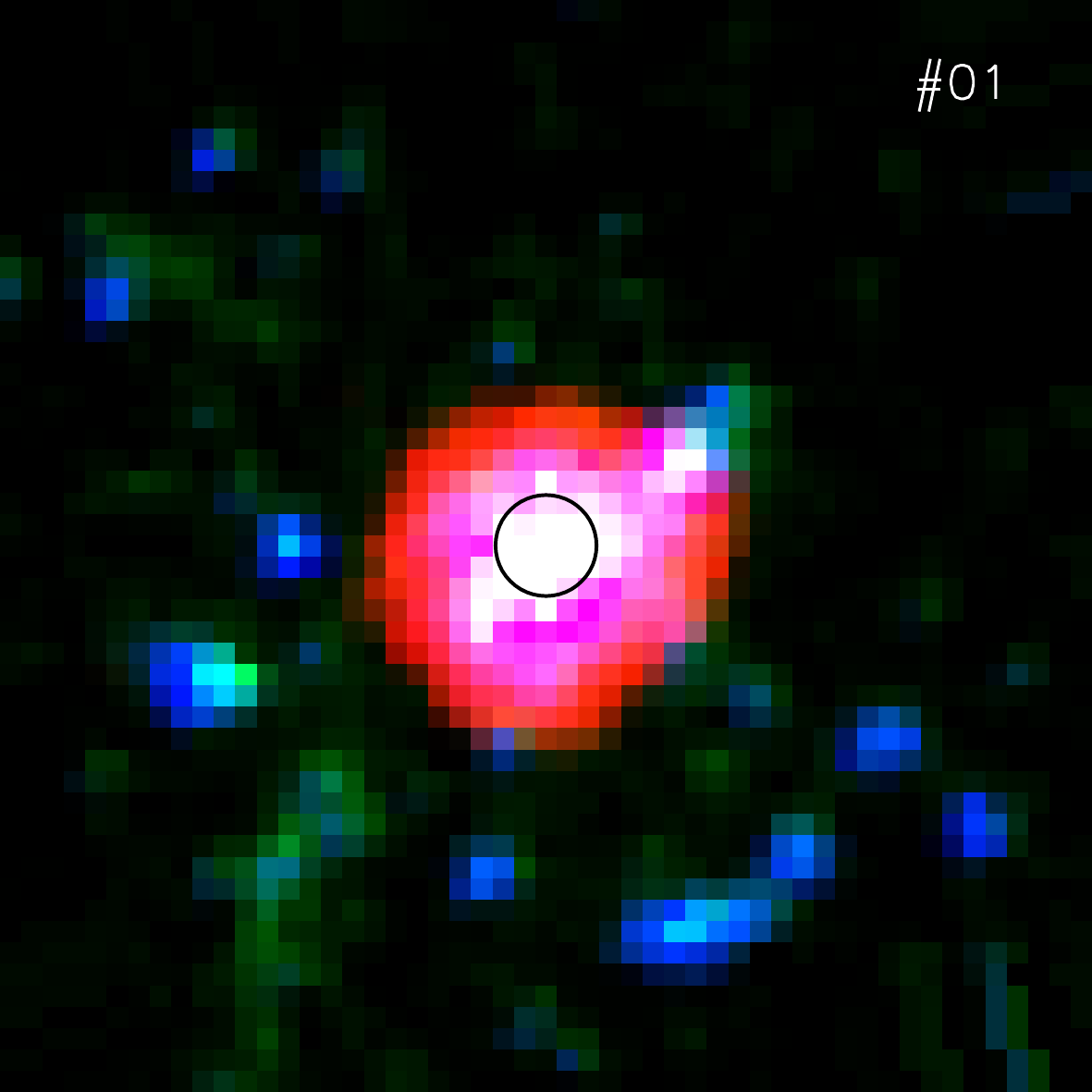}
\includegraphics[scale=0.363]{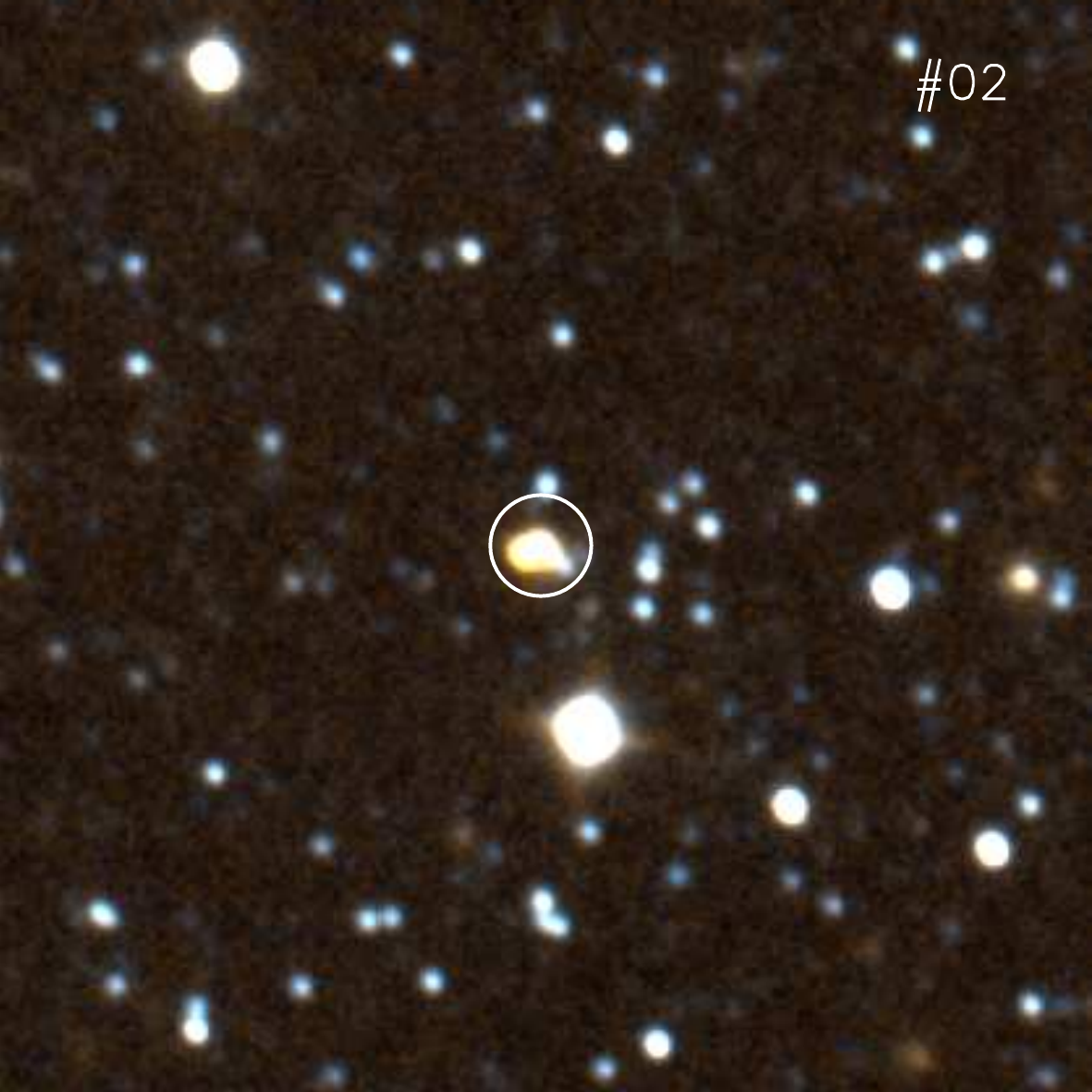}
\includegraphics[scale=0.363]{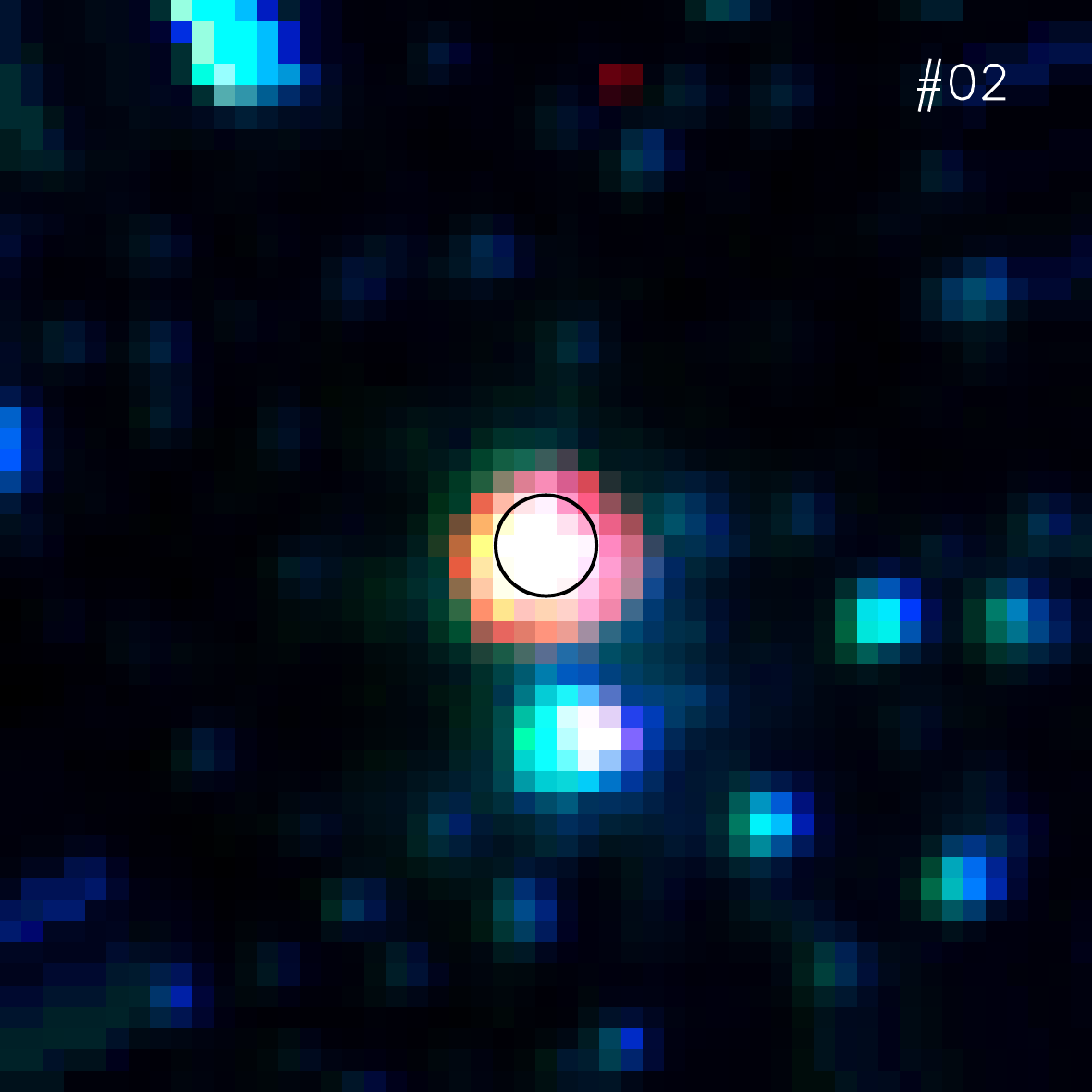}
\includegraphics[scale=0.363]{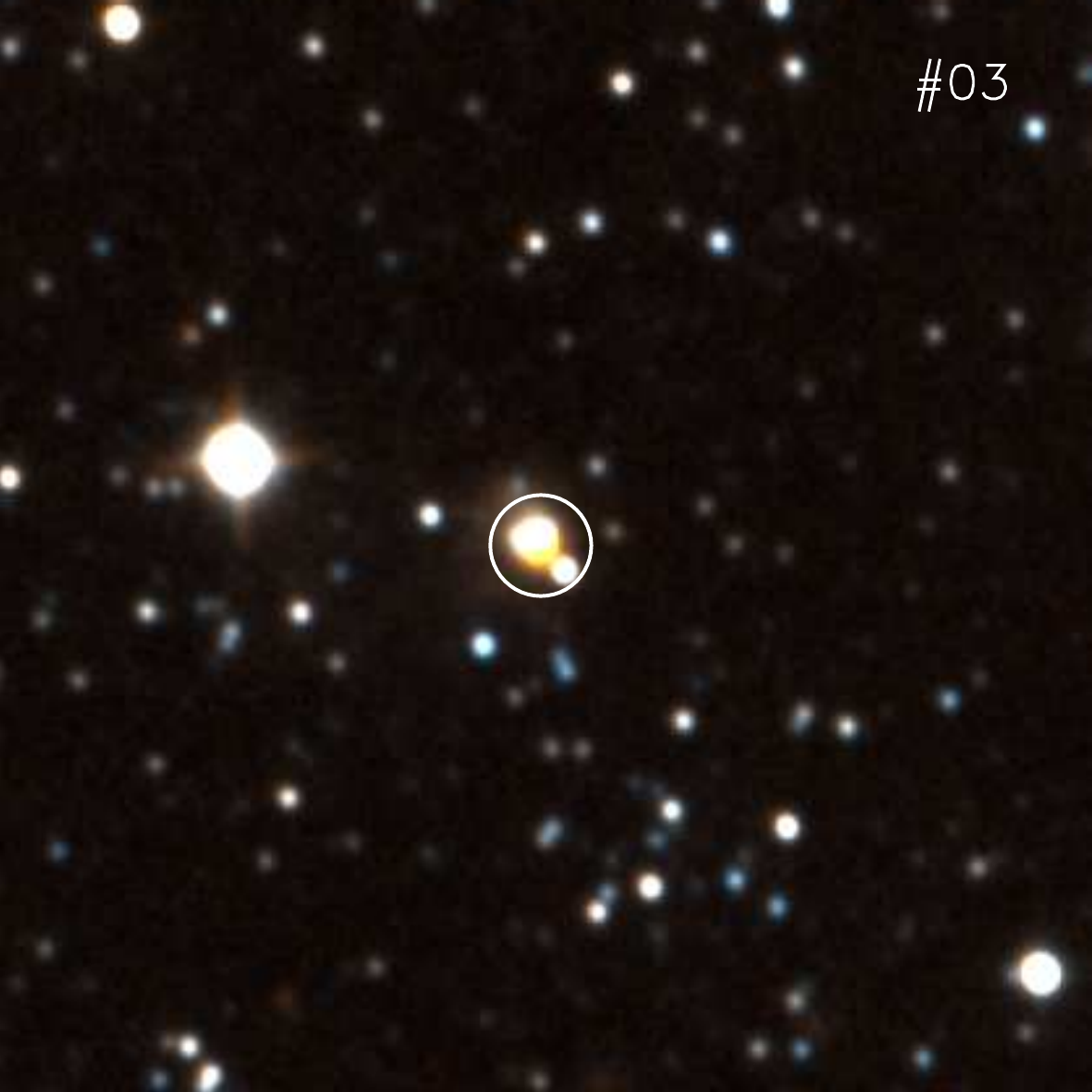}
\includegraphics[scale=0.363]{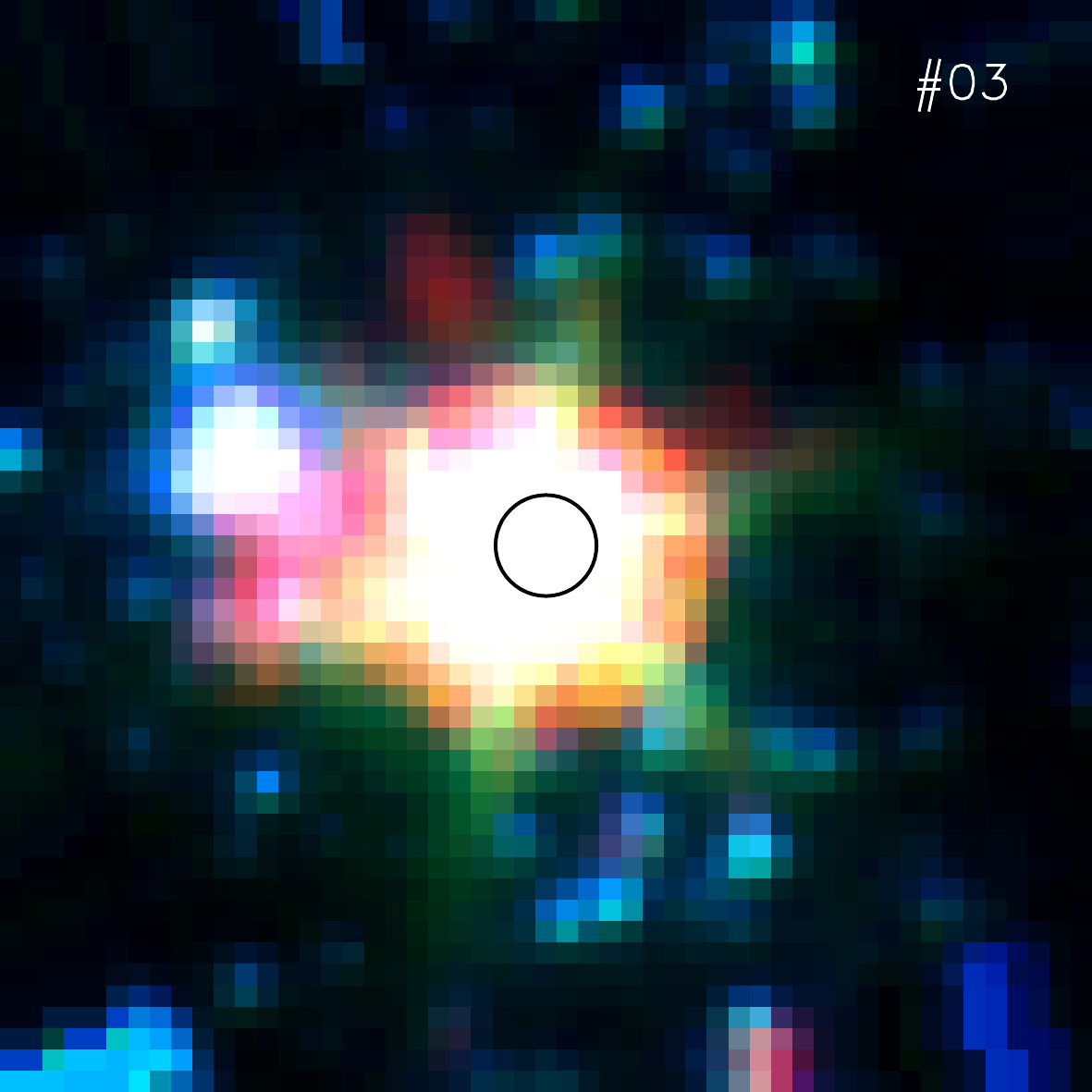}
\includegraphics[scale=0.363]{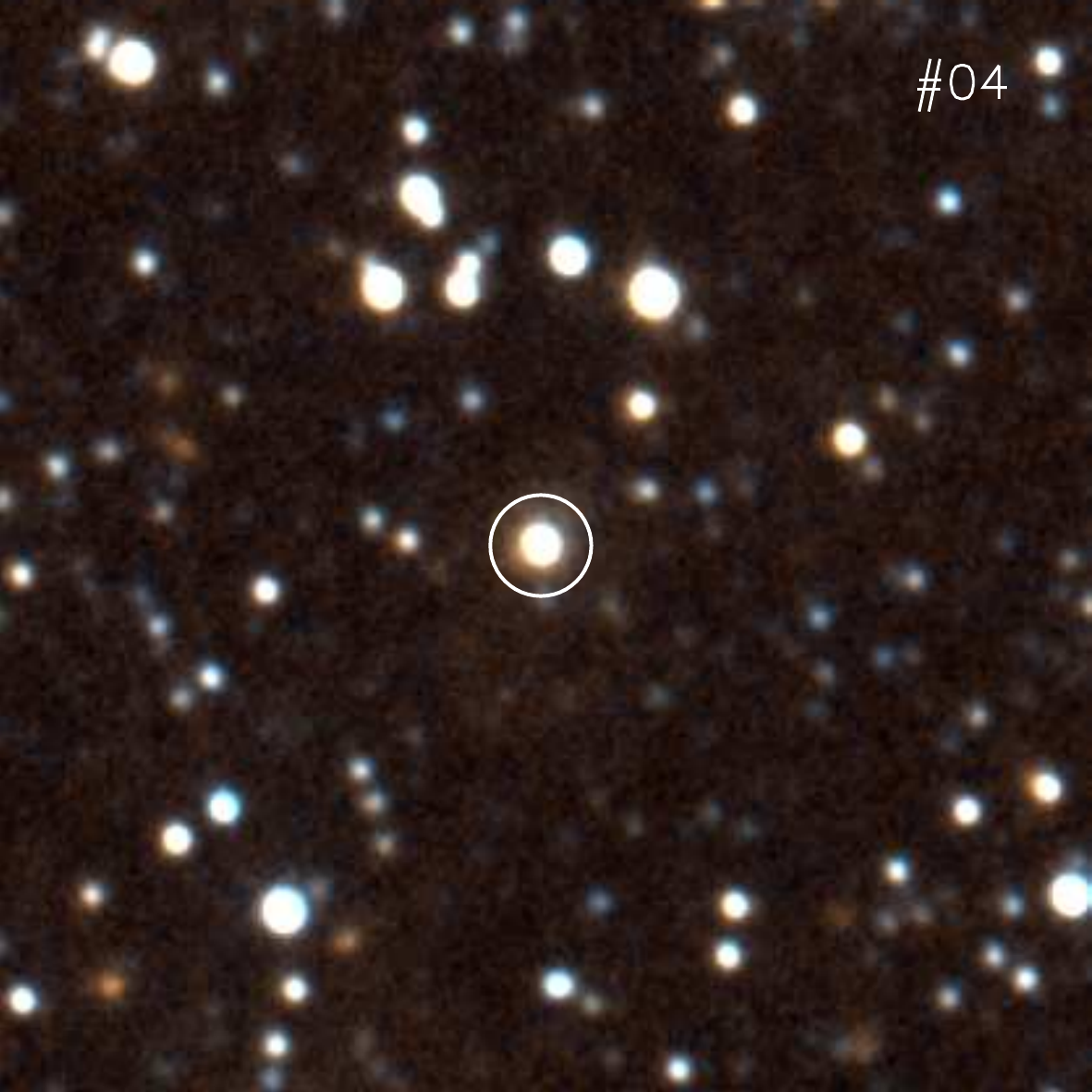}
\includegraphics[scale=0.363]{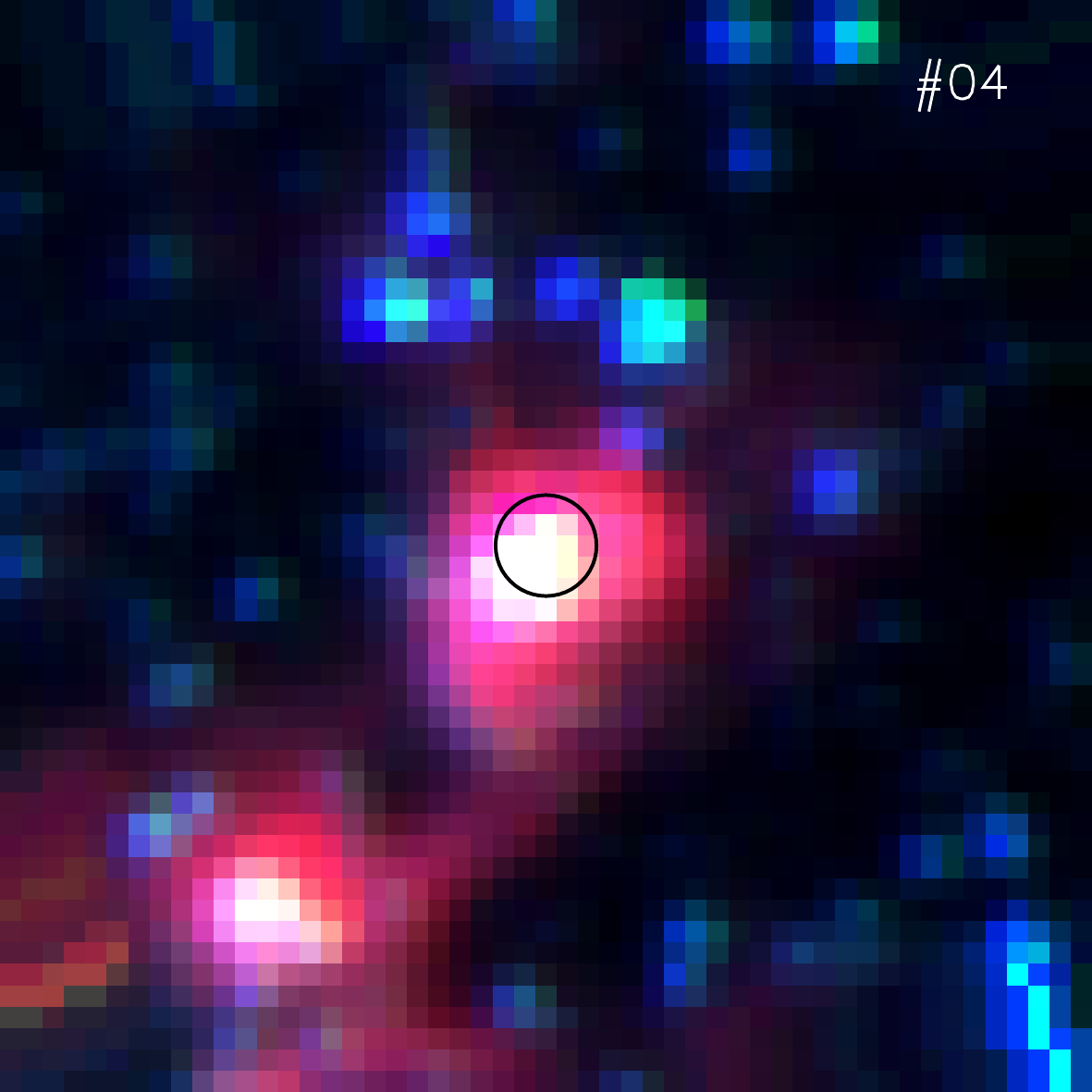}
\includegraphics[scale=0.363]{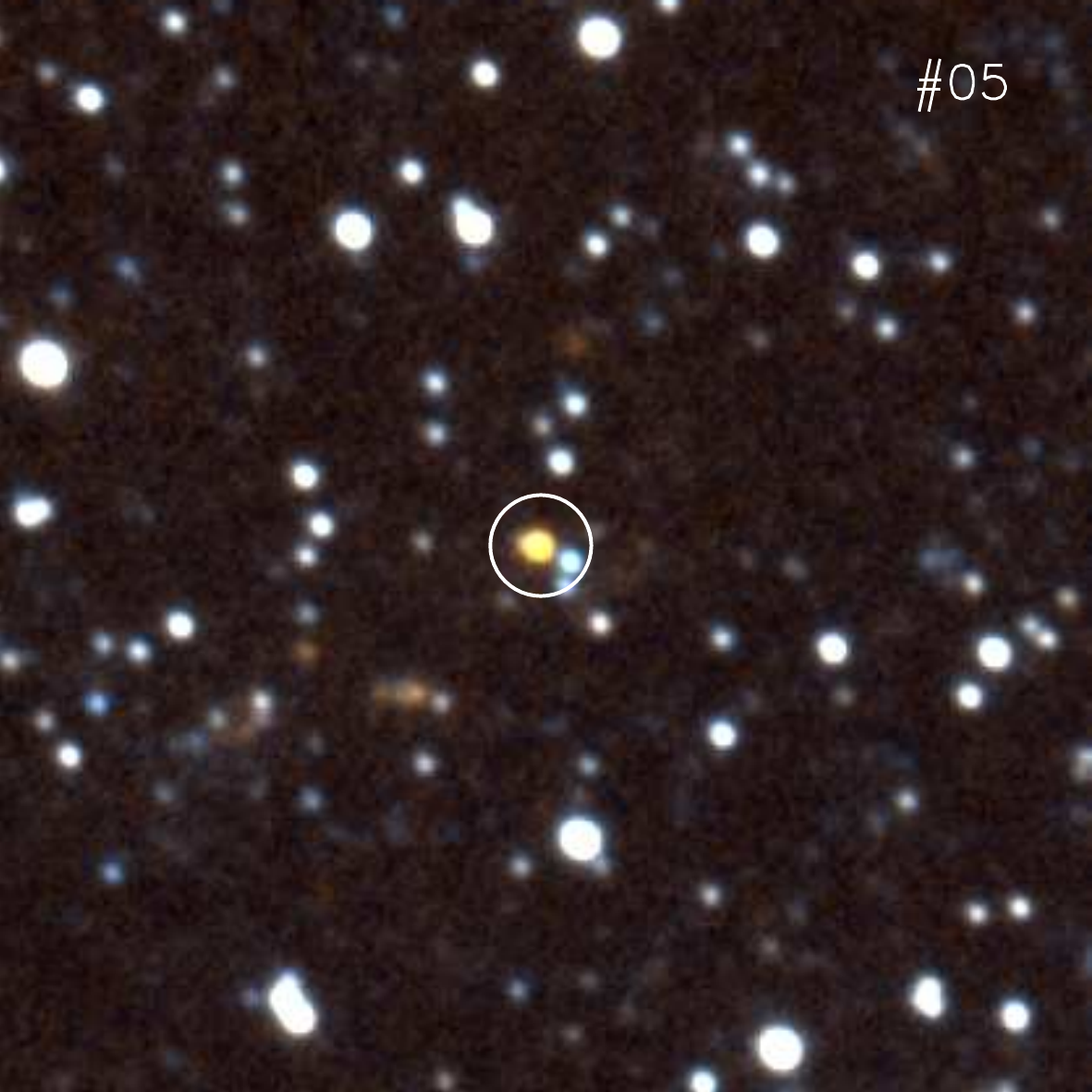}
\includegraphics[scale=0.363]{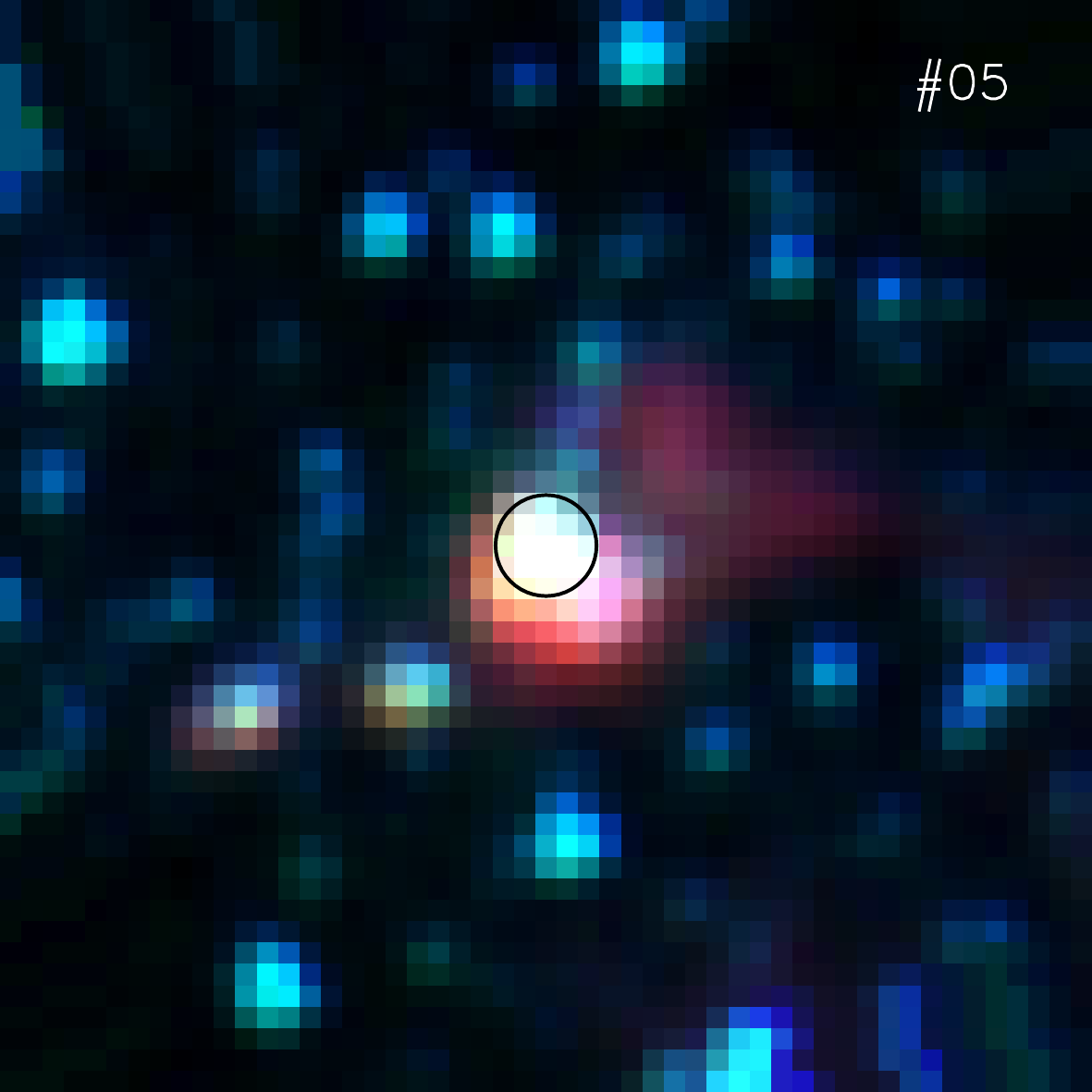}
\includegraphics[scale=0.363]{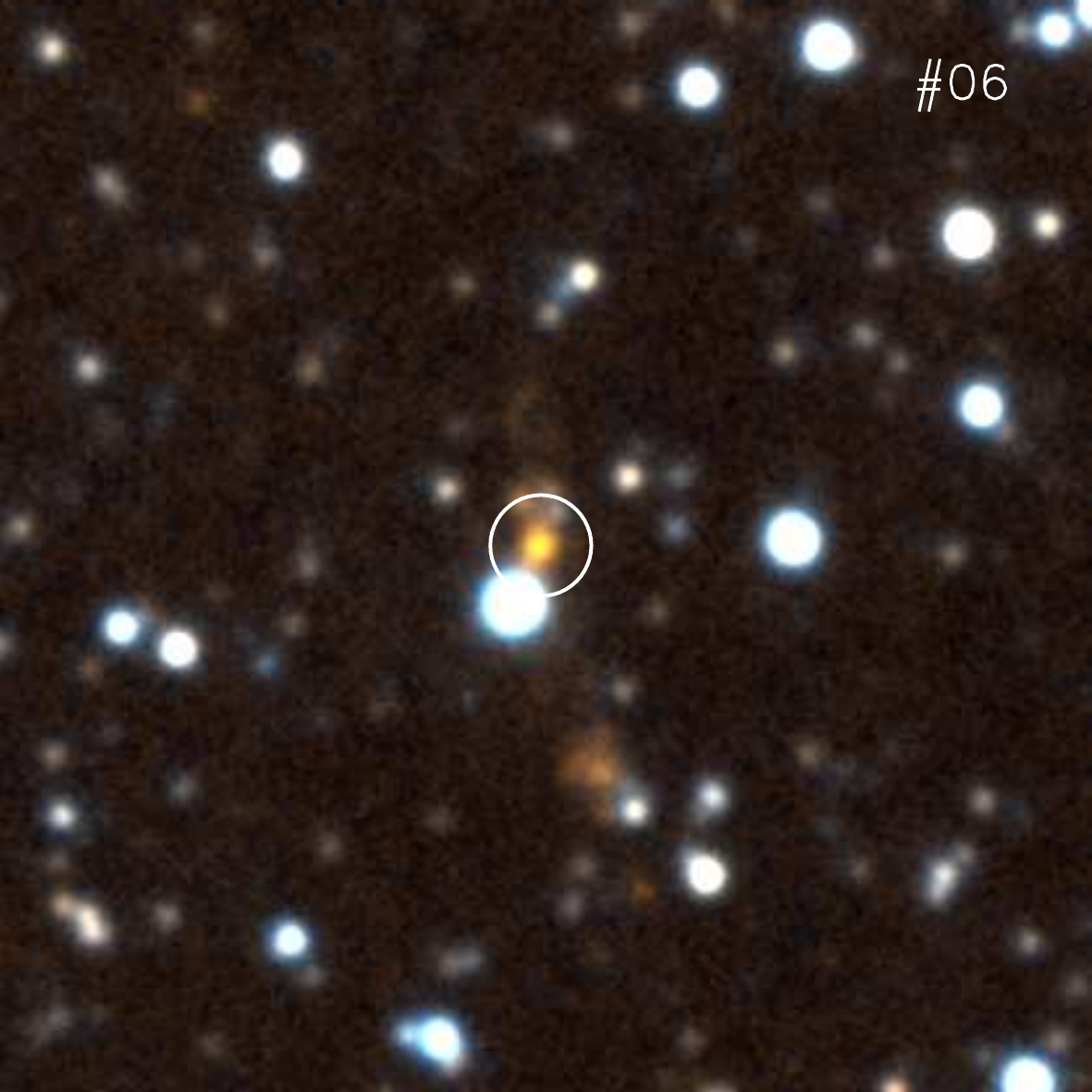}
\includegraphics[scale=0.363]{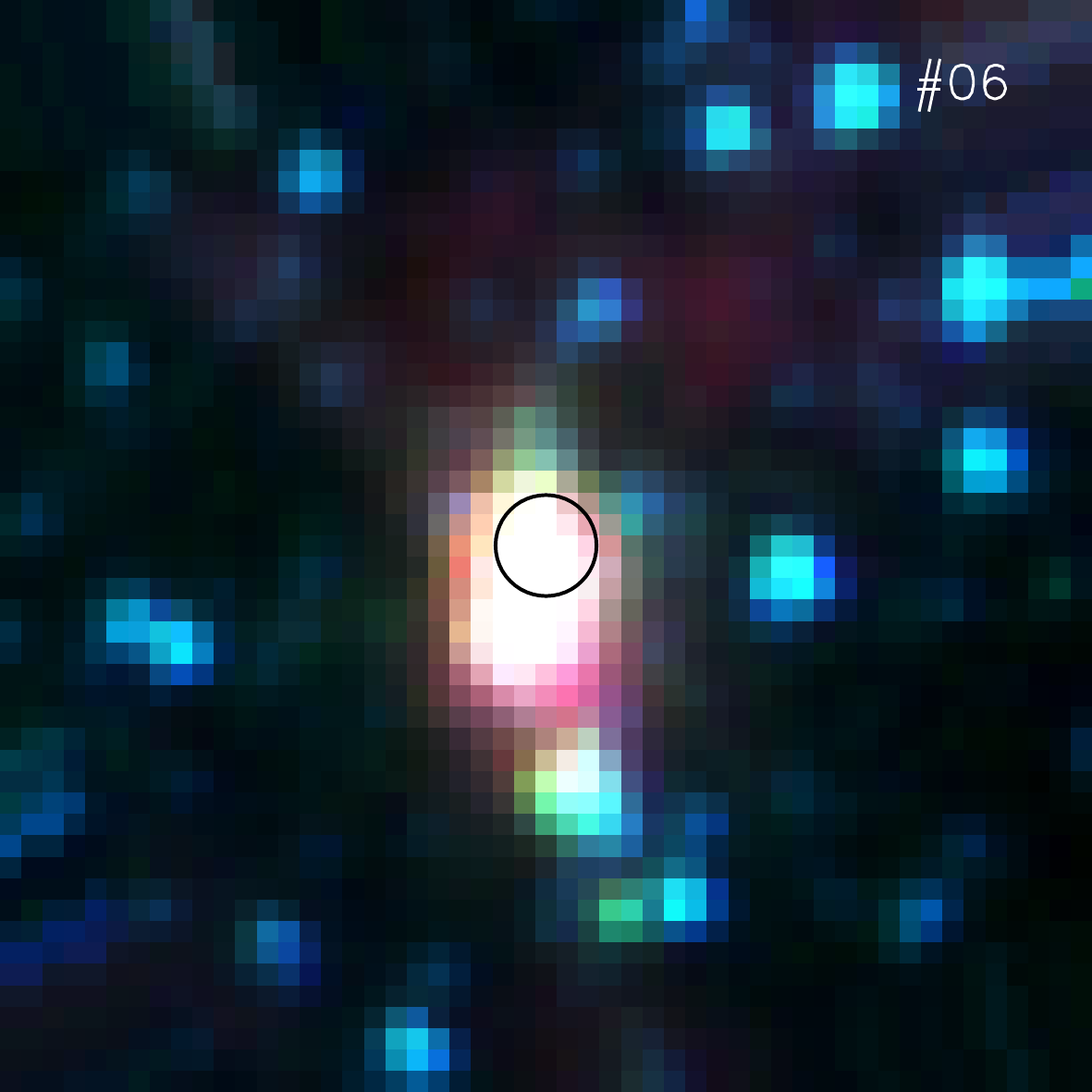}
\includegraphics[scale=0.363]{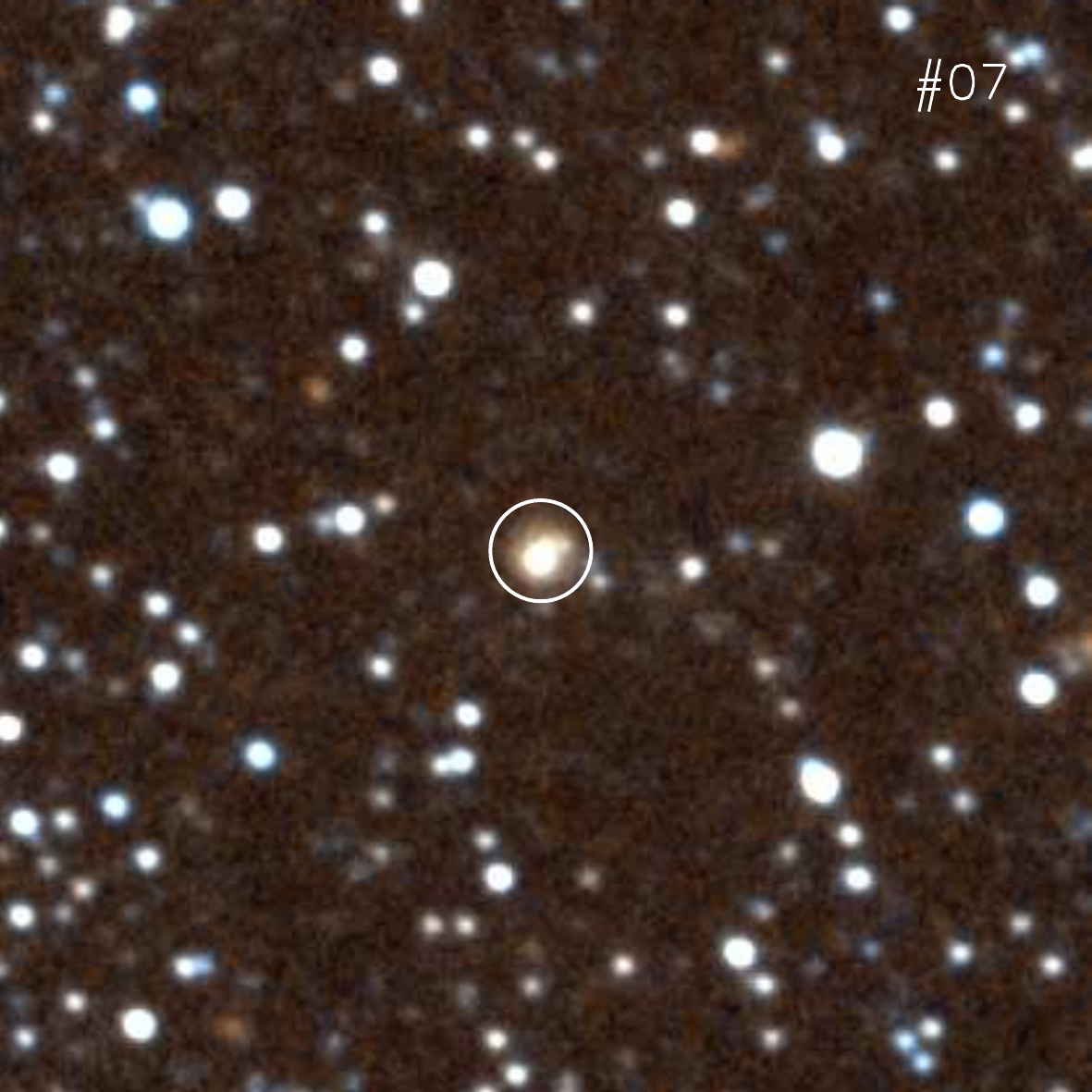}
\includegraphics[scale=0.363]{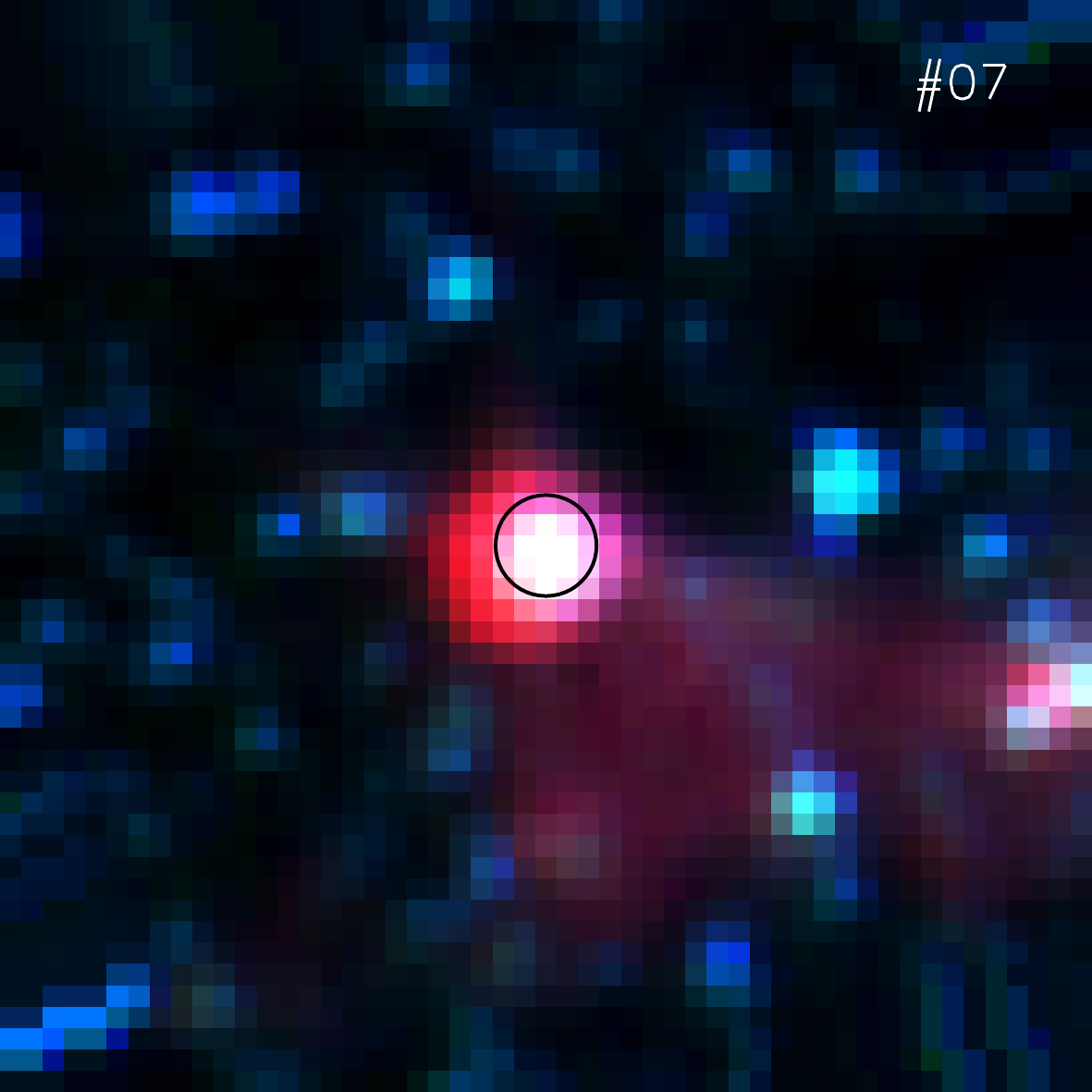}
\includegraphics[scale=0.363]{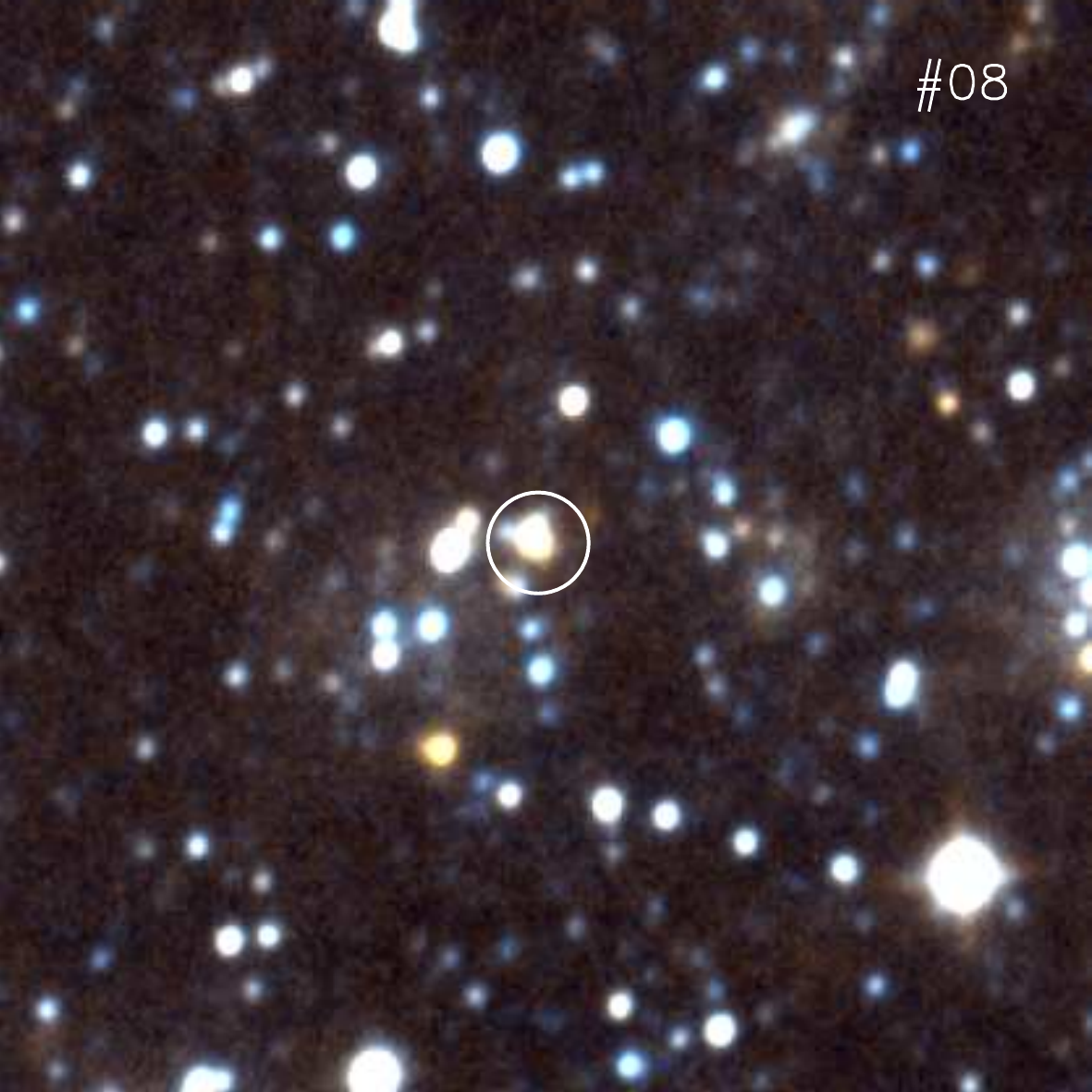}
\includegraphics[scale=0.363]{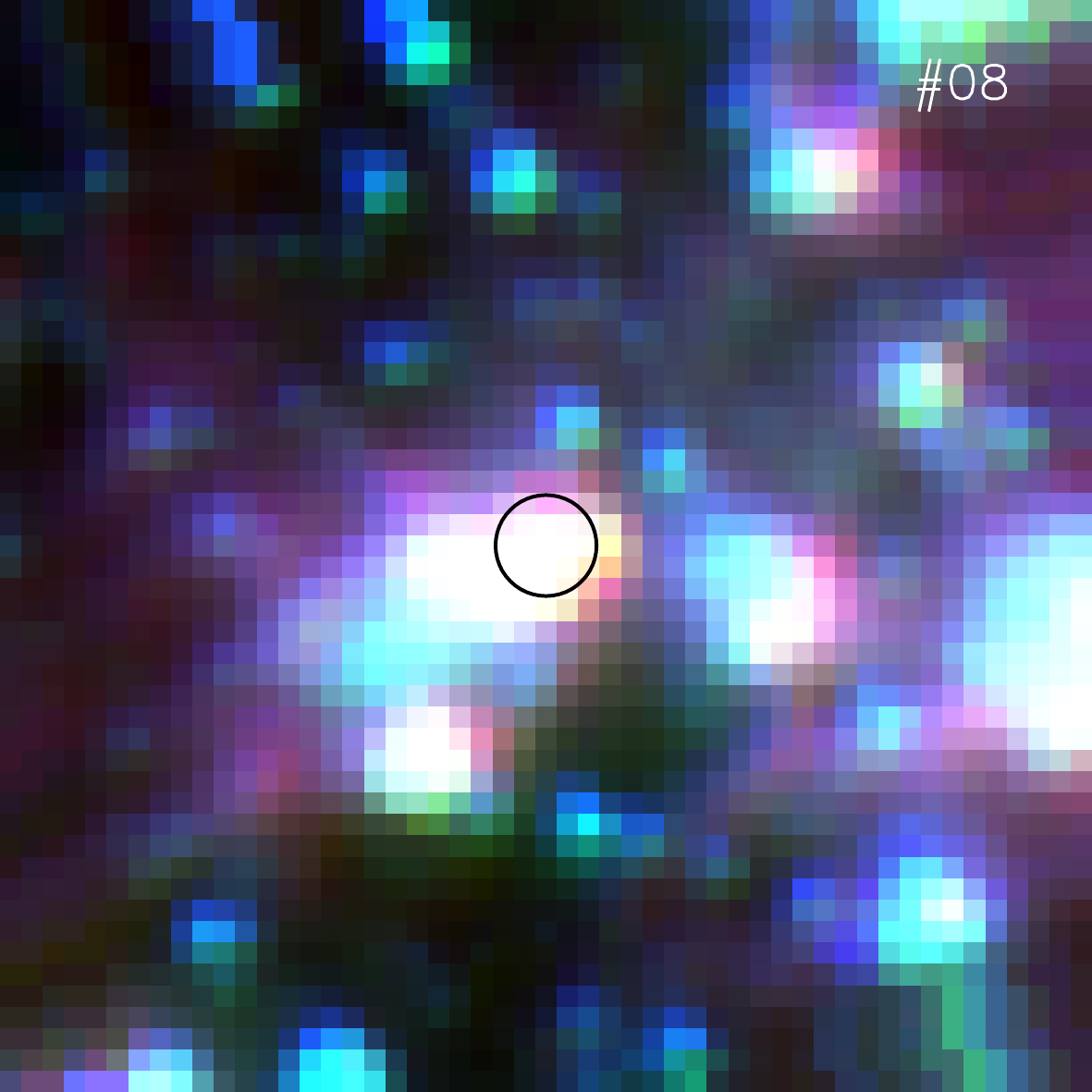}
\includegraphics[scale=0.363]{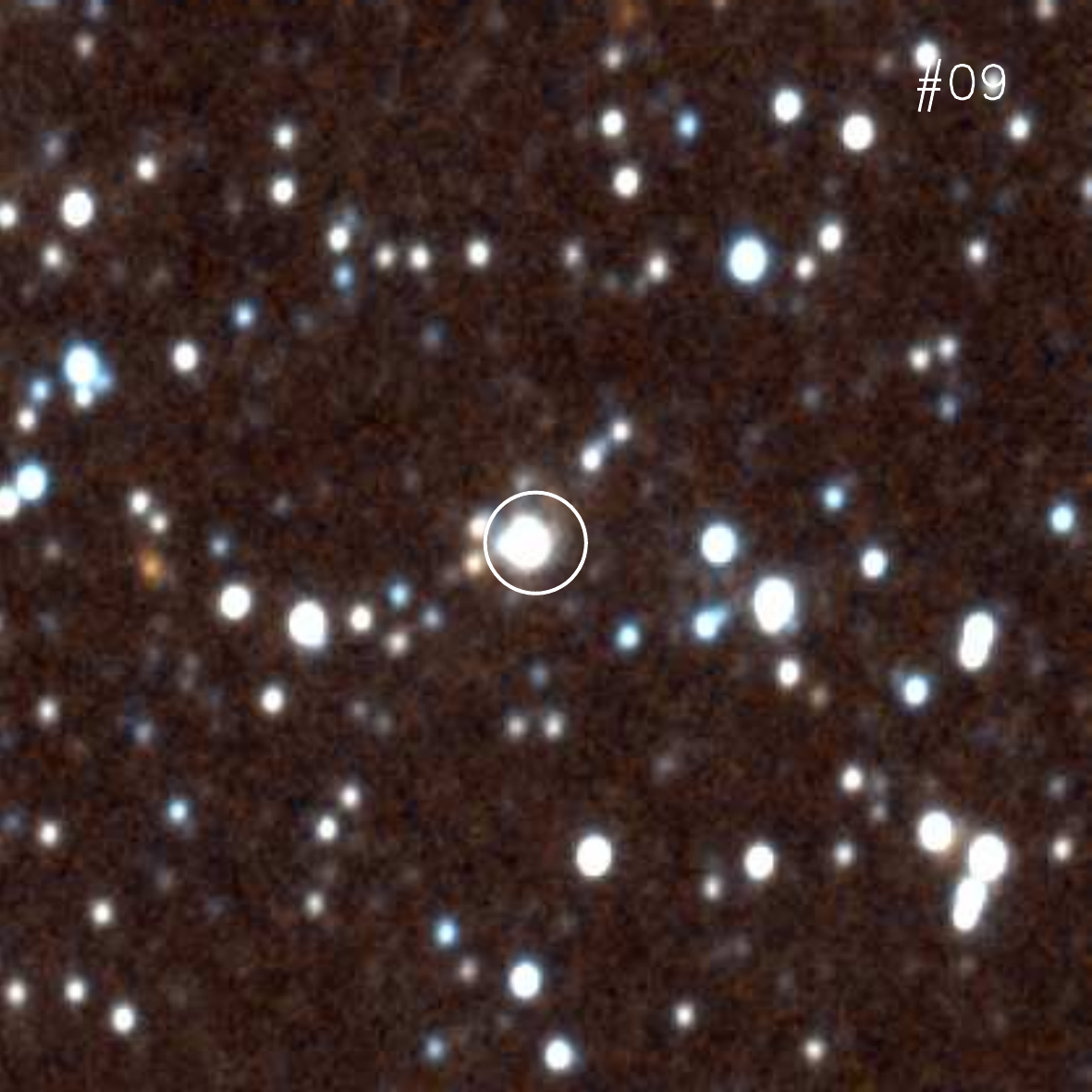}
\includegraphics[scale=0.363]{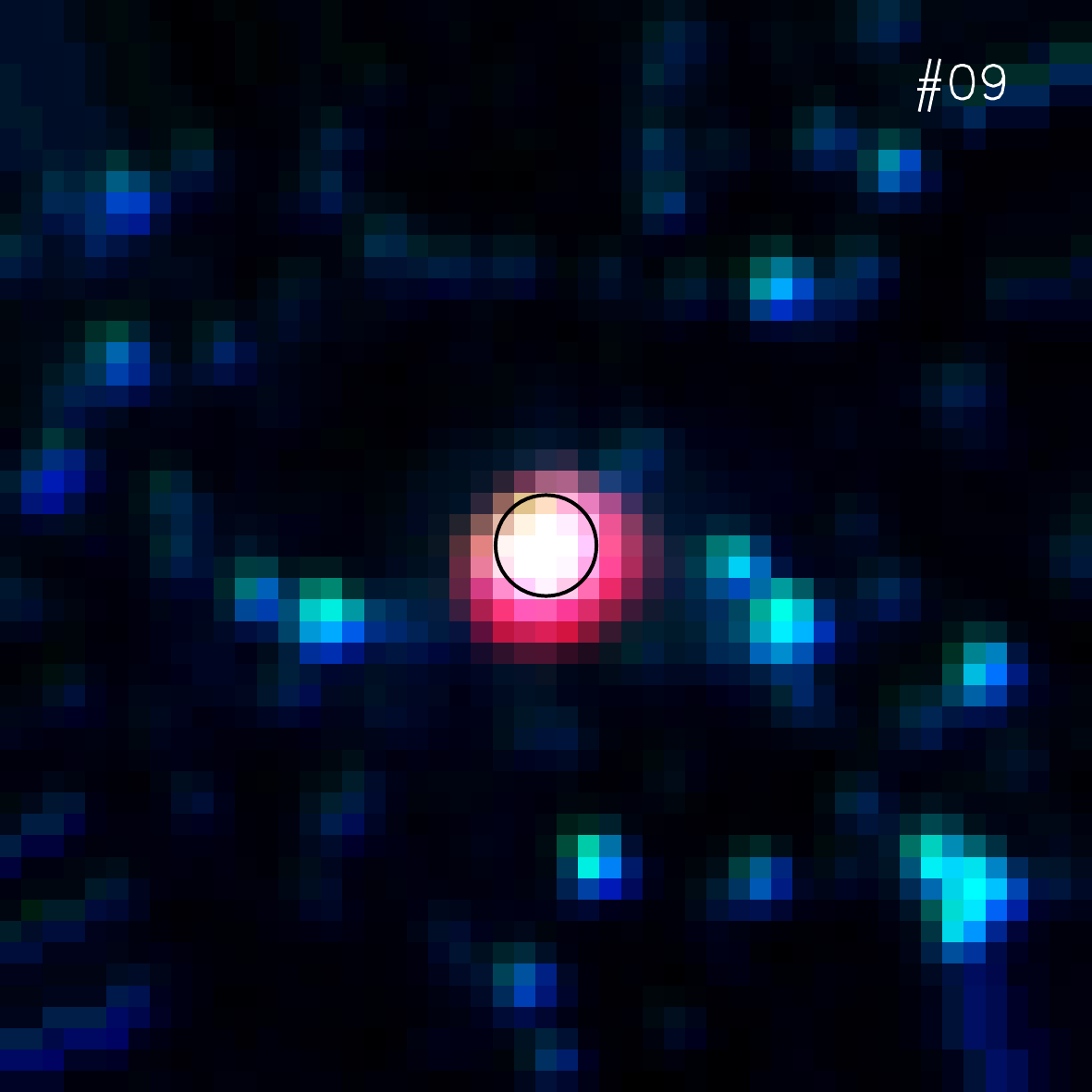}
\includegraphics[scale=0.363]{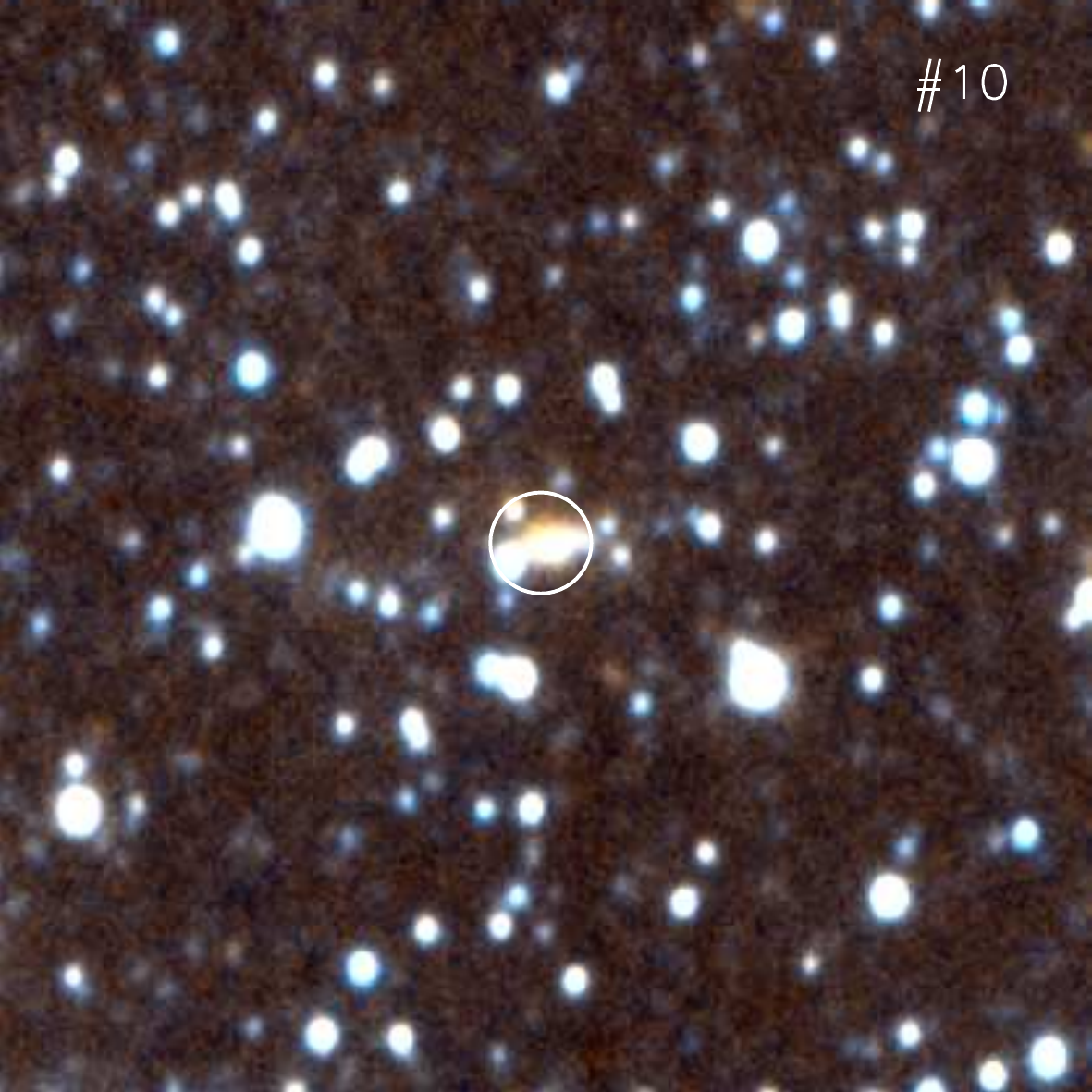}
\includegraphics[scale=0.363]{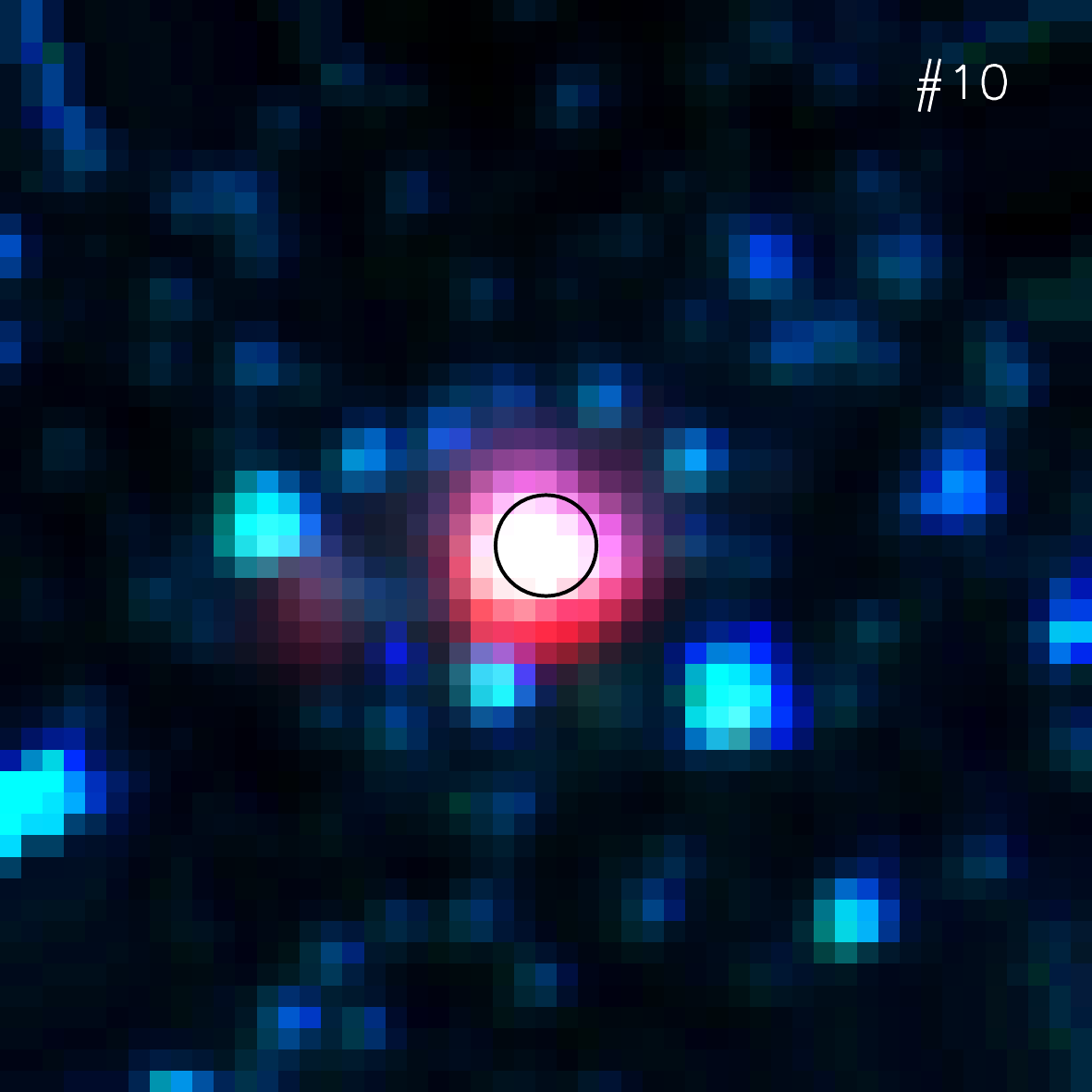}
\caption{Colour-composite images for the YSO candidates in the SMC. For each object 
$J_{\rm s}K_{\rm s}$ (blue/\,red) and \spitzer\ [3.6]-[5.8]-[8] images 
(blue/\,green/\,red) are shown. Each image is $1^\prime \times 1^\prime$ across and 
North is to the top and East to the left. For the last 2 objects we used IRSF images 
and show the $JHK$ colour-composite.}
\label{thumbs}
\end{figure*}
\setcounter{figure}{0}
\begin{figure*}
\includegraphics[scale=0.363]{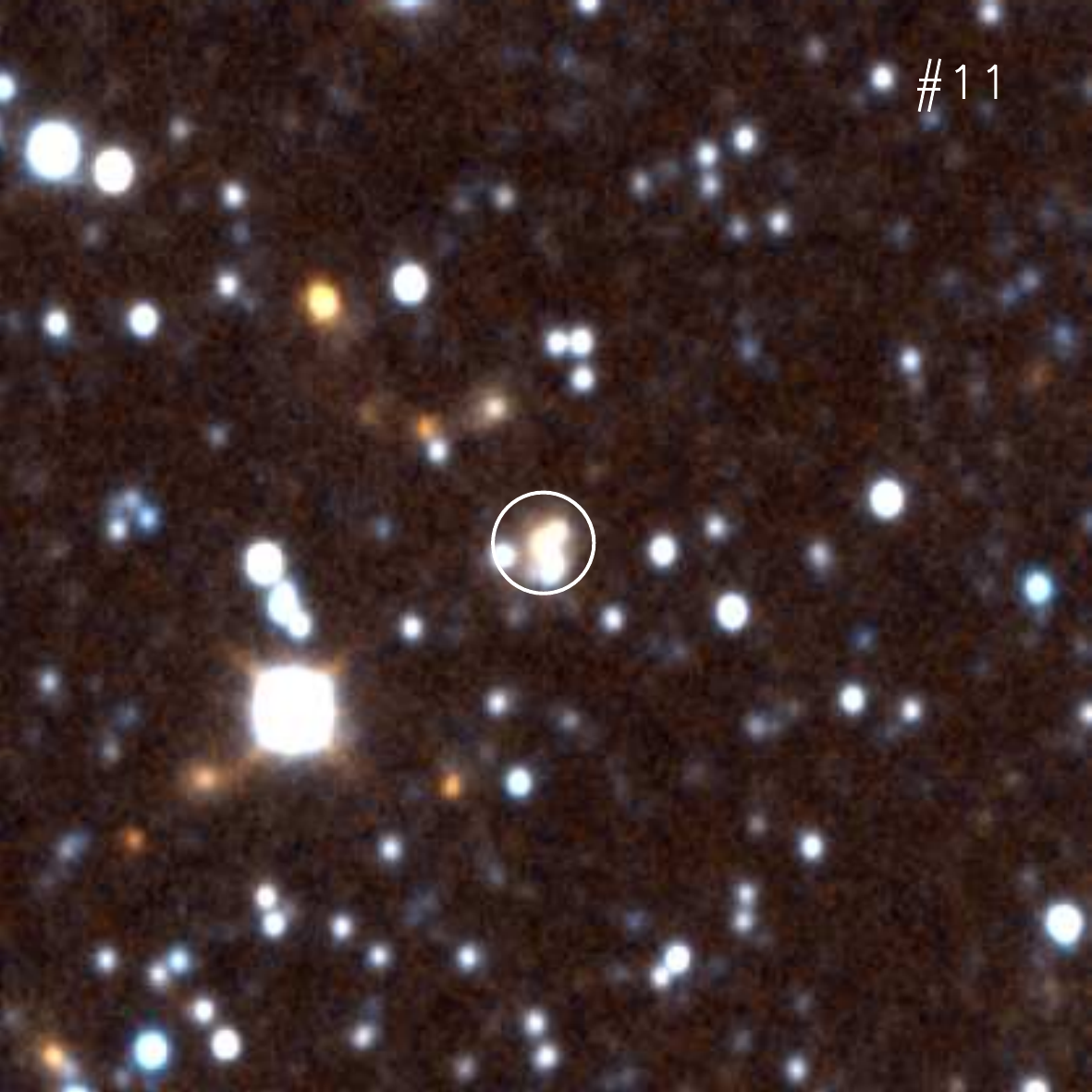}
\includegraphics[scale=0.363]{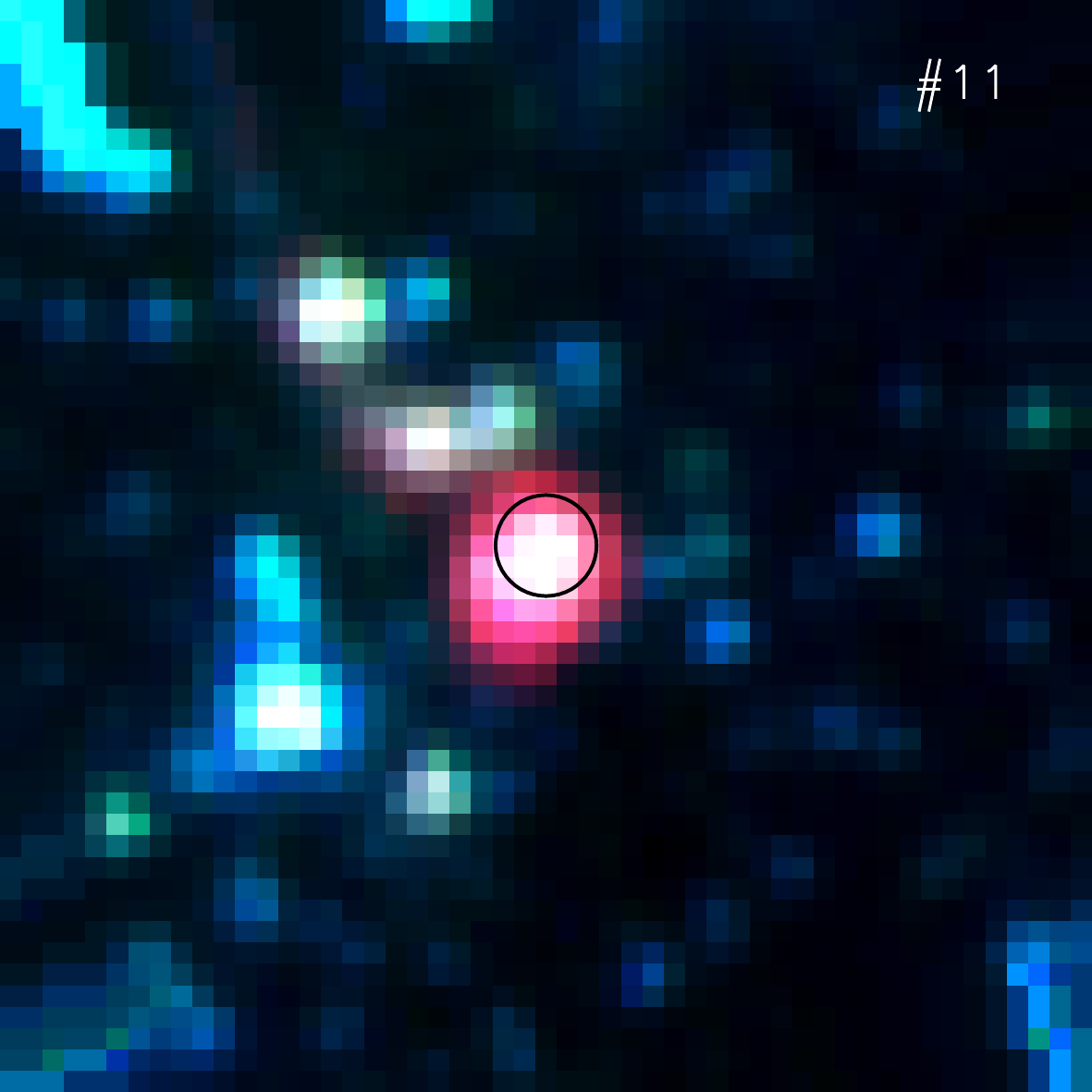}
\includegraphics[scale=0.363]{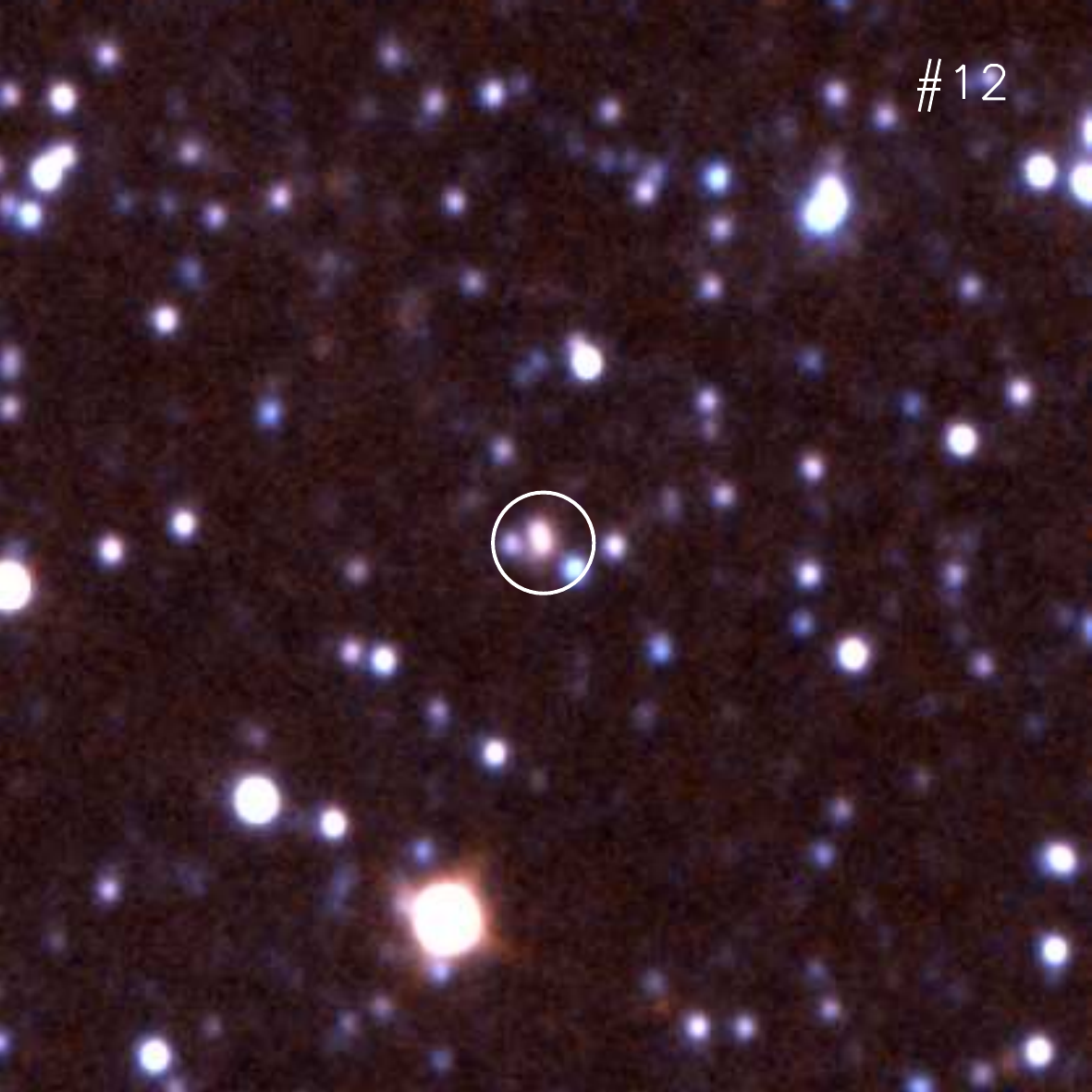}
\includegraphics[scale=0.363]{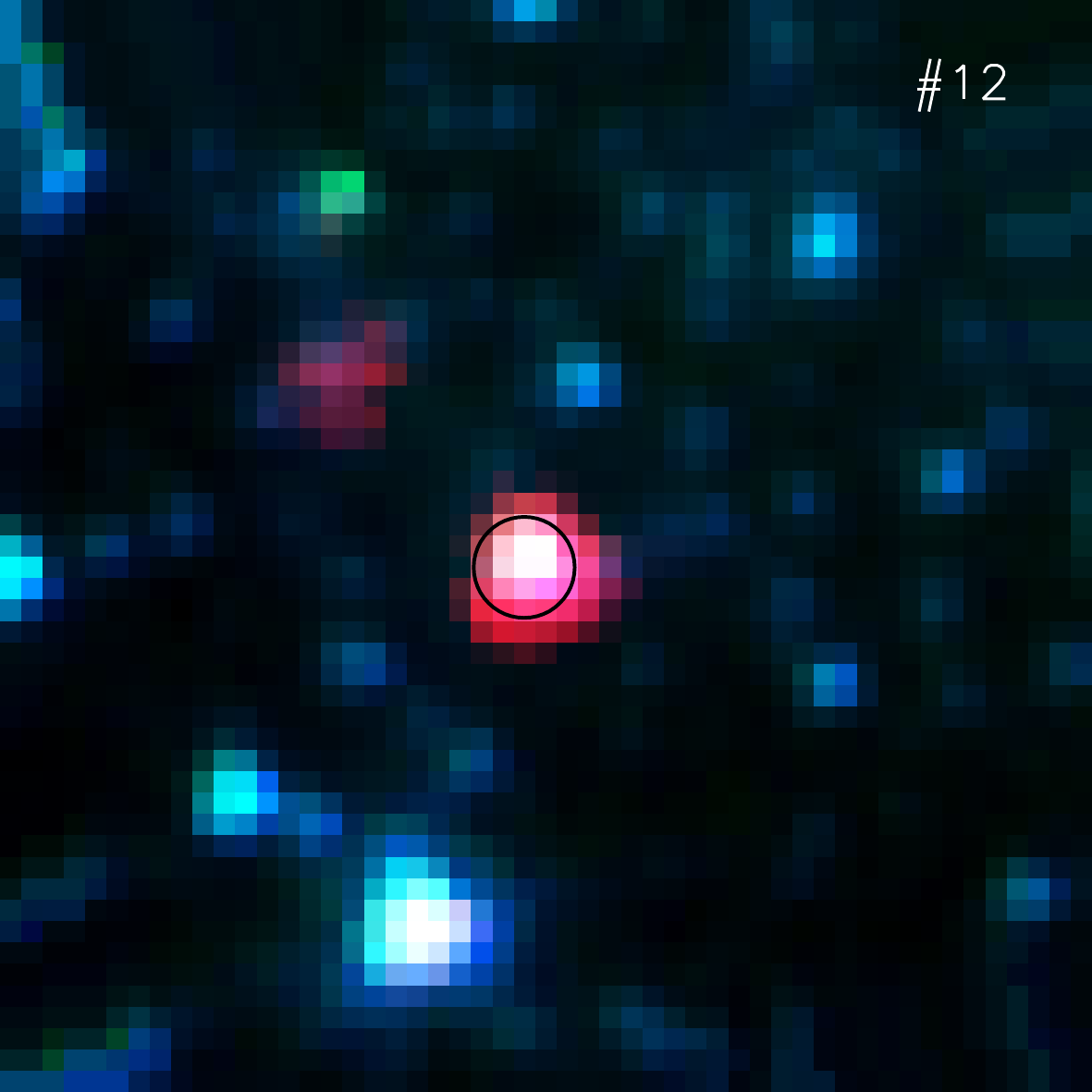}
\includegraphics[scale=0.363]{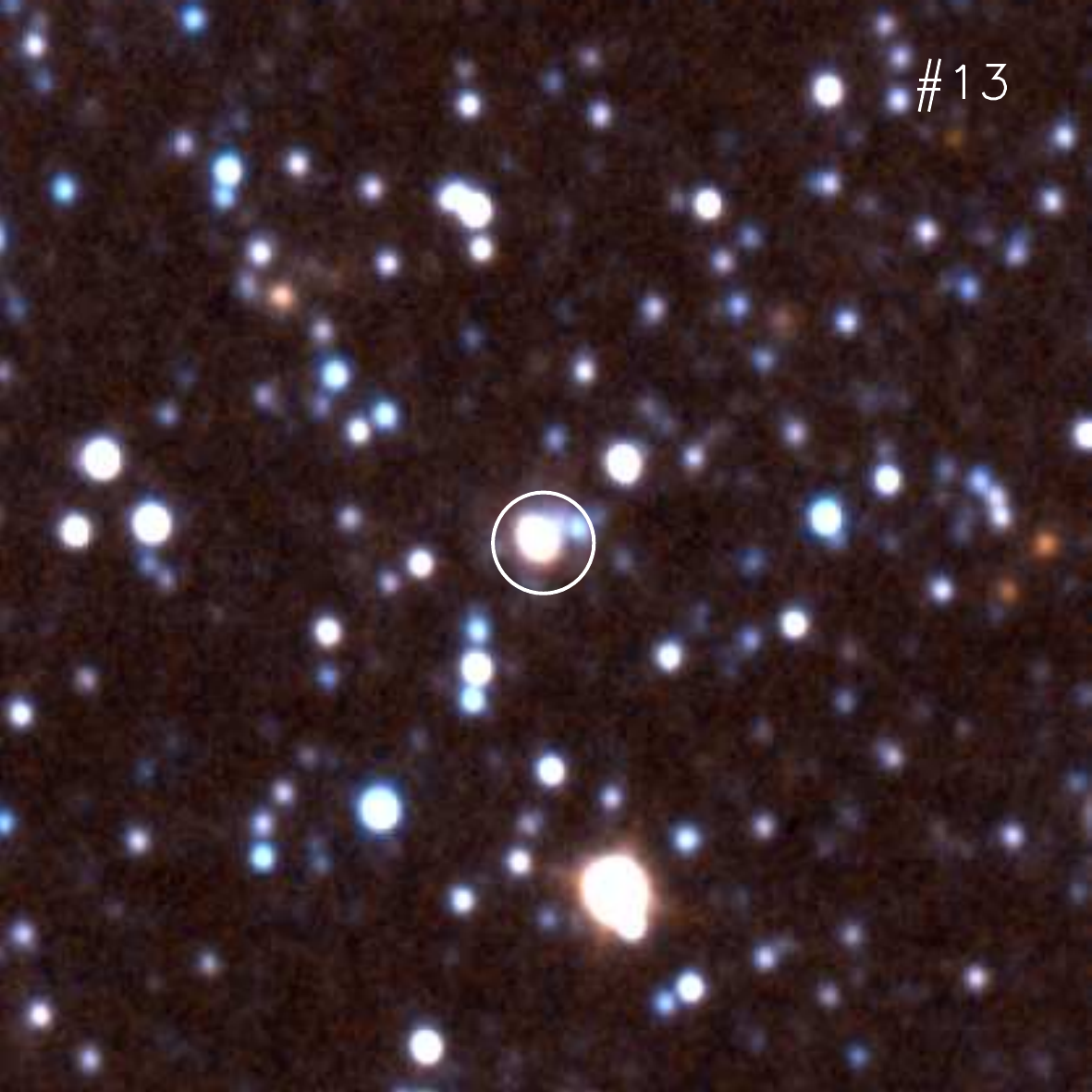}
\includegraphics[scale=0.363]{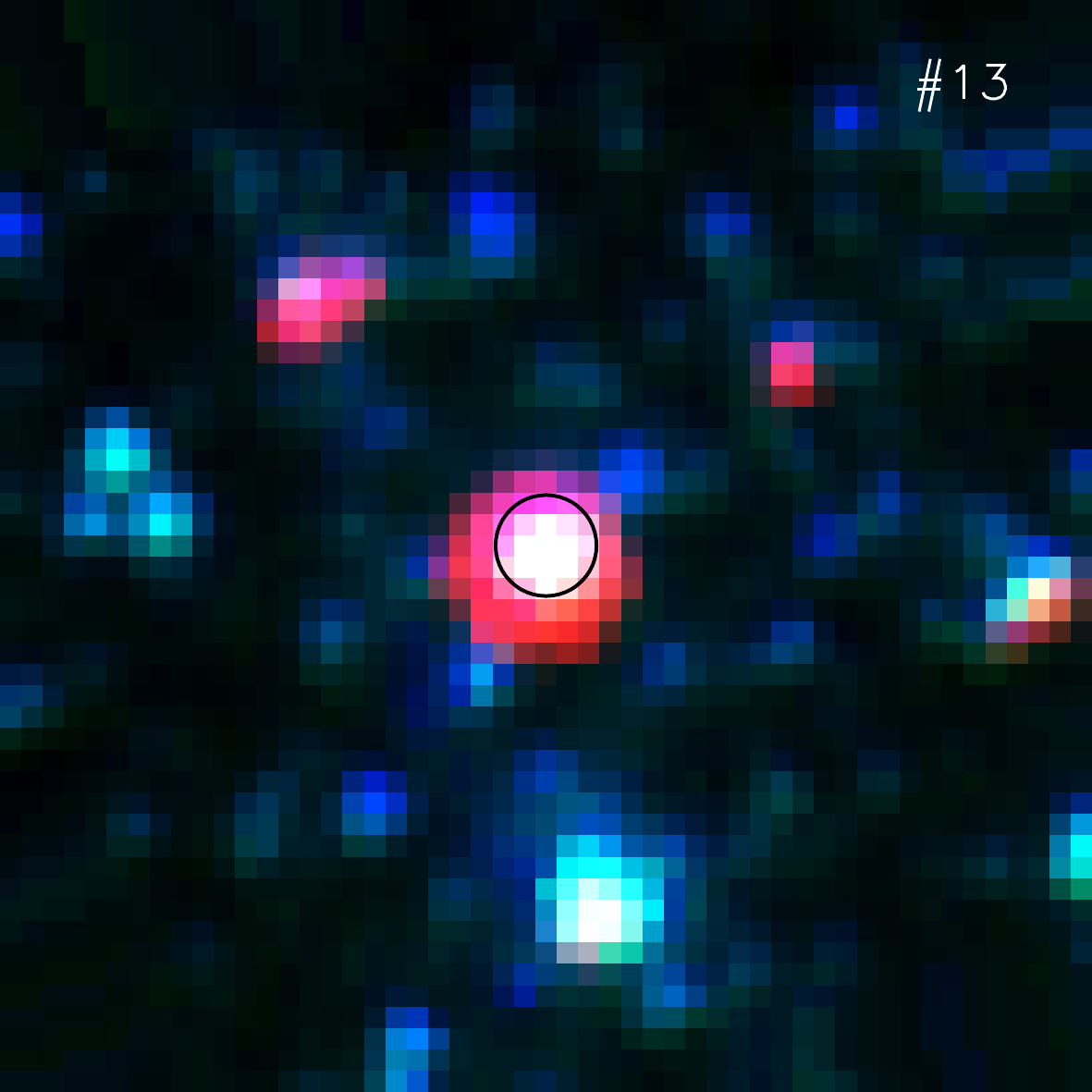}
\includegraphics[scale=0.363]{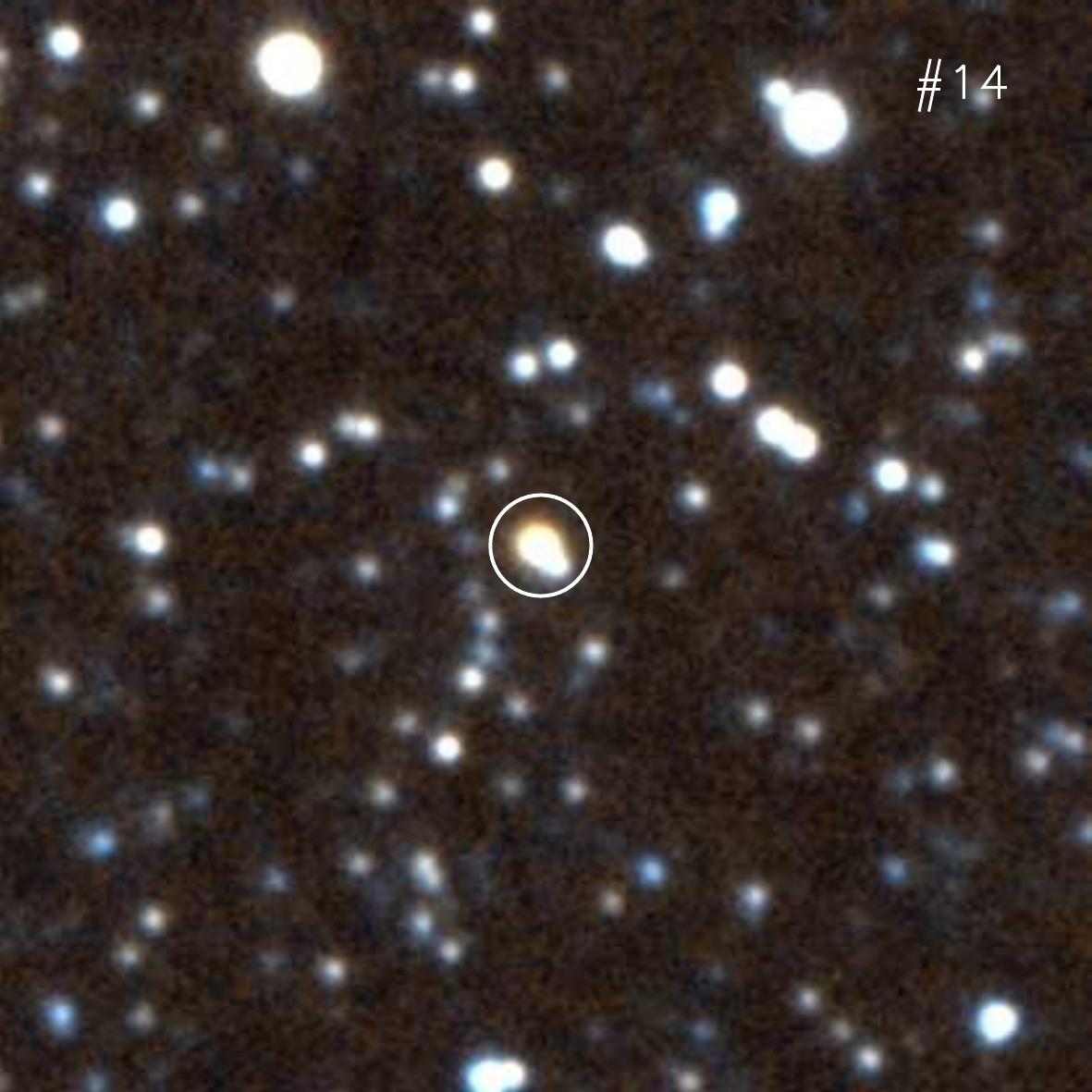}
\includegraphics[scale=0.363]{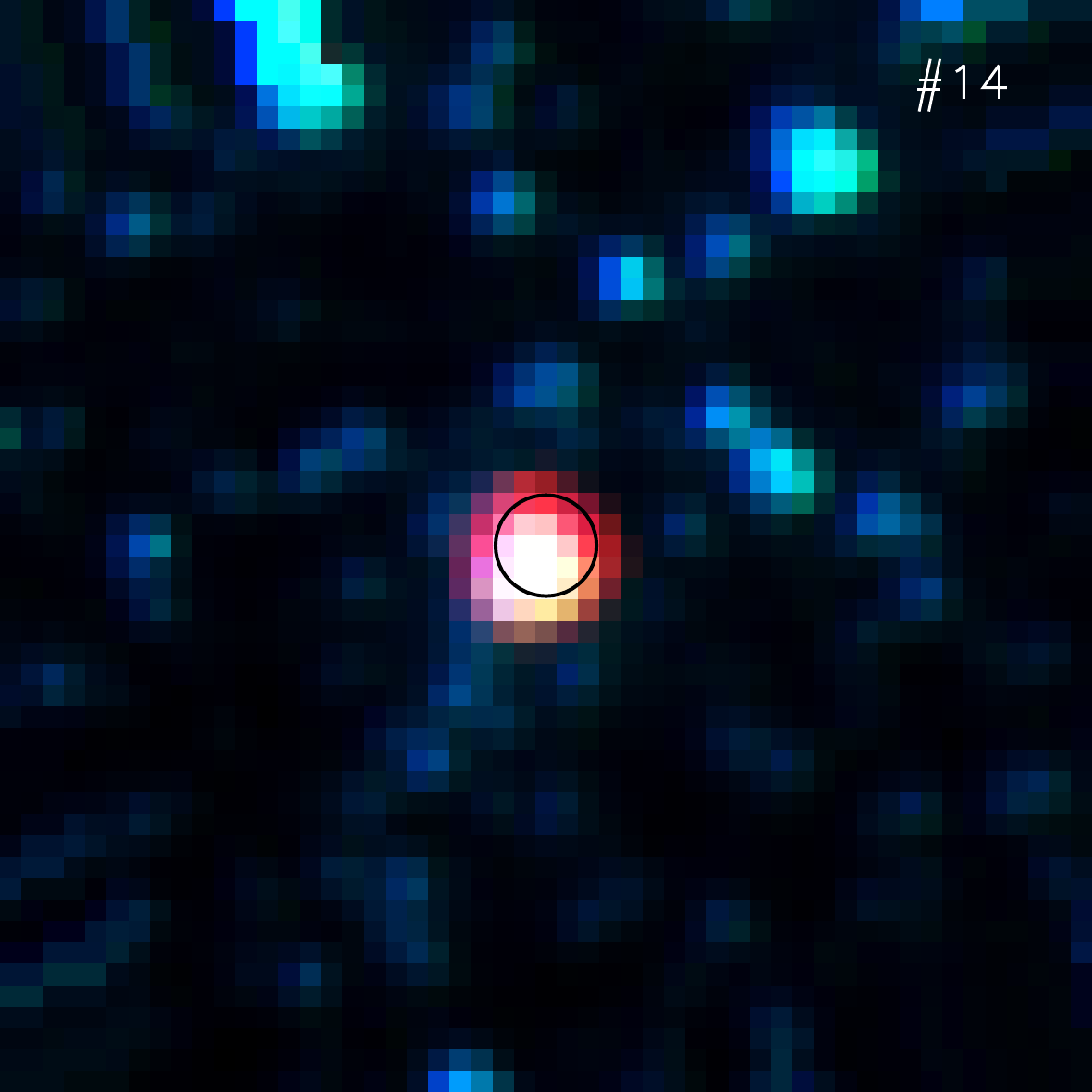}
\includegraphics[scale=0.363]{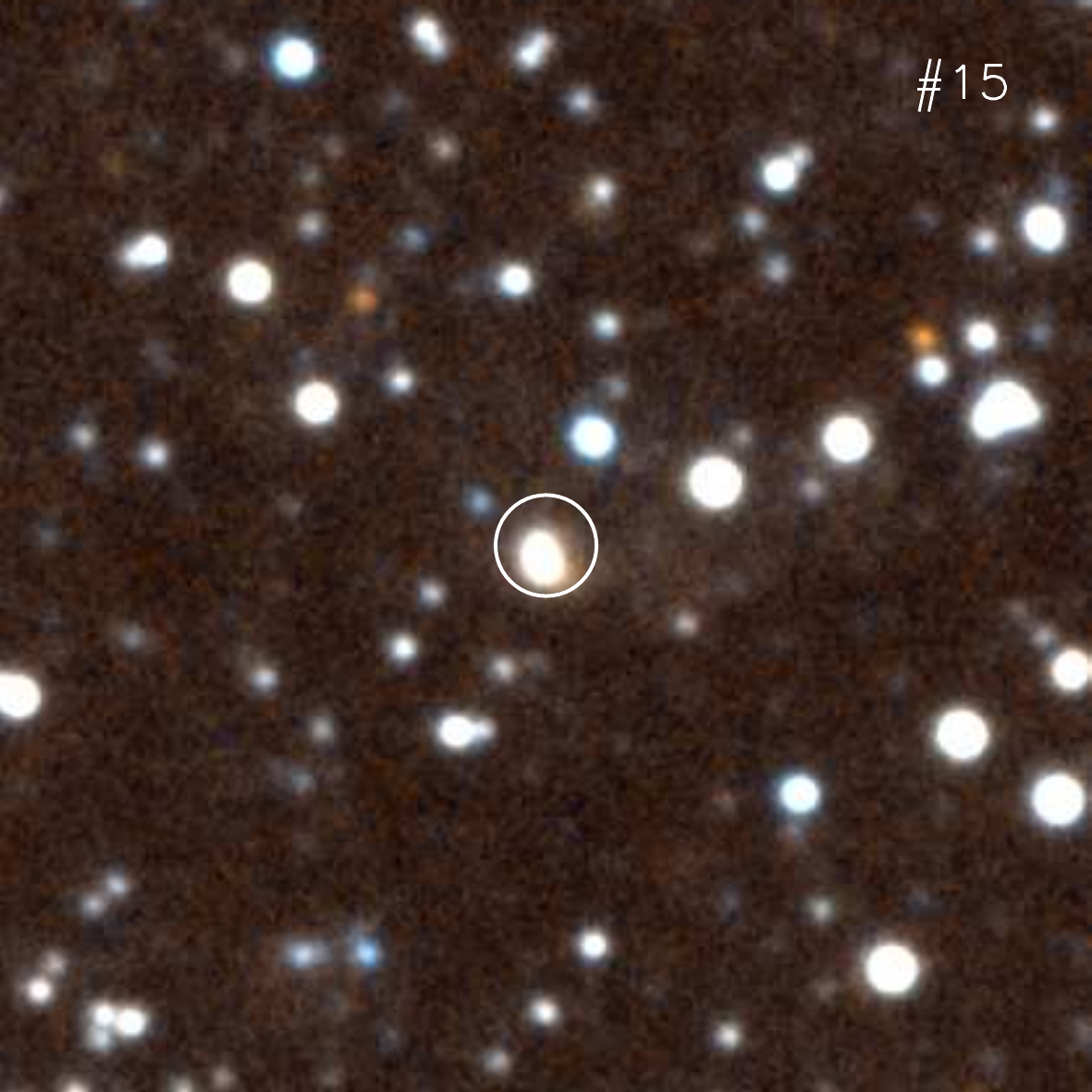}
\includegraphics[scale=0.363]{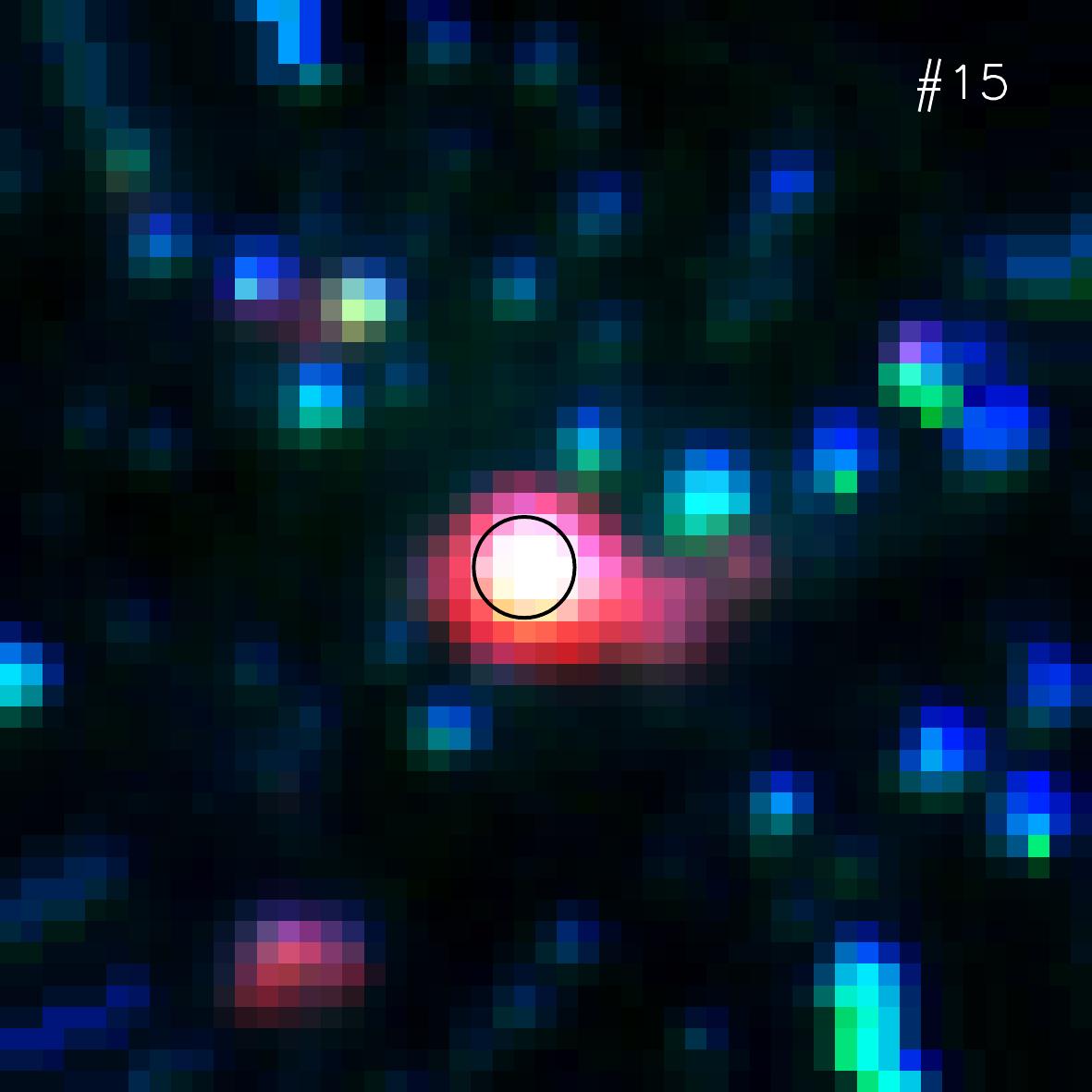}
\includegraphics[scale=0.363]{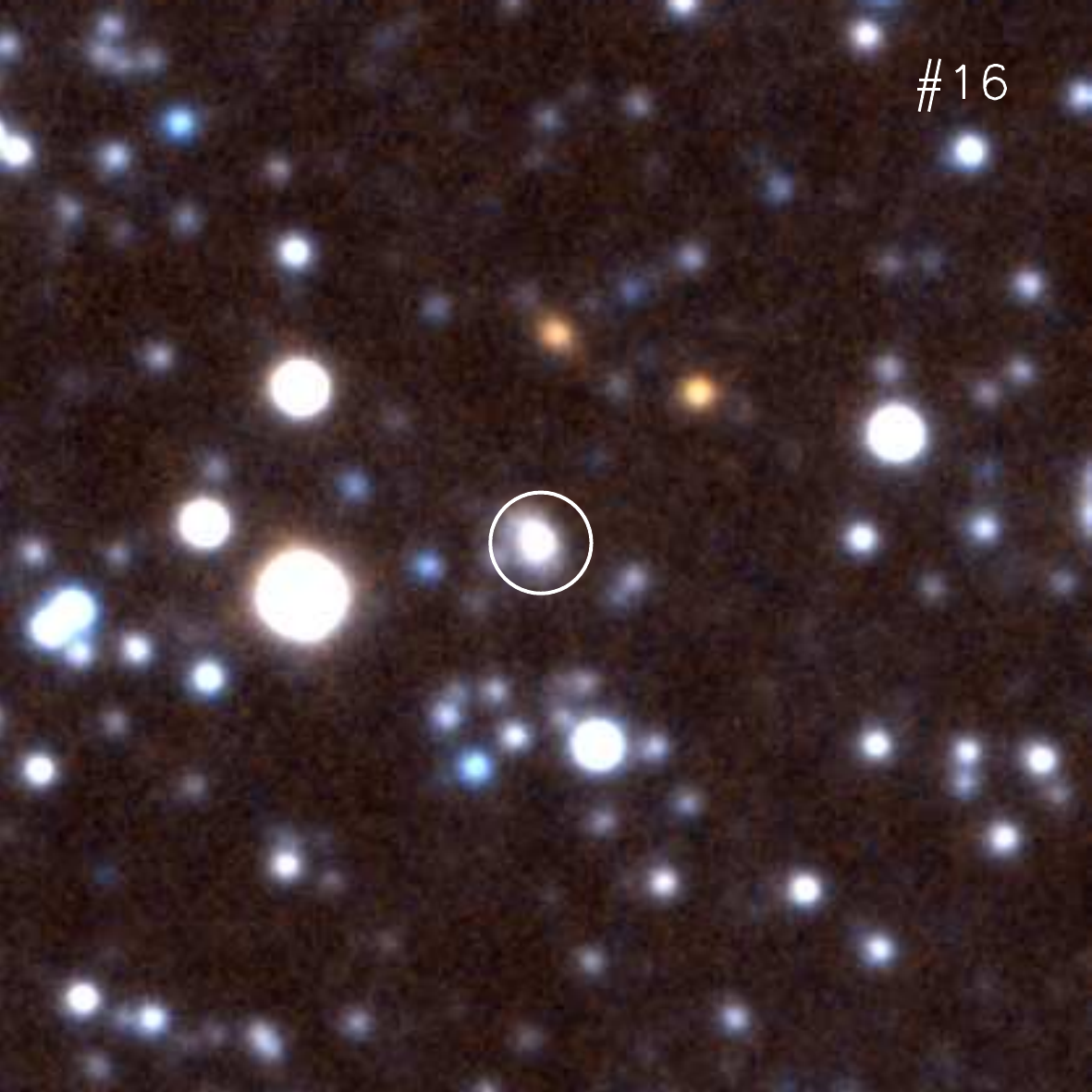}
\includegraphics[scale=0.363]{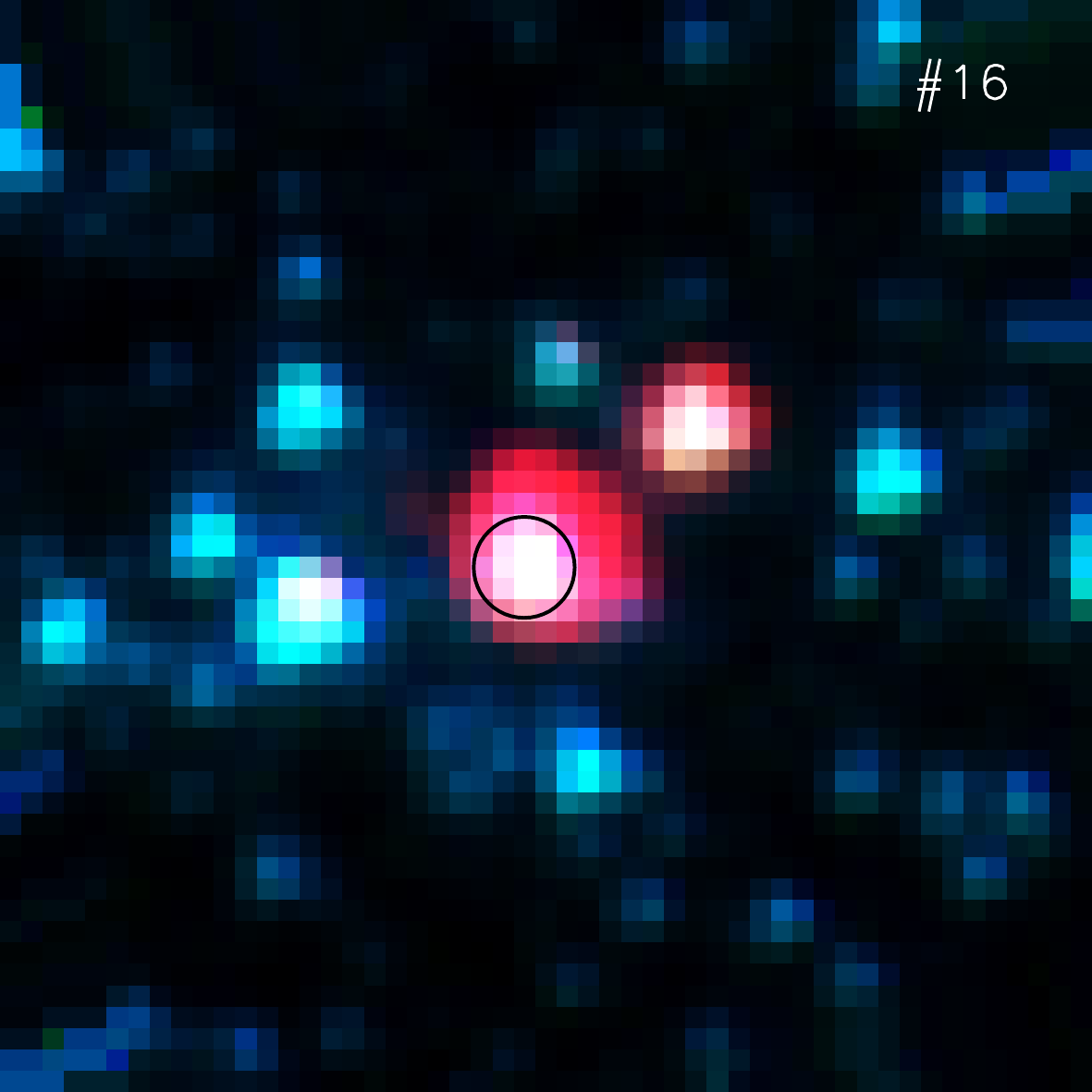}
\includegraphics[scale=0.363]{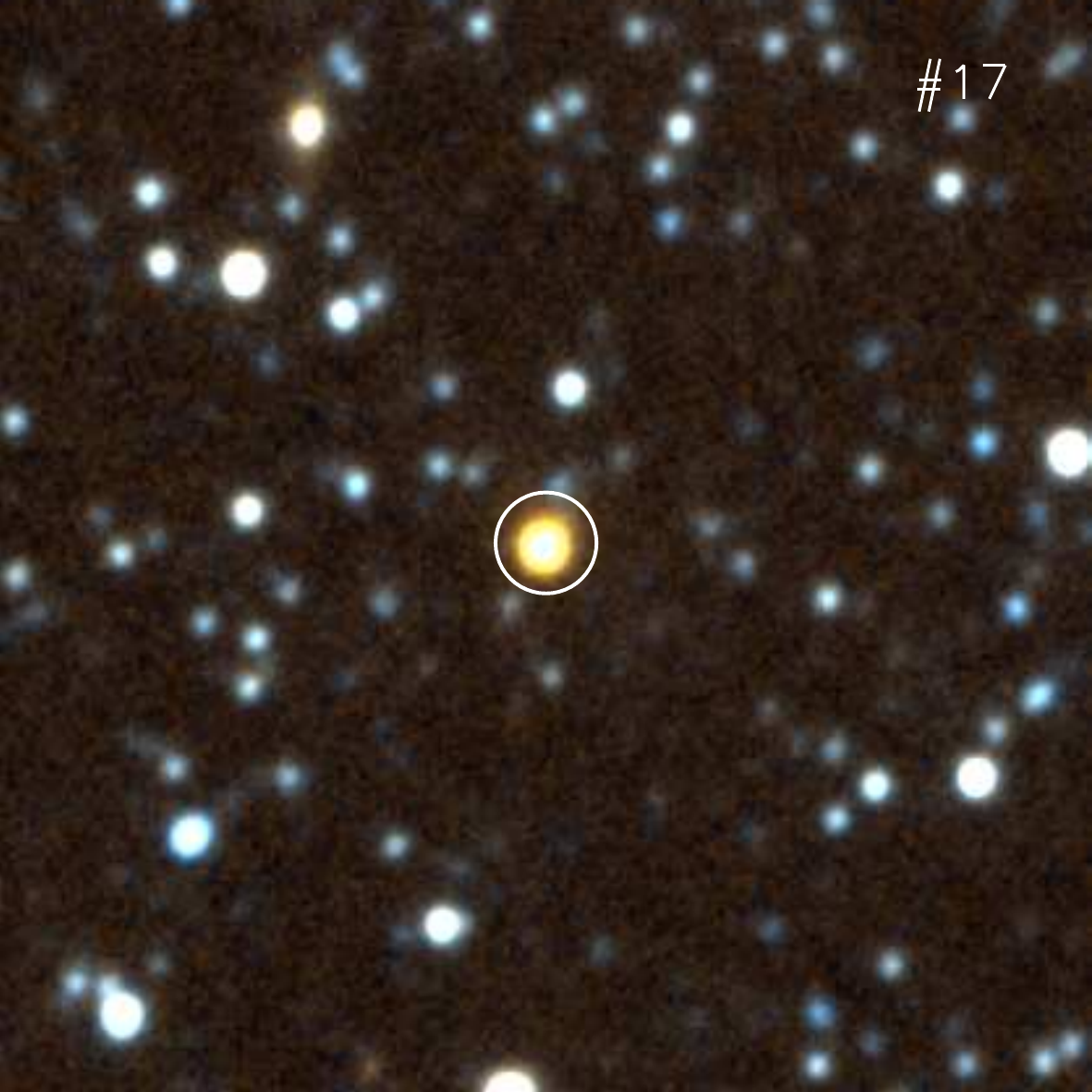}
\includegraphics[scale=0.363]{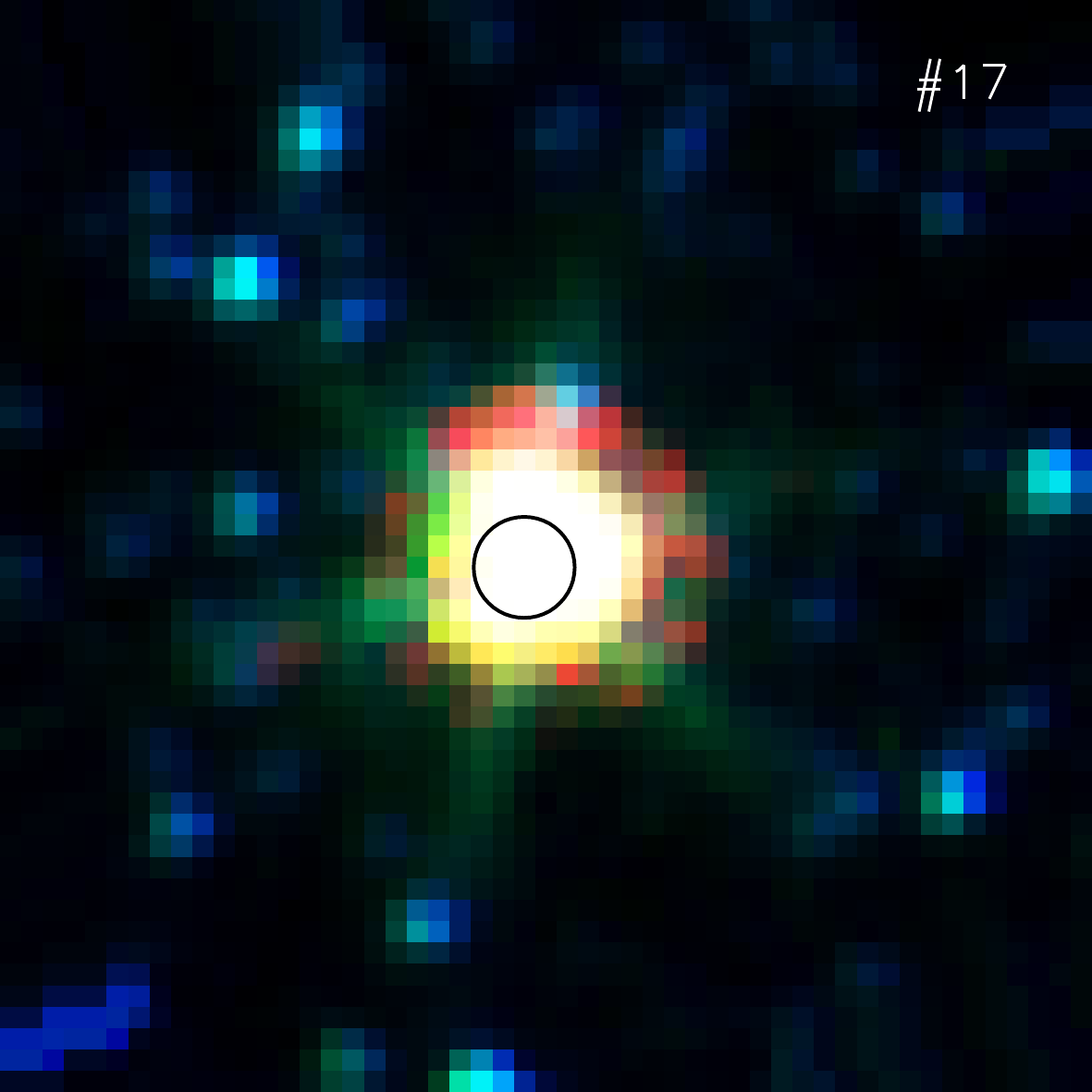}
\includegraphics[scale=0.363]{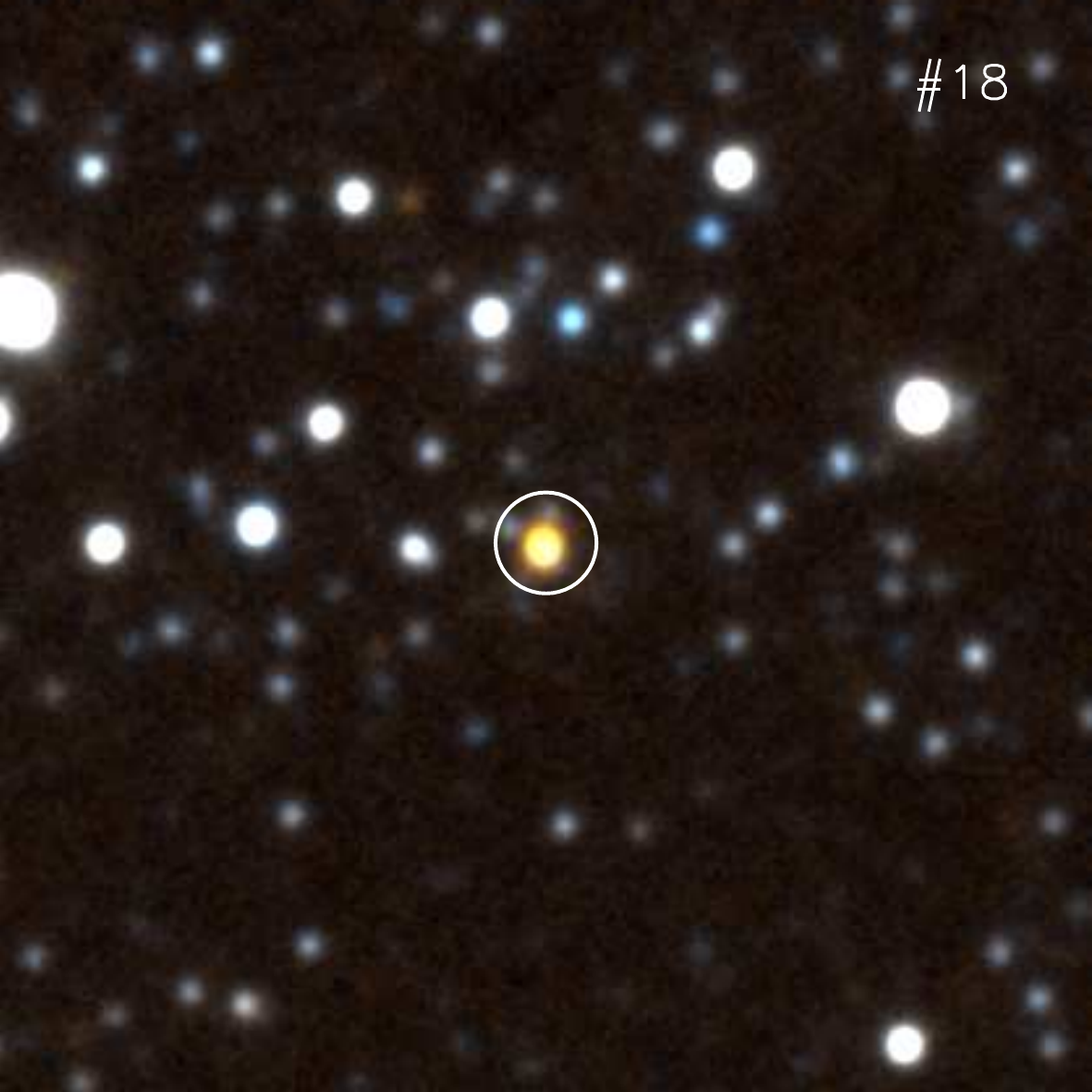}
\includegraphics[scale=0.363]{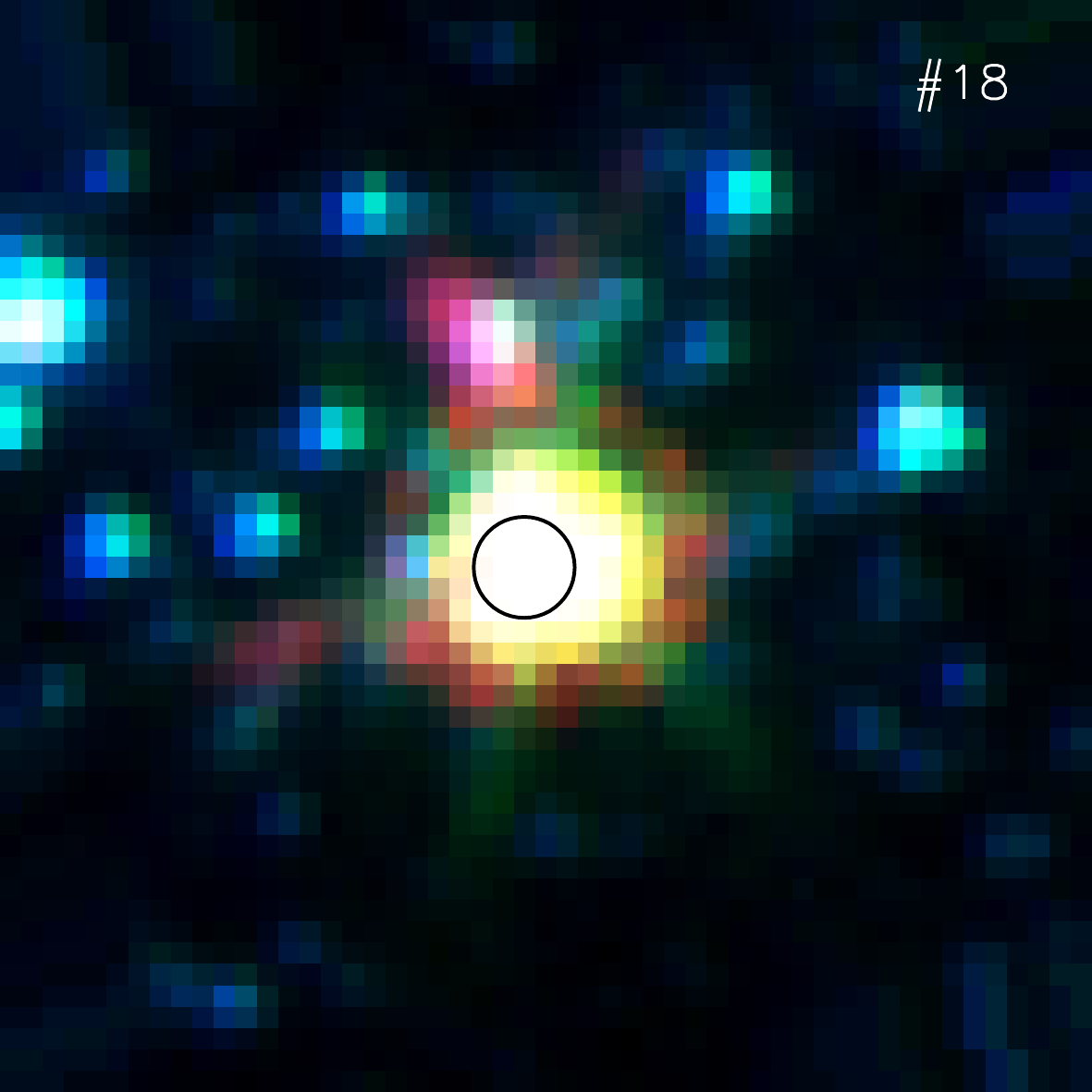}
\includegraphics[scale=0.363]{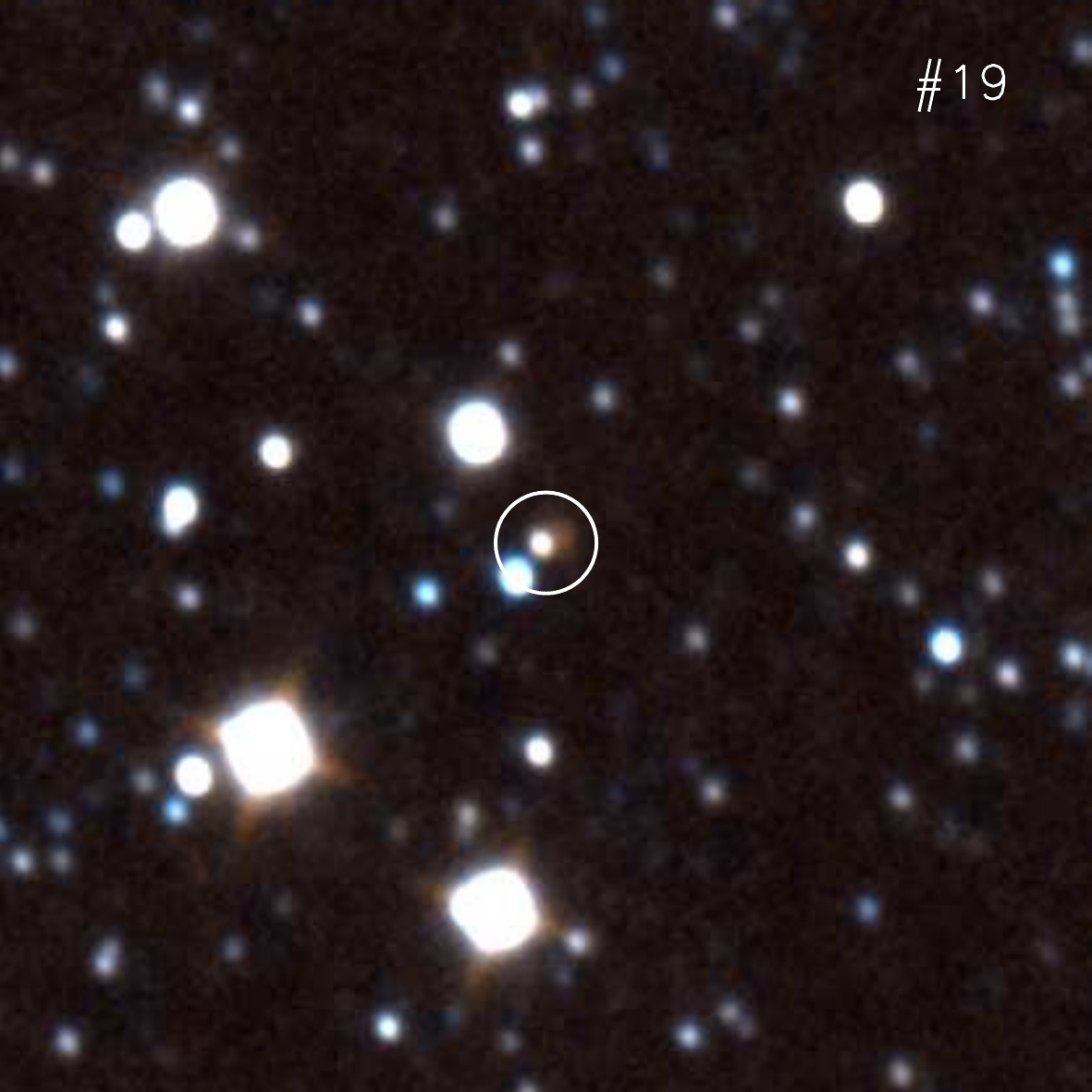}
\includegraphics[scale=0.363]{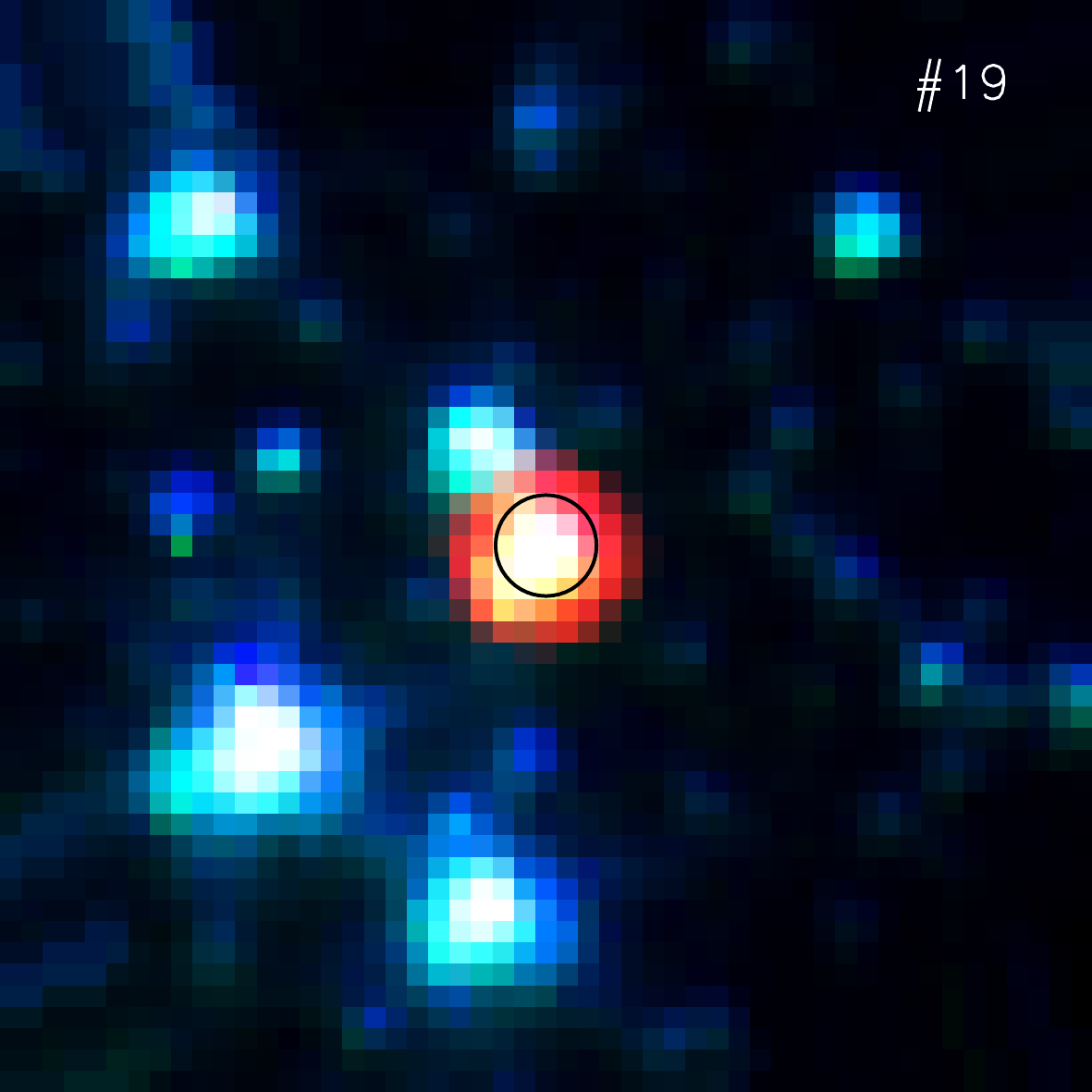}
\includegraphics[scale=0.363]{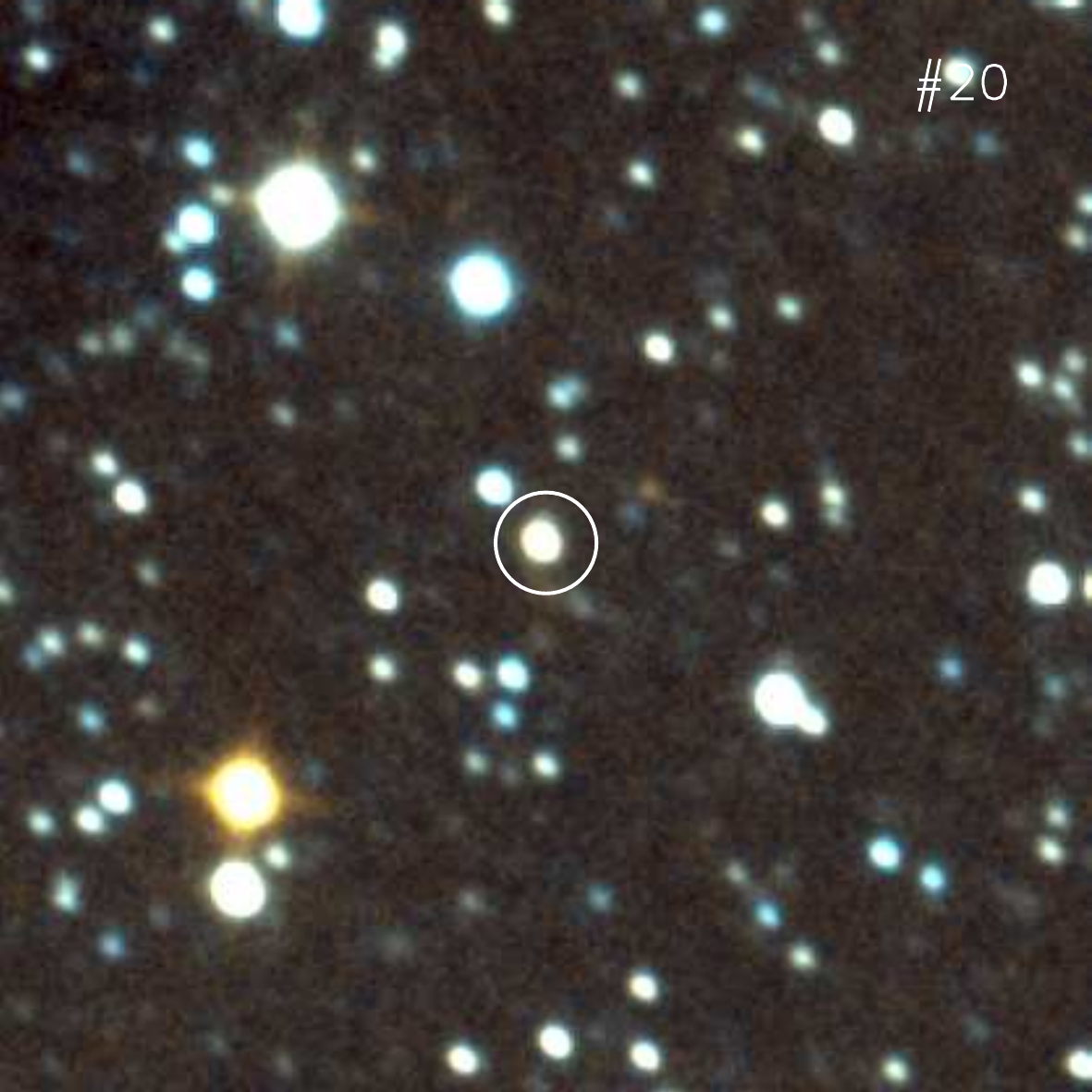}
\includegraphics[scale=0.363]{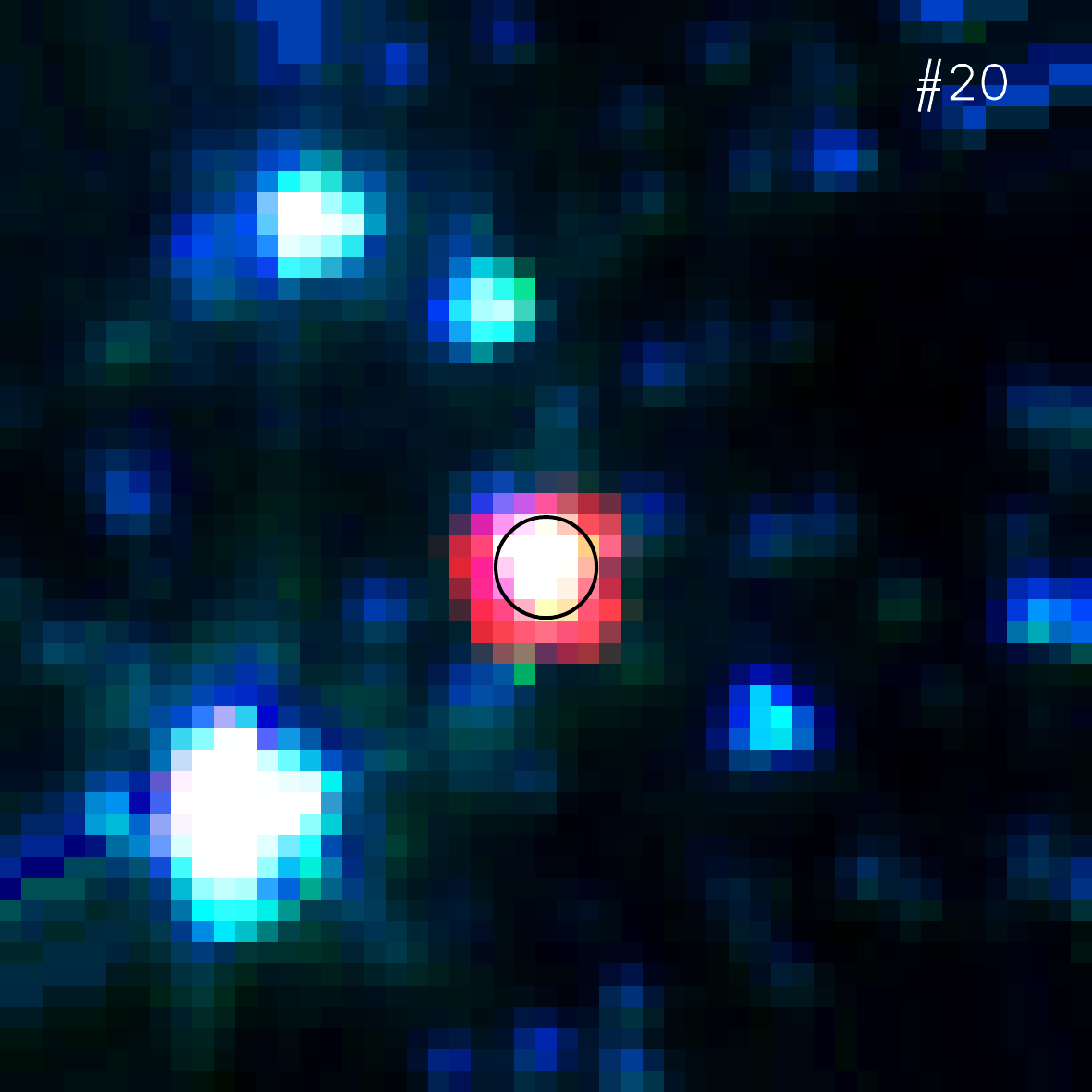}
\caption{Continued.}
\end{figure*}
\begin{figure*}
\includegraphics[scale=0.363]{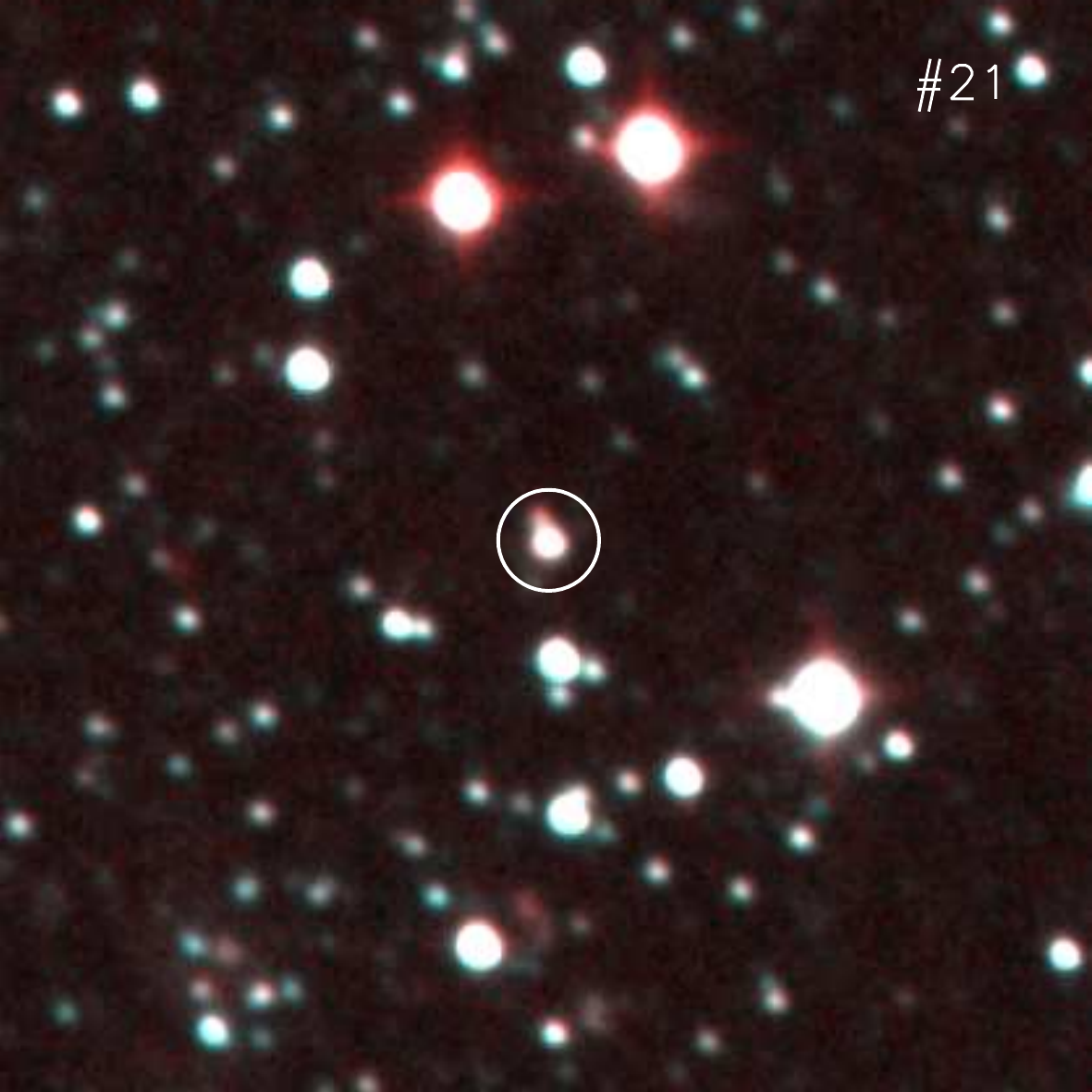}
\includegraphics[scale=0.363]{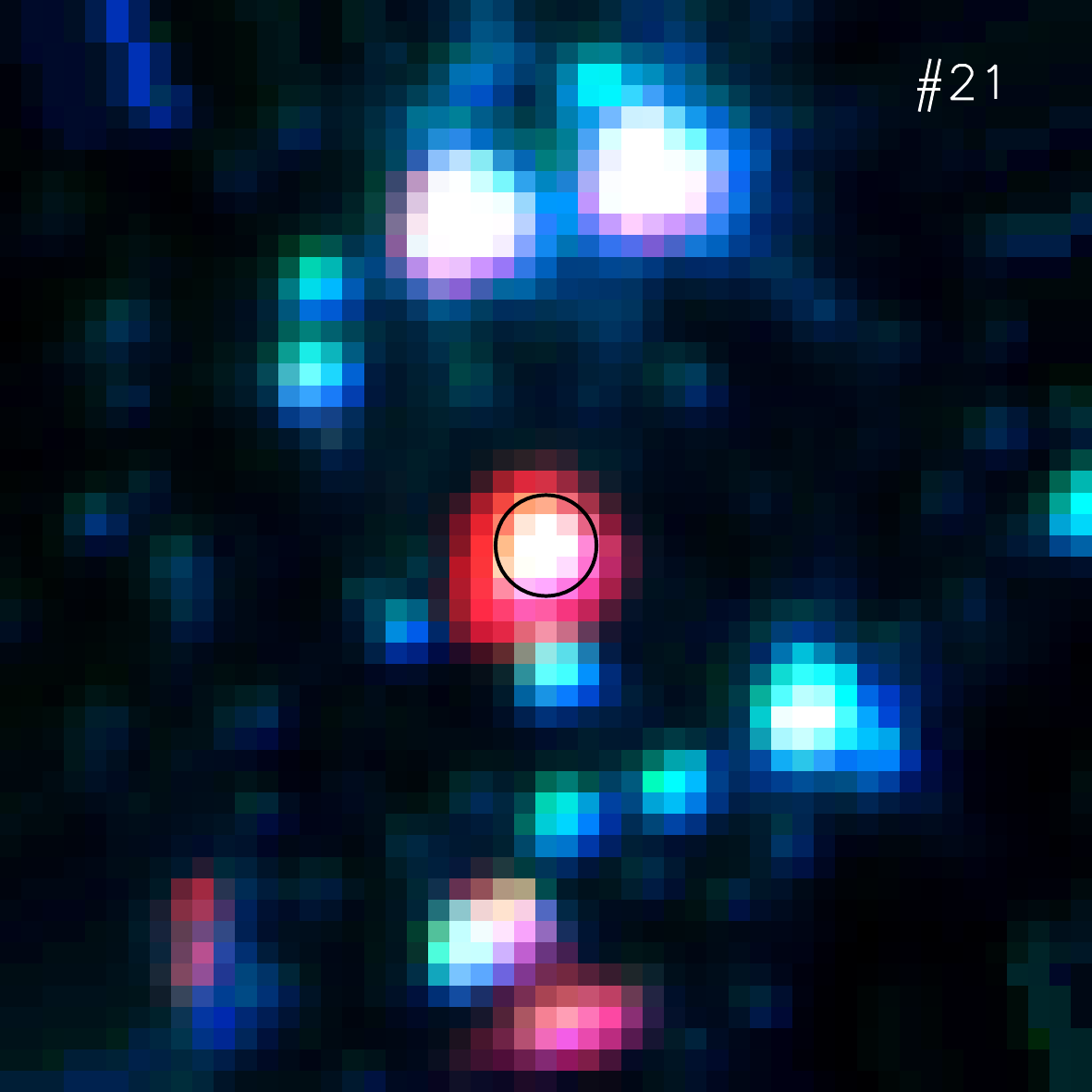}
\includegraphics[scale=0.363]{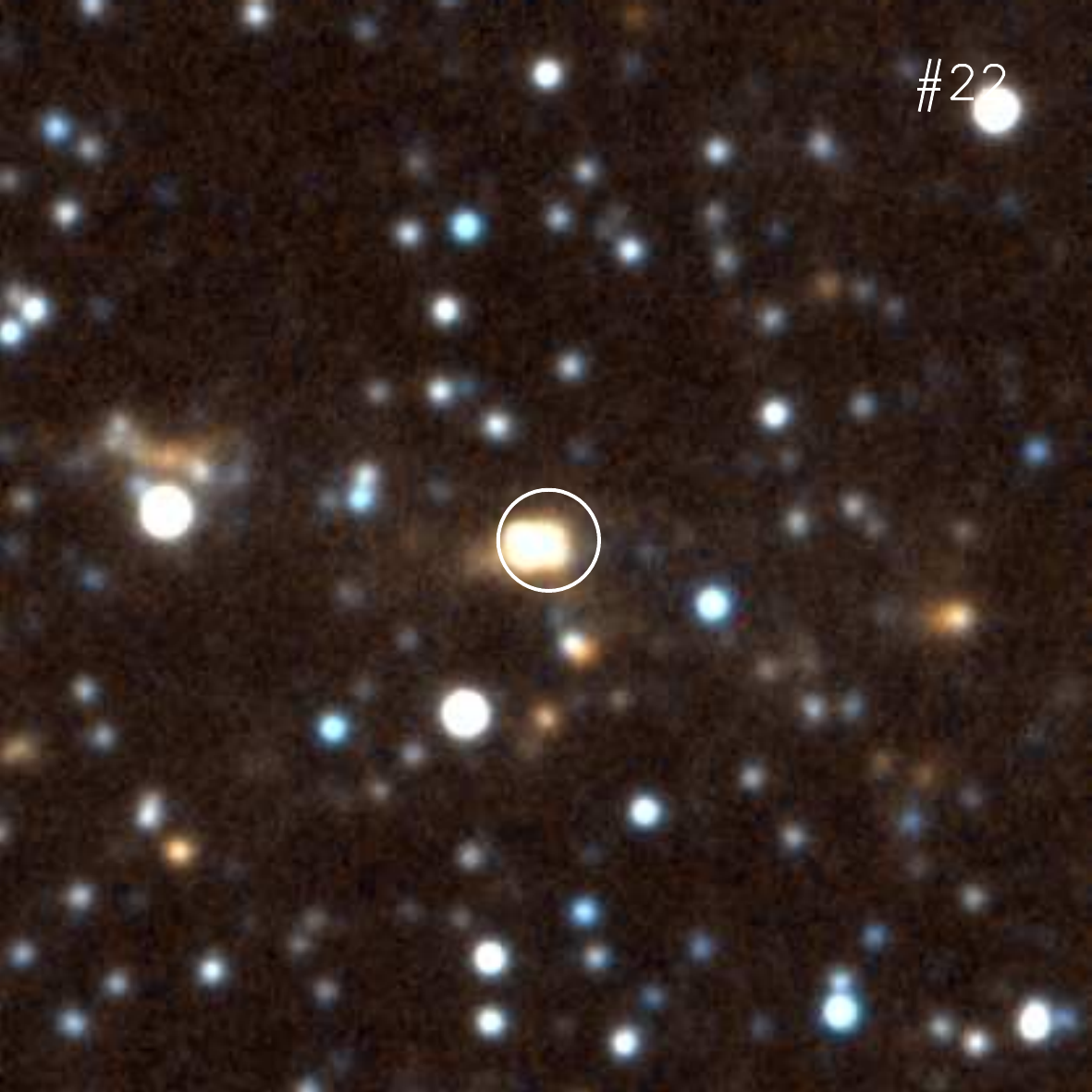}
\includegraphics[scale=0.363]{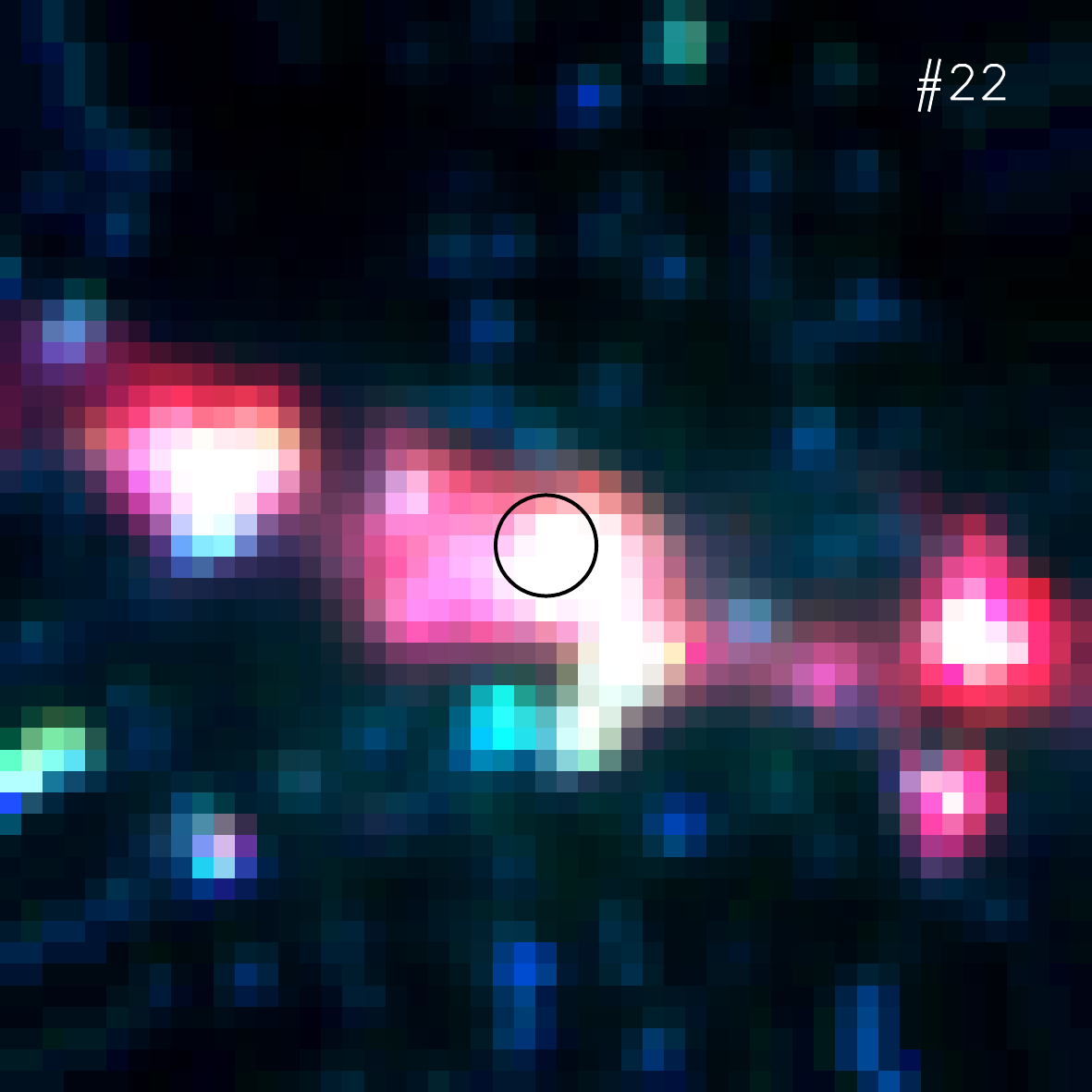}
\includegraphics[scale=0.363]{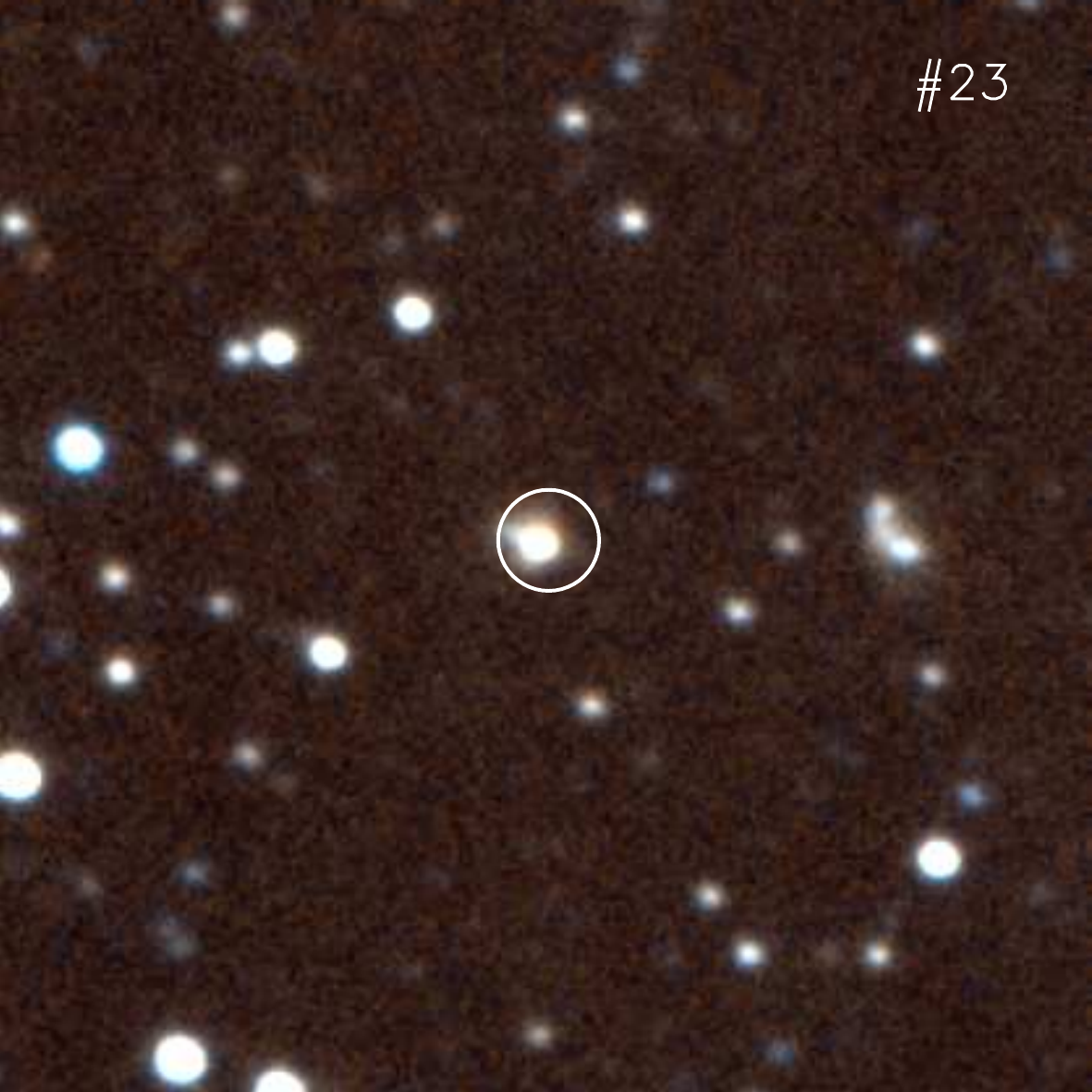}
\includegraphics[scale=0.363]{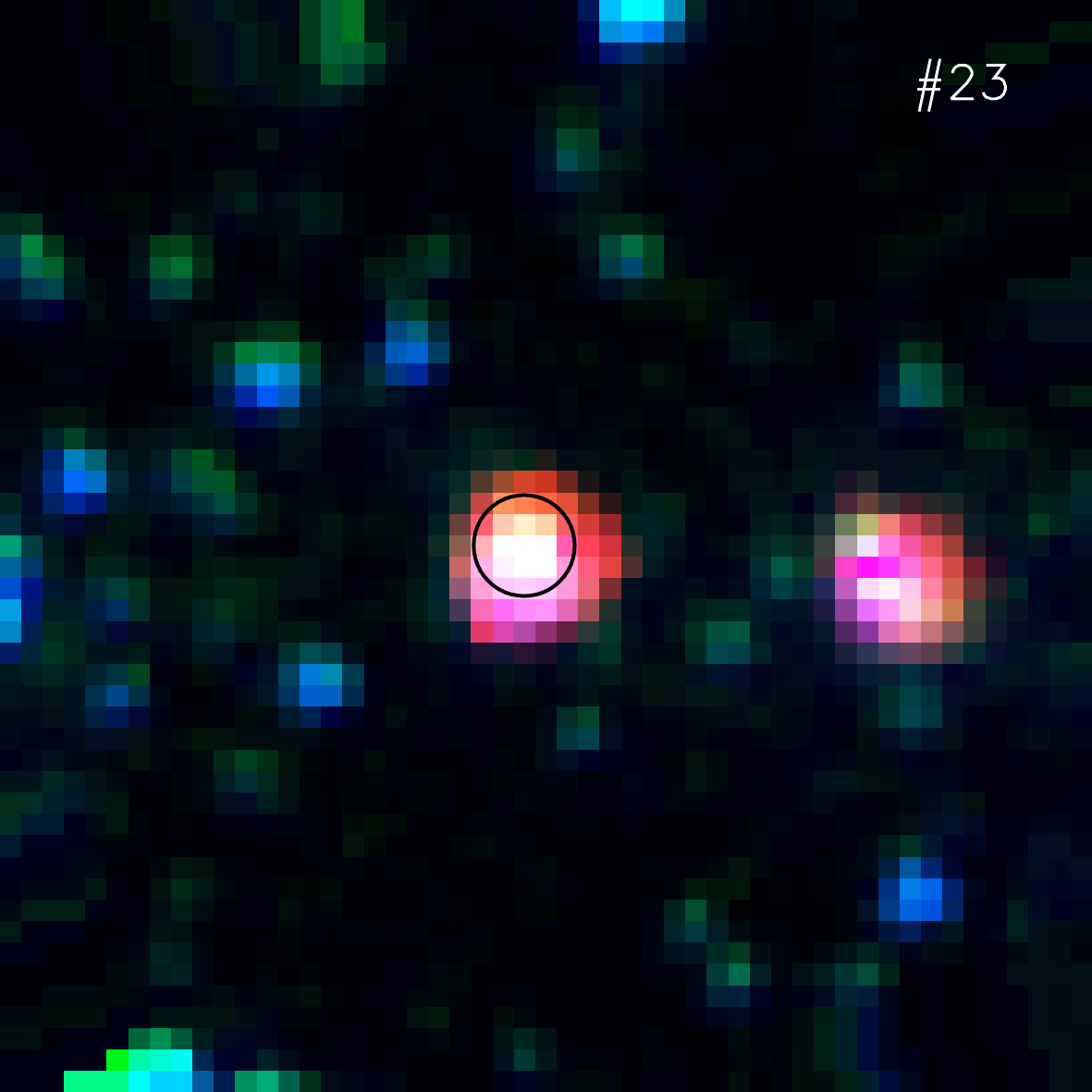}
\includegraphics[scale=0.363]{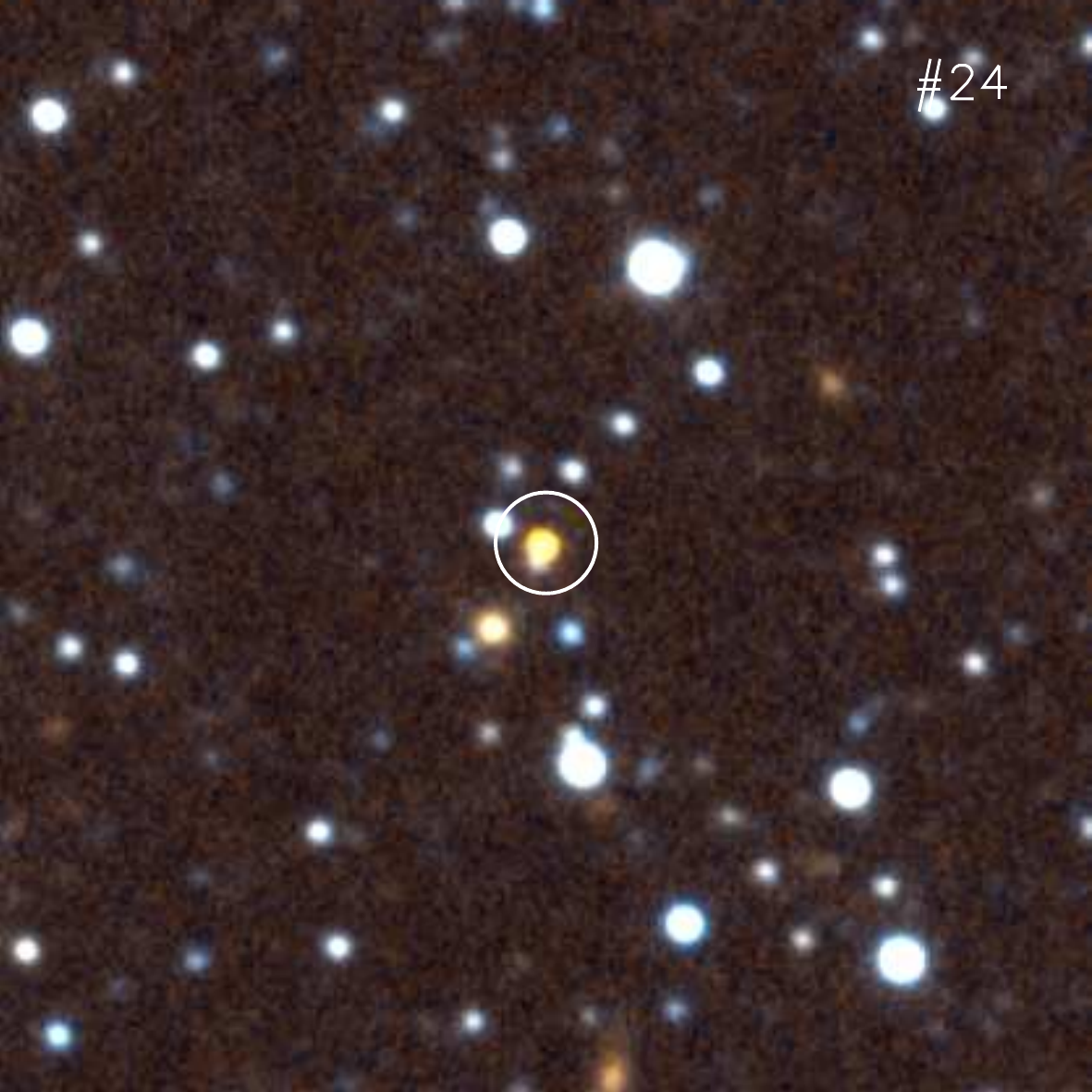}
\includegraphics[scale=0.363]{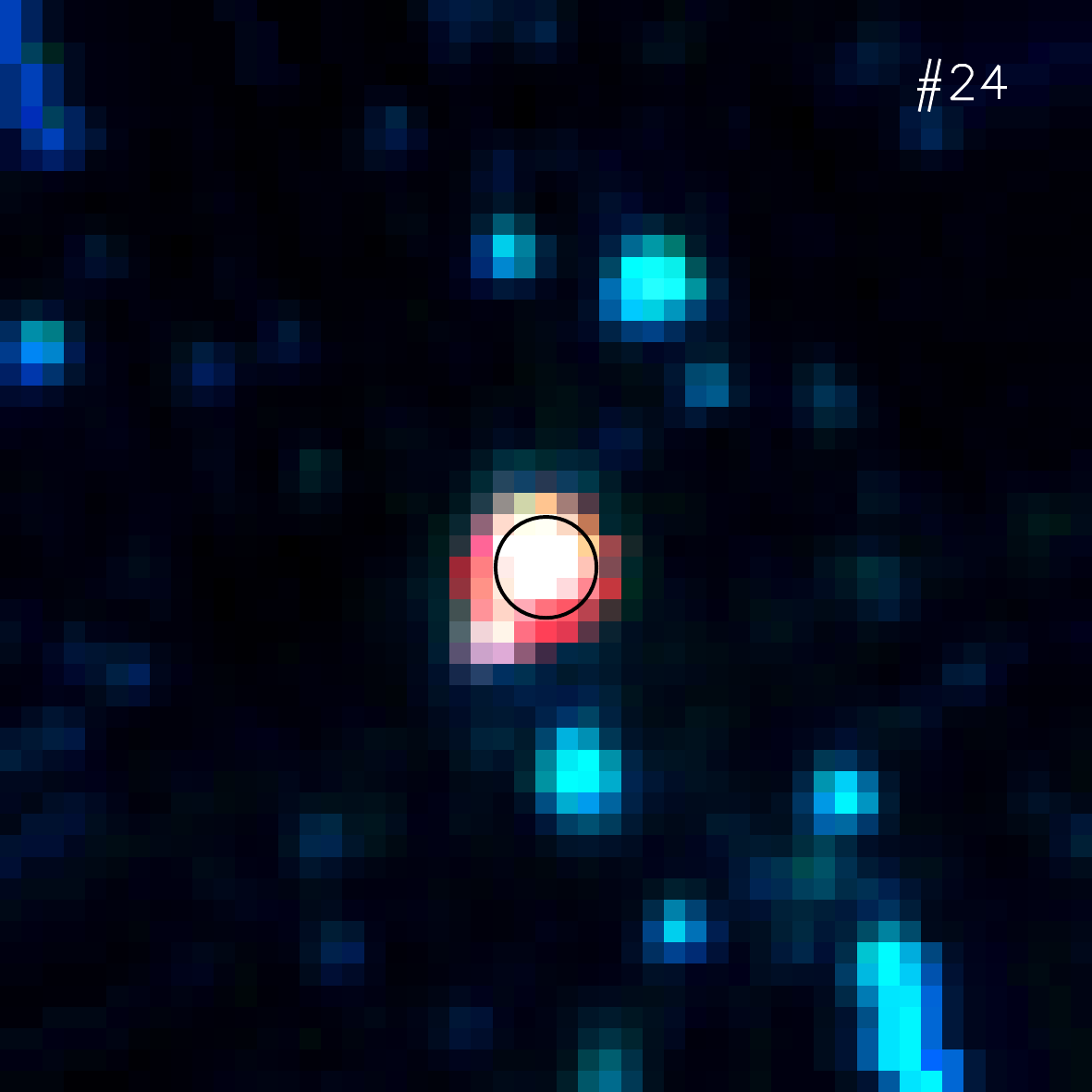}
\includegraphics[scale=0.363]{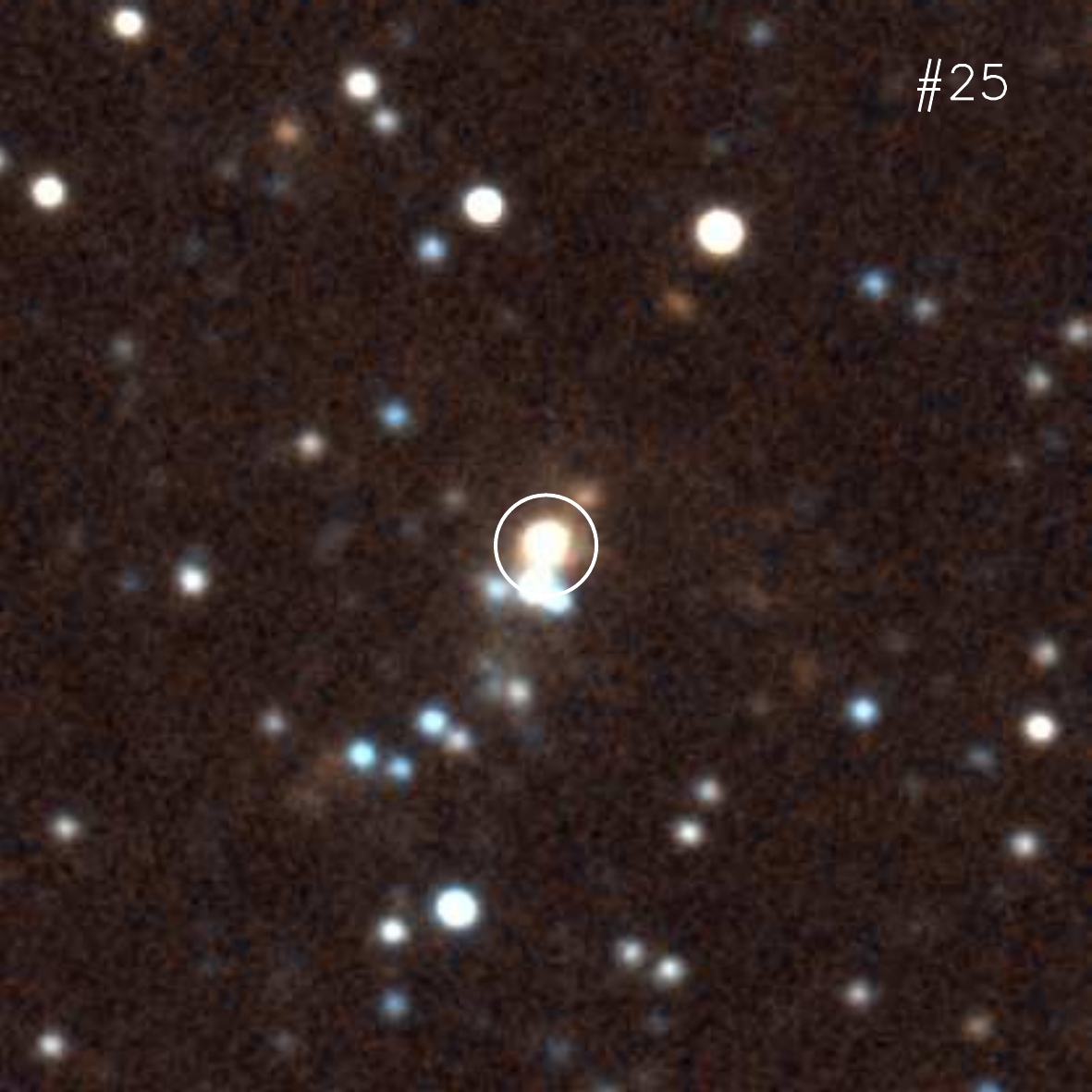}
\includegraphics[scale=0.363]{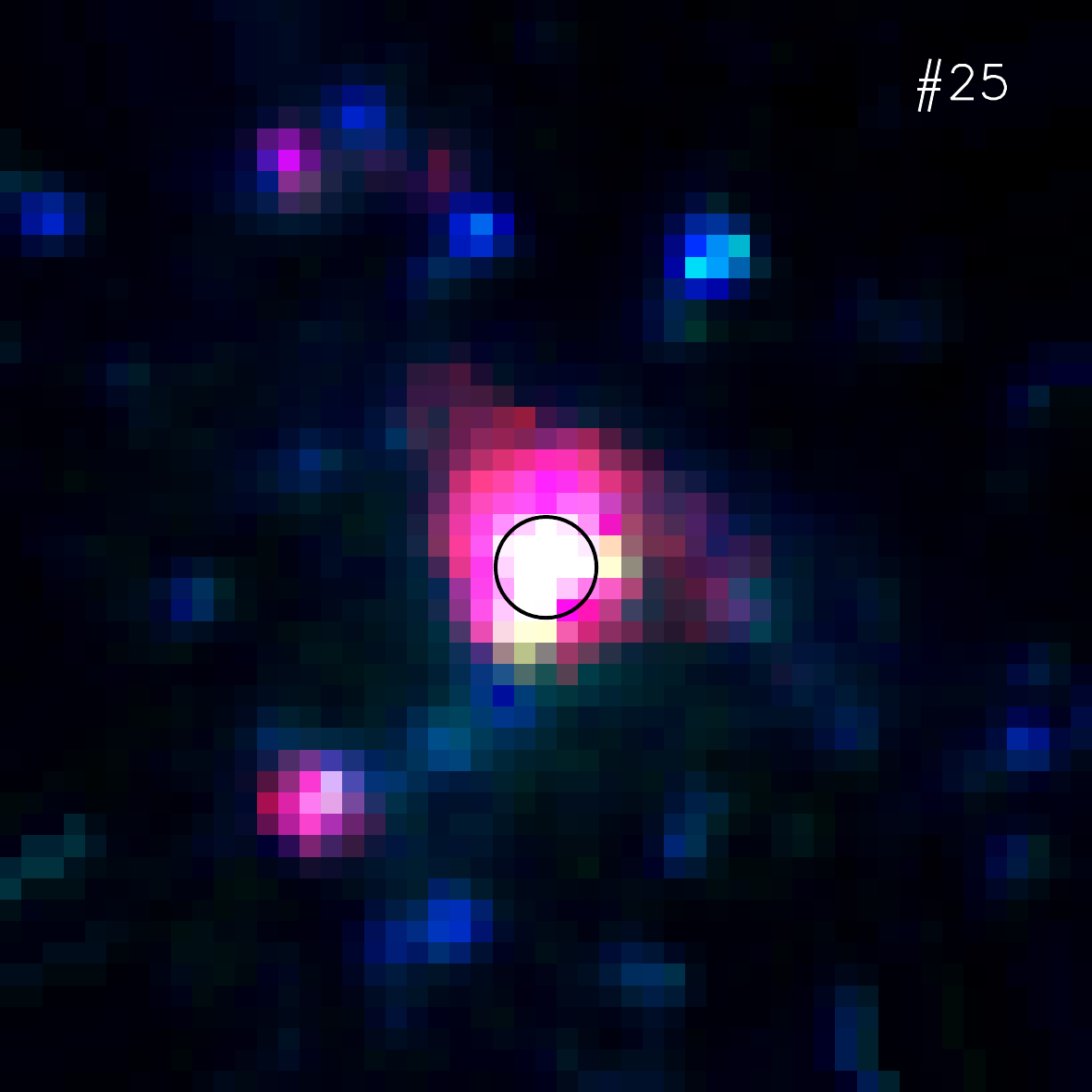}
\includegraphics[scale=0.363]{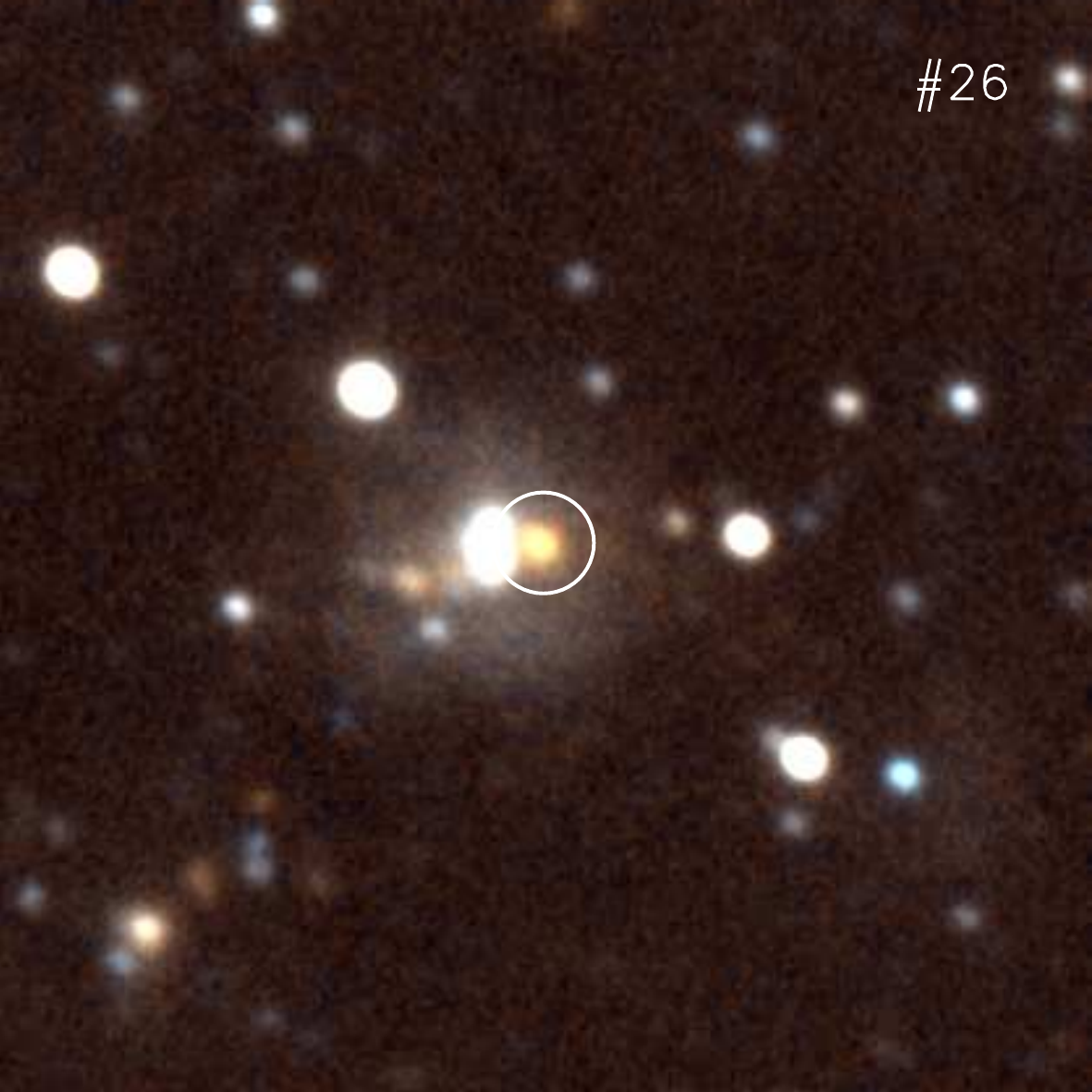}
\includegraphics[scale=0.363]{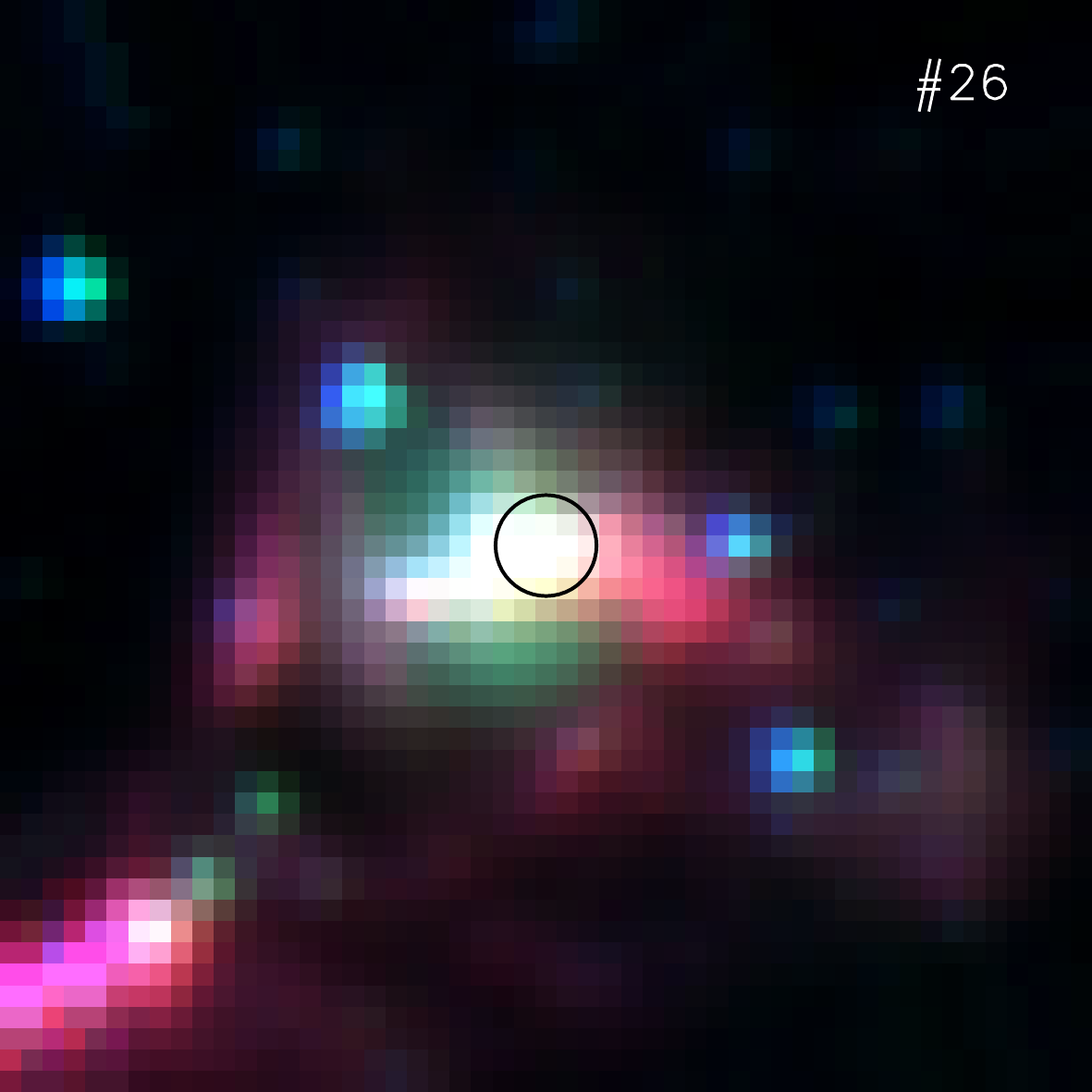}
\includegraphics[scale=0.363]{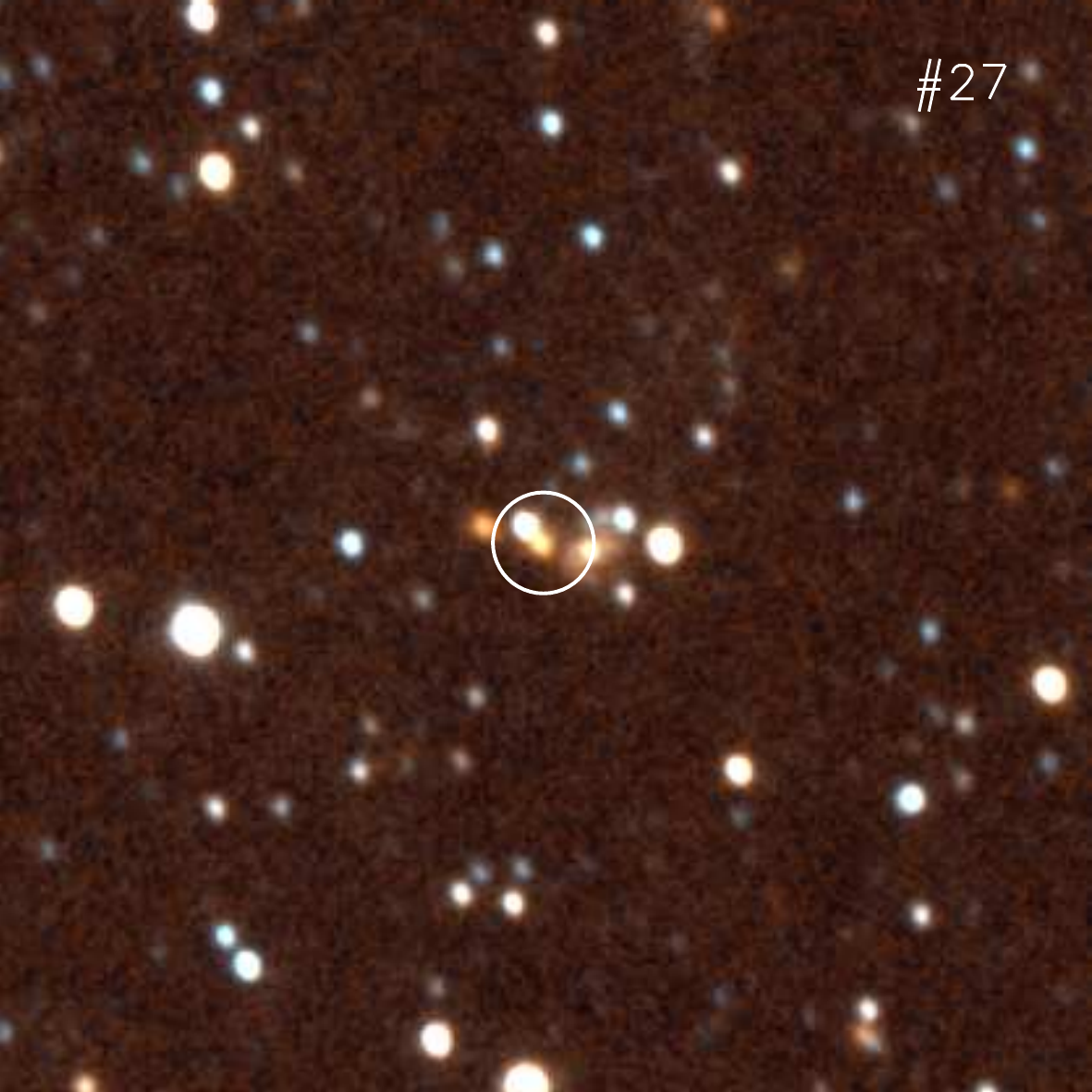}
\includegraphics[scale=0.363]{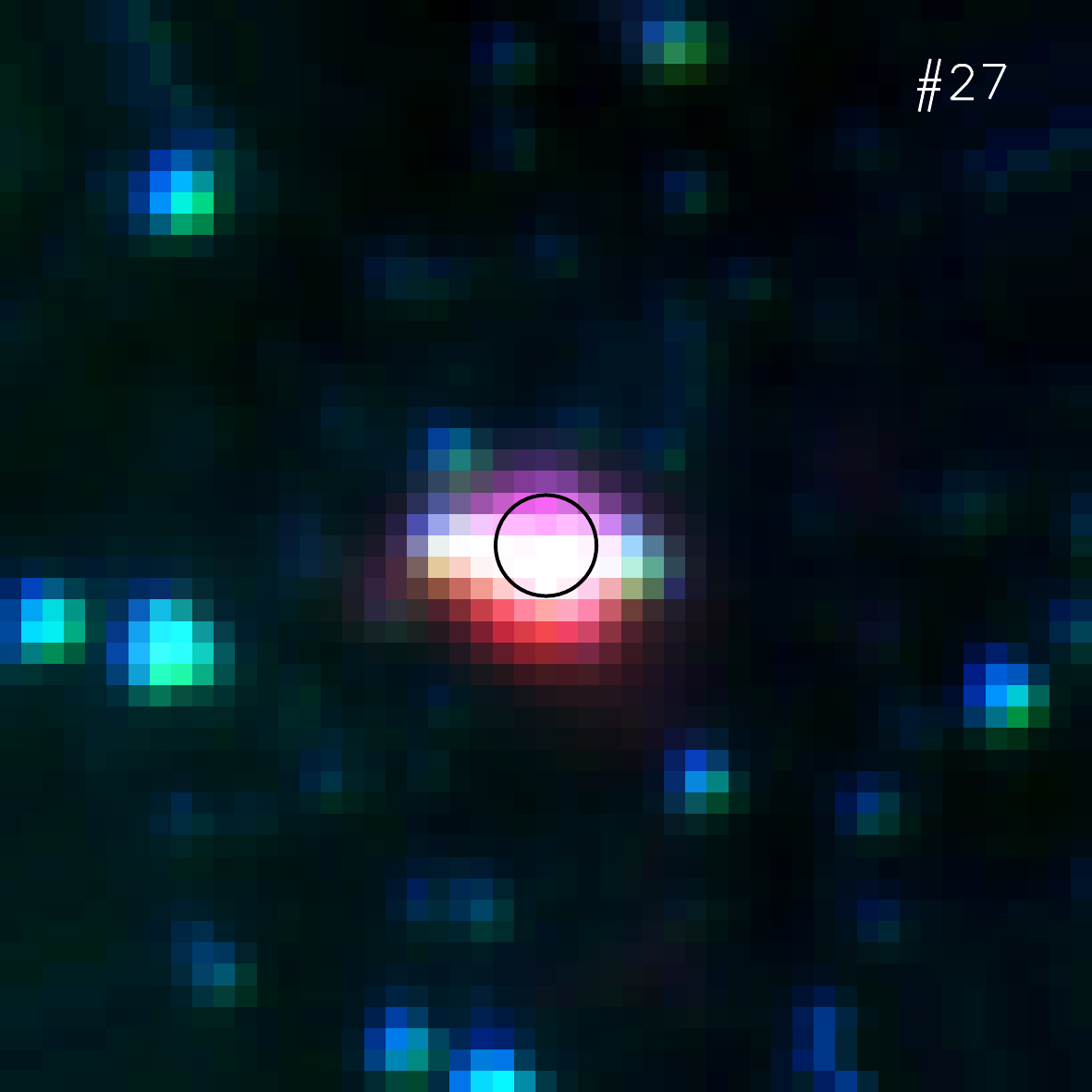}
\includegraphics[scale=0.363]{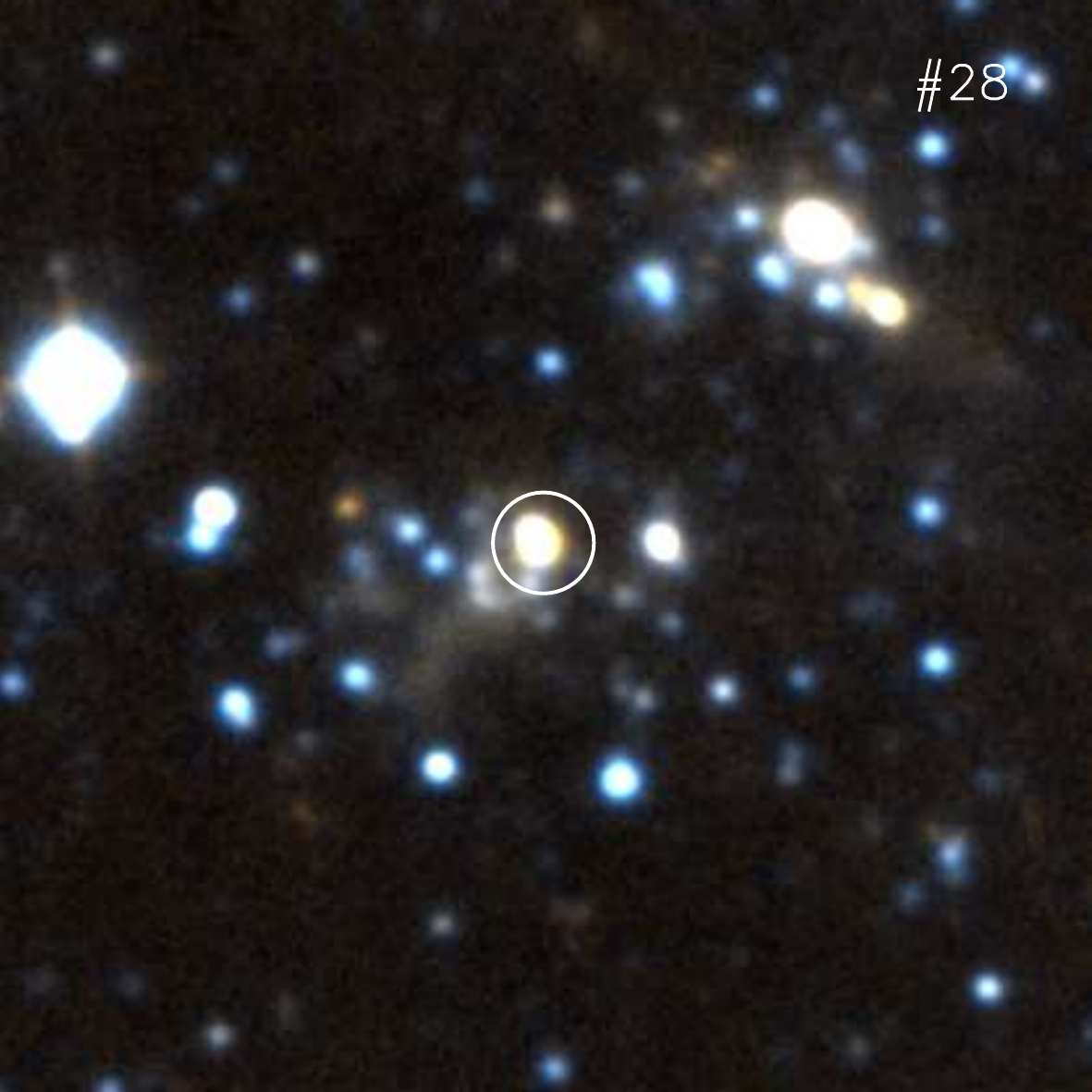}
\includegraphics[scale=0.363]{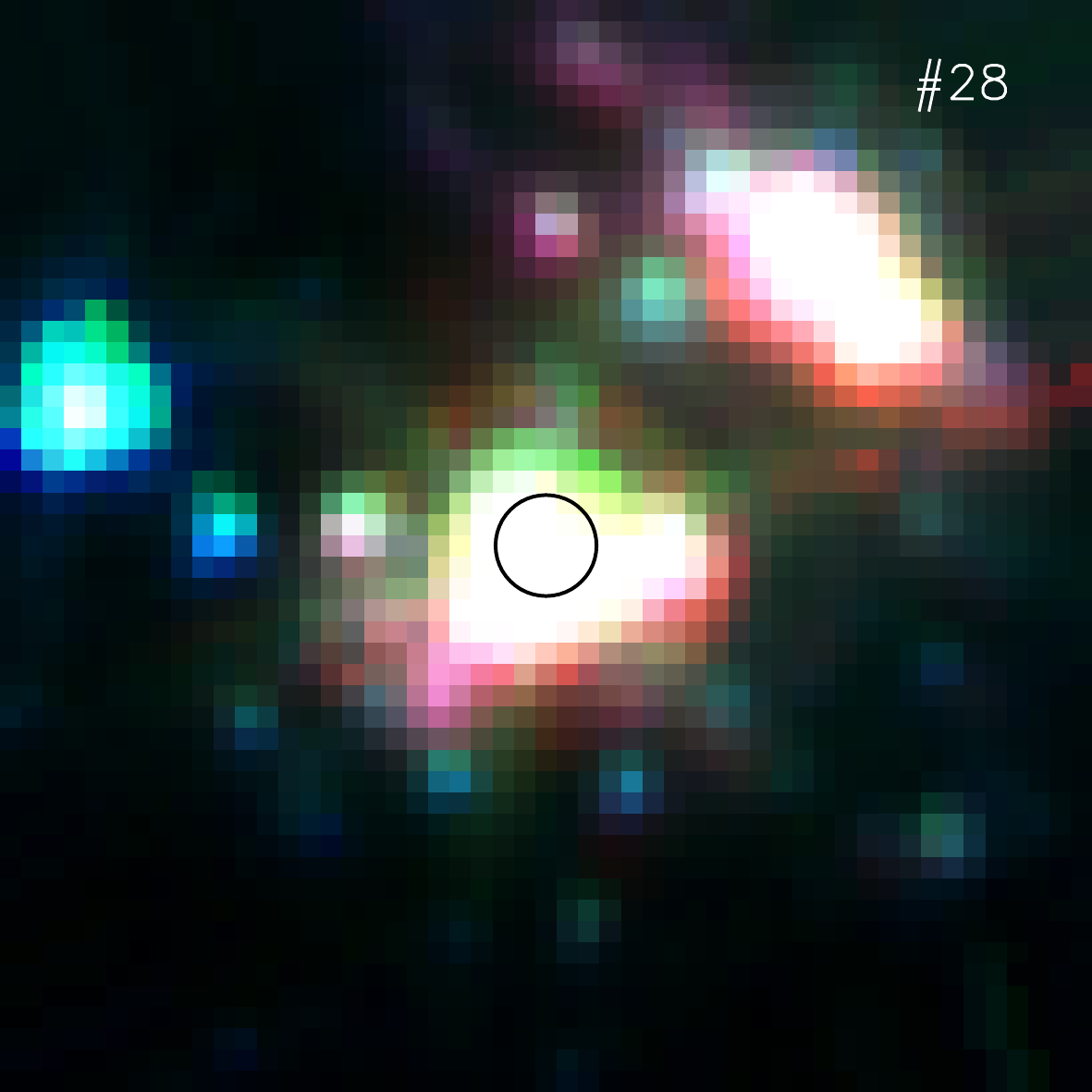}
\includegraphics[scale=0.363]{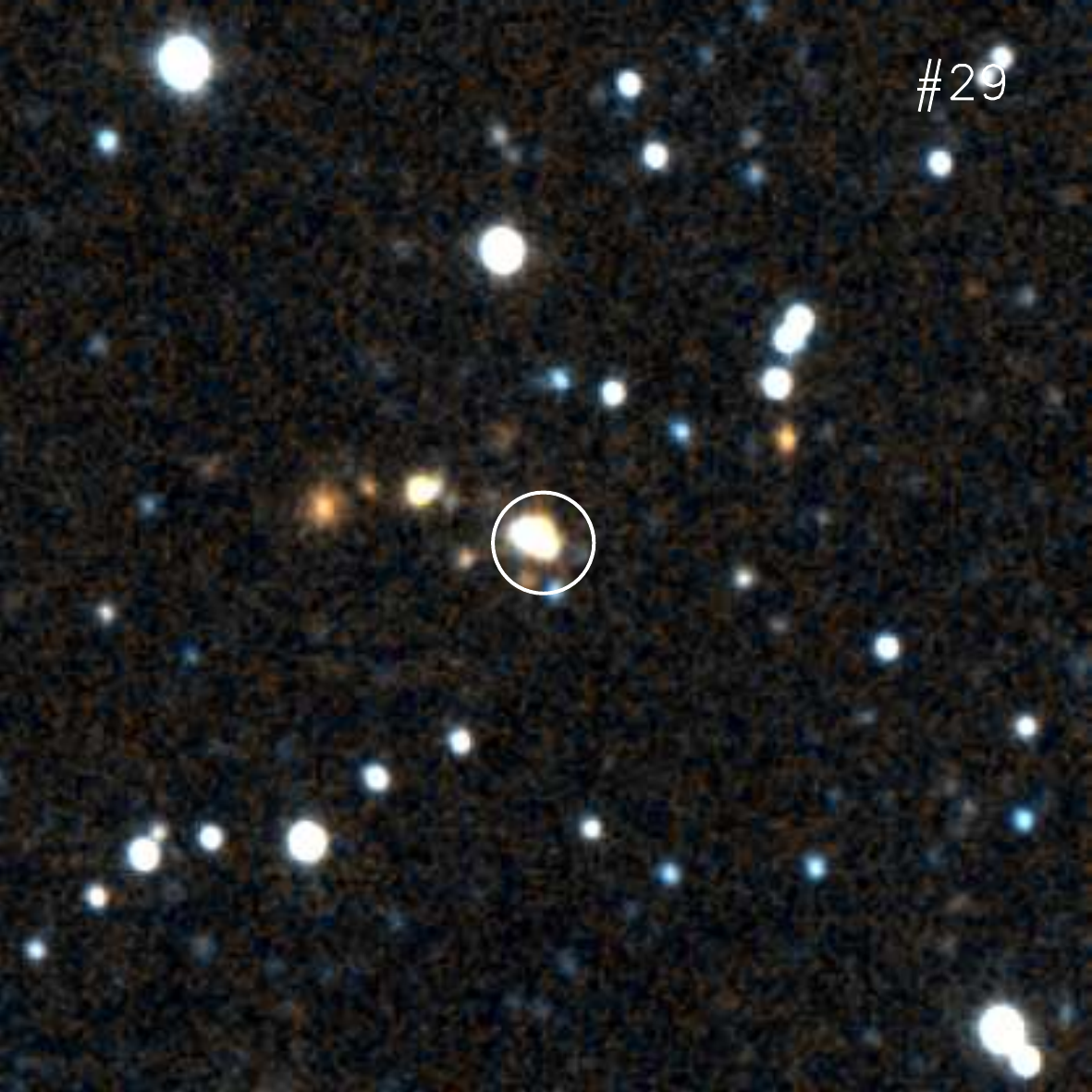}
\includegraphics[scale=0.363]{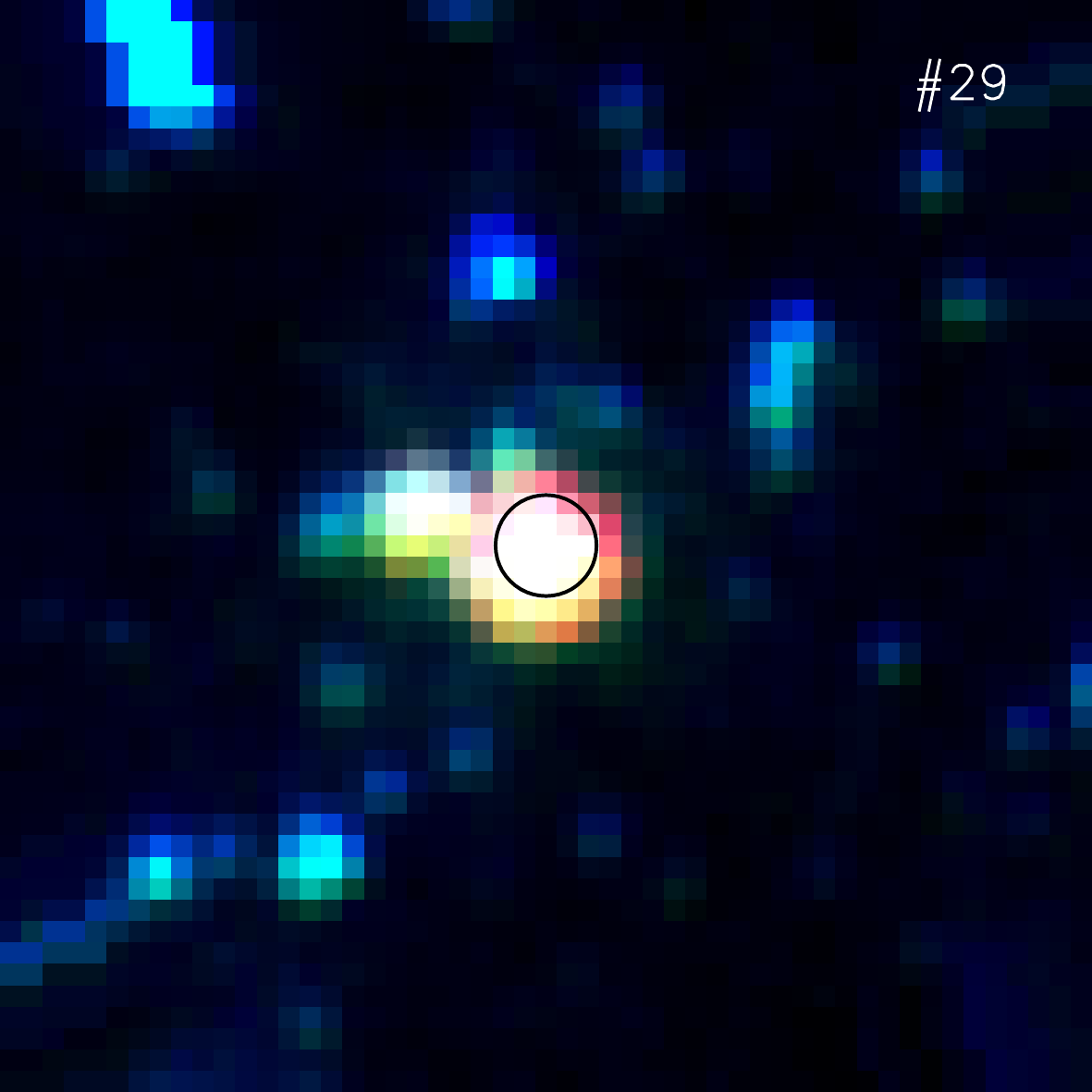}
\includegraphics[scale=0.363]{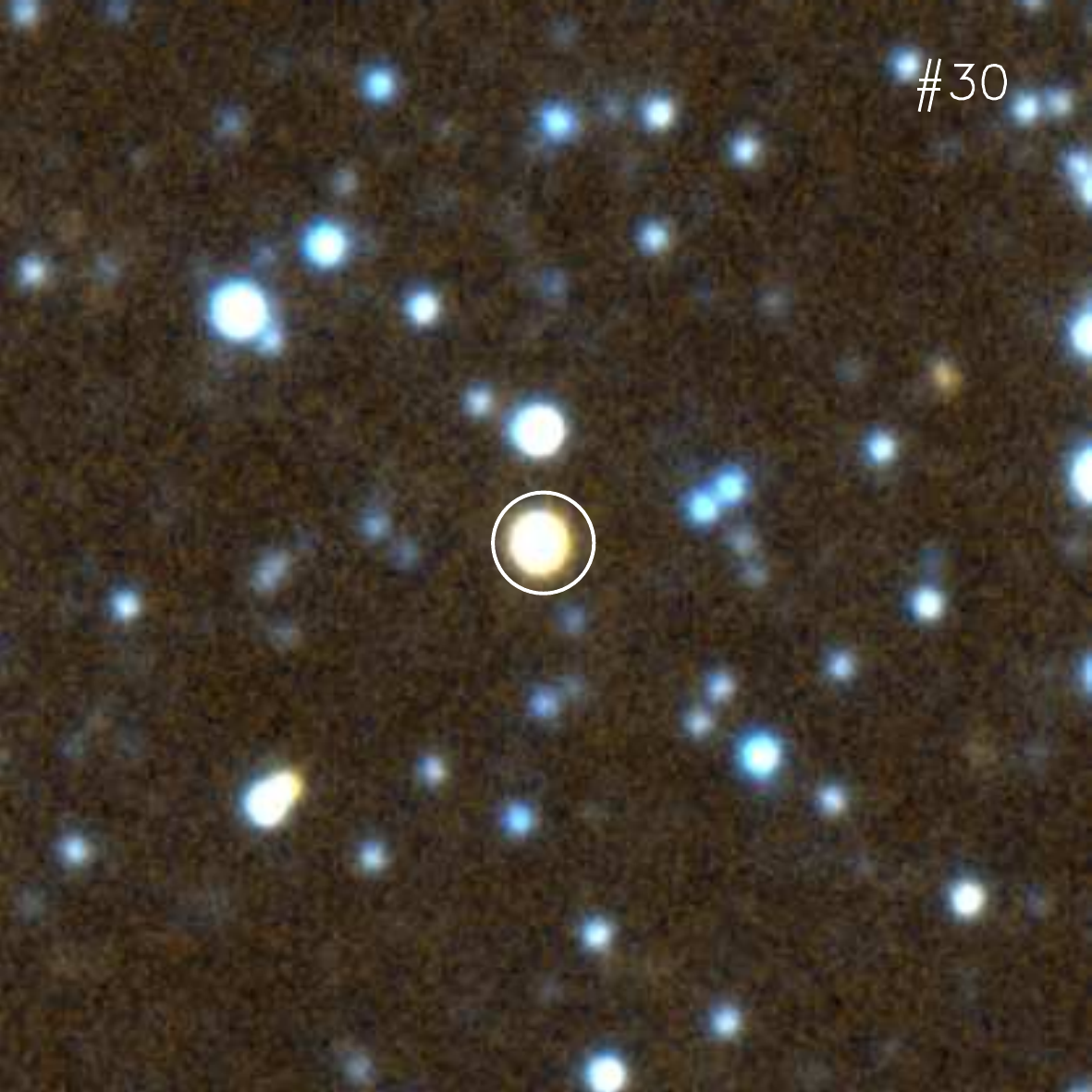}
\includegraphics[scale=0.363]{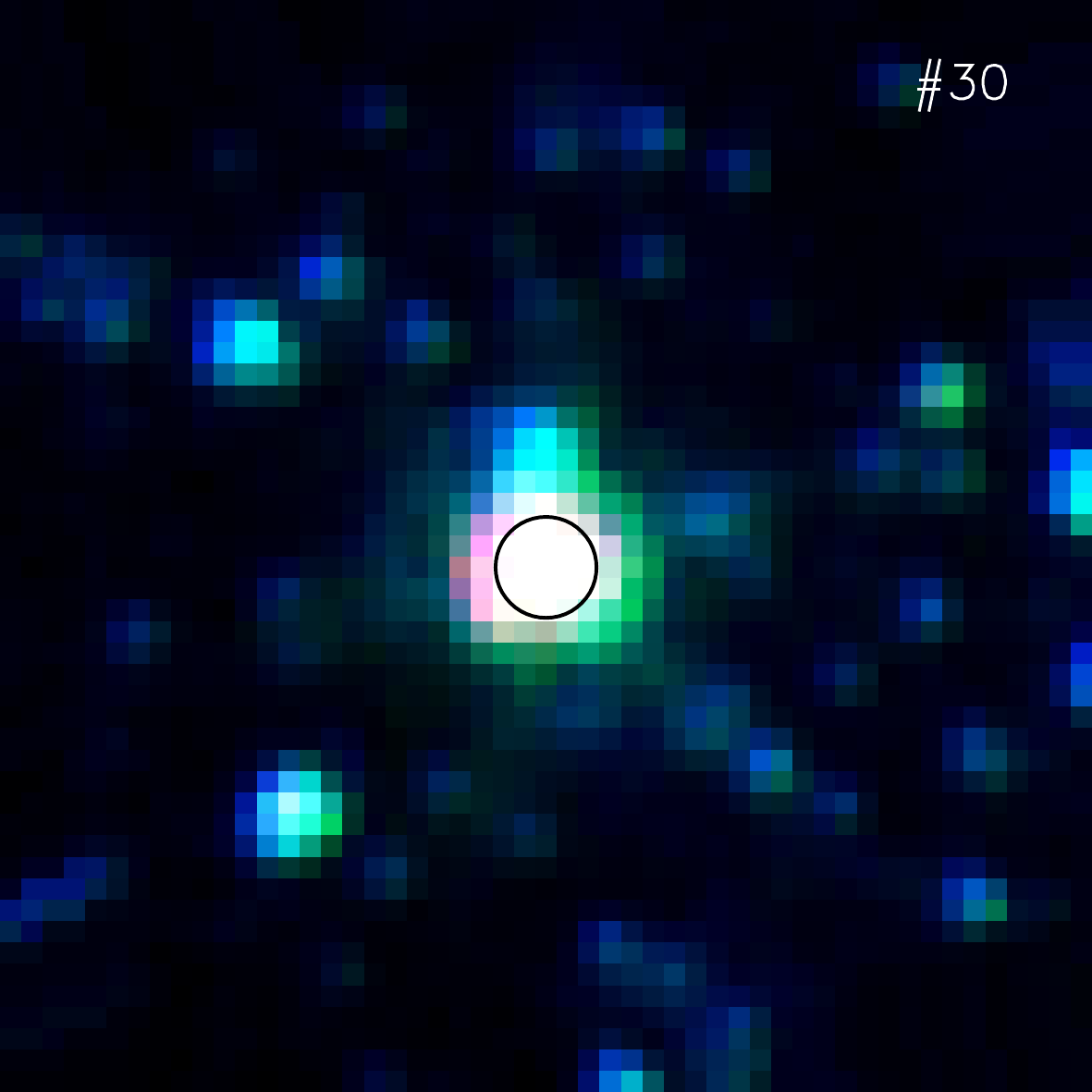}
\contcaption{}
\end{figure*}
\begin{figure*}
\includegraphics[scale=0.363]{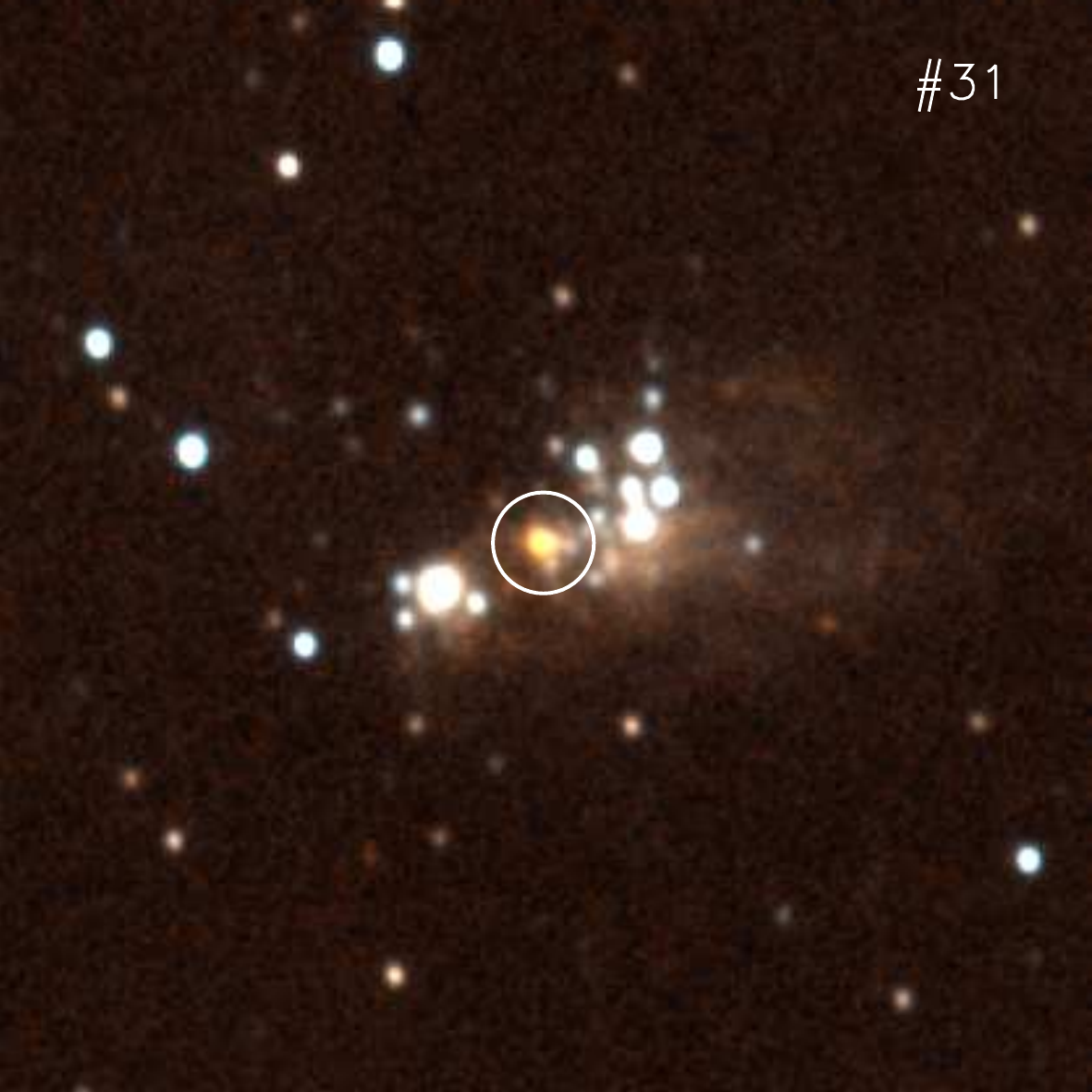}
\includegraphics[scale=0.363]{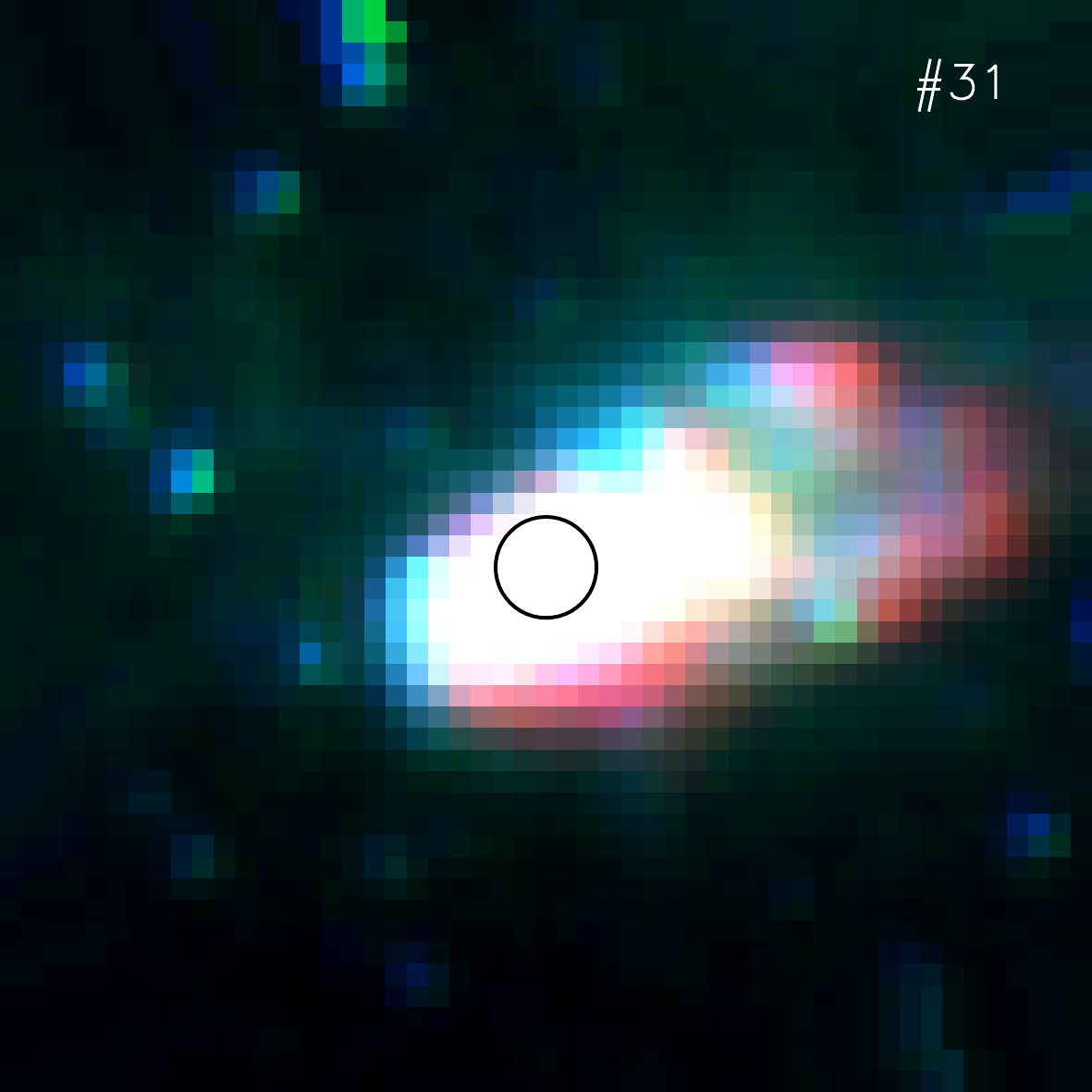}
\includegraphics[scale=0.363]{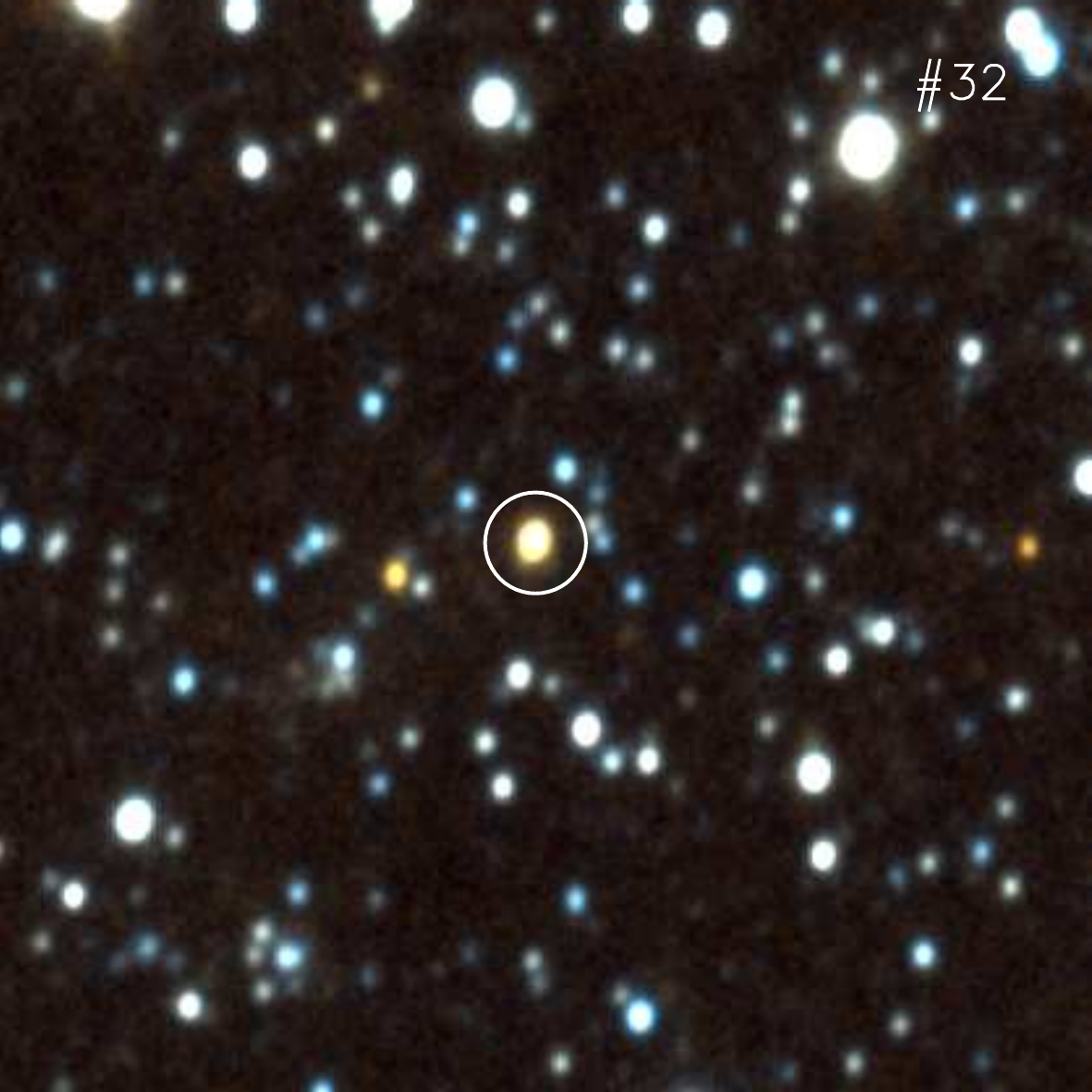}
\includegraphics[scale=0.363]{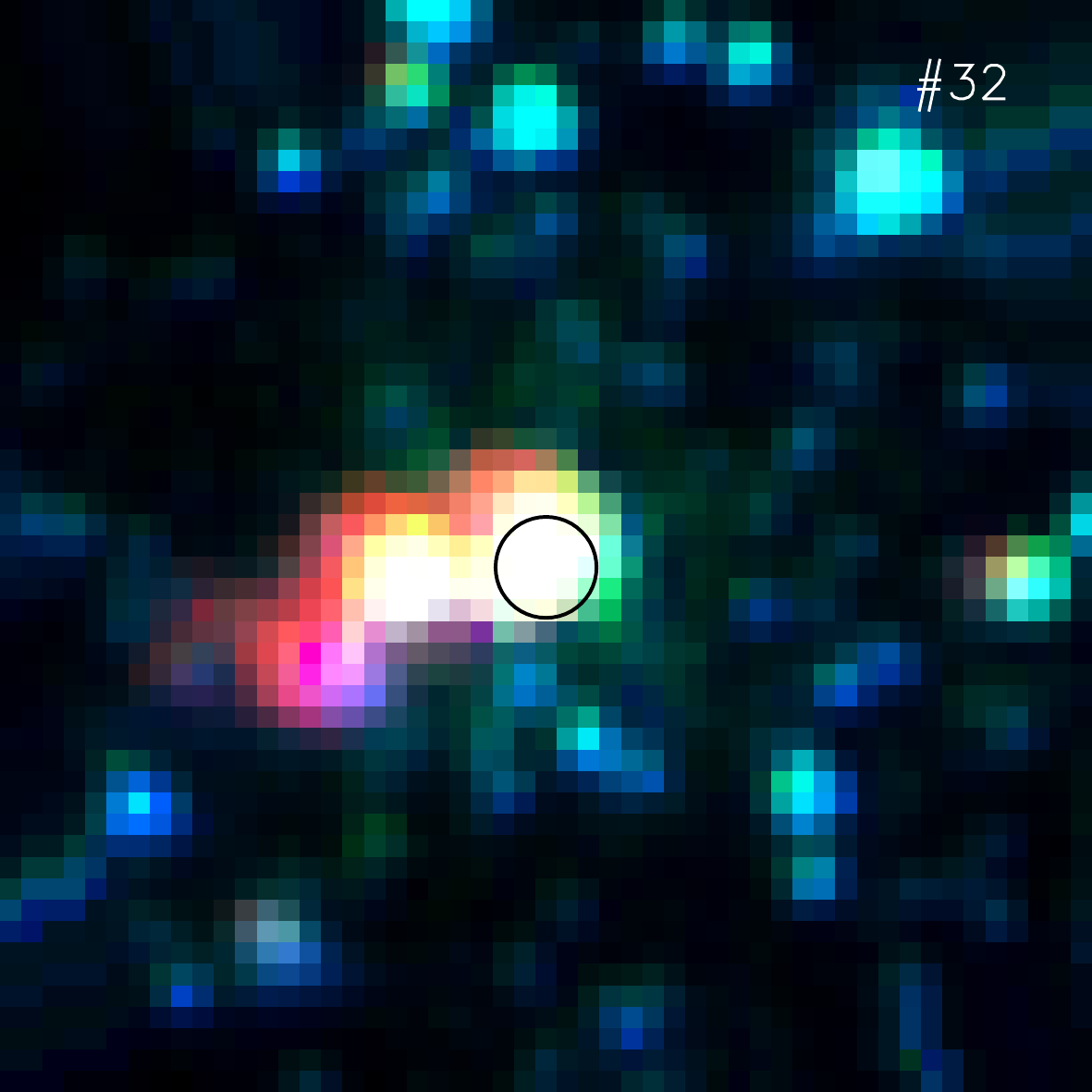}
\includegraphics[scale=0.363]{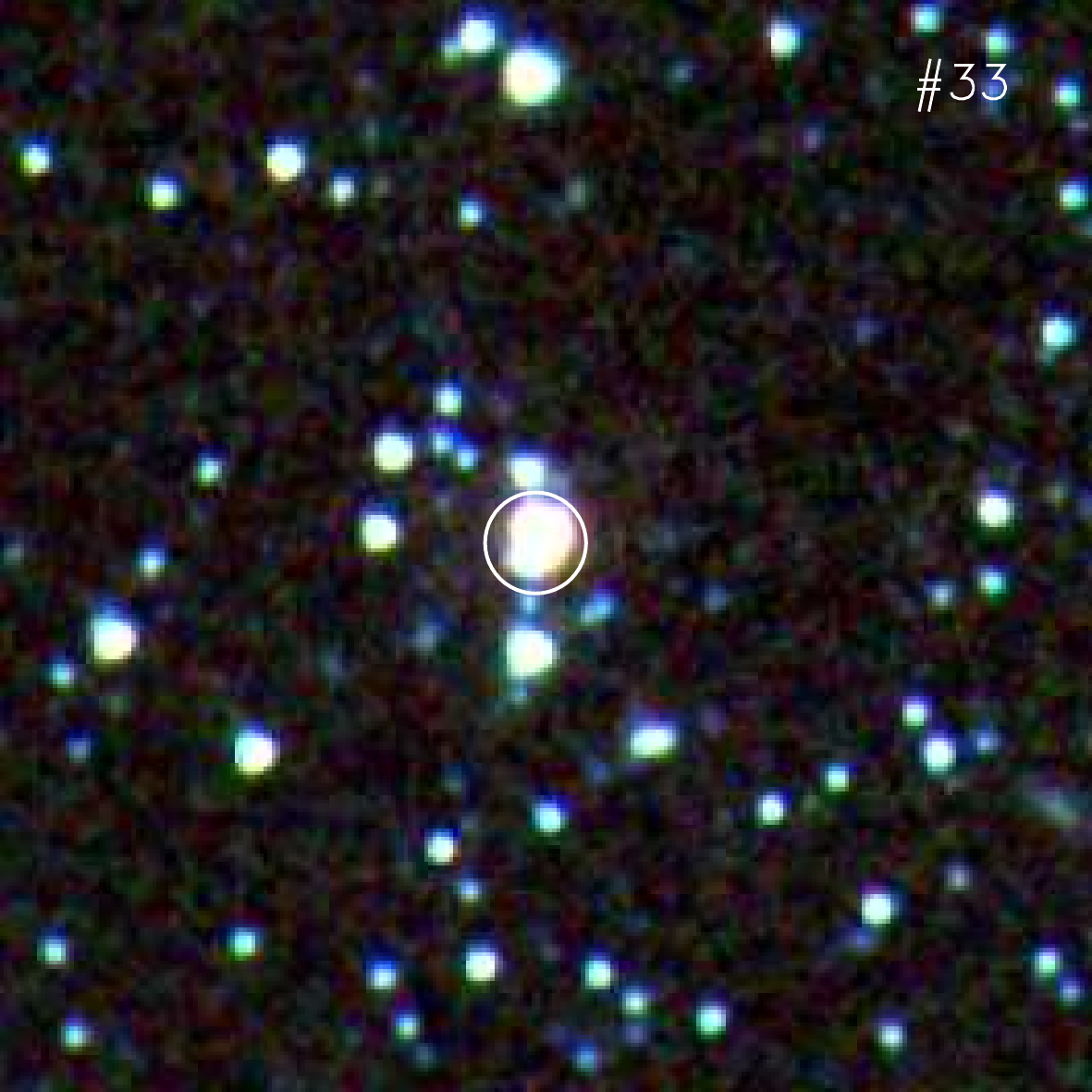}
\includegraphics[scale=0.363]{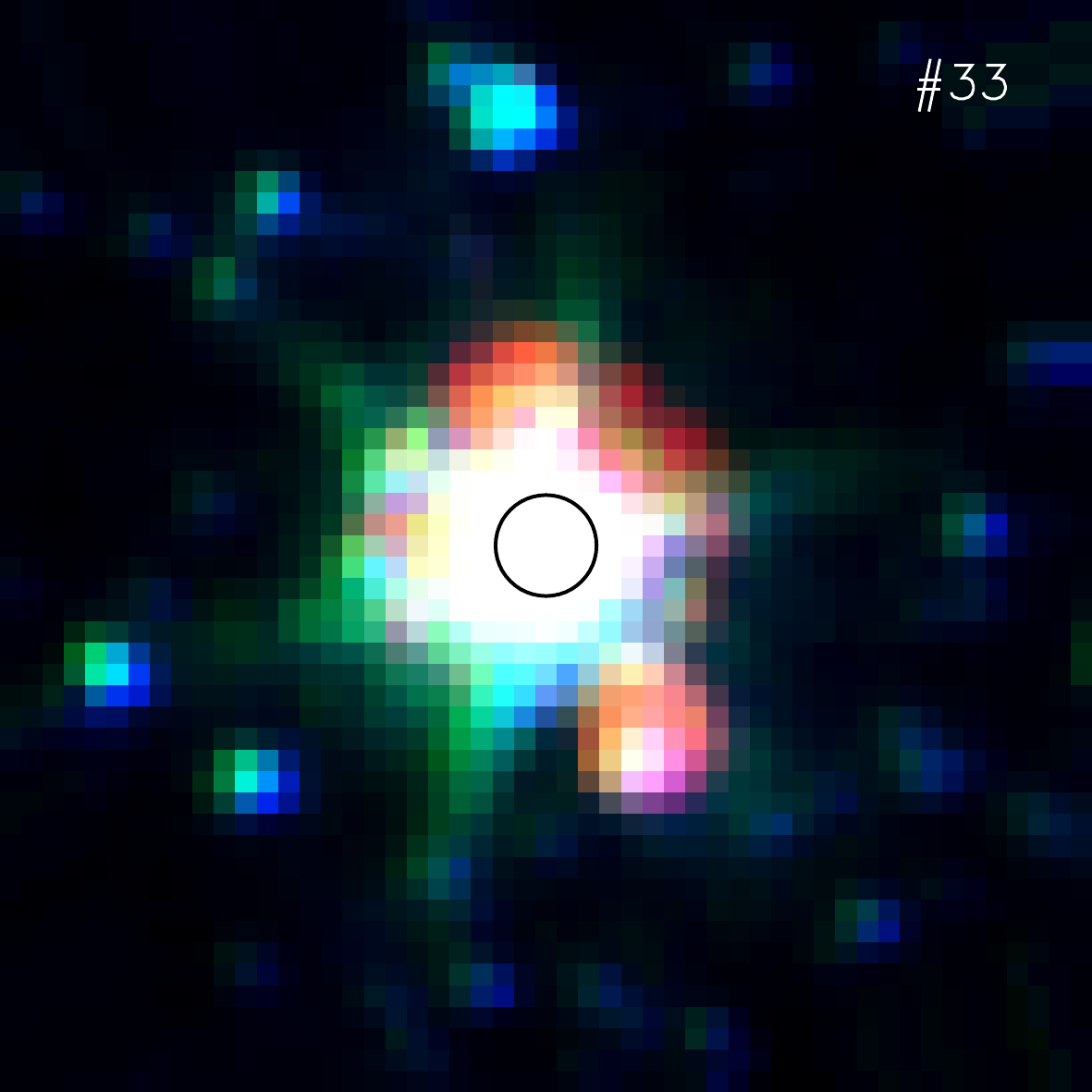}
\includegraphics[scale=0.363]{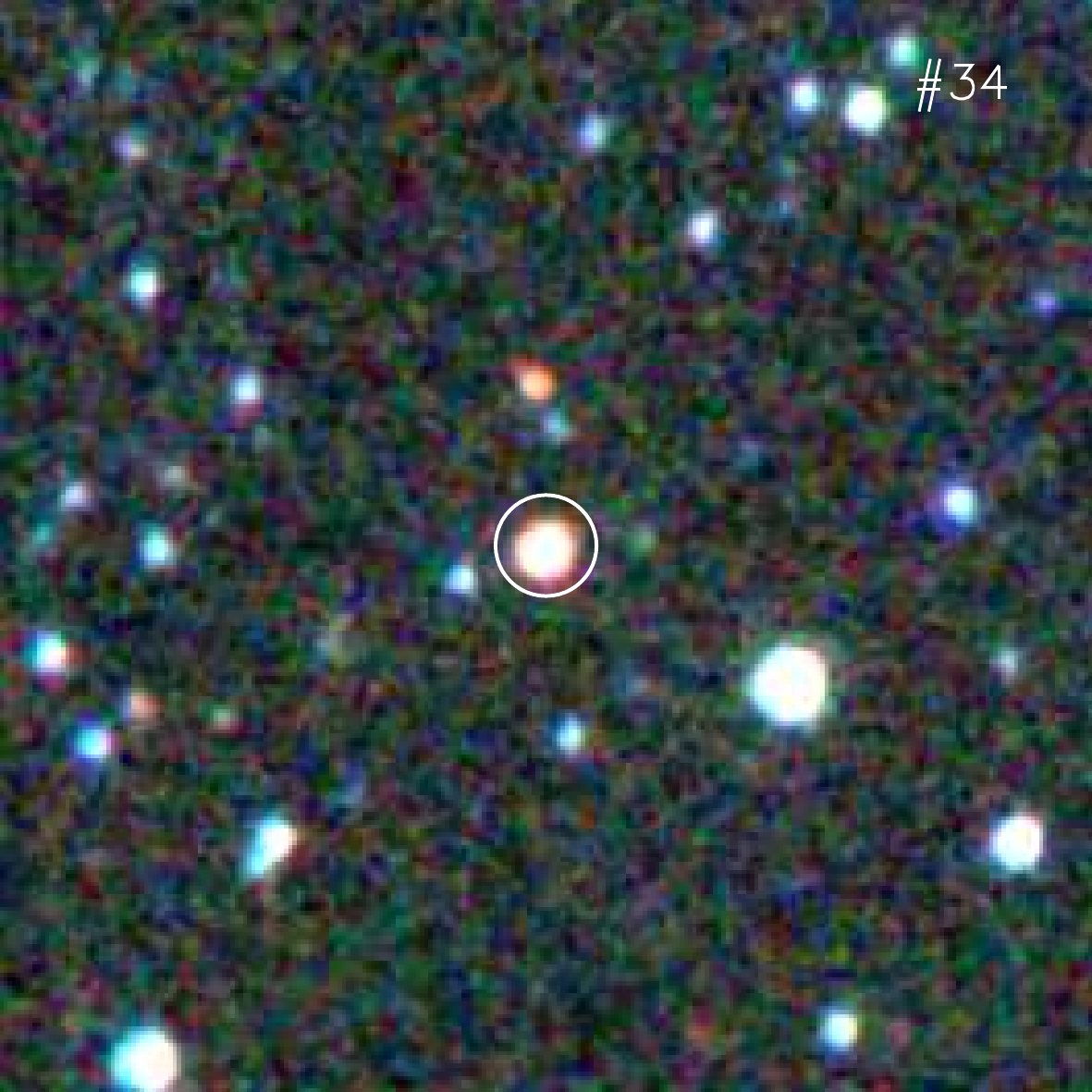}
\includegraphics[scale=0.363]{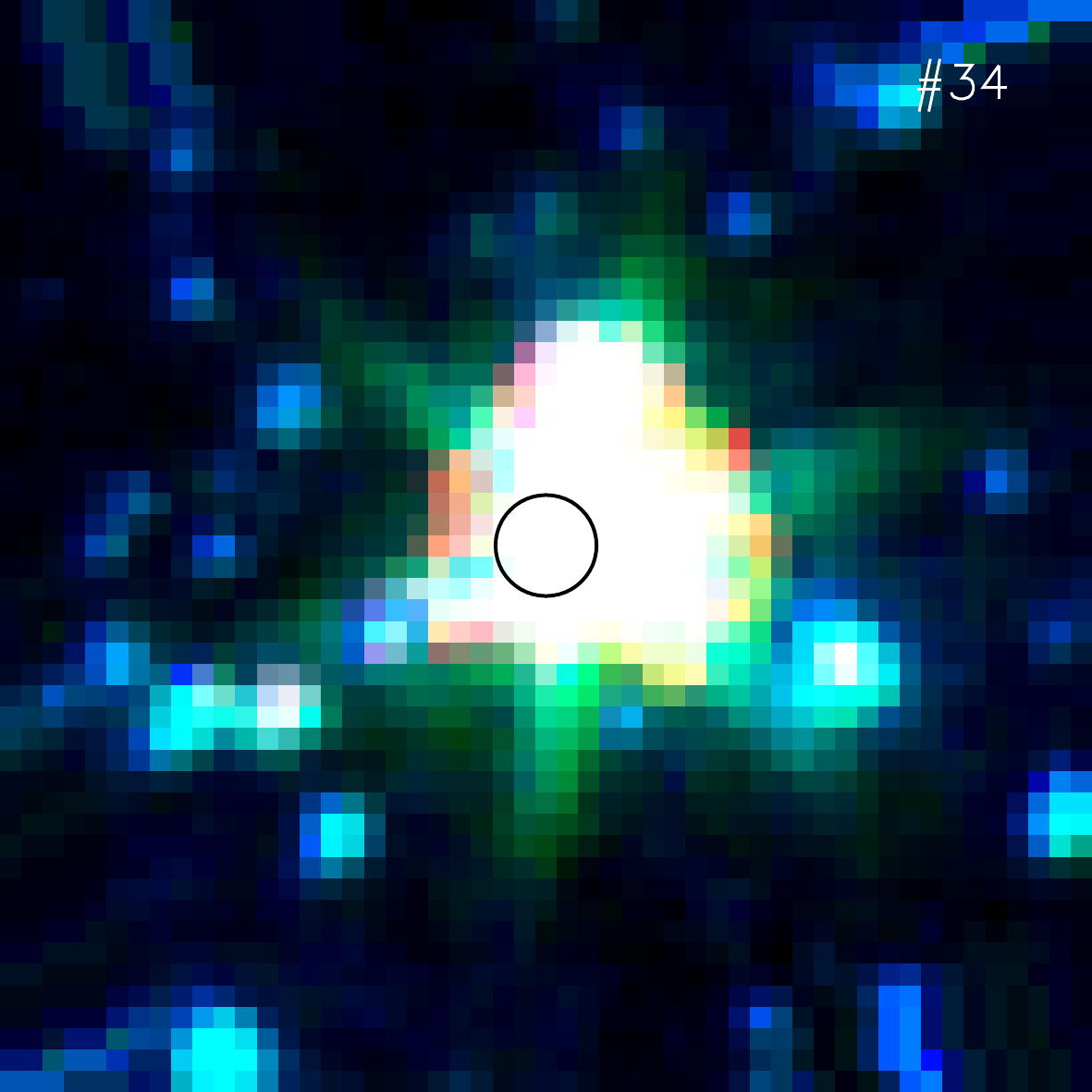}
\contcaption{}
\end{figure*}

\section{Optical spectra}

In this appendix we show all the optical spectra obtained as possible counterparts of
the 34 IR sources in the SMC. As described in Section 3.5.1, the observed optical source 
is not always the correct counterpart for the IR source. No optical counterparts were 
found for sources \#05, 06, 24, 29 and 31. The spectra form the basis of our optical 
classification scheme, using several emission lines. The last panel of
Fig.\,\ref{optical} shows the identifications of the lines discussed in this work for 
source \#26.

\begin{figure*}
\includegraphics[scale=0.9]{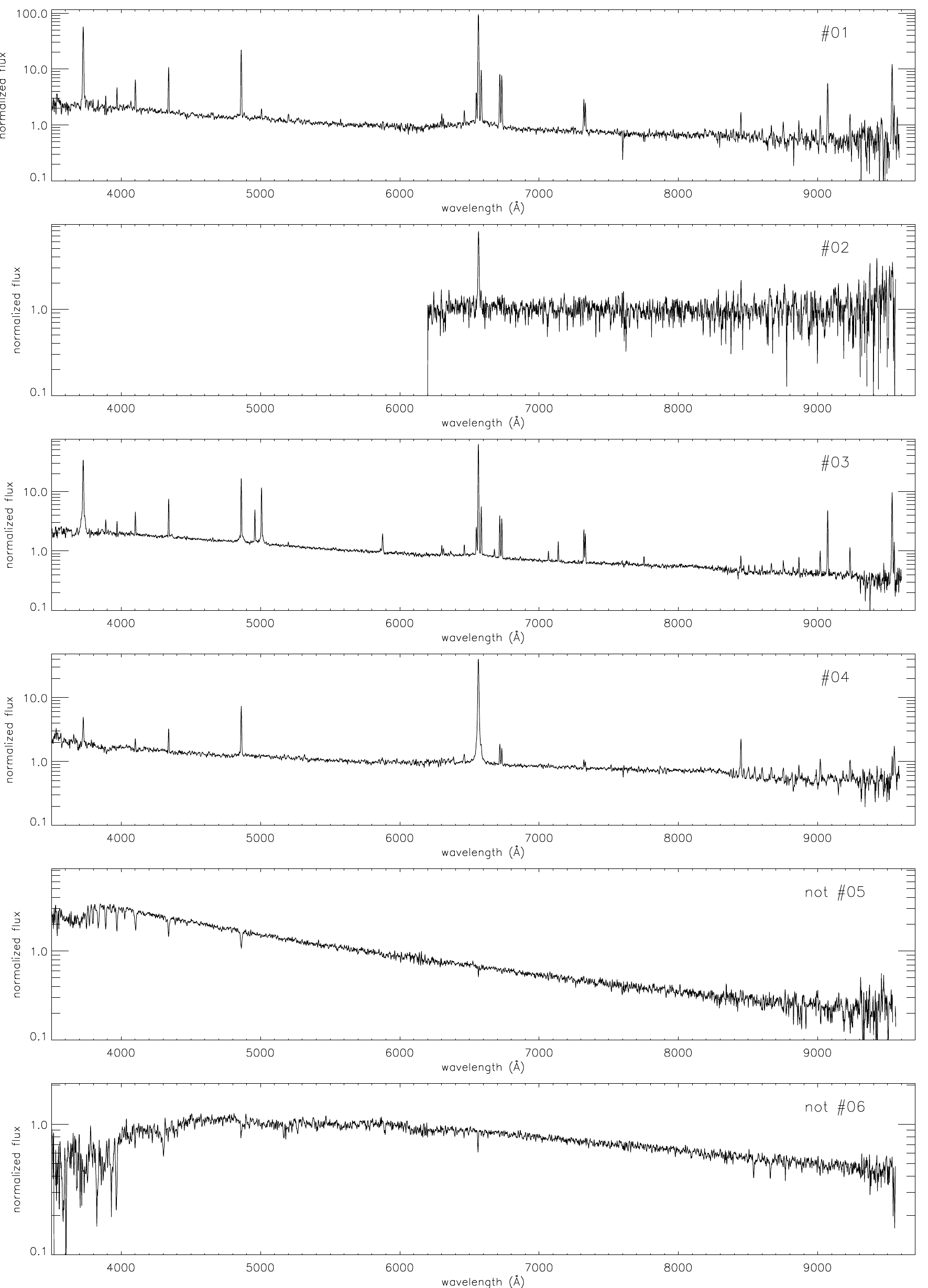}
\caption{Optical spectra of possible optical counterparts for the IR sources in the 
SMC. No spectra were obtained for sources \#24 and 31, and for sources \#05, 06 and 29
the optical spectra are not associated with the IR sources. The most conspicuous 
spectral features are labelled in the last panel for source \#26.}
\label{optical}
\end{figure*}
\begin{figure*}
\includegraphics[scale=0.9]{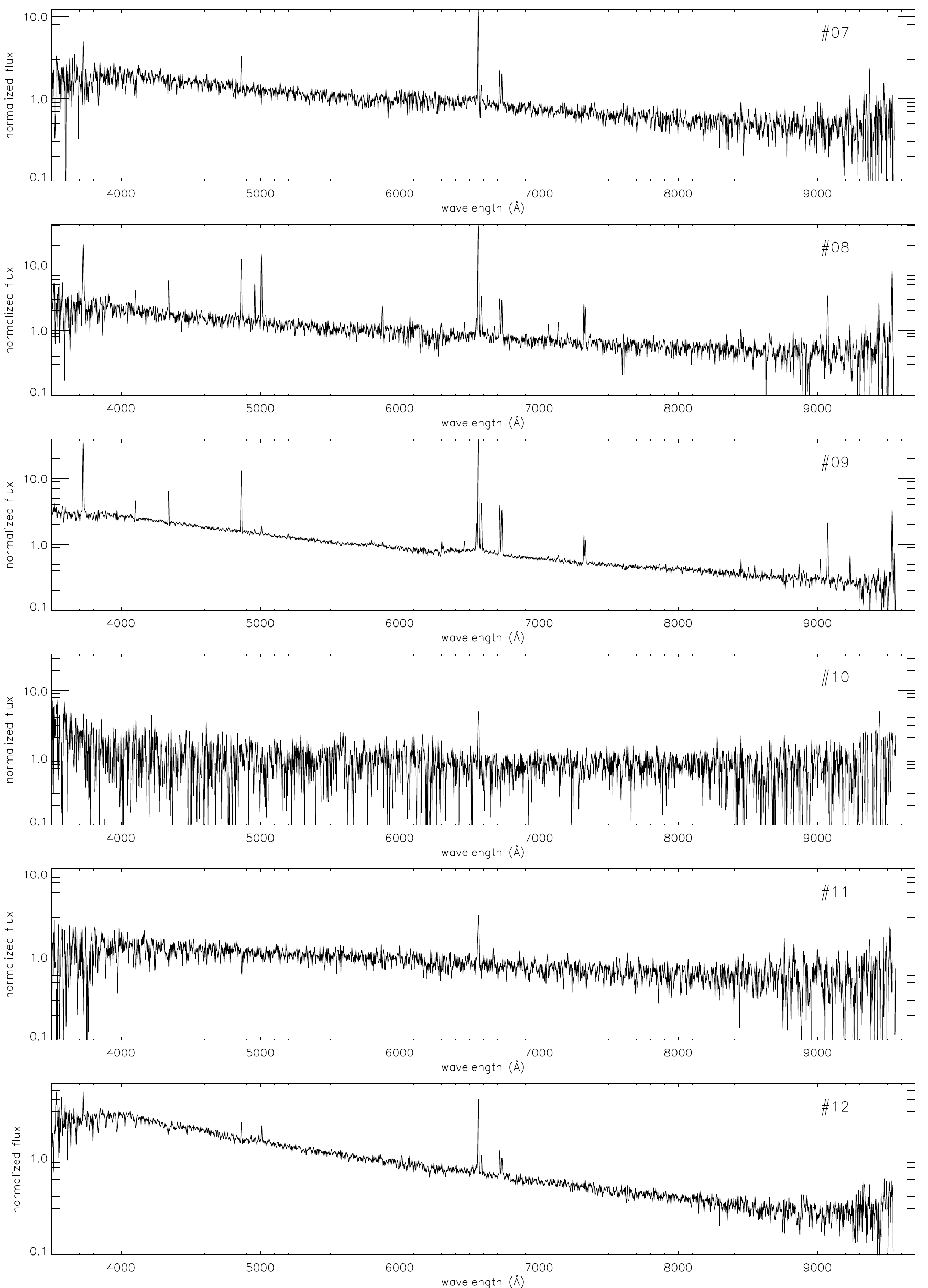}
\contcaption{}
\end{figure*}
\begin{figure*}
\includegraphics[scale=0.9]{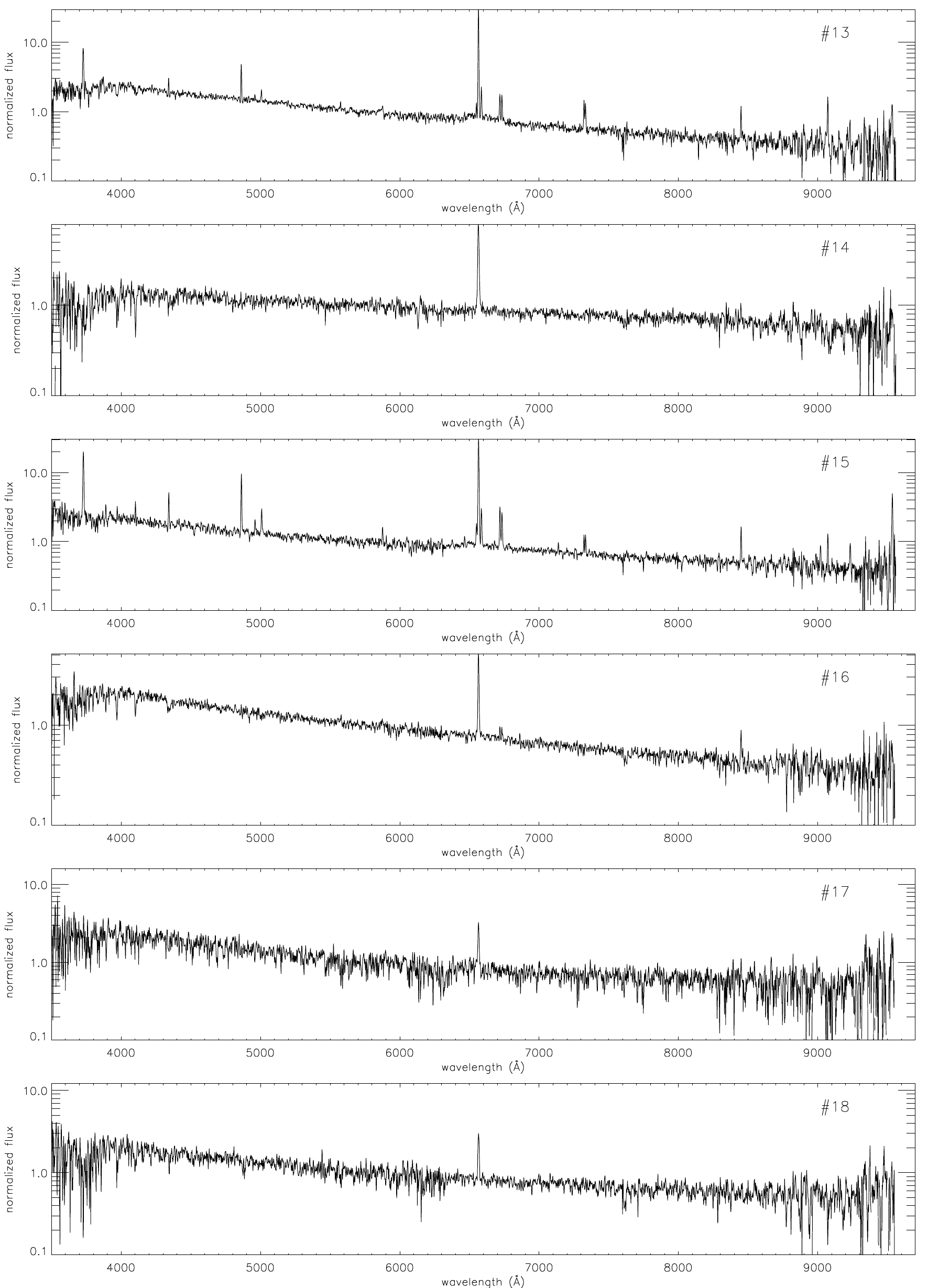}
\contcaption{}
\end{figure*}
\begin{figure*}
\includegraphics[scale=0.9]{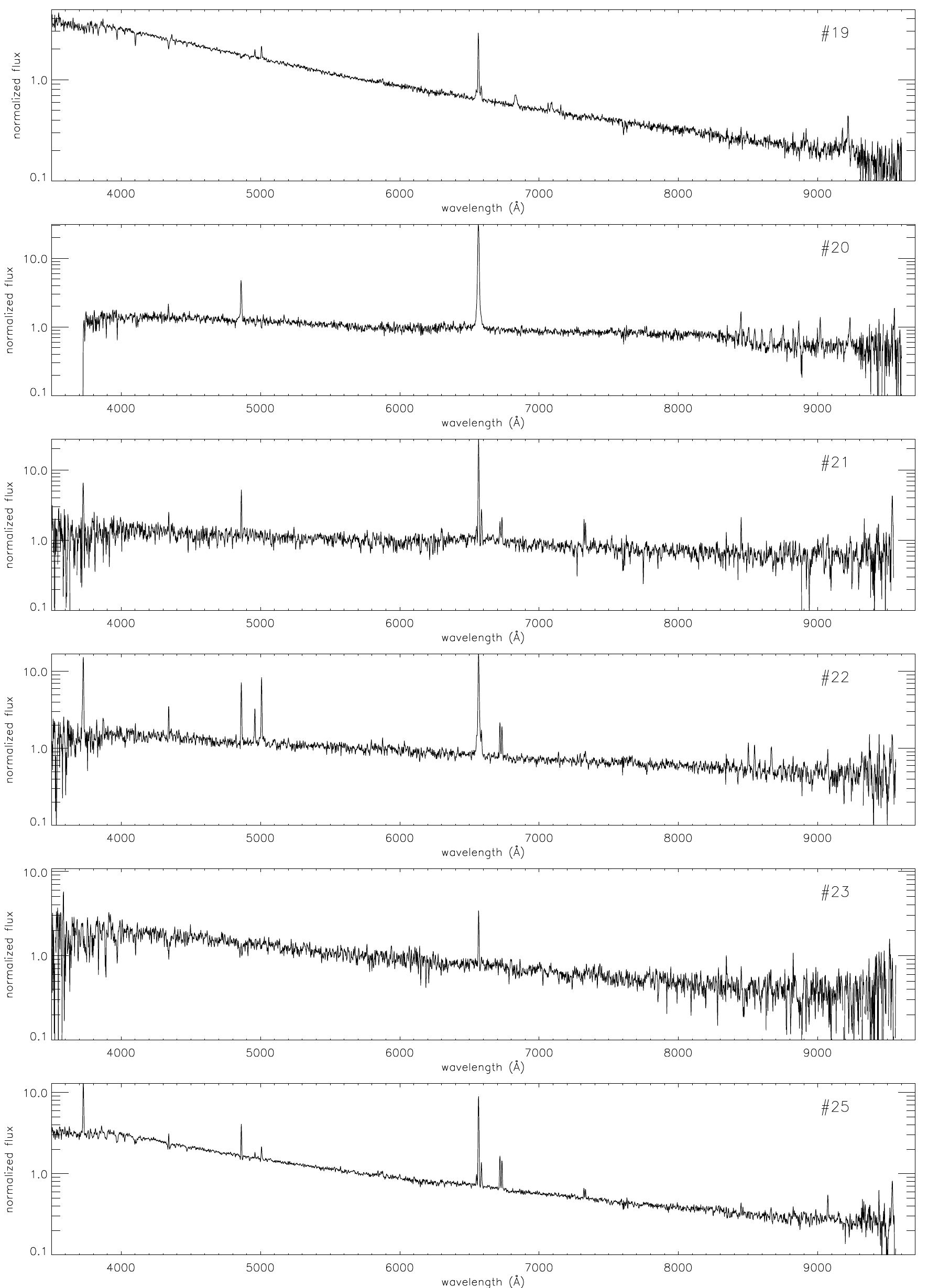}
\contcaption{}
\end{figure*}
\begin{figure*}
\includegraphics[scale=0.9]{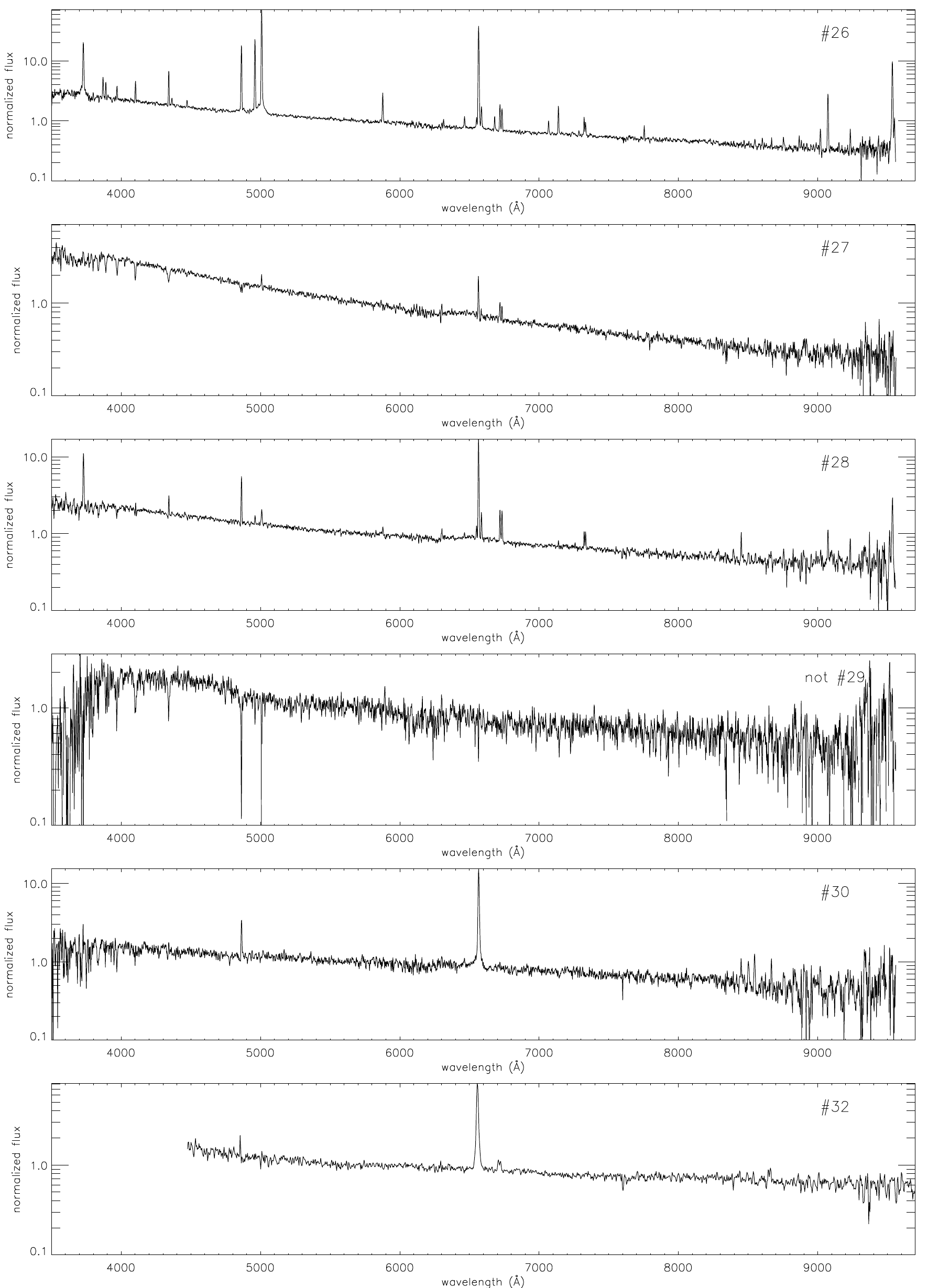}
\contcaption{}
\end{figure*}
\begin{figure*}
\includegraphics[scale=0.9]{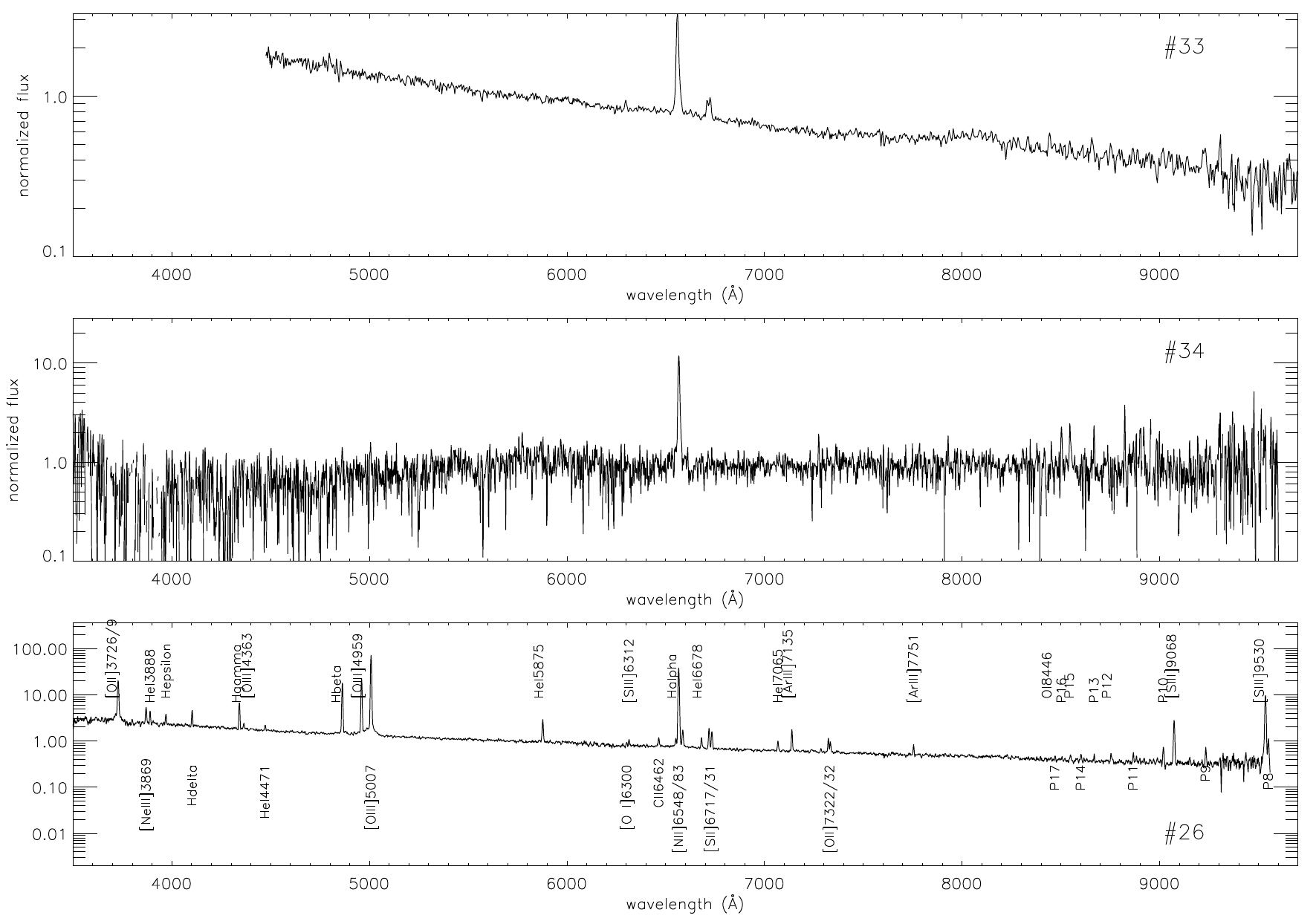}
\contcaption{The most conspicuous spectral features are labelled for source \#26.}
\end{figure*}

\bsp

\label{lastpage}


\begin{thebibliography}{99}
	
\bibitem[\protect\citeauthoryear{Andr\'e \& Montmerle}{1994}]{andre94}
Andr\'e P., Montmerle T., 1994, ApJ, 420, 837

\bibitem[\protect\citeauthoryear{Angeloni et al.}{2007}]{angeloni07}
Angeloni R., Contini M., Ciroi S., Rafanelli P., 2007, AJ, 134, 205

\bibitem[\protect\citeauthoryear{Banerji et al.}{2009}]{banerji09}	
Banerji M., Viti S., Williams D.A., Rawlings J.M.C., 2009, ApJ, 692, 283

\bibitem[\protect\citeauthoryear{Barsony et al.}{2010}]{barsony10}
Barsony M., Wolf-Chase G.A., Ciardi D.R., O'Linger J., 2010, ApJ, 720, 64
	
\bibitem[\protect\citeauthoryear{Belczy\'{n}ski et al.}{2000}]{belczynski00} 
Belczy\'{n}ski K., Miko{\l}ajewska J., Munari U., Ivison R.J., Friedjung M., 2000, 
A\&AS, 146, 407

\bibitem[\protect\citeauthoryear{Bernard-Salas \& Tielens}{2005}]
{bernard-salas05}
Bernard-Salas J., Tielens A.G.G.M., 2005, A\&A, 431, 523

\bibitem[\protect\citeauthoryear{Bolatto et al.}{2007}]{bolatto07}
Bolatto A.D. et al., 2007, ApJ, 655, 212

\bibitem[\protect\citeauthoryear{Boogert et al.}{2008}]{boogert08}
Boogert A.C.A. et al., 2008, ApJ, 678, 985

\bibitem[\protect\citeauthoryear{Boyer et al.}{2011}]{boyer11}
Boyer M.L. et al., 2011, AJ, 142, 103
	
\bibitem[\protect\citeauthoryear{Corradi et al.}{2010}]{corradi10}
Corradi R.L.M. et al., 2010, A\&A, 509, 41

\bibitem[\protect\citeauthoryear{Crawford et al.}{2011}]{crawford11}
Crawford E.J., Filipovi\'c M.D., de Horta A.Y., Wong G.F., Tothill N.F.H., 
Draskovi\'c D., Collier J.D., Galvin T.J., 2011, SerAJ, 183, 95

\bibitem[\protect\citeauthoryear{Draine}{2011}]{draine11}	
Draine B.T., 2011, ``Physics of the Interstellar and Intergalactic Medium'', 
Princeton University Press

\bibitem[\protect\citeauthoryear{Draine \& Li}{2001}]{draine01}	
Draine B.T., Li A., 2001, ApJ, 551, 807

\bibitem[\protect\citeauthoryear{Eisenhardt, Stern \& Brodwin}{Eisenhardt et al.}{2004}]{eisenhardt04}
Eisenhardt P.R., Stern D., Brodwin M., 2004, ApJS, 154, 48

\bibitem[\protect\citeauthoryear{Fazio et al.}{2004}]{fazio04}
Fazio G.G. et al., 2004, ApJS, 154, 10 
	
\bibitem[\protect\citeauthoryear{Filipovi\'c et al.}{1998}]{filipovic98}
Filipovi\'c M.D., Haynes R.F., White G.L., Jones P.A., 1998, A\&AS, 130, 421

\bibitem[\protect\citeauthoryear{Forbrich et al.}{2010}]{forbrich10}
Forbrich J. et al., 2010, ApJ, 716, 1453

\bibitem[\protect\citeauthoryear{Furlan et al.}{2006}]{furlan06}
Furlan E. et al., 2006, ApJS, 165, 568

\bibitem[\protect\citeauthoryear{Furlan et al.}{2008}]{furlan08}
Furlan E. et al., 2008, ApJS, 176, 184

\bibitem[\protect\citeauthoryear{Galliano et al.}{2008}]{galliano08}
Galliano F., Madden S.C., Tielens A.G.G.M., Peeters E., Jones A.P., 2008, ApJ, 
679, 310

\bibitem[\protect\citeauthoryear{Gerakines et al.}{1995}]{gerakines95}
Gerakines P.A., Schutte W.A., Greenberg J.M., van Dishoeck E.F., 1995, A\&A, 
296, 810

\bibitem[\protect\citeauthoryear{Gerakines et al.}{1999}]{gerakines99}	
Gerakines P.A. et al., 1999, ApJ, 522, 357

\bibitem[\protect\citeauthoryear{Gibb et al.}{2004}]{gibb04}
Gibb, E.L., Whittet D.C.B., Boogert A.C.A., Tielens A.G.G.M., 2004, ApJS, 151, 
35

\bibitem[\protect\citeauthoryear{Gordon et al.}{2011a}]{gordon11}
Gordon K. et al., 2011a, AJ, 142, 102

\bibitem[\protect\citeauthoryear{Gordon et al.}{2011b}]{gordon11b}
Gordon K. et al., 2011b, http://data.spitzer.caltech.edu/popular/\\
sage-smc/20110429\_enhanced/documentation/sage-smc\_delivery\_apr11.pdf

\bibitem[\protect\citeauthoryear{Gruendl \& Chu}{2009}]{gruendl09}
Gruendl R.A., Chu Y.-H., 2009, ApJS, 184, 172

\bibitem[\protect\citeauthoryear{Habart et al.}{2005}]{habart05}	
Habart E., Walmsley M., Verstraete L., Cazaux S., Maiolino R., Cox P., 
Boulanger F., Pineau des Forêts G., 2005, SSRv, 119, 71

\bibitem[\protect\citeauthoryear{Haynes et al.}{2010}]{haynes10}
Haynes K., Cannon J.M., Skillman E.D., Jackson D.C., Gehrz R., 2010, ApJ, 724, 215

\bibitem[\protect\citeauthoryear{Hoare et al.}{2007}]{hoare07}
Hoare M.G., Kurtz S.E., Lizano S., Keto E., Hofner P., 2007, in ``Protostars 
and Planets'' V, eds. B. Reipurth, D. Jewitt, K. Keil, University of Arizona 
Press, Tucson, p.\ 181

\bibitem[\protect\citeauthoryear{Houck et al.}{2004}]{houck04}
Houck J.R. et al., 2004, ApJS, 154, 18	

\bibitem[\protect\citeauthoryear{J{\o}rgensen et al.}{2006}]{jorgensen06}
J{\o}rgensen J.K. et al., 2006, ApJ, 645, 1246

\bibitem[\protect\citeauthoryear{Kato et al.}{2007}]{kato07}
Kato D. et al., 2007, PASJ, 59, 615

\bibitem[\protect\citeauthoryear{Keller et al.}{2008}]{keller08}
Keller L.D. et al., 2008, ApJ, 684, 411

\bibitem[\protect\citeauthoryear{Kemper	et al.}{2010}]{kemper10}
Kemper F. et al., 2010, PASP, 122, 683

\bibitem[\protect\citeauthoryear{Kessler-Silacci et al.}{2006}]{kessler06}
Kessler-Silacci J. et al., 2006, ApJ, 639, 275

\bibitem[\protect\citeauthoryear{Lada}{1987}]{lada87}
Lada C.J., 1987, ``Star Forming Regions'', eds. Peimbert M., Jugaku J., 
Proc. IAU Symp. 115, Dreidel, Dordrecht , p.\ 1
	
\bibitem[\protect\citeauthoryear{Lahuis et al.}{2010}]{lahuis10}
Lahuis F., van Dishoeck E.F., J{\o}rgensen J.K., Blake G.A., Evans N.J., 2010, 
A\&A, 519, 3

\bibitem[\protect\citeauthoryear{Lebouteiller et al.}{2010}]{lebouteiller10}
Lebouteiller V., Bernard-Salas J., Sloan G.C., Barry D.J., 2010, PASP, 122, 231

\bibitem[\protect\citeauthoryear{Lee et al.}{2006}]{lee06}
Lee J.-E. et al., 2006, ApJ, 648, 491

\bibitem[\protect\citeauthoryear{Ngeow \& Kanbur}{2008}]{ngeow08}
Ngeow C., Kanbur S.M., 2008, ApJ, 679, 76

\bibitem[\protect\citeauthoryear{Martayan et al.}{2007}]{martayan07}
Martayan C., Floquet M., Hubert A.M., Gutierrez-Soto J., Fabregat J., Neiner C., 
Mekkas M., 2007, A\&A, 472, 577 

\bibitem[\protect\citeauthoryear{Meixner et al.}{2006}]{meixner06}
Meixner M. et al. 2006, AJ 132, 2268

\bibitem[\protect\citeauthoryear{Miko{\l}ajewska}{2004}]{mikolajewska04}
Miko{\l}ajewska J., 2004, RMxAC, 20, 33
	
\bibitem[\protect\citeauthoryear{Muench et al.}{2007}]{muench07}
Muench A.A., Lada C.J., Luhman K.L., Muzerolle J., Young E., 2007, AJ, 134, 411

\bibitem[\protect\citeauthoryear{Oliveira et al.}{2009}]{oliveira09}	
Oliveira J.M. et al., 2009, ApJ, 707, 1269

\bibitem[\protect\citeauthoryear{Oliveira et al.}{2011}]{oliveira11}	
Oliveira J.M. et al., 2011, MNRAS, 411, L36

\bibitem[\protect\citeauthoryear{Parmar, Lacy \& Achtermann}{Parmar et
al.}{1991}]{parmar91}	
Parmar P.S., Lacy J.H., Achtermann J.M., 1991, ApJ, 372, 25

\bibitem[\protect\citeauthoryear{Pontoppidan et al.}{2008}]{pontoppidan08}
Pontoppidan K.M. et al., 2008, ApJ, 678, 1005

\bibitem[\protect\citeauthoryear{Rho et al.}{2006}]{rho06}
Rho J., Reach W.T., Lefloch B., Fazio G.G., 2006, ApJ, 643, 965

\bibitem[\protect\citeauthoryear{Rieke et al.}{2004}]{rieke04}
Rieke G.H. et al., 2004, ApJS, 154, 25

\bibitem[\protect\citeauthoryear{Robinson, Smith \& Maldoni}{Robinson et
al.}{2012}]{robinson12}
Robinson G., Smith R.G., Maldoni M.M., 2012, MNRAS, 424, 1530

\bibitem[\protect\citeauthoryear{Robitaille et al.}{2007}]{robitaille07}
Robitaille T.P., Whitney B.A., Indebetouw R., Wood K., 2007, ApJS, 169, 328

\bibitem[\protect\citeauthoryear{Robitaille et al.}{2006}]{robitaille06}
Robitaille T.P., Whitney B.A., Indebetouw R., Wood K., Denzmore P., 2006, 
ApJS, 167, 256

\bibitem[\protect\citeauthoryear{Rodgers, Conroy \& Bloxham}{Rodgers et al.}{1988}]
{rodgers88}
Rodgers A.W., Conroy P., Bloxham G., 1988, PASP, 100, 626

\bibitem[\protect\citeauthoryear{Russell \& Dopita}{1992}]{russell92}
Russell S.C., Dopita M.A., 1992, ApJ, 384, 508
	
\bibitem[\protect\citeauthoryear{Sandstrom et al.}{2012}]{sandstrom12}
Sandstrom K.M. et al., 2012, ApJ, 744, 20

\bibitem[\protect\citeauthoryear{Sargent et al.}{2009}]{sargent09}
Sargent B.A. et al., 2009, ApJS, 182, 477

\bibitem[\protect\citeauthoryear{Schmid}{1989}]{schmid89}
Schmid H.M., 1989, A\&A, 211, 31	

\bibitem[\protect\citeauthoryear{Seale et al.}{2009}]{seale09}
Seale J.P., Looney L.W., Chu Y.-H., Gruendl R.A., Brandl B., Chen R.C.-H., Brandner
W., Blake G.A., 2009, ApJ, 699, 150

\bibitem[\protect\citeauthoryear{Seale et al.}{2011}]{seale11}	
Seale J.P., Looney L.W., Chen C.-H.R., Chu Y.-H., Gruendl R.A., 2011, ApJ, 727, 36

\bibitem[\protect\citeauthoryear{Shimonishi et al.}{2010}]{shimonishi10}
Shimonishi T., Onaka T., Kato D., Sakon I., Ita Y., Kawamura A., Kaneda H., 
2010, A\&A\, 514, 12
	
\bibitem[\protect\citeauthoryear{Smith et al.}{2007}]{smith07}
Smith J.D.T. et al., 2007, ApJ, 656, 770

\bibitem[\protect\citeauthoryear{Spoon et al.}{2002}]{spoon02}
Spoon H.W.W., Keane J.V., Tielens A.G.G.M., Lutz D., Moorwood A.F.M., Laurent O., 
2002, A\&A 385, 1022
	
\bibitem[\protect\citeauthoryear{Sternberg \& Neufeld}{1999}]{sternberg99}
Sternberg A., Neufeld D.A., 1999, ApJ, 516, 371

\bibitem[\protect\citeauthoryear{Stetson}{1987}]{stetson87}
Stetson P.B., 1987, PASP, 99, 191

\bibitem[\protect\citeauthoryear{St\"orzer \& Hollenbach}{2000}]{storzer00}
St\"orzer H., Hollenbach D., 2000, ApJ, 539, 751

\bibitem[\protect\citeauthoryear{Szewczyk et al.}{2009}]{szewczyk09}
Szewczyk O., Pietrzy\'{n}ski G., Gieren W., Ciechanowska A., Bresolin F., Kudritzki
R.-P., 2009, AJ, 138, 1661

\bibitem[\protect\citeauthoryear{Torres et al.}{2012}]{torres12}
Torres A.F., Kraus M., Cidale L.S., Barb\'{a} R., Borges Fernandes M., Brandi E., 2012,
MNRAS in press, arXiv:1209.2397

\bibitem[\protect\citeauthoryear{van den Ancker}{1999}]{vandenancker99}
van den Ancker M.E., 1999, PhD Thesis, University of Amsterdam

\bibitem[\protect\citeauthoryear{van den Ancker, Tielens \& Wesselius}
{van den Ancker et al.}{2000}]{vandenancker00}
van den Ancker M.E., Tielens A.G.G.M., Wesselius P.R., 2000, A\&A, 358, 1035

\bibitem[\protect\citeauthoryear{van Dishoeck}{2004}]{vandishoeck04}
van Dishoeck E.F., 2004, ARA\&A, 42, 119

\bibitem[\protect\citeauthoryear{van Loon et al.}{2005}]{vanloon05}
van Loon J.Th. et al., 2005, MNRAS, 364, 71
	
\bibitem[\protect\citeauthoryear{van Loon et al.}{2008}]{vanloon08}
van Loon J.Th. et al., 2008, AJ, 487, 1055

\bibitem[\protect\citeauthoryear{van Loon et al.}{2010a}]{vanloon10a}
van Loon J.Th. et al., 2010a, AJ, 139, 68

\bibitem[\protect\citeauthoryear{van Loon et al.}{2010b}]{vanloon10b}
van Loon J.Th., Oliveira, J.M., Gordon K.D., Sloan G.C., Engelbracht C.W., 
2010b, AJ, 139, 68

\bibitem[\protect\citeauthoryear{Werner et al.}{2004}]{werner04}
Werner M.W. et al., 2004, ApJS, 154, 1

\bibitem[\protect\citeauthoryear{White \& Hillenbrand}{2004}]{white04}
White R.J., Hillenbrand L.A., 2004, ApJ, 616, 998

\bibitem[\protect\citeauthoryear{Whitney et al.}{2008}]{whitney08}
Whitney B.A. et al., 2008, AJ, 136, 18

\bibitem[\protect\citeauthoryear{Wong et al.}{2011a}]{wong11a}	
Wong G.F., Filipovi\'c M.D., Crawford E.J., de Horta A.Y., Galvin T.,
Draskovi\'c D., Payne J.L., 2011a, SerAJ, 182, 43

\bibitem[\protect\citeauthoryear{Wong et al.}{2011b}]{wong11b}	
Wong G.F. et al. 2011b, SerAJ, 183, 103

\bibitem[\protect\citeauthoryear{Wong et al.}{2012}]{wong12}
Wong G.F. et al., 2012, SerAJ, 184, 93

\bibitem[\protect\citeauthoryear{Woods et al.}{2011}]{woods11}	
Woods Paul M. et al., 2011, MNRAS, 411, 1597

\bibitem[\protect\citeauthoryear{Zasowski et al.}{2009}]{zasowski09}
Zasowski G., Kemper F., Watson D.M., Furlan E., Bohac C.J., Hull C., Green J.D.,
2009, ApJ, 694, 459

\end{thebibliography}
\end{document}